\documentclass[a4paper,12pt,times,numbered,print,index]{Classes/PhDThesisPSnPDF}
\usepackage{bibentry}
\usepackage{xstring}
\usepackage{etoolbox}
\usepackage{doi}
\usepackage{amsmath}

\nobibliography*
\input{Preamble/preamble}
\usepackage{multirow}
\usepackage{colortbl}
\usepackage{hhline}
\usepackage{hyperref}

\usepackage{pgfplots}
\usetikzlibrary{calc}
\usepackage{tikz}

\usepackage{filecontents}

\title{Computer-Aided Assessment of Tuberculosis with Radiological Imaging}

\subtitle{From rule-based methods to Deep Learning}

\author{Pedro Macías Gordaliza}

\dept{Multimedia and Communication }

\university{Universidad Carlos III de Madrid}

\advisor{ \textcolor{white}{A} \newline Arrate Muñoz Barrutia \newline Juan José Vaquero López}

\supervisor{\textcolor{white}{A} \newline \textcolor{white}{A} Arrate Muñoz Barrutia}
\supervisorrole{Tutor:}
\supervisorlinewidth{0.5\textwidth}

\advisorrole{Advisors: }

\advisorlinewidth{0.5\textwidth}

\subject{LaTeX} \keywords{{LaTeX} {PhD Thesis} {Engineering} {University of
Cambridge}}

\ifdefineAbstract
\fi

\ifdefineChapter
\fi

\begin{document}
\makeatletter
\def\myfnt{\ifx\protect\@typeset@protect\expandafter\footnote\else\expandafter\@gobble\fi}
\makeatother
\frontmatter

\newcommand\blankpage{%
    \null
    \thispagestyle{empty}%
    \addtocounter{page}{-1}%
    \newpage}

\maketitle
\textcolor{white}{None}\\
\vspace{150mm}
\begin{figure*}[b]
\centering
\captionsetup{justification=centering}
\caption*{This thesis is distributed under license “Creative Commons \textbf{Attribution – Non Commercial – Non Derivatives”}.} 
\includegraphics[width=0.25\textwidth,height=0.07\textheight]{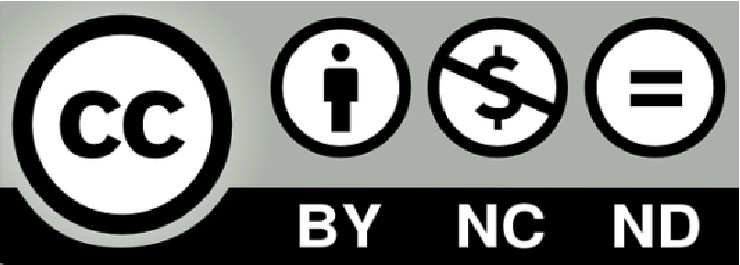}

\end{figure*}

\begin{Publications}
\label{publications}

\subsection*{\underline{Journals Articles}}
\begin{enumerate}
    \item \textbf{Pedro M. Gordaliza}, Arrate Muñoz-Barrutia, Mónica Abella, Manuel Desco, Sally Sharpe, and Juan José Vaquero. \textit{Unsupervised CT Lung Image Segmentation of a Mycobacterium Tuberculosis Infection Model}. Scientific Reports, 8(1), 12 2018. ISSN 2045-2322. doi:\href{https://doi.org/10.1038/s41598-018-28100-x}{10.1038/s41598-018-28100-x}.
\begin{description}
    \item Author contributions: Pedro M. Gordaliza was responsible for conception of the study and the design of the experimental framework, software development, data analysis, writing, editing and reviewing the paper. A.M.B and  J.J.V contribute to the conception of the study and interpretation of the results. S.S. provided the CT data and contributed to the revision of the manuscript. All authors were responsible for  writing, editing and reviewing the paper.
    \item Contribution completely included in Chapter \ref{ch2:chapter2}.
\end{description}
    \item \textbf{Pedro M. Gordaliza}, Arrate Muñoz-Barrutia, Laura E. Via, Sally Sharpe, Manuel Desco, and Juan José Vaquero. \textit{Computed Tomography-Based Biomarker for Longitudinal Assessment of Disease Burden in Pulmonary Tuberculosis}. Molecular Imaging and Biology, 21(1):19–24, 2 2019. ISSN
1536-1632. doi: \href{https://doi.org/10.1007/s11307-018-1215-x}{10.1007/s11307-018-1215-x}.
    \begin{description}
    \item Author contributions: Pedro M. Gordaliza was responsible for conception of the study and the design of the experimental framework, software development, data analysis, writing, editing and reviewing the paper. A.M.B, J.J.V and L.E.V contribute to the conception of the study and interpretation of the results. S.S. and L.E.V provided the data. All authors were responsible for  writing, editing and reviewing the paper.
    \item Contribution completely included in Chapter \ref{ch2:chapter2}.
\end{description}
\item Paula Martin-Gonzalez, Estibaliz Gomez-de-Mariscal, M. Elena Martino, \textbf{Pedro M.
Gordaliza}, Isabel Peligros, Jose Luis Carreras, Felipe A. Calvo, Javier Pascau, Manuel
Desco, and Arrate Muñoz-Barrutia. \textit{Association of visual and quantitative heterogeneity of
18F-FDG PET images with treatment response in locally advanced rectal cancer: A
feasibility study}. Plos One, 15(11):e0242597, 11 2020. ISSN 1932-6203. doi:\href{https://doi.org/10.1371/journal.pone.0242597}{10.1371/journal.pone.0242597}.
\begin{description}
    \item Author contributions: Paula M. G. was responsible for conception of the study, the design of the experimental framework and data analysis. Pedro M. Gordaliza was responsible for code implementation and data analysis. All authors were responsible for  writing, editing and reviewing the paper. 
    \item The code employed is the same developed for Chapter \ref{ch3:Radiomics}.
\end{description}

\item Esperanza Naredo, Javier Pascau, Nemanja Damjanov, Gemma Lepri, \textbf{Pedro M Gordaliza},
Iustina Janta, Juan Gabriel Ovalles-Bonilla, Francisco Javier López-Longo, and Marco
Matucci-Cerinic. \textit{Performance of ultra-high-frequency ultrasound in the evaluation of skin involvement in systemic sclerosis: a preliminary report}. Rheumatology, 59(7):1671–1678, 10, 2019. ISSN 1462-0324. doi:\href{https://doi.org/10.1093/rheumatology/kez439}{10.1093/rheumatology/kez439} .
\begin{description}
    \item Author contributions: E.N. was responsible for conception of the study, the design of the experimental framework and data analysis. Pedro M. Gordaliza was responsible for code implementation and data analysis. All authors were responsible for  writing, editing and reviewing the paper. 
     \item The code employed is the same developed for Chapter \ref{ch3:Radiomics}.
\end{description}

\item Covadonga M. Diáz-Caneja, Clara Alloza, \textbf{Pedro M. Gordaliza}, Alberto Fernández-Pena,
Luciá De Hoyos, Javier Santonja, Elizabeth E.L. Buimer, Neeltje E.M. Van Haren, Wiepke
Cahn, Celso Arango, René S. Kahn, Hilleke E. Hulshoff Pol, Hugo G. Schnack, and Joost
Janssen. \textit{Sex Differences in Lifespan Trajectories and Variability of Human Sulcal and Gyral Morphology}. Cerebral cortex, 31(11):5107–5120, 11 2021. ISSN 1460-2199. doi:\href{https://doi.org/10.1093/cercor/bhab145}{10.1093/cercor/bhab145}.
\begin{description}
    \item Author contributions:  C.M.D was responsible for conception of the study and the design of the experimental framework. Pedro M. Gordaliza was responsible for data analysis. All authors were responsible for  writing, editing and reviewing the paper. 
\end{description}

\item Verónica Aramendía-Vidaurreta, \textbf{Pedro M. Gordaliza}, Rebeca Echeverria-Chasco, Gorka Bastarrika, A Muñoz-Barrutia, and María A. Fernández-Seara. \textit{Reduction
of motion effects in myocardial arterial spin labeling.}. Magnetic Resonance in Medicine, vol. 87, n. 3, pp. 1261-1275, March 2022. doi:\href{https://doi.org/10.1002/mrm.29038}{10.1002/mrm.29038}
\begin{description}
    \item Author contributions: V.A.V. was responsible for conception of the study, the design of the experimental framework, code implementation and data analysis. Pedro M. Gordaliza was responsible for code implementation and data analysis. All authors were responsible for  writing, editing and reviewing the paper. 
\end{description}

\item Joost Janssen, Covadonga M. Díaz-Caneja, Clara Alloza, Anouck Schippers, Lucía de
Hoyos, Javier Santonja, \textbf{Pedro M. Gordaliza}, Elizabeth E.L. L Buimer, Neeltje E.M. van
Haren, Wiepke Cahn, Celso Arango, René S. Kahn, Hilleke E. Hulshoff Pol, Hugo G. Schnack. \textit{Dissimilarity in sulcal width patterns in the cortex can be used to identify
patients with schizophrenia with extreme deficits in cognitive performance}. Schizophrenia
Bulletin, page 2020.02.04.932210, 2020. ISSN 0586-7614. doi: \href{https://doi.org/10.1093/schbul/sbaa131}{10.1093/schbul/sbaa131}
\begin{description}
    \item Author contributions: J.J was responsible for conception of the study, the design of the experimental framework and data analysis. Pedro M. Gordaliza was responsible for data analysis. All authors were responsible for writing, editing and reviewing the paper. 
\end{description}

\item Joost Janssen, Clara Alloza, Covadonga M. Díaz-Caneja, Javier Santonja, Laura
Pina-Camacho, \textbf{Pedro M. Gordaliza}, Alberto Fernández-Pena, Noemi Lois, Elizabeth E.L.
Buimer, Neeltje E.M. Van Haren, Wiepke Cahn, Eduard Vieta, Josefina Castro-Fornieles,
Miquel Bernardo, Celso Arango, René S. Kahn, Hilleke E. Hulshoff Pol, and Hugo G.
Schnack. \textit{Longitudinal allometry of sulcal morphology in health and schizophrenia}. Journal
of Neuroscience, (In press), 2022. Preprint: \href{https://www.biorxiv.org/content/10.1101/2021.03.17.435797v1}{https://www.biorxiv.org/content/10.1101/2021.03.17.435797v1}
\begin{description}
    \item Author contributions: J.J. was responsible for conception of the study, the design of the experimental framework and data analysis. Pedro M. Gordaliza was responsible for data analysis. All authors were responsible for writing, editing and reviewing the paper. 
\end{description}

\item Alberto Fernández-Pena, Daniel Martín-Blas, Luis Marcos-Vidal, \textbf{Pedro M. Gordaliza}, Joost Janssen, Susanna Carmona, Manuel Desco, and Yasser Alemán-Gómez. \textit{ABLE: Automated
Brain Lines Extraction Based on Laplacian Surface Collapse}. Scientific Reports, (Accepted),
2022. Preprint: \href{https://www.biorxiv.org/content/10.1101/2022.01.18.476370v1}{https://www.biorxiv.org/content/10.1101/2022.01.18.476370v1}
\begin{description}
    \item Author contributions: A.F.P. was responsible for conception of the study and the design of the experimental framework, software development and data analysis. All authors were responsible for writing, editing and reviewing the paper.
\end{description}

\end{enumerate}

\subsection*{\underline{Conference Proceedings:}}
\begin{enumerate}
    \item \textbf{Pedro M. Gordaliza}, Juan José Vaquero, Sally Sharpe, Manuel Desco, and Arrate
Munoz-Barrutia. \textit{Towards an informational model for tuberculosis lesion discrimination on
X-ray CT images}. In 15th International Symposium on Biomedical Imaging, Washington, DC, USA, 4-7 April, 2018. doi:\href{https://doi.org/10.1109/ISBI.2018.8363570}{10.1109/ISBI.2018.8363570} 
    \begin{description}
        \item Author contributions: Pedro M. Gordaliza was responsible for conception of the study and the design of the experimental framework, software development, data analysis, writing, editing, reviewing the paper and preparing the presentation. A.M.B  and  J.J.V contribute to the conception of the study and interpretation of the results. S.S. provided the CT data and contributed to the revision of the manuscript. All authors were responsible for  writing, editing and reviewing the paper.
        \item Contribution completely included in Chapter  \ref{ch3:Radiomics}.
    \end{description}
    
    \item \textbf{Pedro M. Gordaliza}, Juan José Vaquero, Sally Sharpe, Manuel Desco, and Arrate Munoz-Barrutia. \textit{Radiomics for the Discrimination of Tuberculosis Lesions}. In European Molecular Imaging Meeting, Donostia, Spain, 20-23 March, 2018. \href{http://eventclass.org/contxt_emim2018/online-program/session?s=PS-12}{http://eventclass.org/contxt\_emim2018/}
    \begin{description}
        \item Author contributions:  Pedro M. Gordaliza was responsible for conception of the study and the design of the experimental framework, software development, data analysis, writing, editing, reviewing the paper and preparing the oral presentation. A.M.B  and  J.J.V contribute to the conception of the study and interpretation of the results. S.S. provided the CT data and contributed to the revision of the manuscript. All authors were responsible for  writing, editing and reviewing the paper.
        \item Contribution completely included in Chapter \ref{ch3:Radiomics}.
    \end{description}
    
    \item \textbf{Pedro M. Gordaliza}, Juan José Vaquero, Sally Sharpe, Fergus Gleeson, and Arrate Muñoz-Barrutia. \textit{A Multi-Task Self-Normalizing 3D-CNN to Infer Tuberculosis
Radiological Manifestations}. In Medical Imaging with Deep Learning (MIDL), London, UK, 8-10 July, 2019. \href{http://arxiv.org/abs/1907.12331}{http://arxiv.org/abs/1907.12331} 
    \begin{description}
        \item Author contributions: Pedro M. Gordaliza was responsible for conception of the study and the design of the experimental framework, software development, data analysis, writing, editing, reviewing the paper and preparing the presentation. A.M.B  and  J.J.V contribute to the conception of the study and interpretation of the results. S.S. and F.G. provided the data and contributed to the revision of the manuscript. All authors were responsible for  writing, editing and reviewing the paper.
        \item Contribution completely included in Chapter  \ref{ch4:Intro}.
    \end{description}
    
    \item \textbf{Pedro M. Gordaliza}, Juan José Vaquero, Sally Sharpe, Fergus Gleeson, and Arrate Muñoz-Barrutia. \textit{Tuberculosis
lesions in CT images inferred using 3D-CNN and multi-task learning}. In 16th International
Symposium on Biomedical Imaging, Venice, Italy, 8-11 April, 2019. doi: \href{https://doi.org/10.1109/ISBI.2019.8759321}{10.1109/ISBI.2019.8759321}
    \begin{description}
        \item Author contributions: Pedro M. Gordaliza was responsible for conception of the study and the design of the experimental framework, software development, data analysis, writing, editing, reviewing the paper and preparing the oral presentation. A.M.B perform the oral presentation  and together with  J.J.V contribute to the conception of the study and interpretation of the results. S.S. and F.G. provided the data and contributed to the revision of the manuscript. All authors were responsible for  writing, editing and reviewing the paper.
        \item Contribution completely included in Chapter  \ref{ch4:Intro}.
    \end{description}

    \item \textbf{Pedro M. Gordaliza}, Juan José Vaquero, and Arrate Muñoz-Barrutia. \textit{Translational Lung Imaging Analysis Through Disentangled Representations}. Radiology: Artificial Intelligence (Under Review), pages 1–13, 12 2021. \href{https://arxiv.org/abs/2203.01668}{https://arxiv.org/abs/2203.01668}
        \begin{description}
        \item Author contributions: Pedro M. Gordaliza was responsible for conception of the study and the design of the experimental framework, software development, data analysis, writing, editing, reviewing the paper and preparing the presentation. A.M.B  and J.J.V contribute to the conception of the study and interpretation of the results. All authors were responsible for  writing, editing and reviewing the paper.
        \item Contribution completely included in Chapter \ref{ch5:Intro}.
    \end{description}

    \item \textbf{Pedro M. Gordaliza}, Verónica Aramendía-Vidaurreta, J.J. Vaquero, Gorka Bastarrika, María A. Fernández-Seara, and A. Muñoz-Barrutia. \textit{Automatic Segmentation Of The Myocardium in Cardiac Arterial Spin Labelling Images Using a Deep Learning Model Facilitates Myocardial Blood Flow}. In 27th International Symposium Magnetic Resonance and Medicine (ISMRM), Montreal, 11-16 May, 2019. \href{https://index.mirasmart.com/ISMRM2019/PDFfiles/4794.html}{index.mirasmart.com/ISMRM2019/PDFfiles/4794.html} 
    \begin{description}
        \item Author contributions: Pedro M. Gordaliza was responsible for conception of the study and the design of the experimental framework, software development, data analysis, writing, editing, reviewing the paper and preparing the presentation. V.A.V. was responsible code implementation and data analysis. All authors were responsible for  writing, editing and reviewing the paper.  
    \end{description}

    \item Verónica Aramendía-Vidaurreta, \textbf{Pedro M. Gordaliza}, Rebeca Echeverria-Chasco, Gorka Bastarrika, A Muñoz-Barrutia, and María A. Fernández-Seara. \textit{Groupwise Non Rigid Registration For Temporal Myocardial Arterial Spin Labeling Images}. In 27th International Symposium Magnetic Resonance and Medicine (ISMRM), Montreal, 11-16 May, 2019. \href{https://index.mirasmart.com/ISMRM2019/PDFfiles/4484.html}{index.mirasmart.com/ISMRM2019/PDFfiles/4484.html}
    \begin{description}
        \item Author contributions: V.A.V. was responsible for conception of the study, the design of the experimental framework, code implementation and data analysis. Pedro M. Gordaliza was responsible for code implementation and data analysis. All authors were responsible for  writing, editing and reviewing the paper. 
    \end{description}
    
    \item Verónica Aramendía-Vidaurreta, \textbf{Pedro M. Gordaliza}, Marta Vidorreta, Rebeca Echeverria-Chasco, Gorka Bastarrika, Arrate Muñoz-Barrutia, and María A. Fernández-Seara. \textit{Comparison of Myocardial Blood Flow Measurements with Arterial Spin Labeling in Breathhold and Synchronized Breathing Acquisitions}. In International Symposium Magnetic Resonance and Medicine (ISMRM), Virtual Conference, 8-14 Agust, 2020. \href{http://archive.ismrm.org/2020/2208.html}{archive.ismrm.org/2020/2208.html}

    \begin{description}
        \item Author contributions:  V.A.V. was responsible for conception of the study, the design of the experimental framework, code implementation and data analysis. Pedro M. Gordaliza was responsible for code implementation and data analysis. All authors were responsible for  writing, editing and reviewing the paper. 
    \end{description}

    \item Leandro A. Hidalgo-Torres, David Pérez-Benito, Rigoberto Chil, \textbf{Pedro M. Gordaliza}, and Juan José Vaquero. \textit{Predicting 3D Photon Interaction in a Hexagonal Positron Emission Tomography Detector: A Deep Learning Approach}. Congreso Anual de la Sociedad Española de Ingeniería Biomédica (CASEIB), Virtual Conference, 25-27 November 2020, \href{https://dialnet.unirioja.es/servlet/articulo?codigo=8207320}{dialnet.unirioja.es/articulo} 
    \begin{description}
        \item Author contributions: L.H.T was responsible for conception of the study and the design of the experimental framework, software development and data analysis. All authors were responsible for writing, editing and reviewing the paper.
    \end{description}
    
\end{enumerate}

\end{Publications}


\begin{dedication} 
\textbf{1870}\\
\vspace{5mm}
Hay una pereza activa \\		
que mientras descansa piensa, \\		
que calla porque se vence,\\		
que duerme pero que sueña.\\		
\vspace{5mm}
Es como un leve reflejo \\		
de la majestad suprema,	\\	
que eternamente tranquila, \\		
sobre el universo reina. \\		
\vspace{5mm}
¡Oh asilo del pensamiento \\		
errante, dulce pereza;	\\	
mil veces feliz el hombre \\		
que de ti goza en la tierra \\
\vspace{5mm}
\textbf{La pereza. Augusto Ferrán}

\end{dedication}


\tableofcontents

\listoffigures

\listoftables


\printnomenclature


\mainmatter
\begin{abstract}\addcontentsline{toc}{chapter}{Abstract}

Tuberculosis (TB) is an infectious disease caused by \textit{Mycobacterium tuberculosis} (\Mtb) that produces pulmonary damage due to its airborne nature.
This fact facilitates the disease fast-spreading, which, according to the \textit{World Health Organization} (WHO), in $2021$ caused $1.2$ million deaths and $9.9$ million new cases.

Traditionally, TB has been considered a binary disease (latent/active) due to the limited specificity of the traditional diagnostic tests. Such a simple model causes difficulties in the longitudinal assessment of pulmonary affectation needed for the development of novel drugs and to control the spread of the disease. 

Fortunately, \textit{X-Ray Computed Tomography} (CT) images enable capturing specific manifestations of TB that are undetectable using regular diagnostic tests, which suffer from limited specificity. In conventional workflows, expert radiologists inspect the CT images.
However, this procedure is unfeasible to process the thousands of volume images belonging to the different TB animal models and humans required for a suitable (pre-)clinical trial.\\

To achieve suitable results, automatization of different image analysis processes is a must to quantify TB. It is also advisable to measure the uncertainty associated with this process and model causal relationships between the specific mechanisms that characterize each animal model and its level of damage. Thus, in this thesis, we introduce a set of novel methods based on the state of the art \textit{Artificial Intelligence} (AI) and \textit{Computer Vision} (CV).

Initially, we present an algorithm to assess \textit{Pathological Lung Segmentation} (PLS) employing an unsupervised rule-based model which  was traditionally considered a needed step before biomarker extraction. This procedure allows robust segmentation in a \Mtb~infection model (\textit{Dice Similarity Coefficient}, DSC, $94\% \pm 4\%$, \textit{Hausdorff Distance}, HD, $8.64\: \mbox{mm} \pm 7.36\:\mbox{mm}$) of damaged lungs with lesions attached to the parenchyma and affected by respiratory movement artefacts. 

Next, a Gaussian Mixture Model ruled by an \textit{Expectation-Maximization} (EM) algorithm is employed to automatically quantify the burden of \Mtb using biomarkers extracted from the segmented CT images. This approach achieves a strong correlation ($R^2 \approx 0.8$) between our automatic method and manual extraction.
 
Consequently, Chapter \ref{ch3:Radiomics} introduces a model to automate the identification of TB lesions and the characterization of disease progression. To this aim, the method employs the \textit{Statistical Region Merging} algorithm to detect lesions subsequently characterized by texture features that feed a \textit{Random Forest} (RF) estimator. The proposed procedure enables a selection of a simple but powerful model able to classify abnormal tissue.

The latest works base their methodology on \textit{Deep Learning} (DL).
Chapter \ref{ch4:Intro} extends the classification of TB lesions. Namely, we introduce a computational model to infer TB manifestations present in each lung lobe of CT scans by employing the associated radiologist reports as ground truth. We do so instead of using the classical manually delimited segmentation masks. The model adjusts the three-dimensional architecture, \textit{V-Net}, to a multi-task classification context in which loss function is weighted by homoscedastic uncertainty. Besides, the method employs \textit{Self-Normalizing Neural Networks} (SNNs) for regularization. Our results are promising with a \textit{Root Mean Square Error} of $1.14$ in the number of nodules and $F_1$-scores above $0.85$ for the most prevalent TB lesions (i.e., conglomerations, cavitations, consolidations, trees in bud) when considering the whole lung.

 In Chapter \ref{ch5:Intro}, we present a DL model capable of extracting disentangled information from images of different animal models, as well as information of the mechanisms that generate the CT volumes. The method provides the segmentation mask of axial slices from three animal models of different species employing a single trained architecture. It also infers the level of TB damage and generates counterfactual images. So, with this methodology, we offer an alternative to promote generalization and explainable AI models.
 
 To sum up, the thesis presents a collection of valuable tools to automate the quantification of pathological lungs and moreover extend the methodology to provide more explainable results which are vital for drug development purposes. Chapter \ref{ch6:Intro} elaborates on these conclusions.

\end{abstract}
\markboth{Abstract}{Abstract}

\begin{Motivation and Objectives}\addcontentsline{toc}{chapter}{Motivation and Objectives}

Tuberculosis (TB) is an infectious disease caused by \textit{Mycobacterium tuberculosis} (\Mtb)  that produces pulmonary damage. TB causes around three thousand deaths per day around the globe; thousand more than the \textit{Coronavirus} disease (COVID-19) at its uprising, a fact even more dramatic considering that such a terrible number maintains steady during the last decades. Thus, numbers in $2021$ amount to $9.9$ million new cases and 1.2 million deaths, according to the World Health Organization (WHO) \cite{WorldHealthOrganization2021Global2021}.

As part of the efforts to control this devastating epidemic, proper modelling of TB as a continuous spectrum between latent and active stages is urgently needed \cite{Pai2016Tuberculosis}. To this aim, several projects are ongoing worldwide, being this thesis part of one of them, the European Accelerator of Tuberculosis Regime Project, \href{https://era4tb.org/}{ERA4TB} \cite{ERA4TBconsotium2021ERA4TB}.

This thesis presents a set of methods based on the \textit{Artificial Intelligence} (AI) state of the art to enrich the ability of \textit{x-ray Computed Tomography} (CT) images to depict specific manifestations of TB, contrarily, to regular diagnostic tests which suffer from limited specificity. By exposing such specific patterns, CT images allow proper modelling of TB, which is essential for disease prognosis and its longitudinal monitoring, and consequently at the development of more effective drugs.

The common practice is that experts meticulously examine CT images manually for various purposes. Namely, to delimit regions/volumes of interest (ROIs/VOIs) in the images, the lungs in this work context, to enable "post-hoc" studies. Also, to examine abnormal regions within the ROIs to establish whether these are characteristic disease manifestations. With this information, they can subsequently provide longitudinal descriptions to enable hypotheses about the mechanisms that rule the disease and its interactions with in-development drugs.

Nevertheless, this approach is highly time-consuming, prone to human errors and therefore, infeasible at large studies. Namely, at (pre-)clinical trials in which thousands of images belonging to different animal, disease burden and strains models are involved, as in the ERA4TB project.

Therefore, the development of new tools capable of automating expert tasks entails the fundamental goal of this work. It particular, we aim for the methods to capture extra information from CT images that capture new findings of disease inner-workings that could contribute to for TB eradication. To this aim, the following specific objectives are proposed in this thesis: 

\begin{itemize}
\item Develop methods to automatically segment lungs damaged by Tuberculosis. 
\item Develop methods for the automatic quantification of TB burden to characterize the disease progression and response to therapy.
\item Develop methods for the automatic detection and characterization of the main TB manifestations to assess the development of the disease and the effectivity of the new drugs.
\item Perform automatic analysis of translational animal models, namely, the mammal models usually employed in clinical trials (i.e., mouse, macaque, human).
\item Provide analysis tools able to yield valuable information about the disease physiopathology under speculative scenarios to leverage interdisciplinary experts hypothesis. 
\end{itemize}

\end{Motivation and Objectives}
\markboth{Motivation and Objectives}{Motivation and Objectives}
\begin{Thesis outline}\addcontentsline{toc}{chapter}{Thesis Outline}
The rest of the document comprehend the following chapters:
\begin{enumerate}
    \item \nameref{ch1:intro}, consists of a comprehensive review of essential background concepts. Thus, briefly introduces both: a) medico-social aspects of TB (e.g., the impact of Tuberculosis in the world population, the importance of radiological imaging in its eradication) and, b) concepts related to the Artificial Intelligence methodology (e.g., advantages and limitations of the frameworks that enable task automation, the appearance of biases in the algorithms due to data scarcity or dataset shifts) used in the different works presented in the document. Finally, both topics joint in the state-of-the-art portray for the automatic analysis of pathological lung images. 
    
    \item \nameref{ch2:chapter2},  presents an algorithm for Pathological Lung Segmentation for a macaque model of Tuberculosis and a Gaussian Mixture based method for quantification.  
    
    \item \nameref{ch3:Radiomics}, introduces a Radiomics approach for Tuberculosis manifestations classification by extracting texture features and applying a Random Forest classifier.
    
    \item \nameref{ch4:Intro}, presents the Deep Learning model yielding a multi-task architecture. It is a self-normalized set-up with weights computed from an estimation of uncertainty. It is designed to identify lesion from complete three-dimensional Computed Tomography volumes.
    
    \item \nameref{ch5:Intro}, presents the model able to disentangle significant factors. We show its application on image synthesis and the automatic and robust delimitation of pathological lungs.  
    
    \item \nameref{ch6:Intro}

\end{enumerate}

\end{Thesis outline}
\markboth{Thesis outline}{Thesis outline}
\chapter{The Key Role of Artificial Intelligence in Tuberculosis Assessment } \label{ch1:intro}  

\ifpdf
    \graphicspath{{Chapter1/Figs/Raster/}{Chapter1/Figs/PDF/}{Chapter1/Figs/}}
\else
    \graphicspath{{Chapter1/Figs/Vector/}{Chapter1/Figs/}}
\fi

\nomenclature[z-Mtb]{\textit{Mtb.}}{Mycobacterium tuberculosis}                                
\nomenclature[z-CT]{CT}{Computed Tomography}
\nomenclature[a-F]{$F$}{complex function}   
\nomenclature[z-SOTA]{SOTA}{State Of The Art}
\nomenclature[z-CADx]{CADx}{Computer Aided Diagnosis}
\nomenclature[z-CADe]{CADe}{Computer Aided Detection}
\nomenclature[z-AI]{AI}{Artificial Intelligence}
\nomenclature[z-ML]{ML}{Machine Learning}
\nomenclature[z-DL]{DL}{Deep Learning}
\nomenclature[g-p]{$\pi$}{ $\simeq 3.14\ldots$}                                             
\nomenclature[r-j]{$j$}{superscript index}                                                       
\nomenclature[s-i]{$i$}{subscript index}                                                        

The bulk of this work lies in the study and subsequent proposal of several solutions based on different branches of Artificial Intelligence (AI).  Similar to the temporal evolution of AI itself, the first work presented in the manuscript (Chapter \ref{ch2:chapter2}) initially applies, adapts and develops classical AI methods (based on coding human-defined rules). Then, it presents solutions employing some of the more common Machine Learning (ML) techniques developed in the last 20 years, which are based on statistical dependence within the study dataset but with few or minimal assumptions about the mechanisms causing such dependencies. Finally, it ends with more complex models that aim to add the predictive power and explainability to the previous ones.

These novel methodologies can be adapted to a wide range of problems and hopefully benefit a large part of the scientific community. However, this scientific work has a strong engineering and therefore translational character. Namely, it is oriented to a specific application such as the automatic evaluation of lungs damaged by Tuberculosis as verbalized in its title.\\ 

Thus, while in subsequent chapters the particular objective, the mathematical machinery, the performance analysis or the implementation of the different methodologies are presented in a self-contained manner\footnote{Each chapter consists of adaptations of individual works shown in \nameref{publications}. These extend those sections that help to synthesize the entire manuscript.}, this opening chapter aims to justify the need for their development.\\  
For this reason, initially (see \nameref{ch1:sec:TBNumbers}) a summary of the updated epidemiological data\footnote{Based on 2020 WHO report. ~\cite{WorldHealthOrganization2020GLOBAL2020}} of Tuberculosis (TB) worldwide is provided showing its pandemic status. 

Next, the biomedical framework of TB diagnosis based on microbiological, pathological and immunological studies and their interactions with treatments is introduced. The main biological causes of the pandemic are exposed together with the different ways to dampen them by employing the strategies presented in the Section \ref{ch1:sec:errad} (\nameref{ch1:sec:errad}).\\ 
The role of the different medical imaging modalities depending on the disease markers critical at the several framework components presented in the previous section are introduced subsequently in Section \ref{ch1:sec:MITB}. 
Finally, the main approaches existing in the literature to carry out the objective of quantifying damaged lungs, obviously with the focus on those devoted to Tuberculosis damage, and which have served as inspiration and support for the development of the different methodologies introduced in the subsequent chapters, are presented in Section \ref{ch1:sec:CAD}. Since the approaches rely on the aforementioned AI principles, Section \ref{ch1:sec:CAD} is divided in two, namely, a first part which describes  \nameref{ch1:ssec:list:rulesvsML} and a second with the specific application's summary.  

\section{Tuberculosis in numbers}\label{ch1:sec:TBNumbers} 

The data shown below is not intended to simplify the harm caused by Tuberculosis throughout the history of humanity and especially nowadays. The mere expectation is to show the unaware reader the reality about the pandemic that causes the most deaths daily at the time of this writing. We would like also to alert them to the need to tackle it globally, in a similar way to how unfortunately and abruptly the society had done since the outbreak of COVID-19; especially given the neglected status of Tuberculosis.

\begin{figure}
\centering
\captionsetup{justification=justified}
\includegraphics[width=\textwidth,height=0.5\textheight]{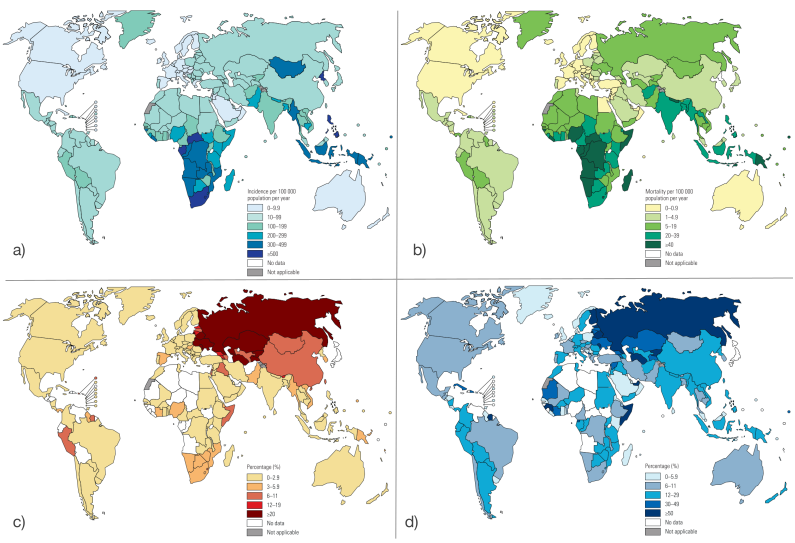}
\caption[Tuberculosis Epidemiology]{Tuberculosis (TB) epidemiology data per country: \textbf{a)} TB prevalence per $100000$ population per year; \textbf{b)} TB mortality per $100000$ population per year; \textbf{c)} Multidrug-resistant TB (MTB-TB) percentage; \textbf{d)} Rifampicin-Resistant TB (RR-TB). Extracted from the WHO (World Health Organization) TB report 2020 \cite{WorldHealthOrganization2020GLOBAL2020}.}\label{ch1:fig:num_Maps} 
\end{figure}

According to the World Health Organization  (WHO) estimations~\cite{WorldHealthOrganization2021Global2021}, in 2021, there were 9.9 million (range, 8.9–11.0 million) million incident cases and 1.3 million (range, 1.2–1.4 million) deaths caused by tuberculosis (TB). The incidence and mortality rates per $100.000$ population per year and country are within the upper part of \figurename \ref{ch1:fig:num_Maps} (maps $a$ and $b$). More strikingly, latent TB (see Section \nameref{ch1:ssec:TBspectrum}) is present in about a quarter of the world’s population. Within this infected population, \textit{Mycobacterium tuberculosis} (\Mtb), the causative agent of TB, becomes active in 10\% of the cases and mainly damages the lungs owing to its airborne nature.

\begin{figure}
\centering
\captionsetup{justification=justified}
\includegraphics[width=\textwidth,height=0.25\textheight]{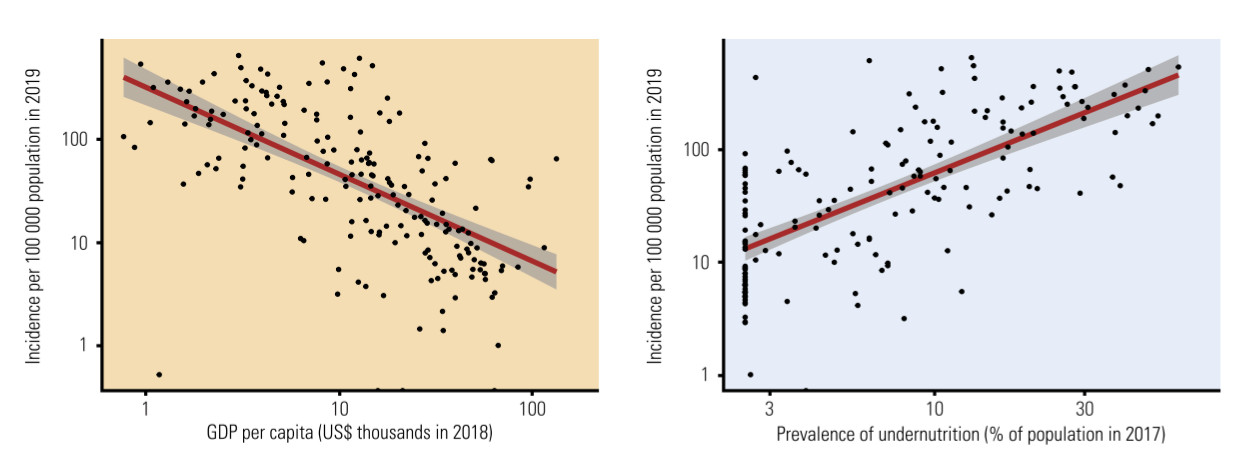}
\caption[Tuberculosis vs GDP \& Undernutrition]{Social cofounders of TB. \textbf{Left)} Relationship between TB prevalence and Gross Domestic Product (GDP); \textbf{Right)} Relationship between TB prevalence and undernutrition. Extracted from the WHO TB report 2020 ~\cite{WorldHealthOrganization2020GLOBAL2020}.}\label{ch1:fig:TB_GDP} 
\end{figure}

As can be seen in \figurename \ref{ch1:fig:num_Maps}, the TB burden distribution is far from homogeneous around the globe, being low and middle-income countries much more hit by the pandemic; mainly due to the well-known relationship between undernutrition and a depressed immunological system~\cite{Odone2014TheTargets, Romanowski2019Long-termMeta-analysis}; \figurename \ref{ch1:fig:TB_GDP} shows this trend clearly. Meanwhile, rich countries understood TB as a disease of the past, turning it into a neglected disease associated with poverty, marginalization and social exclusion of individuals suffering from it.

This fact does not benefit the fight against the pandemic at all. However, a terrible game-changer has become crucial in the last 25 years. Resistance to TB drugs~\cite{Barry3rd2009TheProphylaxis, Gandhi2010Multidrug-resistantTuberculosis,Diacon2009TheTuberculosis,Young2009EliminatingTuberculosis,   Zignol2006GlobalTuberculosis} has increased markedly during this time, especially in those westernized countries that considered the disease eradicated. The data corresponding to new cases in 2020 with MDR-TB (Multidrug-Resistant Tuberculosis) and RR-TB (Rifampicin-Resistant Tuberculosis), as well as the percentage of cases that had already been treated for TB and developed MDR and RR are shown by country in maps $C$ and $D$ of \figurename \ref{ch1:fig:num_Maps}.
This situation has shifted the projections for TB eradication from around 2030~\cite{Barry3rd2009TheProphylaxis,Young2009EliminatingTuberculosis} to 2050~\cite{Houben2016TheModelling}.

Thus, it is evident that in the current context of globalization, TB has become a global concern. Due to undesirable reasons and because they are directly involved in the problem, the wealthy countries have finally been forced to intervene much more active in recent years, not least the United Nations (UN) has included the WHO "End TB Strategy" among its SGDs (Sustainable Development Goals) for the period 2015-2035. \nomenclature[z-SGD]{SGD}{Sustainable Development Goals}
This strategy is deployed through several multisectoral projects with different objectives, some of which aimed at improving the aforementioned social risk factors, while others, such as the work carried out in this project framework (see \nameref{ch1:sec:Erad:ERA4TB}), focus on developing new and more effective treatments for TB (e.g., new drugs, regimes, vaccines), as shown in the following sections.

\nomenclature[z-MDR]{MDR}{Multidrug-Resistant}
\nomenclature[z-RR]{RR}{Rifampicin-Resistant}

\section{Eradicating Tuberculosis: The need for continuous assessment}\label{ch1:sec:errad}

In 1882, Robert Koch discovered that TB was caused by the \textit{Mycobacterium tuberculosis} complex, while the symptoms were already well known since earlier dates. Concretely, molecular evidence of ancient TB-related clades has been found in Egyptian mummies ($~1550-1080$ BC)~\cite{Nerlich1997MolecularMummy} and recent studies based on the molecular clock of \Mtb~\cite{Menardo2019TheTuberculosis} confirm the hypothesis that all \Mtb~lineages go back around $2500$ years, while the Most Common Recent Ancestors (MRCAs) of the \Mtb, such as \textit{Mycobacterium bovis}, \textit{Mycobacterium pinnipedii} or \textit{Mycobacterium canettii} are dated back $11.000$ years. 

While the antiquity of the disease may be surprising to less familiar readers, the literature point to the fact that the emergence of new clades is not particularly prolific for \Mtb~when comparing to other diseases of bacteriological or virological origin (\Mtb~mutation rate is estimated around $1x10^{-8}$ and $5x{10^-7}$ nucleotide changes per site per year)~\cite{Menardo2019TheTuberculosis}. Thus, producing efficient drugs in a reasonable time before the appearance of new clades is feasible in this context. However, the new infections and reinfections numbers show that the design of previously used drugs has been insufficient, mainly due to the lack of knowledge of the molecular events of  \Mtb~itself and its interaction with social risk factors which has led to the recent emergence of new, more adapted and therefore resistant clades and lineages of the disease~\cite{Gygli2021PrisonsTuberculosis, Ngabonziza2020ARegion}.

Partly due to the urgency of the situation given by TB pandemic status and to technical limitations, drug design has been mainly based on a binary interpretation of TB (e.g., active or non-active, infected or non-infected), complicating the understanding of the molecular mechanisms causing the clinical stages traditionally described for TB (see \figurename \ref{ch1:fig:TB_tests}). As shown in the next section, this fact highlights the need for a paradigm shift in assessing the disease.

\subsection{From a binary perspective to continuous spectrum of diagnosis}\label{ch1:ssec:TBspectrum}

Depending on multiple factors (i.e., the viral load, the strain of \Mtb~complex, the attacked immune system condition after transmission), bacteria can either be eliminated from the organism through an innate or adaptive immune response (T-cell mechanisms) or remain in a latent (LTBI, Latent TB Infection) or active state~\cite{Barry3rd2009TheProphylaxis, Kaufmann2011FactLater,Pai2008SystematicUpdate}.

LTBI is not transmissible, and infected subjects present no symptoms. Active TB patients suffer from persistent cough, fever, weight loss, haemoptysis, among other maladies, and they can be transmitters of the bacteria. Due to this fact, from the traditional clinical and public health point of view, TB is understood as a binary disease~\cite{Nachiappan2017PulmonaryManagement, Pai2016Tuberculosis}.
This binary conceptualization is reflected in the most popular tests for disease assessment, which lack the specificity and sensitivity to provide a correct non-dual result. Illustratively, the main tests are shown below, divided according to whether they are used to find the presence of LTBI or for active TB.
\nomenclature[z-LTBI]{LTBI}{Latent Tuberculosis Infection}

\begin{description}
\item [Tests to detect LTBI] \textcolor{white}{:} 

\begin{description}
\item [Tuberculin Skin Test (TST):] Also known as Mantoux test or Mendel–Mantoux test or Purified Protein Derivative (PPD)~\cite{10.2307/4457498}. In this test, a protein of \Mtb~is injected intradermically, usually on the left forearm. The test is read around 72 hours later (48-96 hours). The evaluation consists of measuring the local inflammation caused. If there is no inflammation, the subject diagnostic is negative. In the event of inflammation, the diagnosis is given from the diameter of the inflammation and the patient's risk factors. It is important to note that the test can be positive both when the patient has acquired the corresponding antibodies and has previously eliminated the bacteria (adaptive immune response) and when the TB persists as latent.\\
The subjectivity test reading leads to false positives and false negatives, making complementary diagnostic tests such as X-rays or a second binary test such as Interferon-$\gamma$ needed~\cite{Auguste2017ComparingMeta-analysis}.
\nomenclature[z-TST]{TST}{Tuberculin Skin Test}

\item [Interferon-$\gamma$ Release Assays (IGRA):] Employed as an alternative to TST~\cite{Doan2017Interferon-gammaAnalysis}. The test involves exposing a blood sample from a subject to \Mtb~antigens and measuring the amount of interferon released by T-lymphocytes.
As TST, positive results do not directly link to the persistence of latent infection~\cite{Abubakar2018PrognosticStudy}. Those subjects who have overcome the infection and maintain T-cells from the adaptive immune response will also release interferon.
\nomenclature[z-IGRA]{IGRA}{Interferon $\gamma$ Release Assays}
\end{description}

\item [Tests to detect Active TB] \textcolor{white}{:}
\begin{description}
\item [Sputum smear:] The traditional test analyze the sputum using a conventional microscope to locate the bacteria. The minimum sensitivity is over the threshold of 5000 bacteria per millimeter~\cite{Datta2017ComparisonMeta-analysis, Nachiappan2017PulmonaryManagement}. Thus, it is impossible to detect active TB at early stages. Besides, the test is not very specific since \Mtb~appears the same as  \textit{non-mycobacterium tuberculosis}, which is added once again to the lack of anatomic location of the TB burden.
\item [Culture:] The culture substantially improves the sensitivity of the sputum smear, finding the \Mtb~from 10 mycobacteria per millimetre~\cite{Datta2017ComparisonMeta-analysis, Hobby1973EnumerationTuberculosis}. In return, with the best and most expensive culture methods, the results are obtained in two weeks, although generally, the duration lasts up to six. The test cannot also characterize the longitudinal evolution of the sources of infection beyond deciding whether they are still active or not. 
\end{description}
\end{description}

Besides the mentioned particular features, it is important to remark the tests relatively low cost. However, they cannot provide the continuous longitudinal characterization of the disease (from latent to active) that arises in modern literature~\cite{Arroyo-Ornelas2012ImmuneTechnologies,Ernst2012TheTuberculosis,Kaufmann2011FactLater,Pai2016Tuberculosis}. This longitudinal characterization is fundamental to understand the mechanisms of the immunological life cycle of the disease to facilitate the development of new drugs and the characterization of the resistance to them. Namely, it helps to understand under what circumstances do a latent infection reactivate or what markers are significant in remission~\cite{Barry3rd2009TheProphylaxis,Pai2016Tuberculosis}. This is especially important in the current global scenario in which therapies pretend to shorten their duration to 4 months or less~\cite{Yang2021OneChemotherapeutics}.

\begin{figure}
\centering
\captionsetup{justification=justified}
\includegraphics[width=0.95\textwidth,height=0.5\textheight]{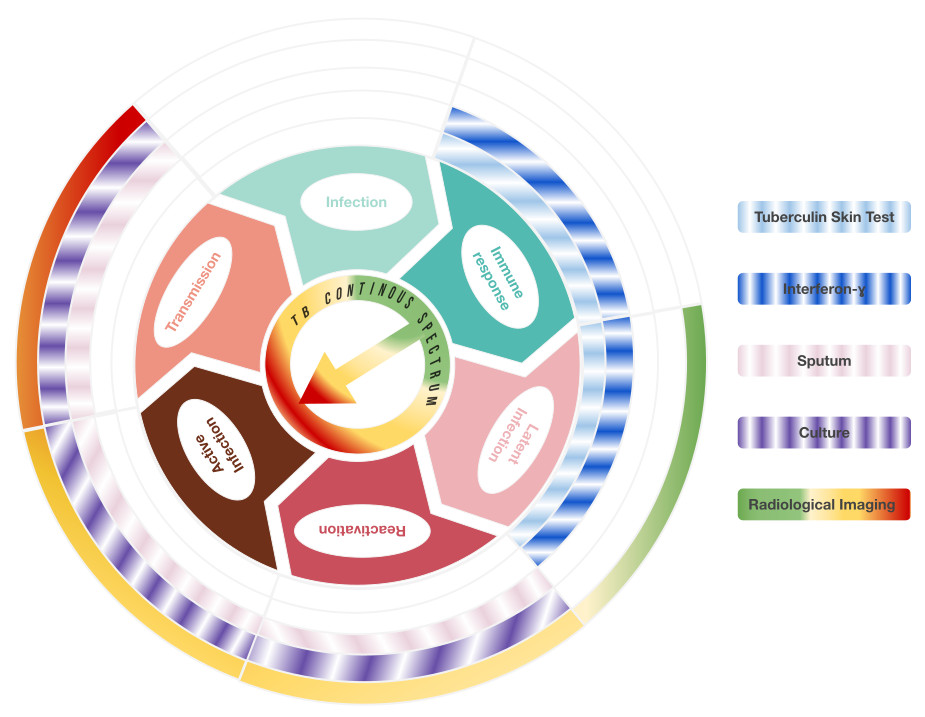}
\caption[Life cycle of \Mtb and main tests]{Life cycle of \Mtb~and main tests to characterise the entire disease spectrum. The inner cycle names the traditional categorical clinical stages of the continuous spectrum of TB immunological life cycle. Each outer circle represent each TB assessment tests capability. Blank spaces for lack of sensibility, bicolour ones represent the binary character of the test, while gradient representation represents the ability to provide a continuous value.} 
\label{ch1:fig:TB_tests}
\end{figure}

The \figurename~\ref{ch1:fig:TB_tests} illustrates this idea showing TB as a continuous spectrum  \cite{Pai2016Tuberculosis,Young2009EliminatingTuberculosis} divided into six main phases (infection, immune response, latent infection, reactivation, active infection and transmission). 
Such an approach is usual in clinical practice and essential during clinical assays. Subject identification in this scale depends on the symptoms but more on the infection burden for which the tests act as proxies. However, the traditional tests lack enough specificity/sensitivity to characterize the whole spectrum undermining the drug discovery process.  
   
For obvious ethical reasons, the development of new treatments avoids to perform tests in humans until its very last stage. So, the hypothesized molecular mechanisms fighting the disease at each phase are extrapolated from different animal models (see Section \ref{ch1:ssec:animals} for further details)~\cite{Yang2021OneChemotherapeutics}.

When employing binary or categorical tests, the relationship between the drug mechanisms and the outcome could be due to confounded correlations between pairs of outputs (i.e., marker $M$ is employed as the output of treatment $T$ for two animal models $X$ and $Y$. However, output $M$ could be due to the specific T-cell mechanism in model $X$ while in $Y$ is just a product of the risk factors). Causative mechanisms could more easily remain unknown under the binary scenario~\cite{Hernan2020CausalIfb,Lim2020DistinguishingVaccines,vanSmeden2020ReflectionResearch}. So, attributing an effect to a specific molecule in a new drug or to some variation in the composition of a vaccine, would be impossible or very expensive. Contrarily, when continuous TB characterization models are employed, the model translation results less uncertain.

Different techniques indirectly allow for the needed continuous assessment of TB. Blood tests and some culture techniques such as those mentioned above allow obtaining the Colony Forming Units (CFUs) from a sample~\cite{Barry3rd2009TheProphylaxis,Pai2016Tuberculosis}. CFUs represent the infectious burden of a subject but present some limitations: a) In cases of LTBI, the sensitivity is quite low or nonexistent, b) samples must be taken from several anatomical regions because c) the measurement is not a unilateral marker and lacks anatomical location. For example, the same number of units can represent a subject with a large localized infection or a pattern of a small source of infections in the different pulmonary lobes and even extrapulmonary. Another alternative is given by the extraction of the lungs (or the region of interest) for their subsequent dissection and complete evaluation. Even if the method is very exhaustive and allows to characterize the different manifestations of the disease in relation to their anatomical location. It is obvious that: a) the longitudinal evaluation of an excised subject is infeasible; b) it is costly; c) it implies bacteriological risks and d) it does not allow direct comparison with the disease in humans. The limitations of the different methods highlight the need to use techniques that: 1) leverage a longitudinal analysis of the subjects; 2) provide sufficient specificity and sensitivity for the detection and characterization of the different TB markers and 3) be translational between the different animal models.\\
\nomenclature[z-CFU]{CFU}{Colony Forming Unit}

Without forgetting that active TB disease requires a microbiological diagnosis, in this context, the use of \textit{in vivo} medical images is an almost perfect fit~\cite{Chen2014PET/CTTuberculosis,Nachiappan2017PulmonaryManagement}. Different imaging modalities allow the longitudinal study of the characteristic radiological manifestations. Longitudinal follow ups provide the required continuous character to the evaluation.  The screening helps to leverage the most accurate TB infection status that is essential to accomplish the objective of eradicating TB by 2035\footnote{$2050$ attending to the new models ~\cite{Houben2016TheModelling}}~\cite{Barry3rd2009TheProphylaxis,Young2009EliminatingTuberculosis}.
Since each medical image technique present advantages and disadvantages and given their relevance to this work, an introduction to the modalities employed for TB assessment are presented in a subsequent section, \nameref{ch1:sec:MITB}.

\subsection{Project Framework: ERA4TB}\label{ch1:sec:Erad:ERA4TB}

The previous sections clearly show how in addition to evolution of the social factors, the eradication of TB involves characterizing the mechanisms of bacterium propagation and its interaction with different drugs longitudinally for the continuous improvement of the compounds. This process is enabled by clinical trials that study infected subjects using "in-vivo" imaging. 
The process requires a highly interdisciplinary environment with diverse scientific profiles ranging from chemists to engineers, biologists and physicians. Namely, chemist develop new drugs. Engineers work on automatic extraction of imaging biomarkers. Biologists and physicians come to an understanding of the biological processes involved. Thus, given the magnitude of the task, it is organized into large projects, especially after the inclusion of \textit{"End TB Strategy"} among the SGDs. Specifically, the work presented in this thesis has been developed within the framework of the  European Accelerator of Tuberculosis Regime Project (\href{https://era4tb.org/}{ERA4TB}) project. ERA4TB is \textit{"a public-private initiative devoted to accelerate the development of new treatment regimens for tuberculosis"} through a \textit{"platform based on a progression pipeline that can cater for a variety of molecules at different stages of development"}~\cite{ERA4TBconsotium2021ERA4TB}. To this aim, ERA4TB is divided into modules or Work Packages (WPs) briefly summarised below, which articulate the needed interfaces during trials, being this thesis subject framed under WP4: \nomenclature[z-WP]{WP}{Work Package}
\nomenclature[z-ERA4TB]{ERA4TB}{European Accelerator of Tuberculosis Regime Project}

\begin{itemize}
    \item \textbf{WP1, Data and Pipeline Management:}
   The activities in WP1 support the development and implementation of a data management (Drug Development Information Management (DDIM)) platform supporting project efforts. A Graphical User Interface (GUI) as a portal to the clinical and preclinical data and an image storage repository. Through the platform, data will be curated, standardized and accessible to researchers. Besides, the platform will provide a plug-and-play infrastructure to employ the software analysis implemented, as presented in this thesis.

    \item \textbf{WP2, \textit{In Vitro} Profiling:}
 As the first stage for the characterization of the interactions between the drug and the bacteria at cell-level, WP2 aims to provide  \textit{in vitro} profiling capacity needed for both: (1) the preclinical profiling of single drugs and (2) the knowledge generation pathway of preclinical combos.
 
    \item \textbf{WP3, \textit{In Vivo} Profiling\label{ch1:ssec:animals}:}
    After \textit{in vitro} validation through the methodology instilled in the platform, WP3 will investigate the efficacy of the identified compounds, alone and in combination, in experimental animal models, focusing initially on relevant mouse models mimicking TB pathogenesis in humans. Subsequently, promising regimens will move to Non-Human Primate (NHP) models for evaluation.\\
    
    The works presented in this manuscript focus on CT volumes analysis of NHP and mice models
     to detect similar imaging biomarkers as those described in \nameref{ch1:sec:MITB}. While the mice model is worth it for the initial characterization of the compound effects, it does not completely recapitulate the full range of characteristics of the pulmonary pathology in humans ~\cite{Yang2021OneChemotherapeutics}. Experts cannot visually distinguish between different lesions in the mice model, so they cannot inject the knowledge into the automation systems. Consequently, the drug evaluation is approximated by analyzing the lungs as a whole.
     
    Fine evaluation requires \nomenclature[z-NHP]{NHP}{Non-Human Primates} NHP models that have been proven to recapitulate relevant clinical characteristics of the human disease. 
    This is due to the high level of gene homology, which underlies anatomical, physiological and immunological similarities~\cite{Kaushal2012TheTuberculosis,Pena2015MonkeyLearned,Scanga2014ModelingPrimates}. These similarities lead to the development of comparable disease pathology, clinical signs and immune features following \Mtb ~infection.
    Animal models are fundamental for developing novel treatments, as they provide a platform in which the efficacy of new interventions can be evaluated against infectious challenges. Longitudinal images of the TB macaque model can be acquired from live animals using medical imaging systems~\cite{Dennis2015ALaboratories,Lewinsohn2006HighMacaque,Scanga2014InLaboratory} – e.g., chest radiographs (CXR), computed tomography (CT) and position emission tomography (PET) – and employed to visualize the evolution of pulmonary disease.
    
\item \textbf{WP4, Imaging:} Imaging technologies are instrumental, enabling translational tools for drug development. Different modalities are employed to characterize the disease evolution from single cells to tissue specimens and in vivo subjects; (1) Single-cell imaging using microfluidic systems will be used to quantify responses to dynamic exposure to single molecules and drug combinations and to determine the PK (pharmacokinetic) driver; (2) MALDI-MS (Matrix-Assisted Laser Desorption/Ionization - Mass Spectrometry) imaging in infected tissues will provide quantitative information about drug penetration and distribution at the site of action in different types of TB lesions; (3) PET/CT on infected mice and NHP will provide non-invasive PK/PD (pharmacodynamic) assessments that will be integrated into response prediction models, in which imaging biomarkers will be incorporated, in close cooperation with WP5.
\nomenclature[z-MALDI-MS]{MALDI-MS}{Matrix-Assisted Laser Desorption/Ionization - Mass Spectrometry}

\item \textbf{WP5, Modelling and Simulation:} WP5 aims to ensure effective translation and extrapolation of experimental findings into clear criteria for selecting candidate molecules for combination therapy. WP5 includes world-leading partners with expertise in mathematical and statistical modelling and simulation in the field of PK/PD of anti-infective drugs, ensuring data integration and translation from WPs 1–4 and 6 with the ultimate goal of ranking suitable candidate compounds for progression into clinical trials. These actions facilitate the implementation of an appropriate database /data management system within WP1 and an efficient workflow for a rational dose selection of single compounds (monotherapy) and combination therapy to be evaluated in Phase I studies.

\item \textbf{WP6, Preclinical Development:} In charge of developing suitable synthetic route for manufacturing each molecule, and its formulation into Drug Products for later use in WP7 to study their safety and to evaluate their PK in humans ensuring their rapid transfer to in-patien trials (Phase II readiness).

\item \textbf{WP7, Phase I. First Time In Humans (FTIH):} For each molecule entering Phase I, WP7 conducts FTIH studies and other Phase I trials on healthy volunteers as needed (i.e., dose-ranging, single ascending dose, multiple ascending doses, combination regimen) for the Phase II dossier. Other Phase I trials (i.e., drug-drug and drug-food interaction studies) could be carried out if prioritised by the development team. FTIH and other Phase I trials will be designed and developed following EMA guidelines and to the highest scientific, quality and ethical standards. The final study protocol design to be implemented for each molecule will consider the recommendations arising from the integrated PK/PD modelling and simulation performed by WP5.\nomenclature[z-FTIH]{FTIH}{First Time In Humans}

\item \textbf{WP8, Management, Outreach and Sustainability:} This WP provides scientific guidance and professional project management for solving trade-offs between scope, time, quality and cost to ensure adequate progress and successful project completion. Additionally, this WP aims at developing outreach and sustainability strategies for the long-term maintenance of the ERA4TB platform.

\item \textbf{WP9, Ethics and Data Privacy:} This WP aims to define and follow up all the Ethical and Data Privacy implications to be considered throughout the project in a cross-sectional way. It also ensures that research integrity is followed across the project. It also takes care that all its activities fulfil ethical and regulatory requirements for preclinical experimentation and clinical trials under the applicable local and EU regulation on ethics and data privacy. A dedicated chapter in the Consortium Agreement ensures that Ethics and Data Privacy is be a matter for discussion included in all the routine meetings and decisions of ERA4TB's governing bodies.
\end{itemize}

\section{Medical Imaging for Tuberculosis Assessment}\label{ch1:sec:MITB}

The main \textit{in vivo} imaging modalities employed to provide the needed TB longitudinal and continuous assessment under different scenarios are listed below.
Their application is limited by the quality of the lung images acquired (i.e., SNR, resolution, artefacts). In the most aggressive cases, TB can attack the majority of the organs (extrapulmonary tuberculosis) ~\cite{ Rodriguez-Takeuchi2019ExtrapulmonaryFindings,Sharma2012PerformanceAnalysis,Wallis2013TuberculosisChallenges}. However, the lung parenchyma is often the most and main damaged region. Therefore, most studies attempt to characterize the disease progression (or remission) based on their hypothesis about the evolution of the TB manifestations within the respiratory system.
\nomenclature[z-SNR]{SNR}{Signal-to-Noise-Ratio}

\begin{description}
\item [Ultrasound (US):] It is usually used in paediatrics or in the extended clinical environment where patients cannot remain immobile during the time that image acquisition with scanners requires. Ultrasound sensitivity is enough for the detection of Pleural Effusion (see Section \ref{ch1:ssec:manifestList}) and extrapulmonary manifestations (i.e., hepatosplenomegaly and abdominal lymphadenopathy) but lacks the power to show for other main findings (i.e, difference among granulomas, conglomerations or milary TB, ground glass opacities, cavities walls)~\cite{Gennaro2018PotentialReview, Heuvelings2019ChestTuberculosis}.
\nomenclature[z-US]{US}{Ultrasound Imaging}

\item[Magnetic Resonance Imaging (MRI):] It is prescribed for extrapulmonary tuberculosis~\cite{ Wallis2013TuberculosisChallenges,Young2009EliminatingTuberculosis}. In addition, due to its non-ionizing nature, it is usually performed for pregnant women and very young patients.
However, the air-like structure of the lung parenchyma yields a low (received) signal from the tissue, which translates into a poor image contrast. Still, it is beneficial in lymph nodes study, abnormalities in the pleura and caseation (see Section \ref{ch1:ssec:manifestList})~\cite{Rodriguez-Takeuchi2019ExtrapulmonaryFindings, Sharma2021ExtrapulmonaryTuberculosis}.
\nomenclature[z-MRI]{MRI}{Magnetic Resonance Imaging}

\item[Chest X-Ray (CXR):] Together with the previously mentioned sputum tests and cultures (see Section \ref{ch1:ssec:TBspectrum}) form the triumvirate for the initial diagnosis of the disease~\cite{Bhalla2015ChestRecommendations, Nachiappan2017PulmonaryManagement}. Although a wide variety of disease manifestations can be detected and sometimes localized by CXR images~\cite{VanGinneken2002AutomaticAnalysis}, the modality does not provide enough information to accurately measure them. So, the longitudinal follow up of the disease using CXR is difficult as the images just provide a planar projection of a 3D volume being highly dependent on subjective interpretation. Therefore, CXRs just provide a quick, cheap and rough assessment. 
\nomenclature[z-CXR]{CXR}{Chest X-Ray}

\item[Computed Tomography (CT):] CT allows the acquisition of volumetric images (see \figurename \ref{ch1:fig:TB_CT_cortes}), providing a more reliable representation of the different tissues, unlike the CXR. Current high-resolution CTs (HRCT) leverage the characterization of structures at submillimetre resolution~\cite{Galban2012ComputedProgression}. This allows the detection and quantification of manifestations that remain hidden under other modalities; for example, Ground Glass Opacities (GCO), miliary and conglomerated nodules, necrosis in Lymphatic Nodules (LNs) or other parenchymal lesions.
\nomenclature[z-HRCT]{HRCT}{High-Resolution Computed Tomography}
\nomenclature[z-LN]{LN}{Lymphatic Nodule}

\begin{figure}
\centering
\includegraphics[width=0.8\textwidth, height= 0.15\textheight]{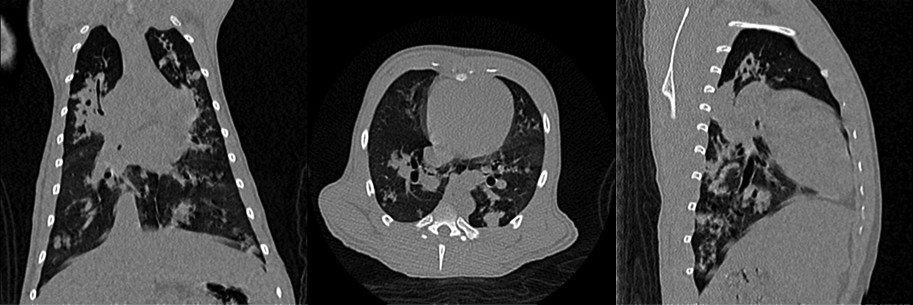}
\caption[CT image of a TB infected subject]{The abdomen of a macaque infected with TB: a) Coronal view; b) Axial view; c) Sagittal view. }\label{ch1:fig:TB_CT_cortes}
\end{figure}

\item[Positron Emission Tomography-CT (PET-CT):] PET addition, normally fluorodeoxyglucose - PET (FDG-PET), to the CT scanner adds to the virtues of the second the possibility of detecting inflammation and infection through the use of Standardized Uptake Values (SUVs). Although the use of this technique in TB is currently under consideration~\cite{Mota2020RadiosynthesisTomography, Ordonez2019RadiosynthesisTuberculosis}, mainly due to the lack of a specific PET radiotracer for TB, the results to date suggest that PET could be of great help for better measurement of the \Mtb~activity since the additional insights suggest that some manifestations as calcifications are not static~\cite{ Ordonez2020DynamicLesions, Urbanowski2020CavitaryTransmission,Yang2021OneChemotherapeutics}. This recent discovery could be a breakthrough for PK/PD modelling which would be fundamental to characterize the response to drugs administration ~\cite{Ordonez2020DynamicLesions, Urbanowski2020CavitaryTransmission}.
\nomenclature[z-PET]{PET}{Positron Emission Tomography}
\nomenclature[z-PK]{PK}{Pharmacokinetic}
\nomenclature[z-PD]{PD}{Pharmacodynamic}

\end{description}

\begin{figure}
\centering
\captionsetup{justification=justified}
\includegraphics[width=0.95\textwidth,height=0.55\textheight]{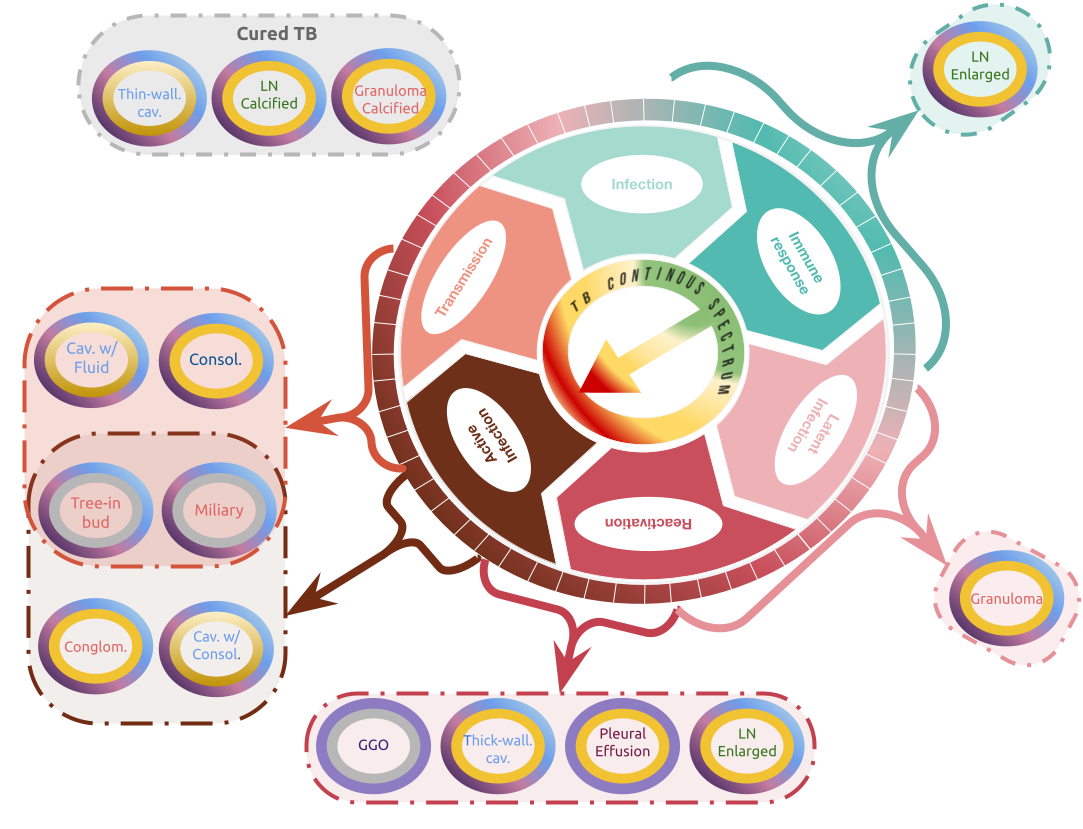}
\caption[Assessment power for Computed Tomography (CT) and Chest X-Ray (CXR) at main TB manifestation  ]{Assessment power for Computed Tomography (CT) and Chest X-Ray (CXR) at main TB manifestations through the continuous epidemiological cycle of TB. Main TB manifestations, namely, \textcolor{teal}{Lymph Nodes (LN)}: Enlarged and Calcified, \textcolor{violet}{Pleural Effusion}, \textcolor{magenta}{Ground Glass Opacities (GGO)}, \textcolor{cyan}{Consolidation}, \textcolor{blue}{Cavities}: Thin-walled Cavity, Thick-walled Cavity, Cavity with Consolidation and Cavity with Fluid and \textcolor{red}{Nodules}: Granuloma, Calcified Granuloma, Conglomeration, Tree-in-bud and Miliary nodules, are displayed across the TB spectrum range where they are usually most common. CT and CXR scan quantification power per manifestation is showed by the outer and inner toroids surrounding them respectively; purplish for CT and yellowish for CXR. Gradient colouring toroids yield for detection, localization and size measurement capacity, solid colour for detection and localization and solid grey for lack of sensibility of the imaging modality. }\label{ch1:fig:Radiology_test} 
\end{figure}

Under the listed principles, it is clear that CXR and especially CT, enable the study of the disease in the most detailed way from its macroscopic lung manifestations (shape, size, texture, localization, rate of change) to obtain suitable biomarkers which able to provide a complete spectrum for TB. \figurename \ref{ch1:fig:Radiology_test} illustrates this idea. Each manifestation is positioned at the interval of the TB immunological cycle where they typically appear. The detection and quantification power of CXR and CT turn up by the colours of their inner and outer circumferences. A brief description of these radiological manifestations is provided below~\cite{Burrill2007Tuberculosis:Review, Nachiappan2017PulmonaryManagement}. Similarly to a clinical radiological description, the standard units of CT images, \textit{Hounsfield Units} (HUs)~\cite{Knechel01042009} are referred to describe contrast differences within the TB findings and the rest of the tissues.    

\begin{figure}[htpb]
 \begin{subfigure}[t]{\textwidth}
    \centering
        \includegraphics[width=0.7\linewidth, height= 0.15\textheight]{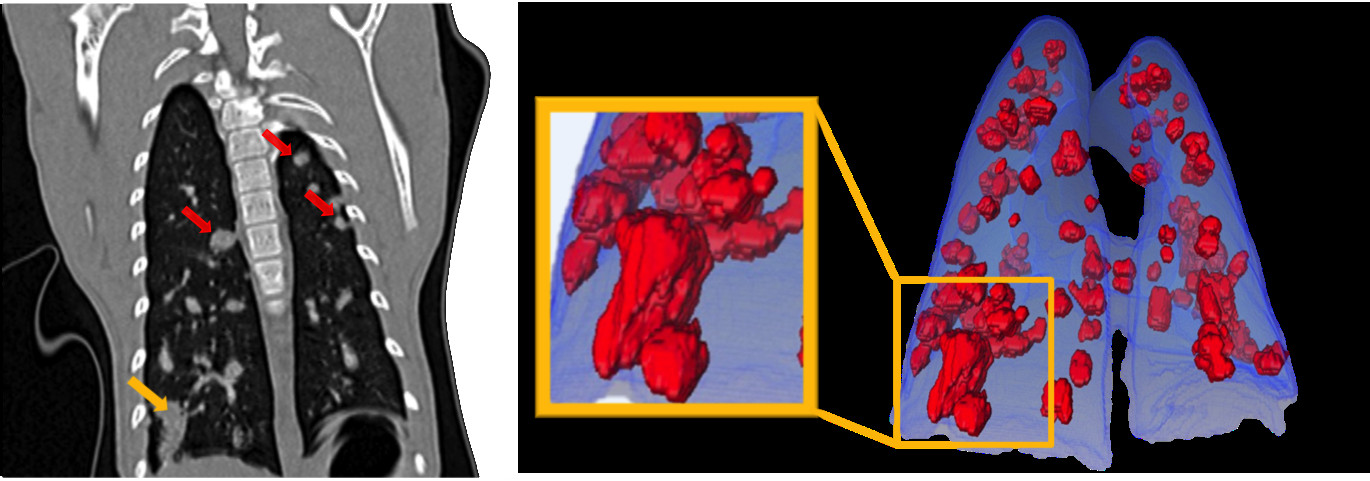} 
        \caption[Granulomas and conglomerations caused by TB ]{left) CT image of a subject 13 weeks after being infected. Granulomas (red arrows). Conglomeration (yellow arrow) in the lower part of the right lobe; right) 3D image of the infected lung.}\label{ch1:fig:TB_lesions_image}
    \end{subfigure}
    
    \centering
    \vspace{1cm}
    \begin{subfigure}[t]{0.3\textwidth}
        \centering
        \includegraphics[width=\linewidth, height= 0.125\textheight]{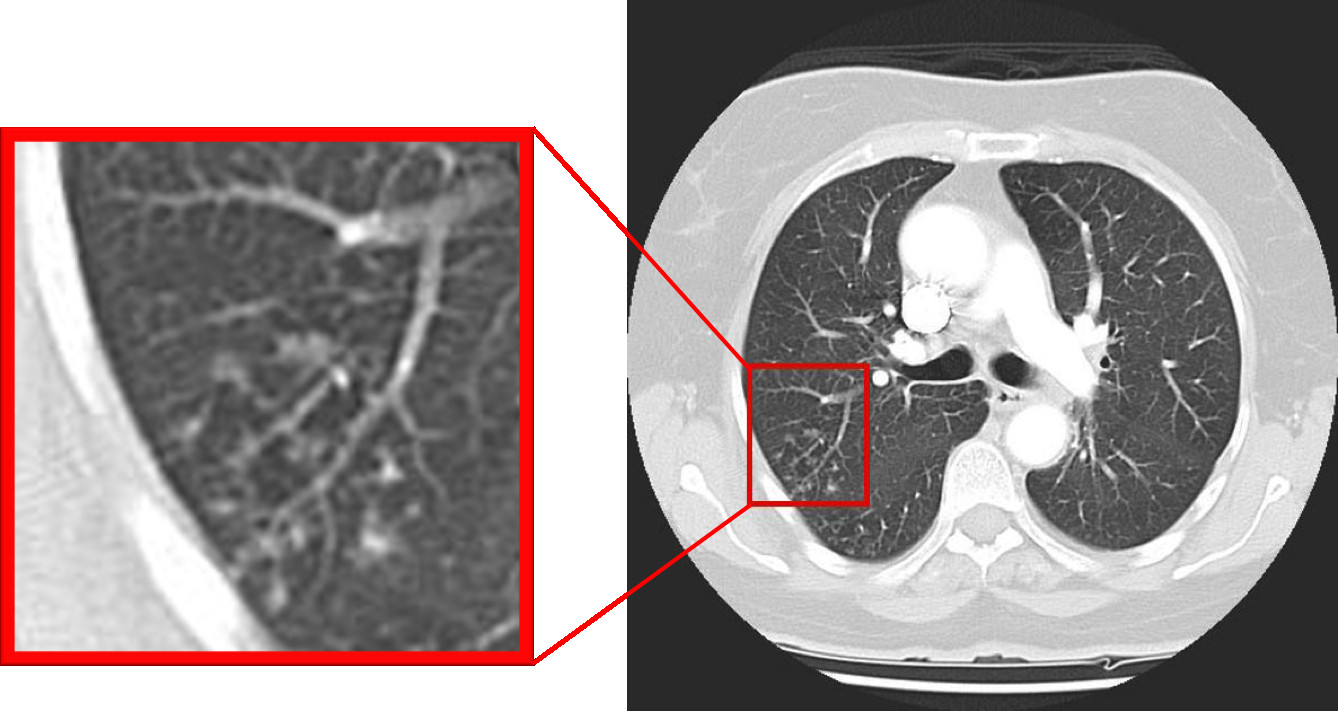} 
        \caption[Tree-in-bud manifestation CT image]{Sample of tree-on-bud pattern caused by TB: (Left) Zoomed area; (Right) Axial CT slice.}\label{ch1:fig:TB_lesions_TreeB}
    \end{subfigure}
    \hspace{0.3cm}
    \begin{subfigure}[t]{0.3\textwidth}
        \centering
        \includegraphics[width=\linewidth, height= 0.125\textheight]{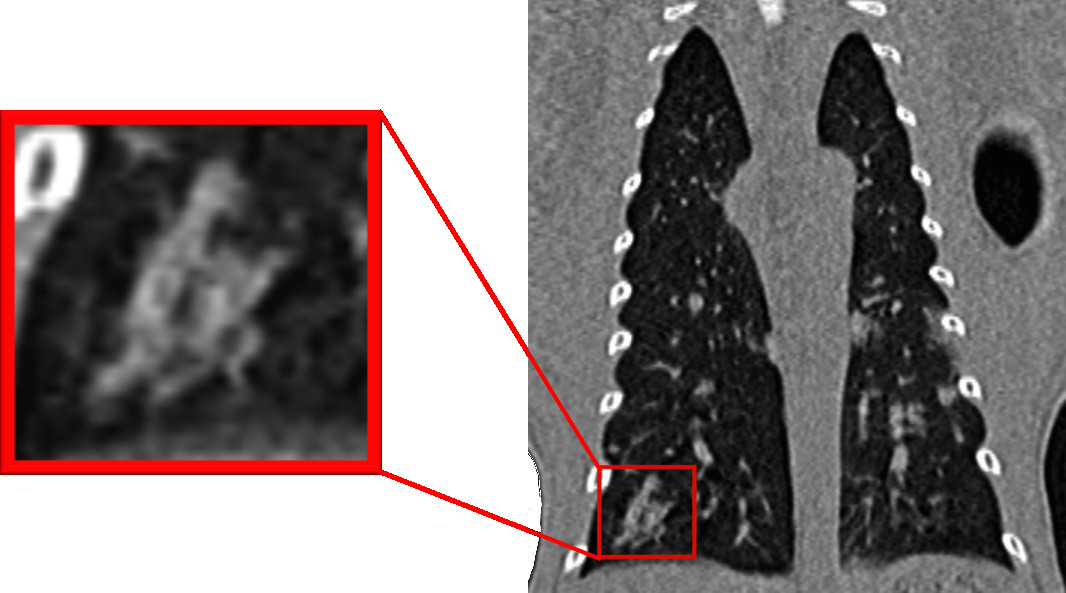} 
        \caption[Consolidation caused by TB ]{Sample of the consolidation caused by TB: (Left) Zoomed area; (Right) Sagittal slice.}\label{ch1:fig:consolidacion}
    \end{subfigure}
    \hspace{0.3cm}
    \begin{subfigure}[t]{0.3\textwidth}
        \centering
        \includegraphics[width=\linewidth, height= 0.125\textheight]{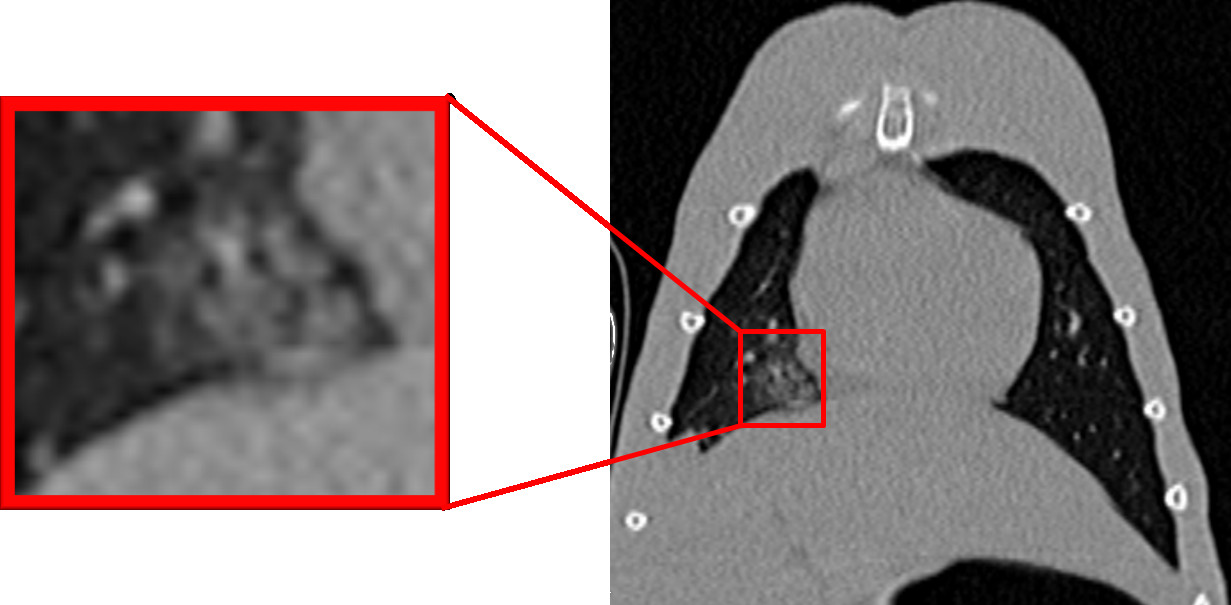} 
        \caption[Ground Glass Opacity caused by TB]{Ground Glass Opacity (GGO) caused by TB: (Left) Zoomed area; (Right) Sagittal slice.}\label{ch1:fig:TB_lesions_gco}
    \end{subfigure}

    \vspace{1cm}
    \begin{subfigure}[t]{\textwidth}
    \centering
        \includegraphics[width=0.7\textwidth, height= 0.15\textheight]{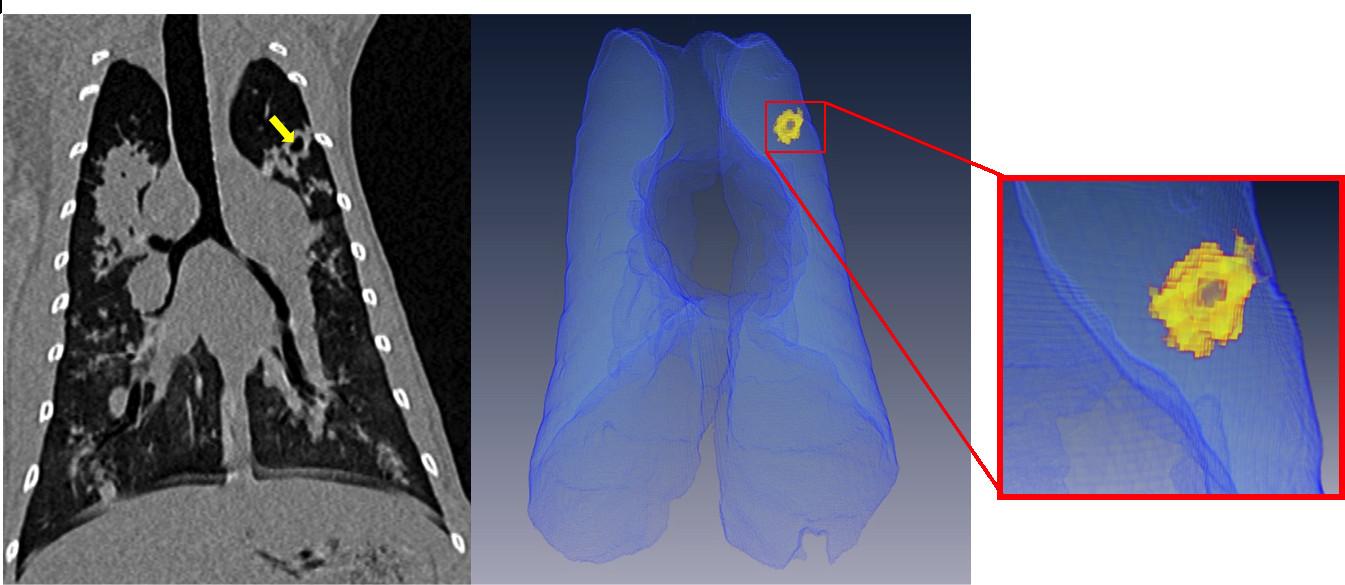} 
        \caption[Cavity caused by TB. ]{Sample of the cavity caused by TB: (Left) Sagittal slice. The cavity is pointed with a yellow arrow; (Middle) 3D visualization of the segmented lung (blue) and the cavity (yellow); (Right) Zoom showing the segmented cavity in yellow.}\label{ch1:fig:TB_lesions_cavidad}
    \end{subfigure}
    
    \caption[Radiological Tuberculosis Manifestations]{Radiological Tuberculosis Manifestations}\label{ch1:fig:radiological_manis}
\end{figure}

\begin{itemize}\label{ch1:ssec:manifestList}

\item Granuloma or tuberculoma: They are the most characteristic lesions produced by TB and the essential biomarker during the latent stage. Granulomas are spherical and present high homogeneous values on the HU scale. Due to their similar intensity on this scale to blood vessels and mediastinal tissue, it is easy to classify them incorrectly (\figurename\ref{ch1:fig:TB_lesions_image}).

\item Nodule Conglomeration: Adhesion of granulomas usually occurs when the disease is advanced and not treated. Its structure can be seen in \figurename  \ref{ch1:fig:TB_lesions_image}. Alternatively, tuberculomas could appear as randomly distributed micronodules, named miliary TB\footnote{The Miliary term refers to the random distribution of small nodules which is not exclusive of TB disease}.


\item Tree-in-bud: It appears when the infection begins to occupy air-like structures (i.e., alveoli, airways) during medium and high active stages. For this reason, they are identified by observing opacities (very bright areas) within the structures, as can be seen in \figurename \ref{ch1:fig:TB_lesions_TreeB}.


\item Infiltrate or consolidation: Opacification of air spaces within the lung parenchyma. The consolidation may be dense and may have irregular, poorly defined, or hazy margins (see \figurename \ref{ch1:fig:consolidacion}).


\item Cavity: They are air-filled areas inside granulomas. It originates at the beginning of the disease in immunosuppressed subjects, in whom the defence mechanisms cannot contain the infection. First, a proliferative lesion is formed. It has a reactive inflammatory component around the infectious focus that tends to evolve to necrosis in the central part. Necrosis is called caseosis because of the whitish appearance reminiscent of cheese. Cavities are formed when the foci of caseosis are emptied (see \figurename \ref{ch1:fig:TB_lesions_cavidad}). In the most active periods of the disease, the cavity can fill with liquid.  


\item Ground Glass Opacity (GGO): Refers to an area of greatest attenuation in the lung with bronchial and vascular damage (see \figurename \ref{ch1:fig:TB_lesions_gco}). 
\nomenclature[z-GGO]{GGO}{Ground Glass Opacity}

\end{itemize}

The CT capacity to represent specific radiological TB manifestations (as shown in the \figurename~\ref{ch1:fig:radiological_manis}) makes this modality a workhorse for damaged lung assessment  \cite{Galban2012ComputedProgression, Nachiappan2017PulmonaryManagement}.

However, volumetric CT image evaluation is complex. The CT scanners high level of detail comes at the cost of generating large images. 
Traditionally, the CT volume assessment is done manually by experts radiologists who face a tedious and time-consuming task, prone to errors and with a wide intra- and inter-expert variability being their interpretation subjective~\cite{Udupa2006AAlgorithms} (\textit{"With great power comes great responsibility"}\footnote{ \href{https://www.youtube.com/watch?v=b23wrRfy7SM}{Spider-Man phrase} widely attributed to the character Uncle Ben. \href{https://en.wikipedia.org/wiki/Amazing_Fantasy}{\textit{Amazing Fantasy. vol.15}, 1962.}}).

Aforementioned, the development of new drugs requires massive clinical trials, including different animal models with their corresponding needed lung image analysis, which is infeasible manually. 

In this way, the need to develop automatic tools for lung damage assessment, such as those presented in the following chapters, is obvious.

 The following section contextualizes the methodological environment which encapsulates such tools, introducing the main AI methodology principles from medical imaging and the most representative approaches.

\section{Computer Aided Diagnosis: The way to automated quantification}\label{ch1:sec:CAD}

Numerous and varied approaches that use Computer-Aided Diagnosis/Computer-Aided Detection (CADx/CADe) for medical imaging analysis can be found in the literature, obviously twinned with the advances and trends of AI methods. Indeed, the papers presented in this thesis redirect AI general-purpose principles to the specific problem of pathological lung CT image analysis, relying on works with a similar goal developed over the last three decades. Therefore, this section presents the relevant AI concepts in different applications that act as scaffolding for the more relevant literature subsequently introduced. The literature described is extended in each particular chapter on specific goals. In this manner, each chapter maintains its self-contained organization.


\subsection{AI Learning Principles with Emphasis in Lung Analysis}

From a general perspective, the particular lung imaging analysis question can be treated as an inverse problem\footnote{Although many works do not make explicit the resolution of an inverse problem, this framework allows to characterize them} that encompasses it within a mathematical framework that provides answers through different AI approaches (i.e., rule-based models, statistical learning, machine learning, causal learning).  The solutions provided intend to mathematically skeletonize a physical system from a limited set of $N$ observations:

\begin{equation}\label{ch1:eq:observations}
    (x_1,y_1), (x_2, y_2)\dots (x_i,y_i)\dots(x_N, y_N),
\end{equation}

where $x_i \in \mathcal{X}$ and $y_i \in \mathcal{Y}$ are inputs and outputs draws of the system. In the most frequent case within this thesis, $x_i$ are CT images and $y_i$ the corresponding segmentation or diagnosis labels. The inputs and outputs are traditionally treated as \textit{independent and identically distributed} (i.i.d) variables, that build up the sample of the random variables $(X_1, Y_1) \dots (X_n, Y_n)$ with unknown probability distribution, $P_{XY}$.

Informally contextualized, the physical systems modelled in the works presented in this thesis correspond to different tasks typically performed manually by radiologists. Thus the inputs to the system, $x_i$, would correspond to chest CT volumes.Concurrently, experts in the specific problem field depict the outputs yielded as labels, $y_i$ (e.g., binary masks, tabulated reports), in the best-case scenario. Namely, experiments in which the experts are previously trained to provide annotations in a specific format for a supervised problem (see below).
\nomenclature[a-N]{$N$}{Number of observations}

Formally, the characterization by the function $f$ must accomplish:
\nomenclature[x-X]{$\mathcal{X}, \mathcal{Y}$}{Metric Space}
\nomenclature[x-dx]{$dx$}{Differential respect to $x$}
\nomenclature[x-E]{$\mathbb{E}$}{Expectation function}
\nomenclature[x-F]{$\mathcal{F}$}{Functions Space}
\nomenclature[a-XR]{$X, Y, Z$}{Random Variable}
\nomenclature[a-P]{$P$}{Probability distribution}
\nomenclature[a-p]{$p$}{Probability density function of $P$}
\nomenclature[x-PX]{$P_X$}{Probability distribution of $X$}
\nomenclature[x-px]{$p(x)$}{Probability density function of $P_X$ at point $x$ }
\nomenclature[x-Pyx]{$P_{Y|X}$}{Conditional probability of $y|X$}
\nomenclature[z-iid]{i.i.d.}{independent and identically distributed}
\nomenclature[a-x]{$x$}{Draw from a random variable $X$}
\nomenclature[a-x]{$y$}{Draw from a random variable $Y$}
\nomenclature[a-R]{R}{Risk function}
\nomenclature[a-f]{f}{Function $f$}
\nomenclature[z-ERM]{ERM}{Empirical Risk Minimization}
\nomenclature[z-SSL]{SSL}{Self-Supervised Learning}
\nomenclature[z-OOD]{OOD}{Out-Of-Distribution}
\nomenclature[x-Loss]{$\mathcal{L}$}{Loss Function}

\begin{equation}\label{ch1:eq:XtoY}
    f:\mathcal{X}\longrightarrow\mathcal{Y}
\end{equation}
being $f$ usually found over some set of functions $\mathcal{F}$ under the optimization of a \textit{risk} or error:
\begin{equation}
    \underset{f \in \mathcal{F}}{\text{minimize}} \quad R(f),
\end{equation}
representing $R(f)$ as \footnote{Several derivations of $R$ and $\mathcal{L}$  can be found in the literature depending on the methodological framework (from ML to Bayesian). Indeed, $\mathcal{L}$ arise from the assumptions about $P_{X,Y}$ and $f$} 
\begin{equation}\label{ch1:eq:theoryRisk}
    R(f) = \int \mathcal{L}(f(x), y)p(x,y)dxdy
\end{equation}
where $p(x,y)$, if $P_{X,Y}$ admits a density, is its probability density function. $\mathcal{L}$ is a loss function (e.g., Mean Square Error, Cross Entropy) to measure the difference between the system prediction, $f(x)$ or $\hat{y}$, and the real observed output, $y$.
However, since $P_{XY}$ is unknown, $R$ is approximated in most statistical learning approaches, employing the \textit{Empirical Risk Minimization} (ERM), which converges when $N$ tends $\to \infty$ (\textit{consistent})~\cite{Boser:1992:TAO:130385.130401,vapnik2013nature}. It is formulated as:
\begin{equation}
    ERM(f) = \frac{1}{N}\sum_{i=0}^{N-1} \mathcal{L}(f(x_i), y_i).
\end{equation}

Within this context, from the type of observations and the necessary assumptions to characterize the distribution, $P_{X,Y}$, and the model, $f$, several AI modelling concepts and their limitations, can be distilled.  

Shelling the framework components out, the model, $f$, defined in eq. \eqref{ch1:eq:XtoY}, is naively identified as the kernel. Usually, in modelling, two stages are distinguished. In the first stage, features are extracted from the input data to obtain robust \textit{representations}~\cite{Bengio2013RepresentationPerspectives} of each entity main characteristics. In the second phase, the features are assembled to decide which ones resemble best the system output. Ideally, the function \eqref{ch1:eq:XtoY} must be bijective. 

Thus, the polymorphic mathematical machinery that governs $f$ mutates depending on the application, giving rise to different algorithms instantiated through their parameters within the domain $\mathcal{F}$. 
 Despite being this process consisting in determining the most suitable model, $f^*$ ($f^* \in \mathcal{F}$), referred to as \textit{learning}, highly heterogeneous, some essential principles rule them.

Each learning principle is closely related to primary AI branches~\cite{Pearl1984Heuristics:Solving}, which development and transitory predominance are closely linked to the historical evolution of the computational processing power. The processing gain boost models and their optimization processes to obtain better representations. Focusing on the learning representation process and algorithms complexity, two of the three frameworks for the automation of image processing tasks contemplated in this thesis chapters can be distinguished. Namely, the two most commonly used up to date: 

\begin{enumerate}\label{ch1:ssec:list:rulesvsML}
\item \texttt{Rule-based models:}  Tentatively, the model encodes the input data through code statements to obtain the main features (e.g., in image problems: blobs isolation, blobs roundness, edges orientation in objects). Namely, experts define representative characteristics (input representation) automatically extracted through computer programming (rules). Consecutively, the features are combined through a heuristically parameterized model.

While this modelling is hugely advantageous in terms of a) being an unsupervised process (does not require annotated data for implementation, at least explicitly), b) usual computational simplicity, and c) \textit{explainability} (defined below) since all processes are well-known; the enormous disadvantage is the very low \textit{generalization} capacity since expert knowledge is primarily based on perception. Coding perception based on predetermined rules is highly subjective and data-dependent, which supposes a constant need for model reparametrization even with slight domain changes in the input data. The segmentation method presented in Chapter \ref{ch2:chapter2} employs a rule-based algorithm.  

\item \texttt{Machine Learning Models:} Contrary to rule-based models, the parameters that govern the machine learning models are obtained by explicitly exploiting the statistical dependencies between the features that represent the inputs  of the observed sample (\textit{representations}) (e.q. \eqref{ch1:eq:observations}), modelled as random variables and their outputs (e.g. $\mathbb{E}[y|f(x)]$), usually under i.i.d. conjecture.
Based on different assumptions, many algorithms are available. All of them relate the available features through a large number of parameters which are fitted by exhaustive search computation processes 
to optimize a cost function \eqref{ch1:eq:theoryRisk}.

Traditionally, handcrafted features are employed to feed algorithms, such as~\textit{linear regression} (LR)~\cite{fISHER1936TheProblems}, \textit{Neural Networks}~(NNs)~\cite{LeCun1998TheNetworks}, \textit{Support Vector Machines}~(SVM)~\cite{vapnik2013nature} or \textit{Random Forests} (RF)~\cite{Breiman2001} among others in the \textit{supervised} case. Clustering techniques as \textit{k-means}~\cite{Bishop2006PatternLearning}, latent variable models as \textit{Expectation-Maximization}, etc., are implemented in unsupervised setups. 
The quantification approach of Chapter \ref{ch2:chapter2} implements an EM algorithm (see Section \nameref{ch2:sec:quantication}) (EM)~\cite{Neal1998AVariants}. These features can be similar to those extracted for rule-based models. However, considering their limitations and the capacity of modern ML algorithms to work in high-dimensional domains, it is frequent to use as many features as can be obtained without prior knowledge about their representation capacities. This approach intersects with radiological imaging in what is referred to as \textit{Radiomics}~\cite{Gillies2016Radiomics:Data, Lambin2017Radiomics:Medicine}.

\textit{Radiomics} techniques extract numerous quantitative characteristics (e.g., texture features~\cite{Haralick1979StatisticalTexture}, scale-invariant feature transform (SIFT)) leveraged by digitized radiological images. The extracted features are combined to obtain the predictions, $Y$, that best suit the statistical assumptions aforementioned. The work presented in Chapter \ref{ch3:Radiomics} is an adaptation of radiomics techniques for the detection of the different TB lesions.

These techniques are based on the use of statistical descriptors as features to represent the input entities. They are difficult to interpret by humans but have been proved to yield better results than those based on more human-understandable ones~\cite{Yip2016ApplicationsRadiomics.}. So much that the classical two-phase modelling approaches (feature extraction + mathematical modelling) has become a minority in the recent literature through the introduction of end-to-end models. In this alternative scenario, features are automatically extracted and combined in a single phase employing Deep Neural Networks (DNNs), which give rise to the now hackneyed term \textit{Deep Learning}~(DL)~\cite{LeCun2015DeepLearning}. The results with DL have equalled or overpass human performance for predictive analytics~\cite{Esteva2017Dermatologist-levelNetworks, Rajpurkar2017CheXNet:Learning}. However, DL models are usually understood as \textit{black boxes} \cite{London2019ArtificialExplainability}, which results in trustability issues, as introduced later in this section.

 \nomenclature[z-SVM]{SVM}{Support Vector Machines} 
  \nomenclature[z-LR]{LR}{Linear Regression} 
    \nomenclature[z-EM]{EM}{Expectation Maximization} 
\end{enumerate}

Following with the dissection of the framework terms presented in equations \eqref{ch1:eq:observations}-\eqref{ch1:eq:theoryRisk}, from the role of the outputs (labels) arise the aforementioned concept of \textit{supervised} or \textit{unsupervised} learning (\textit{self-supervised learning}, SSL\footnote{Self-Supervised Learning is more appropriate. As \textit{Le Cun} points out, unsupervised is a "confusing and overloaded" term~\cite{LeCun2021Self-supervisedIntelligence}.}), depending on whether or not the model employs system outputs in its design (or to learn the model). Although ML literature shows better results for supervised techniques, obtaining enough data is not always possible (e.g., it is not ethical to infect humans with TB to rely on a bigger $N$). Besides, novel SSL techniques could be essential for model generalization purposes~\cite{Chen2019Self-supervisedRestoration,Hinton2021HowNetwork,  Taleb2021MultimodalAnalysis}.

Therefore, in the clinical environment with a frequent lack of annotated data, such techniques represent fresh opportunities \cite{Castro2020CausalityImaging, Hussain2017DifferentialTasks, Kayalibay2017CNN-basedData}. Besides, even when is possible to obtain good quality annotations, they require expert work for long periods of time which is an unaffordable cost in many scenarios. Unfortunately, the biomedical environment is paradigmatic in this case since only a few highly qualified people\footnote{Experts' knowledge is the physical system to be modelled by $f$.} can generate annotations (i.e., radiologists, pathologists)~\cite{Maddox2019QuestionsCare, Topol2019High-performanceIntelligence}.

The lack of data or \textit{data scarcity} turns the inverse problem (eq. \eqref{ch1:eq:XtoY}) into an ill-posed one ~\cite{Peters2017ElementsAlgorithmsb, vapnik2013nature}. Namely, for each realization $(x_i, y_i)$ that is not in the observed sample, $P_{XY}$ is not defined. Therefore, untractable and unknown as defined before. Among others (i.e., Variational Bayesian Methods \cite{Blei2017VariationalStatisticians}), a classical way to circumvent this fact fall on  building a more tractable distribution as the conditional probability $P(Y|X)$, relying on probability theory techniques. Usual ML terms as \textit{regression} or \textit{classification} problems arise from the assumption of such approach, namely, $f(x)=\mathbb{E}[Y|X=x]$, being $\mathcal{Y}=\mathbb{R}$ or $f(x)=arg\,max_{y \in \mathcal{Y}}P(Y=y|X=x)$, being $\mathcal{Y}=\mathbb{Z}$. Chapter \ref{ch4:Intro} follows this approach for a multi-task classification problem. Employing just the conditional probability is enough in specific environments, where unseen datasets present conditional distributions similar to those employed during model learning.

This aspect is rarely fulfilled in real-world problems, resulting in a lack of generalization of the proposed model. This concept, which refers to the effectiveness  of $f$ to generalize solutions, is named \textit{capacity}. It usually depends on the model \textit{expressiveness}, in the sense of complexity of functions \cite{Cybenko1989ApproximationFunction}. In addition to \textit{data scarcity}, the model \textit{capacity} is severely affected by \textit{i.i.d.} assumption, generally false, broken due to data mismatches (\textit{distribution shifts} and \textit{selection biases}). They occur naturally in the real world and particularly and very significantly in clinical scenarios, penalizing the \textit{robustness} of the model. To cope with this issue is vitally important to analyze the recognizable mismatches, and for this thesis purpose, how they are extrapolated to clinical datasets. Following the nomenclature given by ~\citet{Castro2020CausalityImaging} (in brackets the nomenclature in traditional ML literature), the following are reckon\footnote{Further details, especially how the shifts arise from the causal learning framework can be found in the remarkable work ~\cite{Castro2020CausalityImaging}.}:

\begin{description} \label{ch1:shifts}
\item \textit{Population shift} (\textit{covariate shift}~\cite{Ioffe2015BatchShift}): It refers to the case when the observed realizations (\ref{ch1:eq:observations}), employed during model implementation, and the realizations features feeding subsequent inference processes (predictions over new clinical data), are in separated regions of their domain (i.e., different age, smoking status, stays at-risk countries, at samples of subjects included in a TB study). 

\item \textit{Annotation shift} (\textit{concept shift}~\cite{Cohen2020OnPrediction, Moreno-Torres2012AClassification}):  It arises when observed instances belonging to the same class are labelled differently due to annotators subjectivity (i.e., annotator experience, class definitions). 

\item \textit{Prevalence shift} (\textit{target shift}~\cite{Zhang2013DomainShift}): It appears in the cases with class balance differences between observed data and new data.  

\item \textit{Acquisition shift} (\textit{domain shift}~\cite{Perone2019UnsupervisedSelf-ensembling, Sankaranarayanan2018LearningSegmentation}):  Usually, it is due to the use of different scanner and imaging protocols that introduces spurious correlations in the datasets. For example, the same ROI in a CT image, even for the same subject acquired at practically the same time looks different under two CT scanners. 

\item \textit{Manifestation shift} (\textit{conditional shift}~\cite{Tsafack2020AThings, Zhang2013DomainShift}): Contrarily to the acquisition shift case, manifestation shift occurs when the system outputs present changes between datasets. 

\item \textit{Sample selection bias}~\cite{Heckman2013SampleError, Stukel2007AnalysisMethods}: Contrarily to the rest of the shifts, in which the mismatch arises at the data generation process, selection bias occurs at the data collection process. It happens when the dataset subsampling is not uniform due to some selection criteria as image quality control or patients admission criteria. 
\end{description}

Given the almost constant need of mitigating distribution shifts, a vast recent literature present methods that favours the model \textit{transfer} to new \textit{Out-Of-Distribution} (OOD) samples (in other words, different datasets). The result is an improved model generalization~\cite{Kamnitsas2017UnsupervisedNetworks, Perone2019UnsupervisedSelf-ensembling, Zhang2017CurriculumScenes, Zhang2019BridgingAdaptation}.

Conceptually, this set of methods is coined under the broad term, \textit{Domain Adaptation}, closely linked to hackneyed \textit{transfer learning}~\cite{Karimi2020CriticalNetworks,Neyshabur2020WhatLearning, Raghu2019Transfusion:Imaging, Tan2018ALearning, Zhuang2021ALearning}. These depend intimately, and their instantiation has produced almost innumerable options captured in the literature.
Reviewing each particular technique is outside the scope of this work. 
Still, among \textit{Domain Adaptation} techniques, it is noteworthy to mention \textit{Data Augmentation} given its widespread use, even before the explosion in the use of DL-based models~\cite{Dyk2001TheAugmentation, Hussain2017DifferentialTasks, Perez2017TheLearning, Shorten2019ALearning, Tanner1987TheAugmentation}.  \textit{Data Augmentation} techniques could not properly perform \textit{Domain Adaptation} but are included since these augment the joint distribution. Namely, \textit{Data Augmentation} intends to alleviate distributions shifts lessening \textit{data scarcity} by generating new artificial images (experiment draws) transforming available samples to increase the dataset size and its quality. The augmented images depend on \textit{a priori knowledge} about the problem (\textit{domain knowledge}~\cite{Lenc2014UnderstandingEquivalence})  since they should simulate real-world data. To this aim, colour space modifications, mixing images, projective transformations\footnote{Generalization framework for DL provided by \textit{Geometric Deep Learning}~\cite{Bronstein2021GeometricGauges} }, noise addition, patches deletion, among other strategies, are taken into account. Even when the data generation process is mostly unknown, the transformations are inferred using dedicated generative models \cite{Kazeminia2020GANsAnalysis}. This fact and the nature of the \textit{Data Augmentation} algorithms, based on classic transformations or deep generative models (i.e., PixelCNN~\cite{Oord2016ConditionalDecoders}, Variational Autoencoders -VAEs-~\cite{Kingma2013Auto-EncodingBayes}, Generative Adversarial Networks -GANs-~\cite{Goodfellow2014GenerativeNets}, Cycle-GANs~\cite{Zhu2017UnpairedNetworks})~\cite{Kingma2014Semi-SupervisedModels, Zhou2021APromises}, make \textit{Data Augmentation} itself an important field inside the AI framework.

\nomenclature[z-CNN]{CNN}{Convolutional Neural Network}
\nomenclature[z-DNN]{DNN}{Deep Neural Network}
\nomenclature[z-VAE]{VAE}{Variational Autoencoder}
\nomenclature[z-GAN]{GAN}{Generative Adversarial Network}
\nomenclature[z-GDL]{GDL}{Geometric Deep Learning}

The previous paragraphs present several principles for automation of the image analysis task  from the different strategies that exist to approximate the members in eq. \eqref{ch1:eq:theoryRisk}. While their development and adaptation to particular problems have recently~\cite{Hinton2018DeepCare,Litjens2017AAnalysis, Zhou2021APromises} yielded exceptional and previously inconceivable results, there is still "an elephant in the room". Namely, how trustable are the predictions and therefore AI for the automation tasks? 
\textit{Trustability}, vital in many complex AI-driven applications, should be a design requirement in healthcare applications~\cite{Maddox2019QuestionsCare,Shaban-Nejad2021GuestHealthcare,Topol2019High-performanceIntelligence} to take a sector with die-hard extended ideas to the next level.
Trustability is an abstract concept that gives rise to various interpretations~\cite{Ignatiev2020TowardsAI}. For the sake of clarity, in this work, \textit{trustability} is understood as the addition of two related but commonly studied independently pillars: \textit{uncertainty quantification} and \textit{explainability}~\cite{Seu2021BridgingTrustabilityb}. 
\textit{Uncertainty} is usually inferred as the measure of confidence in the predictions given by a model~\cite{Gal2016UncertaintyLearning,Krzywinski2013PointsUncertain, Tagasovska2019Single-ModelLearning}. Again, there are many different approaches to quantifying \textit{uncertainty} (specific examples appear throughout the thesis), which intend to characterize the following sources of \textit{uncertainty}:

\begin{description} \label{ch1:desc:uncert}
\item \textit{Epistemic uncertainty:} This type is due to the model structure and parameters chosen to explain the available observed data. Usually, epistemic uncertainty reduces when the number of observed data increases and the complexity of the model decreases.
\item \textit{Aleatoric uncertainty:} This type is caused by noise unexplained from the data, which mainly arise from \textit{homoscedastic/heteroscedastic} noise and labels overlap. Contrary to \textit{epistemic uncertainty}, \textit{aleatoric uncertainty} is not reducible with extra data. Formally, it is referred as:
\begin{enumerate}
    \item \textit{Homoscedastic or task-dependent uncertainty}: Occurs when the predictions errors variance have the same distribution independently of the input data values. If a single model is designed to yield more than one prediction/task simultaneously, each task will have its own homoscedastic uncertainty. Chapter \ref{ch4:Intro} presents an approach which estimates Homoscedastic uncertainty to optimize the proposed model loss function.
    \item \textit{Heteroscedastic or data-dependent uncertainty}: Contrarily to \textit{Homoscedastic uncertainty}, the variance of the prediction errors depends on the input data. 
\end{enumerate}
\end{description}

Finally, \textit{Explainability} qualifies the model by estimating the knowledge humans have about how the model makes a decision \cite{Arya2019OneTechniques, Gunning2019XAI-ExplainableIntelligence}.  This knowledge allows us to deal with \textit{uncertainty} since it enables the following two control procedures: 1) Interventions in choosing and assembling model mechanisms that better mimic the underlying physical system and 2) direct interventions on these mechanisms values, which have pronounced beneficial effects over healthcare applications \cite{Holzinger2019CausabilityMedicine}.
Therefore, it is necessary to understand as best as possible the processes of the DL model. This way, \textit{explainability} comprehend \textit{interpretability}, two terms that are often mistakenly used interchangeably. However, the second one measures humans ability to predict model outputs given different inputs or model parameters variations regardless of "why?". Thus, ML-based models, especially in DL, suffer from the "black box" effect~\cite{London2019ArtificialExplainability, Loyola-Gonzalez2019Black-boxView, Singla2021ExplainingApproach} that can be interpretable but hardly explainable. Not surprisingly, techniques such as \textit{disentangling}~\cite{Montavon2018MethodsNetworks,Zhang2020DisentanglingImages, Zhang2018VisualSurvey} have recently re-emerged in the field to provide explainability in DNNs. All the concepts above are encapsulated within the novel in-development paradigm, Causal Representation Learning \cite{Liu2021ADomain, Scholkopf2021TowardLearning}, which is the third automation framework used in this thesis. The work in Chapter \ref{ch5:Intro} follows a causal approach for disentangling. Namely,

\begin{enumerate}
\setcounter{enumi}{2}
    \item \texttt{Causal Representation Learning Models:} Formally, this paradigm arises from the interaction of  ML and causal inference \cite{Ghahramani2015ProbabilisticIntelligence,Luo2020WhenLearningb,Pearl2019TheLearning,Scholkopf2021TowardLearning}.
As mentioned, the superior performance of current ML/DL models in predictive tasks for medical imaging is 
indisputable. Their success lies in the enormous capacity to extract representations of the input data that are strongly related to the output observations. However, these models "are a victim of their success". When modelled naively, ML/DL methods seek relationships based on a mere statistical correlation on the available data \cite{DeGrave2021AISignal, Roberts2021CommonScans}. Since correlation usually appears in biased environments (see dataset shifts above), generalization and robustness problems emerge, in addition to explainability ones. 

To alleviate these problems and at the same time exploit the high capacity of DL models to extract features, several approaches appear in the literature giving rise to significant subfields (some of them already mentioned) \cite{Goyal2021InductiveCognition}. We highlight here three archetypal approaches: 1) Incorporating new terms to the lost function, either explicit regularizers\footnote{In Chapter \ref{ch5:Intro} a novel regularizing technique, \textit{Self-Normalizing Neural Networks (SNNs) \cite{Klambauer2017Self-NormalizingNetworks}} is adapted to the medical imaging field} or linked to the particular problem (e.g., including overlapping coefficients in a medical image segmentation problem) \cite{Goodfellow2016RegularizationLearning, Kukacka2017RegularizationTaxonomy, Neyshabur2014InLearning}; 2) Including a \textit{Data Augmentation} or \textit{Domain Adaptation} step; 3) Implementing a new architecture \cite{Bronstein2021GeometricGauges,Cicek20163DAnnotation, He2015DeepRecognition, Milletari2016V-Net:Segmentation, Ronneberger2015U-net:Segmentation}.

Regardless of the complexity and usefulness of each approach, these techniques allow us to limit the vast solutions space, $\mathcal{F}$, 
that models with millions of parameters can generate, based on a series of assumptions derived from prior knowledge \cite{Bishop2013Model-basedLearning, Vasudevan2021Off-the-shelfCausality}.
For example, from experiments to validate models, it is well-known that parameters with high specific values are a clear symptom of overfitting. This is controlled by inserting regularization terms as $L^1$ or $L^2$\cite{Bishop2006PatternLearning,Lin2018FocalDetection}\footnote{From a Bayesian perspective, $L^1$ and $L^2$ correspond to adding a Laplacian and Gaussian prior over the parameters distribution, respectively}. Alternatively, the Convolutional Neural Networks (CNNs) assume spatial correlation in the input data \cite{Dumoulin2018ALearning, Zeiler2014VisualizingNetworks}. As expected in the neighbourhoods of the images for which, in the automation of their analysis, the CNN present the best results in the literature. This is said considering that the penetration of Transformers literature is still low \cite{Chen2021TransUNet:Segmentation, Dosovitskiy2020AnScale, Hatamizadeh2021UNETR:Segmentation}.

This set of assumptions or inductive biases \cite{Baxter2000ALearning,Goyal2021InductiveCognition} is fundamental to the success of DL models. They delimit the solution space \cite{Huszar2017IsLearning}. Namely, they force the model to find more general representations which intend to hold across different datasets. Therefore, high-correlated but spurious signals present in shifted datasets used in the learning phases are not prioritized \cite{Arjovsky2020InvariantMinimization, Castro2020CausalityImaging}.
 However, introducing inductive biases in models guided by statistical learning may not be enough. Even with enriched models able to adapt to more environments or different datasets, solutions are build on the correlations between the representations and the observed outputs of the system under this framework. Since "correlation does not imply causation"\cite{Matthews2000Storks0.008}, both the generalization and the explainability of the models are difficult to prove.
 
  Fortunately, this problem is not new in AI. Causal inference theory pioneered by \textit{Judea Pearl} \cite{Pearl2011Causality:Edition,Pearl2018TheWhy, Pearl2016CausalStatistics} and developed during the last 40 years allows finding causal-effect relationships from the statistical characterization of the observed variables. However, its integration with DL models is not trivial and is currently a hot topic in the literature, which adopts the mathematical machinery of causation (\textit{calculus of causation}) \cite{Pearl2012TheCA, Tucci2013IntroductionDo-Calculus} to enable causal DL models. Thus, we can develop automated algorithms formalized through causal models based on \textit{a priori} knowledge (inductive biases) embedded as causal graphs, structural equations and more \cite{Hernan2020CausalIfb,Pearl2018TheWhy, Peters2017ElementsAlgorithmsb}.
  
  This intersectional framework is still under development and important concepts beyond the scope of the present work such as \textit{identifiability} needs to be integrated \cite{Peters2016CausalIntervals, Tian2002AEffects}. Adopting this framework, it is possible to establish causal models guided by graphical diagrams. Their high-dimensional input and output mechanisms are defined by DL architectures. They allow to establish hierarchical models governed by meaningful variables in the image generation process, as in the work presented in Chapter \ref{ch5:Intro}.
  
This configuration intends to build the causal knowledge structure defined by Pearl as "Ladder of Causation" \cite{Pearl2018TheWhy}. This structure enables not only the statistical characterization of particular observations from the available datasets but also perform interventions on the variables that govern the model. Interventions empower counterfactuals that enable imagined spaces \cite{Lewis1973Counterfactuals, Lorenz1973BehindKnowledge.}. They are potentially vital in medical applications \cite{Gordaliza2021TranslationalRepresentations,Pawlowski2020DeepInference, Reinhold2021ASclerosis} to answer questions such as, what would be the evolution of a damaged lung if the patient had followed a different treatment? What would happen to a human lung if the treatment only consisted of clinical trials on non-human primates and mice?

\end{enumerate}

Summarizing, this section briefly presents several basic principles of AI that are fundamental in their intersection with the field of medical imaging and specifically, with the analysis of damaged lungs. The constant evolution of these principles and the adaptation to the problem at hand is presented in the following section, through the fundamental papers published in recent years. 

\subsection{AI in service of CT Pathological Lung Assessment Approaches}\label{ch1:ssec:LungSegSOTA}

The successful use of radiological imaging for diseases assessment falls on the reliability of the information yielded by imaging biomarkers. 
Imaging biomarkers can be of a very different nature; ranging from the segmentation and measurement of the entire Region of Interest (ROI) to those extracted by a DNN, including the measurement of specific manifestations or the characterization of voxel neighbourhoods using statistical descriptors (\textit{radiomics}).

While it is true that we can attempt to quantify the damage caused by TB in the lungs without first isolating them, prior segmentation of damaged lungs is a convenient initial step to limit the area in which lesions will be located ~\cite{Mansoor2015SegmentationTrends}. Following this approach, we drastically avoid extracting spurious correlation from the ROI context, which harms biomarker quantification~\cite{Doel2015ReviewCT, Lee2019EfficientImages,Mansoor2014ASegmentation} (see Section \ref{ch1:ssec:list:rulesvsML}). Indeed, the segmented ROI acts as an inductive bias~\cite{Goyal2021InductiveCognition}.

Moreover, Pathological Lung Segmentation (PLS) is critical for a majority of CADx or CADe applications~\cite{Galban2012ComputedProgression, Setio2016PulmonaryNetworks, Xu2014EfficientFeatures}. While the quantification of disease-specific lesions is more a particular problem. For instance, in the case of lung cancer or Chronic Obstructive Pulmonary Disease (COPD) the characteristic manifestations to quantify are emphysema, fibrosis or specific nodules~\cite{Haruna2010CTCOPD,Henschke2006SurvivalScreening,Jonas2021ScreeningForce, Setio2015AutomaticImages, Tanabe2021AssociationsCOPD}, while for human TB are cavitations, tree in buds, etc. (see Section \nameref{ch1:sec:MITB}).

Besides, the development of new drugs, ERA4TB project motivation (\nameref{ch1:sec:Erad:ERA4TB}), depends on translational biomarkers. Usually there is not a unique correspondence between human and other animal models TB manifestations or sometimes the radiological manifestations are not even defined. For that, the correct segmentation of the whole pathological lung represents a suitable alternative.\\

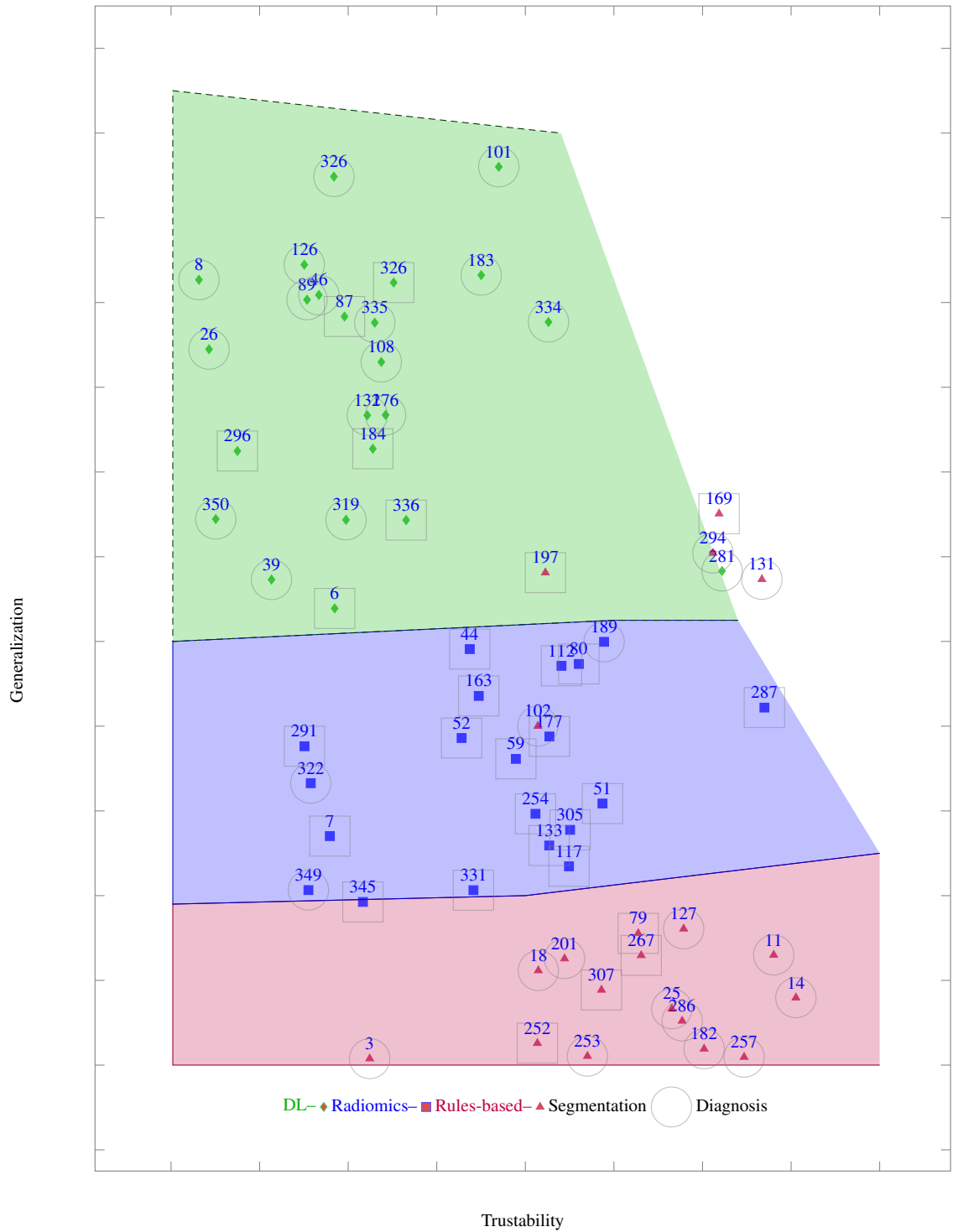
\begin{figure}
\newcommand\TikCircle[1][2.5]{\tikz[]{\draw[thick](0,0.0)circle[radius=4pt];}}

\begin{tikzpicture}
\hypersetup{hidelinks, pdfborder = {0 0 0}}

\begin{axis}[width=150mm, height=200mm, xlabel=Trustability, ylabel=Generalization, legend style={at={(0.8,0.055)}, {draw=none},
	anchor=east,legend columns=-1}, 
	legend entries={
            [black!30!green]DL--\doublespacing,
            [blue]Radiomics--\doublespacing,
            [purple]Rules-based--\doublespacing, 
            Segmentation\space,
            Diagnosis\space%
        }, 
    axis line style={gray}, 
    every axis/.append style={
    font=\scriptsize,
    x axis line style={<->},
    y axis line style={<->},
    xlabel style={below},
    ylabel style={left},
    },xmax=4.4, ymax=2.5, 
    yticklabels={,,}, 
    xticklabels={,,}
    ]

\addplot+[nodes near coords,only marks, mark=None,
point meta=explicit symbolic]
table[x=Trustability, y=General2, meta=mycitess]{Chapter1/Data/papers_summ_review.dat};

\addplot+[scatter, only marks,opacity=0.7,
   scatter/@pre marker code/.append code={%
        \let\pgfplotspointmeta=\modality, 
    }, 
    scatter/classes*={
        A={mark=diamond*, black!30!green},%
        B={mark=square*, blue},%
        C={mark=triangle*, purple}
        },%
    nodes near coords*=,
    %
    visualization depends on={value \thisrow{ModalityCat} \as \modality},
    visualization depends on={value \thisrow{mycitess} \as \mycitess}
    ]
table[x=Trustability, y=General2, meta=ModalityCat]{Chapter1/Data/papers_summ_review.dat};

\addplot+[scatter, only marks, opacity=0.5, mark size=9,
   scatter/@pre marker code/.append code={%
        \let\pgfplotspointmeta=\Aim
    },%
    scatter/classes*={
        Segmetation={mark=o, black!50!white},%
        Diagnosis={mark=square, black!50!white}},%
    nodes near coords*=,
    %
    visualization depends on={value \thisrow{Aim} \as \Aim},
    ]
table[x=Trustability, y=General2,meta=Aim]{Chapter1/Data/papers_summ_review.dat};

\addplot+[purple!80!black,fill=purple,fill opacity=0.25, mark=none] 
coordinates
{(4,0.0)  (0.01,0.0) (0.01,0.38) (2,0.4)  (4,0.5)};
\addplot+[blue!80!black,fill=blue,fill opacity=0.25, mark=none] 
coordinates
{(4,0.5) (2,0.4) (0.01,0.38) (0.01,0.38) (0.01,1.0)  (2.5,1.05) (3.2,1.05)};
\addplot+[black!80!green,fill=black!30!green,fill opacity=0.25, mark=none] 
coordinates
{(3.2,1.05) (2.5,1.05) (0.01,1.0) (0.01,2.3)  (2.2,2.2) };
\end{axis}
\end{tikzpicture}
\caption[SOTA in generalization and trustability terms]{Qualitative categorization of representative pathological lung analysis works (bibliography reference number) depending on their generalization and trustability (uncertainty measure and explainability) characteristics. Red, blue and green depict Rules-based, radiomics and DL models, respectively, for the dominant AI approach of each work. $\TikCircle$ or $\square$ represent the main focus of each entry, namely, Segmentation or Diagnosis (which depends on some previous segmentation algorithm). Shadowed areas conceptualize each AI branch capacity for our purpose.

Note that \cite{Hu2001AutomaticImages} consider only healthy subjects and \cite{Kuhnigk2006MorphologicalScans, Soliman2016AccurateModeling} are Human-in-the-loop (HIL) approaches.}\label{ch1:fig:SOTA1}
\end{figure}

Given the particular importance of the segmentation problem, it is common in the literature to categorize the works into those based solely on segmentation or those based on the extraction of imaging biomarkers for diagnostic, even when the preprocessing includes automatic segmentation. Adding this fact to the significant concepts introduced through the section, the relevant literature is examined with a view in:
 1) the model $f$, namely, a) rule-based models or ML models, meaning, b) radiomics (with or without handcrafted features) and c) DL models; 2) whether the application aims PLS or diagnosis, and 3) their capabilities in terms of trustability and generalization as shown in \figurename  \ref{ch1:fig:SOTA1}. The values of generalization and explainability in \figurename~\ref{ch1:fig:SOTA1} depends on qualitative criteria, i.e.,  the number and diversity of the datasets, the analysis of the ablation experiments, the ability to generate realistic synthetic images from random and intervened models. Therefore, the figure provides an approximate representation of the exact capabilities presented in each paper.
 
 As it is depicted in \figurename \ref{ch1:fig:SOTA1}, rule-based methods generalize poorly, in contrast to their trustability. This fact is explained by both the chance to measure epistemic uncertainty in less complex models and the higher levels of explainability due to the direct injection of expert knowledge to obtain features as inductive biases. Actually, in a counterproductive manner,  it is common to find models excessively biased to fit a small dataset that describes a particular pathology in a specific sample as in ~\citet{Abdillah2017ImageImages}.
 
In general, the complexity of the algorithms grows to reflect the biological heterogeneity within clinical datasets due to both inter/intra-subject and pathological variability.
The initial approaches, mainly rule-based, focus on the segmentation of healthy lungs using thresholding algorithms, the simplest inductive bias. These methods extract objects (blobs) from the distribution of grey values in the image. To this aim, the algorithm attempts to find the threshold value that best separates the objects of interest (the lungs, in this case).

This technique works well with CT images since these have grey levels associated with the different tissues of interest. However, the algorithms are very sensitive to noise and abnormal patterns present in the infected tissue. Thus, it is necessary to apply several morphological operations to obtain a still sub-optimal segmentation. Thresholding methods cannot provide proper segmentation of damaged lungs. However, they represent the first step for many algorithms.  In the literature, we find paradigmatic thresholding works for the segmentation of healthy lungs, such as the one by~\citet{Hu2001AutomaticImages}. This work solves a simple problem employing an iterative thresholding model that is easily interpretable. The model is able to adapt to the different domains of the healthy lung problem much better than most rules-based algorithms mentioned in Section \ref{ch1:fig:SOTA1}, given that they focus on PLS.

To improve the generalization of rule-based models, the algorithms add complexity up by combining thresholding with region growing techniques~\cite{Adams1994SeededGrowing}, as in \citet{Shen2020Quantitative2019}. However, the study presents the aforementioned overfitting issue; parameterization fits a particular dataset (image acquired just from one scanner).

As mentioned, most thresholding techniques focus on the segmentation of the whole lung since inferring the mechanisms to delimit lesions is much more complex. Even so, there are approaches for specific cases where thresholding techniques (i.e., segmenting nodules) are applied for  diagnosis. For example, in the work presented by ~\citet{Roy2006AutomatedIndex}.

Continuing with the description of rule-based techniques, we come across region-based methods such as that of ~\citet{Hojjatoleslami1998RegionApproach}. Their method follows a traditional seeded region growing algorithm. It consists in the evaluation of voxels close to a previously deterministically established voxel, the seed, and setting a criteria to decide whether the evaluated voxels belong to the lung region.

The accuracy of this class of methods depends heavily on: a) the correct identification of significant voxels that can be considered seeds; b) the definition of the neighbourhood, i.e., which voxels are close enough to it to be evaluated, and c) the definition of the neighbourhood resemblance criteria. Thus, human expert intervention controlling all the parameters becomes frequently essential to achieve appropriate segmentation or diagnostic results as in ~\citet{Farag2013ASets}.

Alternatively, employing techniques halfway between rule-based and ML methods to improve model expressiveness and therefore generalization is another explored via, such as ~\citet{Grady2006RandomSegmentation} where growth is ruled through a \textit{Random Walk} algorithm.

The next class of rule-based methods are shape-based models. These methods segment the structure employing an atlas defined or built by experts as~\citet{Li2003EstablishingImages}. An atlas consists of a lung template for CT images in the present case, containing labels of the anatomical structures of the regions of interest. This template is aligned, or in the image processing jargon, registered, with the image to be segmented by optimizing an indicator of similarity between the two (e.g., the mutual information between the images). Once both images are aligned, the region of interest is segmented. These methods work well enough for cases where lesions have not abruptly modified the expected structure. However, in the case of tuberculosis, they are often ineffective because the lungs of damaged subjects differ significantly from healthy lungs.

The last major group of rule-based methods for PLS commonly defined in the literature are neighbouring anatomy-guided methods. This class of methods uses information about the structures expected to appear around the organ of interest. For example, the rib cage, the heart or the diaphragm as in~\citet{Artaechevarria2010}, in the case of the lungs. This procedure is valuable when the area of interest is so damaged that it is impossible to recognize it. Its main problem is that it requires that the neighbouring organs are not damaged or affected by image acquisition artefacts or pathologies to function correctly.

Although the segmentation or diagnosis obtained is better than with less complex methods, most algorithms do not generalize well. Moreover, it is common to use larger datasets to adapt the parameters of the system and sacrifice quality for an improvement in generalization as in the work ~\citet{Kuhnigk2006MorphologicalScans, Soliman2016AccurateModeling}. Thus, most rule-based methods require the introduction of the Human-in-the-loop (HIL) to refine the results \cite{Kuhnigk2006MorphologicalScans}.\\
\nomenclature[z-HIL]{HIL}{Human-in-the-loop}

To avoid this fact, radiomics encompasses both those methods that use a combination of handcrafted features obtained from the \textit{a priori} knowledge of the radiologists and those methods that extract descriptive statistics from the images. In both cases, the features are extracted from the image input through an \textit{ad-hoc} method, oppositely to DL methods which obtain them automatically. Commonly, classifiers use all kinds of features together. 

Thus, the different radiomics methodologies presented in the literature are distinguished by the nature of the features that feed the model, $f$, and the algorithm that governs it. The model choice depends on the study of diverse factors related to the nature of the data available and the capabilities of each model to deal with noise, mismatches, etc.

 In general, as hypothesized by the blue shaded area (\figurename \ref{ch1:fig:SOTA1}), radiomics methods present an intermediate trustability and generalization capacity. Since features are known, their importance can be measured through multivariate analysis as in ~\citet{Coroller2015CT-basedAdenocarcinoma} and  ~\citet{Hawkins2016PredictingScans}. The specific statistical dependencies on each model can also be exploited to estimate the importance of the features by leveraging uncertainty measures and explainability. The explainability is conditioned by the features nature too. Thus, the model design must consider the trade-off between trustability and generalization, as a consequence of employing human-interpretable features and those that are not but provide greater expressiveness.
 
The most up-to-date methodological examples can be found by closely scrutinizing among "all smoke and mirrors" COVID related works~\cite{Roberts2021CommonScans}. Thus, in ~\citet{Shi2021Large-scaleClassification}, Severe Acute Respiratory Syndrome Coronavirus-2 (SARS-CoV-2) correlated features were handcrafted and exploited by a biased version of an RF~\cite{Breiman2001}. It this way, the model was general enough to work with a multicenter sample achieving a good comparison against radiologist scores. 

RF was the most successful algorithm, in performance terms, until the advent of DL. Thus, a large part of the literature employs RF models fed by mixed feature sets (texture, wavelet, handcrafted, etc.)~\cite{Gillies2016Radiomics:Data, Lambin2017Radiomics:Medicine}. As it is the case of~\citet{Tang2020SeverityImages} which exploit both texture features and the volumes of the regions to be classified or~\citet{Wilson2017RadiomicsCancer} which adds wavelets features~\cite{Ricker1953WaveletResolution}. In addition, RF is also often used for feature reduction employing the Gini importance metric~\cite{Menze2009AData}. Different approaches select the prediction most correlated features with an RF in the first stage of the model.  Subsequently, they built a simpler classifier such as an LR, similar to that proposed by~\citet{Qi2020MachineStudy}. This way, the model is more interpretive, and its uncertainty is easier to estimate. Oppositely, for numerous works, the analysis of the features leading to predictions is not the goal. In this context,~\citet{Christodoulidis2017MultisourceAnalysis} presents an hybrid approach between DL and radiomics to classify lung tissue patterns by extracting features with a CNN.\\

There are a reduced number of works in the literature, employing radiomics for PLS. Algorithms can recognize texture patterns and classify them as parenchymal tissue or not by setting a threshold for the model metric. However, the fine delimitation of such a region presents a problem. Most algorithms extract features and assign the same class (tissue, lesion, etc.) from regular fixed-dimensional voxel neighbourhoods  ($3x3$ for 2D images, $3x3x3$ for image volumes), providing a coarse segmentation. Necessarily, a post-processing algorithm to provide a finer level of detail is applied as in the work of~\citet{Liu2020AForest}.

Although RF is the classifier most used in recent years for radiomics, classifiers of a very different nature, such as SVMs based on kernel methods, are also common: ~\citet{Chen2018RadiomicClassification}, ~\citet{Alam2018Multi-StageClassifie} or ~\citet{Singh2018PerformanceHumans}. Since the SVMs performance is similar to other models and the lack of interpretability caused by the use of kernels, SVMs are less common in clinical environments.\\ 
  
However, the reluctance to "black box" models have diminished with the surge of DL models that yield results that far exceed those obtained with previous paradigms. Thus, as already mentioned, the current state-of-the-art (SOTA) is dominated by DL approaches and the incorporation of tools to improve their generalization. As a proof, we have the recent COVID-19 related literature that could serve as a decalogue of the prevailing methodology concerning the automation of image analysis of pathological lung, even despite its pitfalls \cite{Roberts2021CommonScans} given by the pandemic emergency and the annoying publication bias \cite{Ioannidis2005WhyFalse}. \citet{Cao2020LongitudinalCases} paper illustrates the capabilities of DL models to detect and delimit specific disease manifestations using a toy dataset of two cases. Several works employ the ubiquitous U-Net for segmentation as in \citet{Chen2020DeepTomography} or the U-Net three-dimensional counterpart \cite{Milletari2016V-Net:Segmentation,Ronneberger2015U-net:Segmentation} in the search for abnormalities in lung structure \cite{Shan2021AbnormalPrediction}. The U-nets and other architectures, as the \textit{ResNet} \cite{He2015DeepRecognition} in the work of \citet{Li2020ArtificialCT} and \citet{Song2021DeepImages}, are sometimes used as the baseline architecture for detection and classification. Work examples that extend naive DL models such as the above, incorporating new tools that mitigate the negative effect of strongly biased datasets (see \ref{ch1:shifts}), include, among others,  the work of \citet{Zheng2020DeepLabel} using \textit{weak supervision} \cite{Zhou2018ALearning} or those of \citet{Gozes2020CoronavirusLearning} and \citet{Wu2021JCS:Segmentation} adding explainability to the models by the use of \textit{saliency maps} \cite{Simonyan2013DeepMaps}.

Beyond COVID-related literature, we found essential contributions to the intersection between automation of pathological lung analysis and DL. Thus, it is very remarkable the research of \citet{VanTulder2016CombiningMachines}. One of the first works adapting DL-based representation learning theory to lung CT image processing. To do so, it combines generative and discriminative models, named \textit{hybrid models}. This work is before the explosion of deep generative model's \cite{Goodfellow2014GenerativeNets, Kingma2013Auto-EncodingBayes} but still proves the effectiveness of such approaches. 

Alternatively, there exists numerous works pioneering PLS tasks using DL. It is worth mentioning the work of \citet{Gao2018HolisticNetworks} that already employs multiresolution analysis to address pattern recognition in \textit{Interstitial lung diseases} (ILD). Also, the paper of \citet{Alakwaa2017Lung3D-CNN} among so many appeared to tackle the cancer nodule detection problem presented at the \textit{Kaggle Data Science Bowl 2017} \cite{KaggleInc.2017DataKaggle}\footnote{\href{https://www.kaggle.com/c/data-science-bowl-2017}{https://www.kaggle.com/c/data-science-bowl-2017} }. Regarding the generalization problems already mentioned, it is necessary to point out that the \textit{Kaggle} challenge was subsequently won by \citet{Liao2019EvaluateNetwork}. They adopted for the medical imaging field, a 3-D Deep Leaky Noisy-OR Network. Shortly after, Google AI researchers \cite{Ardila2019End-to-endTomography} surpassed these results employing a 3-D Mask R-CNN \cite{He2017MaskR-CNN}. However, both applications turn out to be hardly integrable in the clinical workflow \cite{Jacobs2019GooglesValidation}.

Continuing with significant work for PLS, the paper by \citet{Harrison2017ProgressiveImagesb}, published in 2017, presents a remarkable alternative using "progressive and multi-path holistically nested neural networks" to the U-net architecture.  
 U-Net is still the baseline in the field, probably because the chance of obtaining similar results to those reported depends more on the diversity of the data available during learning than the model choice, as pointed out by \citet{Hofmanninger2020AutomaticProblemb}.

For this reason, this summary includes innovative work on models that encourage this generalization, even employing limited datasets. Thus, \citet{Gerard2020Multi-resolutionSpecies} paper stands out, which shows that training a model in several steps allows the use of images of multiple mammalian species for PLS. A similar strategy to segment human lung lobes in high-resolution CT images follows \citet{Lee2019EfficientImages} work. Likewise, \citet{Xie2020RelationalScans} also gives an alternative to segment lobes on images acquired with a high-resolution protocol. They add structural relationships that act as inductive biases for the DL model. Finally, the work of \citet{Amyar2020Multi-taskSegmentation} enriches the model by proposing multi-task learning. Namely, learning the lung segmentation masks together with the severity degree caused by the pathology. The Chapter \ref{ch4:Intro} follows a multi-task approach for the segmentation of TB radiological manifestations.

\nomenclature[z-PLS]{PLS}{Pathological Lung Segmentation}
\nomenclature[z-TB]{TB}{Tuberculosis}
\nomenclature[z-TB]{WHO}{World Health Organization}

\newcommand\crule[3][black]{\textcolor{#1}{\rule{#2}{#3}}}
\chapter{Lung Segmentation and Quantification with Rule-based Approach}\label{ch2:chapter2}

\ifpdf
    \graphicspath{{Chapter2/Figs/Raster/}{Chapter2/Figs/PDF/}{Chapter2/Figs/}}
\else
    \graphicspath{{Chapter2/Figs/Vector/}{Chapter2/Figs/}}
\fi

\section{Coding human perception and expert knowledge}\label{ch2:sec:heuAIPer}

From a general perspective, the specific goal of segmenting and quantifying a pathological lung image could be seen as the attempt to capture heuristic knowledge from the human experts to encode it as methods, processes or algorithms to perform the task.

Regardless of whether the knowledge comes as a \textit{rule of thumb}, intuitive judgments or educated guesses~\cite{Pearl1984Heuristics:Solving} as in the problem case (from experienced radiologists), the most traditional way to employ Artificial Intelligence (AI) needed for task automation is through a set of encoding rules (\texttt{if A: do B...})~\cite{Clancey1983TheExplanation}; in contrast with Machine Learning (ML), where the rules are learned from labelled data and/or features prescribed (supervised or unsupervised learning) or even just the data (e.g., Deep 
Learning (DL)) employing methods in the intersection between statistics and computer science (see Section \ref{ch1:ssec:list:rulesvsML}).

This primary approach to AI for CADx/CADe presents some limitations.
The encoded models do not just intend to represent a human-defined procedure (for example, decision tree protocols) or a well-known shape (e.g., ellipse detection through the Hough transform~\cite{Duda1972UsePictures, Hough1959MachinePictures}), but also need to encode the complexity of human perception, which is loosely defined for this purpose. DL models are based on NNs \nomenclature[z-NN]{NN}{Neural Network}, which are strongly inspired in the neurological basis~\cite{Hubel1962ReceptiveCortex, LeCun2015DeepLearning}. Taking this into account, it is maybe less surprising that DL methods, instead of rule-based ones, have recently achieved the most successful results for perception related tasks~\cite{Litjens2017AAnalysis, Topol2019High-performanceIntelligence}. 

Besides, the expert knowledge cannot be adequately encoded using a set of classic programming control structures. Usually, users apart from the author are unable to set up the proper parameters for a new environment (e.g., new animal model, new acquisition scanner) (see Section \ref{ch2:sec:OutOfComf}).

However, we can gain much valuable information employing a rule-based approach. We can obtain easily informative models for domain-specific problems that, even with limitations, could be enough in some scenarios.

The use of a traditional methodology at this thesis stage serves a double purpose: 1) to promote a solution to the problems mentioned above in datasets composed of TB-infected lungs for different animal models (see Section \ref{ch1:ssec:LungSegSOTA}) and 2)  to gain further insight into the segmentation problem. In this way, we avoid the usual black box effect of the more optimized but less informative ML/DL most employed models~\cite{Litjens2017AAnalysis,Murphy2022ProbabilisticIntroduction, Topol2019High-performanceIntelligence}, to subsequently enrich them, by injecting in different ways, as far as possible, the knowledge acquired during this first approach.

\section{Rules-based Lung Segmentation Method for a Specific Domain}\label{ch2:ssec:ruleAppr}

\begin{figure}
\centering
\captionsetup{justification=justified}
\includegraphics[width=0.8\textwidth,height=0.27\textheight]{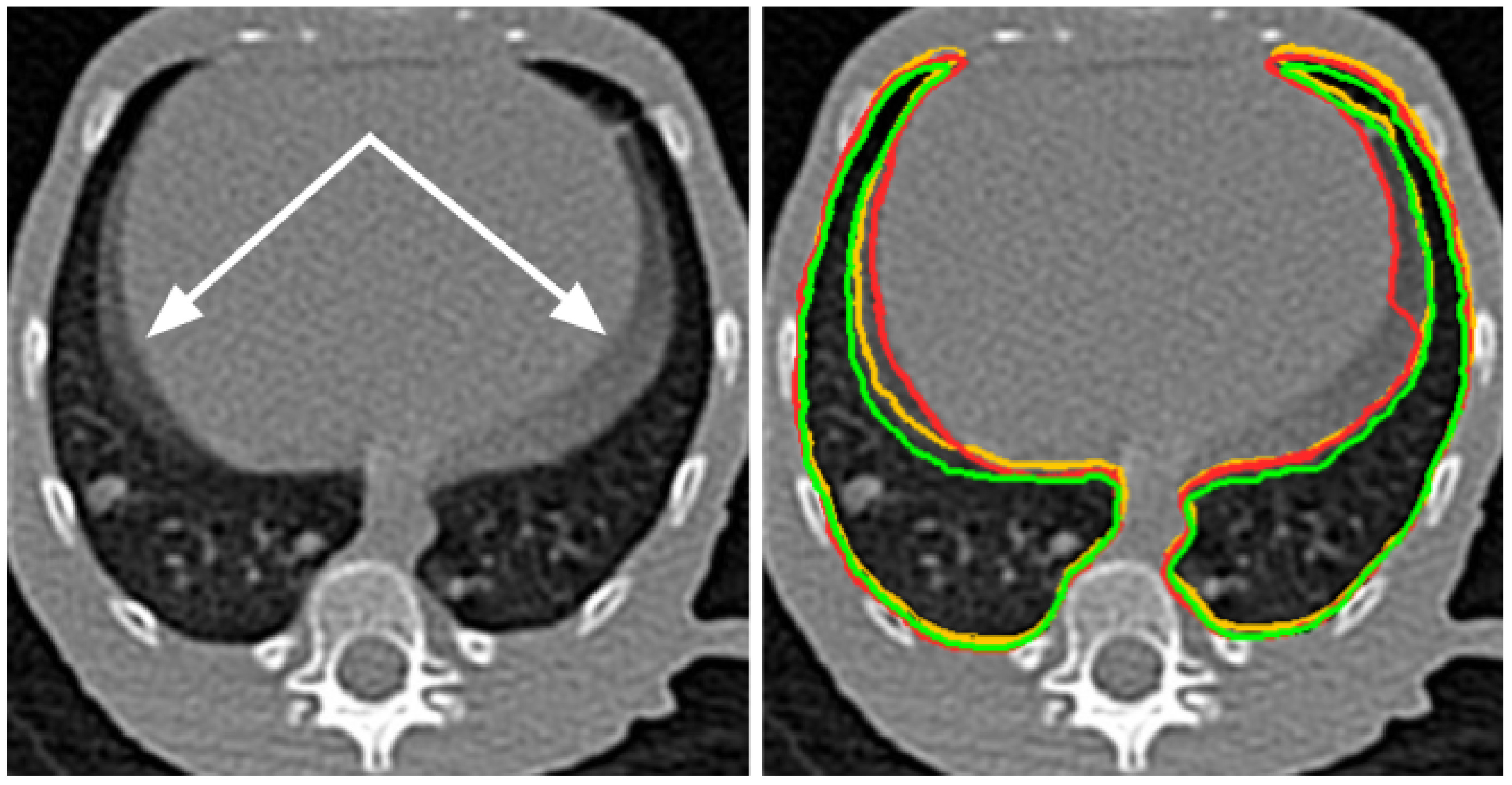}
\caption[Fuzzy Boundaries Example]{(Left) Sample slice from a chest CT volume of a subject infected with \textit{Mycobacterium tuberculosis}. The presence of fuzzy boundaries (white arrow) caused by respiratory movement artifacts makes it difficult to delimit the lung boundary; (Right) The annotations performed by the experts are combined to explicitly illustrate the differences and shown with a red, yellow and green outline, respectively.}\label{fig:chap2/Figure_1} 
\end{figure}

As was mentioned in Section \nameref{ch1:ssec:LungSegSOTA}, segmentation of TB-infected lungs is complex in clinical and preclinical studies. The expected variability of the pulmonary inflation caused by the respiratory cycle is increased and less predictable than the healthy subjects. This is due to the changes in lung compliance caused by the disease and the breathing difficulties experienced by anaesthetized infected animals. Moreover, CT image acquisition in TB animal models is usually performed on free-breathing animals to avoid the additional level of complexity added by the intubation. It results in the presence of significant respiratory motion artefacts. This effect produces fuzzy boundaries, especially in the diaphragm area (\figurename~\ref{fig:chap2/Figure_1}). Thus, it implies an uncertain delimitation of the lungs beyond the segmentation technique used. 

Manual segmentation of the lungs is subject to wide intra-, and inter-expert variability in the presence of those fuzzy boundaries ~\cite{Udupa2006AAlgorithms}. Most of the state-of-the-art methods for automatic lung segmentation are not designed to deal with the specific problems present in \textit{Mtb}-infected lungs under the presence of strong respiratory motion artefacts ~\cite{Mansoor2015SegmentationTrends}, as was mentioned before, even under the DL scope (see \hyperref[ch1:ssec:SOTA]{PLS approaches}). They generally are not able to differentiate between the neighbouring soft tissue and the lesions attached to the pleura since their density (Hounsfield Units, HU) \nomenclature[z-HU]{HU}{Hounsfield Units} is similar ~\cite{Messay2015SegmentationDataset}. Well-known thresholding methods ~\cite{Hu2001AutomaticImages} perform well when extracting healthy tissue but cannot cope with HU variability. Region-based methods  ~\cite{Grady2006RandomSegmentation,Hojjatoleslami1998RegionApproach} fail in the presence of abnormalities and are highly user-dependent. Atlas-based methods ~\cite{Li2003EstablishingImages} fail to obtain a suitable general model able to capture the singularity of the disease. The more recent approaches are primarily based on supervised learning methods ~\cite{Yip2016ApplicationsRadiomics.}. They require a large dataset labelled by an expert to ensure appropriate training and even then, are not free from bias.  

In the remainder of this chapter, an automatic pipeline able to segment lungs infected with \textit{Mtb.} placing considerable importance on the robust and consistent identification of fuzzy boundaries is presented. This is our first approach to the segmentation of infected lungs and follows a traditional rule-based methodology. This method was already published in the paper, \textit{Unsupervised CT Lung Image Segmentation of a Mycobacterium Tuberculosis Infection Model}, \cite{Gordaliza2018}. Therefore, most of the content of the following paragraphs were already presented within it.

\subsection{Materials}
\label{ch2:sec:material}

\subsubsection{Experimental Animals}
\label{ch2:ssec:ExpAnimals}

Male cynomolgus macaques, aged $3$ to $4$ years, were obtained from an established UK breeding colony for these studies. Genetic analysis of this colony has previously confirmed the cynomolgus macaques to be of Indonesian genotype ~\cite{Mitchell2012CharacterisationDiversity}. The absence of previous exposure to mycobacterial antigens was confirmed. All animal procedures and study designs were approved by the Public Health England Animal Welfare and Ethical Review Body, Porton Down, UK, and authorized under an appropriate UK Home Office project license. All animal procedures were performed on a facility with biosafety level 3 laboratories.

\subsubsection{Aerosol Exposure}
\label{ch2:ssec:Aerosol}

Macaques were challenged by exposure to aerosols of \textit{Mtb} as previously described ~\cite{Sharpe2010EstablishmentTesting,Sharpe2016UltraMacaques}. Mono-dispersed bacteria in particles were generated using a 3-jet Collison nebuliser (BGI, Waltham, MA, USA) and, in conjunction with a modified Henderson apparatus, delivered to the nares of each sedated primate via a modified veterinary anaesthetic mask. The challenge was performed on sedated animals. They were placed within a head out plethysmography chamber (Buxco, Wilmington, North Carolina, USA) to enable the aerosol to be delivered simultaneously to measure respired volume. The calculations to derive the presented dose (PD) \nomenclature[z-PD]{PD}{Presented Dose} (the number of organisms that the animals inhale) and the retained dose (the number of organisms assumed to be retained in the lung) have been described previously ~\cite{Harper1953TheSpores.,Sharpe2010EstablishmentTesting,Sharpe2016UltraMacaques}.

\subsubsection{CT Imaging}
\label{ch2:ssec:CTScans} 

Our dataset comprises 63 CT scans of the chest acquired from $9$ different subjects at $7$ time points ($0$, $3$, $12$, $16$, $20$, $24$ and $28$ weeks after aerosol exposure to \textit{Mtb}). The subjects were treated with different combinations of antibiotics (see \tablename~ \ref{ch2:tbl:treatments})~\cite{Sharpe2016UltraMacaques}. The chest CT scans were acquired with a $16$-slice Lightspeed CT scanner (General Electric Healthcare, Milwaukee, WI, USA) with voxel spacing of $0.23$ mm x $0.23$ mm x $0.625$ mm and image size of $512$ pixels x $512$ pixels. 

\begin{figure}
\includegraphics[width=\textwidth]{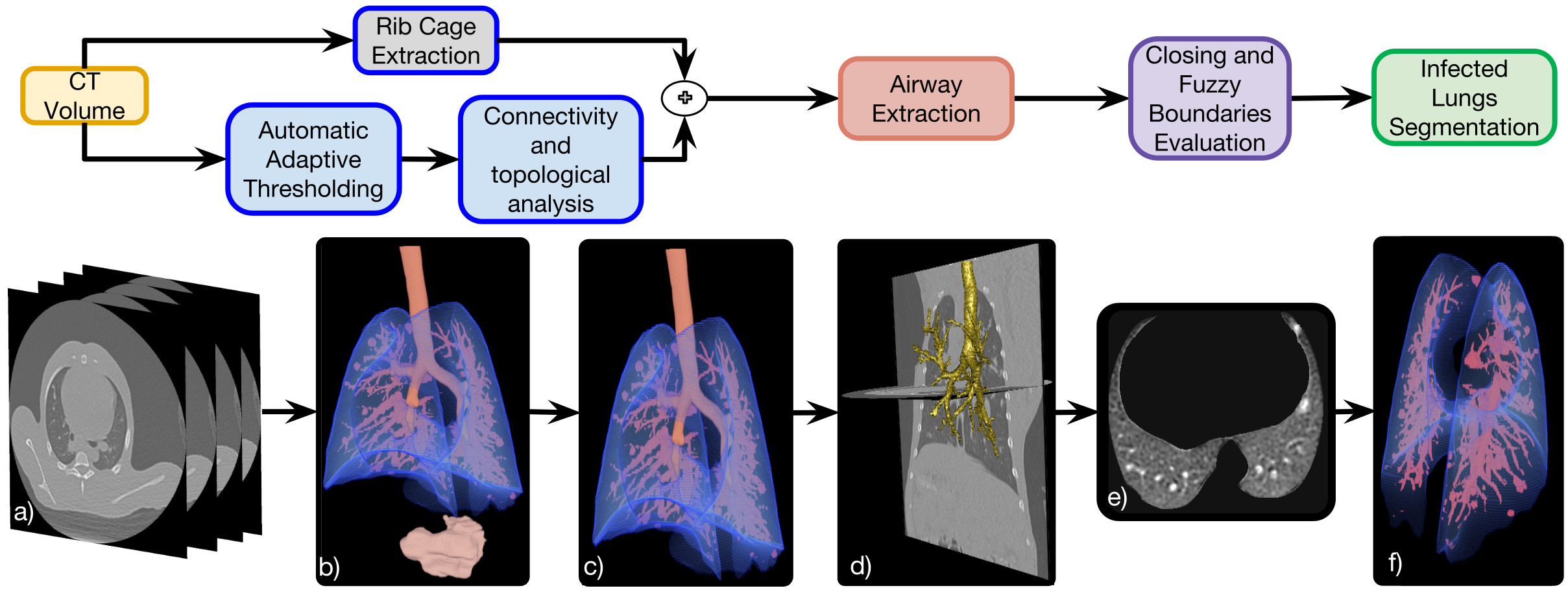}
\captionsetup{justification=justified}
\caption[Automatic lung segmentation pipeline]{Automatic lung segmentation pipeline: (a) Source chest CT volume; (b) 3D rendering of the air-like structures detected in the image using automatic adaptive thresholding; (c) 3D rendering of the preliminary lung and connected airways segmentation obtained using a set of topological operations based on the position of all pre-segmented structures; (d) Isolated airways tree extracted with a propagating wavefront approach; (e) Axial slice of the final lung segmentation in which the lesions caused by \textit{Mtb} and attached to the pleura have been included and the motion artefacts discarded; (f) 3D rendering of the final lung segmentation including healthy parenchyma, the damaged parenchyma and the blood vessels.} 
\label{fig:chap2/Figure_2}
\end{figure}

\subsection{Methods\myfnt{The \texttt{C++/ITK} \cite{Johnson2015TheGuide} code implementation for the methodology description can be found under the Tuberculosis Lung Segmentation (TLS) \href{https://github.com/BIIG-UC3M/TLS-Piped}{GitHub repository}} }
\label{ch2:sec:Methods}

\subsubsection{Automatic Lung Segmentation}
\label{ch2:ssec:lungsegmentationGeneralDescr}

The automatic lung segmentation pipeline is composed of three main steps, as depicted in \figurename~\ref{fig:chap2/Figure_2} and explained in the following sections.

\subsubsection{Preliminary Lung and Airway Tree Segmentation}
\label{ch2:ssec:lungsAndAirwSegm}

\textsl{Automatic Adaptive Thresholding:} The first step goal is to obtain a rough segmentation of the lungs, including the airway tree, similarly as was introduced by \textit{Hu et al.}~\cite{Hu2001AutomaticImages} by separating air-filled structures (i.e., healthy parenchyma, stomach, airways, image background) from more dense tissues in the whole image volume (\figurename~\ref{fig:chap2/Figure_2} (a) and (b)) in which a bi-modal histogram distribution is expected, especially for healthy lungs, by finding a threshold $T$ given by the iterative equation \ref{ch2:eq:hu}.
\begin{equation}\label{ch2:eq:hu}
    T_{i+1} = \frac{\mu_a + \mu_{na}}{2},
\end{equation}
where $\mu_a$ y $\mu_{na}$ are the average intensity of voxels below and above $T$ for the iteration $i$. The find ends once $T_{i+1}$ = $T_i$. While this method have been widely employed in several studies is quite biased as is based in a unreal assumption (inductive prior) for this work environment: volumetric images to segment present big and isotropic Signal-to-Noise ratio (SNR).\\ 
Since the study images present movement artefacts as was mentioned before (see \figurename~\ref{fig:chap2/Figure_1}), \textit{Hu et al.} method was discarded in favour of \textit{Otsu} method~\cite{Otsu1979AHistograms, Vala2013AAlgorithm}. \textit{Otsu}'s method adapt the classic Fisher Linear Discriminant (LD)~\cite{Bishop2006PatternLearning} \nomenclature[z-LD]{LD}{Linear Discriminant} for image thresholding. To this aim, the algorithm assumes the existence of a bimodal (two Gaussians) distribution (air-like and non air-like structures), two tissue classes ($c = [1,2]$), and minimize the within-class variance, e.g: $s_1^2 + s_2^2$, while maximize the separation between the class means ($S_B$), e.g: $(m_2 - m_1)^2$, minimizing the class overlap under an optimized $T$ as follows: 
\begin{equation} \label{ch2:eq:Otsu_opt}
\argmax_T \frac{(m_2(T)-m_1(T))^2}{s_1(T)^2+s_2(T)^2} , 
\end{equation}
being 
\begin{equation}
m_k(T) =\frac{1}{N_k} \sum\limits_{v \in k(T)} I(v) \qquad s_k^2(T) = \sum_{v \in k(T)} (I(v) - m_k)^2,
\end{equation}
where $I(v)$ is the image intensity value (HUs) at voxel $v$ and $N_k$ the number of voxels belonging to class $k$.\\
\nomenclature[a-m]{m}{population sample average}
\nomenclature[a-s]{s}{population sample standard deviation}

\textsl{Rib Cage Extraction:} Although the literature contains robust approaches to rib cage and sternum segmentation \cite{Kang2003AData,Liu2015SegmentationImages,Staal2007AutomaticData}, it was not necessary for our purpose— and beyond the scope of the present study— to implement a highly accurate and time-consuming segmentation. Instead, we used a simple technique, which, although unable to capture each bone's specific shape, was good enough to establish a convex hull for the ribcage.  First, we defined voxels with a value similar to the rib cage bones (over $900$ Hounsfield units (HU)) as seeds. Then, we perform region-growing segmentation using the criteria given by the confidence connected segmentation method ~\cite{Piekos2007ConfidenceITK}.\\ 

\textsl{Connectivity and Topological Analysis:} In order to isolate the lungs from the rest of the segmented air-filled structures, as described in ~\cite{Artaechevarria2010AutomatedTomography,Lehmann2007LabelITK}, we utilized the differences in size and anatomical location of the hidden objects as follows:  a) excluding the objects located outside the convex hull formed by the partially extracted ribcage (\figurename~\ref{fig:chap2/Figure_2}, as those which a volume less than $10\:mm^3$ (c))  and b) selecting as lung tissue, the structures at the minimal Euclidean distance to the ribcage centroid.

\subsubsection{Airway Tree Extraction}
\label{ch2:ssec:AirwayTreeExtraction}

Due to the intricate morphology of the airway tree, a specific algorithm was needed to extract it from the overall lung volume (\figurename~\ref{fig:chap2/Figure_2} (d)). Our approach adapted a method based on modelling a propagating wavefront through the trachea, as introduced by \textit{Schlathoelter et al.}~\cite{Schlathoelter2002SimultaneousBronchoscopy} and extended by \textit{Bulöw et al.}~\cite{Bulow2004AData}. The implementation described in the following sections focuses on leakage detection, which is usually a significant problem for the mentioned approaches.\\

\begin{figure}
\includegraphics[width=\textwidth]{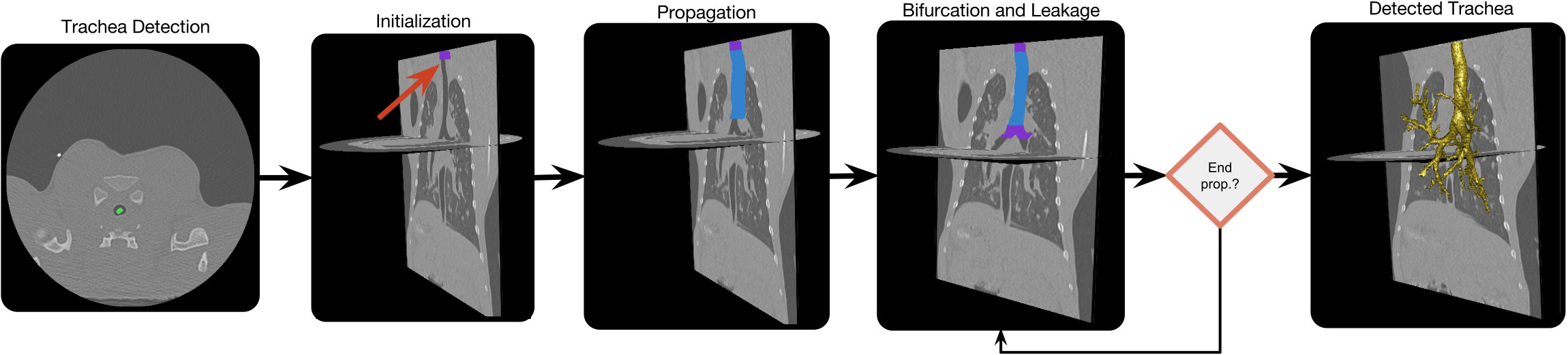}
\caption[Airway tree extraction workflow]{Airway tree extraction workflow. Step $1$: Trachea seed detected by morphological analysis of the \textit{HRCT} slices; Step $2$: The trachea section is initialized by adding the neighboring voxels to the seed, thus creating a dome; Step $3$: Spherical wavefront propagation ruled by the algorithm~\cite{Bulow2004AData,Schlathoelter2002SimultaneousBronchoscopy}; Step $4$: Check for bifurcations and leakages of the wavefront into the lungs; Step $5$: The resultant isolated trachea after propagation.}
\label{ch2:fig:Figure_3_sup}
\end{figure}

\textsl{Trachea detection and initialization:} 
The origin of the trachea is detected using a slice-by-slice search for the first isolated, air-filled area with a diameter from $5.5$ $mm$ to $8.5$ 
$mm$ (depending on the animal's weight ~\cite{Pinkerton2015ArchitectureTree}) and a roundness above $0.9$. The centre of mass (highlighted in green in \figurename~\ref{ch2:fig:Figure_3_sup}-Step $1$) is chosen to be the seed to form a dome, including the surrounding voxels and emulating a spherical wave.

\textsl{Wavefront Propagation:} 
The wavefront propagates, and the decision on whether to add voxels from the neighborhood (segments) is based on a 3D fast marching level set algorithm, which is ruled by the $time\ step$ and two thresholds as defined in \textsl{Artaecheverria et al.}~\cite{Artaechevarria2009AirwayMicro-CT,Deschamps2001FastEndoscopy,Forcadel2008GeneralizedSegmentation}:
\begin{itemize}
\item $T_i = \mu_s + \alpha \cdot \max (\sigma_{s-1},\sigma_{s-2})$ for the similarity between a voxel and its neighborhood, where $\mu_s$ is the mean intensity of the neighborhood voxel, $\sigma_{s-1}$ y $\sigma_{s-2}$ the standard deviation of the voxels added HUs at the two previous iterations and $\alpha$ the propagation factor.
\item and $T_s$ for the intensity gradient given by the evaluated voxel and its neighborhood computed using a three-dimensional Sobel filter.
\end{itemize} 
After each propagation step, the dome shape is checked to detect possible bifurcations and the presence of leakages. If these are detected, the propagation of the current wavefront ends, thus defining a segment. If a bifurcation is found, two new wavefronts are initialized. 

\nomenclature[g-alpha]{$\alpha$, $\beta$, $\gamma$}{factor}
\nomenclature[a-r]{$r_a$, $r_e$}{radius}

\textsl{Bifurcation Detection:} 
The algorithm exploits the expected wavefront shape to determine when a bifurcation occurs. The following rule is used: if $r_a >\beta\cdot r_e$, then mark that a bifurcation has occurred. The parameter $r_a$ is the actual radius of the dome, $r_e$ the expected radius and $\beta$ a scalar factor (commonly chosen between one and two).

\textsl{Leakage Detection:} 
Owing to the presence of partial volume effects, beam hardening, motion artefacts or low-radiation induced noise, the contrast between the airway lumen and the walls might become insufficient to guide the segmentation. Consequently, the wavefront could leak into the lungs. Two control mechanisms are implemented ~\cite{Artaechevarria2009AirwayMicro-CT}: (a) The number of newly generated wavefronts after bifurcation is restricted to two. As a larger number generally indicates that several small segments are growing next to each other; this is a common indicator of leakage; (b) To be accepted, a fully grown segment needs to comply with three restrictions as measured by the growth rate, the compactness and the differences between wavefront sizes. 

The \textit{Growth Rate, $GR$}, indicator evaluates whether the waveform has propagated uniformly. It is defined as:

\begin{equation}
GR = \frac{1}{N}\sum_{i=1}^{N} \frac{|{W_i}|}{|W_{i-1}|}\quad <\quad T_{GR},
\end{equation}

where \(|W_i|\) is the number of wavefront voxels at propagation step $i$, $N$ the number of propagation steps and $T_{GR}$ a threshold. Commonly, $T_{GR}$ is chosen slightly larger than one. 

The \textit{Discrete Compactness}, $C$~\cite{Bribiesca2008AnShapes}, is computed as:

\begin{equation}
	C = \frac{n-\frac{A}{6}}{n-(\sqrt[3]{n})^{2}} > T_{C} 
\end{equation}

where $n$ is the number of voxels of the solid volume, $A$ is the segment surface area and $T_{C}$ is a threshold defined to separate correct from incorrect segments. The typical range for $T_{C}$ is $[0, 1]$.

Finally, the difference between the sizes of the last $(W_{Last})$ and the first $(W_{First})$ wavefronts is also computed and compared with a threshold $T_{W}$ because a large difference (over $10\%$) is a typical sign of leakage:

\begin{equation}
	|W_{last} - W_{first}| < T_W
\end{equation}


\begin{figure}[h]
\includegraphics[width=\textwidth]{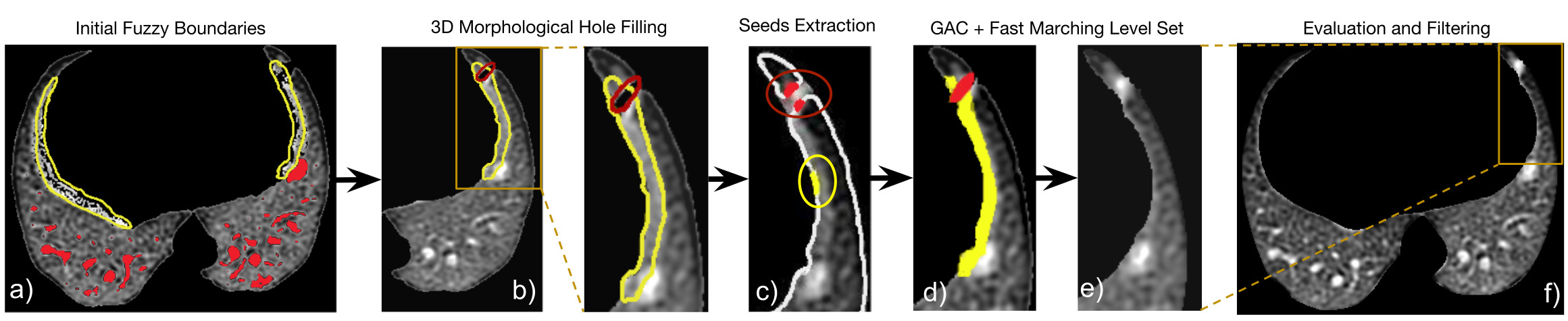}
\captionsetup{justification=justified}
\caption[Lung segmentation pipeline at work]{Lung segmentation evaluation workflow illustrated using a sample sagittal CT slice multiplied by its lung mask: 
(a) Axial slice of the segmented lung obtained after the 
\hyperref[ch2:ssec:lungsAndAirwSegm]{Lung and Airway Segmentation} and \hyperref[ch2:ssec:AirwayTreeExtraction]{Airway Extraction} processes showing holes (black areas inside the parenchyma) and fuzzy boundaries (in yellow); (b) Segmentation after the 3D morphological hole filling process including the holes enclosed by the lung parenchyma; (c) Seeds extracted on the eroded lung surface both in fuzzy boundaries (in yellow) and in TB lesions attached to the pleura (in red); (d) Respiratory motion artefact in the diaphragm area (in yellow) and TB lesion mask (in red) extracted by the combined level set and active contour approach; (e) Final segmentation in which the lesion attached to the pleura has been included and the fuzzy boundaries excluded.}
\label{fig:chapt2/Figure_4}
\end{figure}

\subsubsection{Morphological Closing and Fuzzy Boundaries Evaluation}
\label{ch2:ssec:ClossingAndFuzzyBoundaries}

The last step of the automatic lung segmentation procedure is a refinement process to include missing lesions attached to the pleura and remove the fuzzy boundaries produced by the respiratory motion artefact. 

\textsl{Morphological 3D Hole Filling:} 
Holes, defined as black voxels of the mask that are not connected to the boundaries of the lung segmentation, are removed with an iterative hole-filling filter using the approach described in \textit{Janaszewski et al.} ~\cite{Janaszewski2010HoleObjects} (see \figurename~\ref{fig:chapt2/Figure_4} (b)). At each iteration, a hole neighbourhood ($1mm \ x \ 1mm \ x \ 1mm$) was evaluated to add new voxels to the mask. It is important to remark that the parameters driving the morphological operations are fixed based on the prior knowledge about the subject's anatomy (see \hyperref[ch2:ssec:ExpAnimals]{Experimental Animals}), and its value is kept the same for all CT volumes.\\

\textsl{Fuzzy Lung Border Segmentation and Evaluation:}\label{ch2:sssec:chapt2/fuzzy} 
We specifically propose excluding movement artefacts and including lesions attached to the pleura in our lung segmentation using level sets and active geodesic contours ~\cite{Caselles1997GeodesicContours}, which have proven successful in similar tasks~\cite{Farag2013ASets,Suzuki2010Computer-aidedAlgorithms}. First, the lung surface was extracted from the mask obtained after the morphological 3D hole-filling process: the lung surface was computed as the subtraction of the mask, and an eroded version was computed using a kernel of $1 \ mm$ radius. Then, to obtain the seeds automatically  for the level-sets algorithm, we assumed that the fuzzy regions (lesions or respiratory movement artefacts) had the highest values at the lung boundary (see \figurename~\ref{fig:chapt2/Figure_4} (b)). Therefore, the seeds are chosen to be the outliers (or less probable values) of the intensity distribution at the previously delimited lung boundary (see \figurename~\ref{fig:chapt2/Figure_4} (c)). We set a voxel, $v_{i}$, as the seed based on the following criteria:

\begin{equation}
\centering
\label{eq:seeds}
 v_i \in seeds \iff I(v_i) \geq \mu_{sp} + 2.5\sigma_{sp} \quad \forall v_i \in sp_{border}
\end{equation}
\nomenclature[a-I]{$I$}{Intensity function}
\nomenclature[g-mu]{$\mu$}{population mean}
\nomenclature[g-sd]{$\sigma$}{population standard deviation}
\nomenclature[a-I]{$v$}{voxel}

where $I(\cdot)$ is the voxel intensity, $sp$ represents the segmented lung parenchyma obtained as the output from the morphological hole-filling routine, $sp_{border}$ corresponds to the boundary voxels, and $\mu_{sp}$ and $\sigma_{sp}$ are the mean and standard deviation of the intensities of the voxels within $sp_{border}$, respectively. Assuming a Gaussian distribution of the intensities and setting the number of standard deviations from the mean to $2.5$, we retain $0.65\%$ of the voxels (one-tail, highest values), to capture just a few reliable outliers. 

These seeds were used to create the initial contours for the fast marching level sets. Several seeds could be placed at a given fuzzy boundary, but the level sets will expand, evolving into complex shapes and merging since the intensity gradient is smooth. However, the level sets placed on the fuzzy boundaries do not merge with those placed on the lesion areas, as can be observed in \figurename~\ref{fig:chapt2/Figure_4}  (c), where the intensity gradient was too large. 

Several coarse level sets were obtained as output. These were used as initial contours ($x_{0}$) for the geodesic active contour algorithm~\cite{Caselles1997GeodesicContours}. 
Namely, a contour was fitted to the region ruled by the following partial differential equation (PDE): \nomenclature[z-PDE]{PDE}{Partial Differential Equation}

\begin{equation}
\label{eq:GAC}
\frac{\partial \Psi}{\partial t} = -\alpha \textbf{A}(\textbf{x})\cdot \nabla\Psi - \beta P(\textbf{x})|\nabla\Psi| + \gamma Z(\textbf{x})\kappa |\nabla\Psi|,
\end{equation}

where $\Psi$ is the level set, \textbf{x} is a point of the contour, $\textbf{A}(\textbf{x})$ controls the advection, $P(\textbf{x})$ is the propagation and $Z(\textbf{x})$ is the spatial modification of the mean curvature $\kappa$; $\alpha$,$\beta$ and $\gamma$ are scalars which module each term of the contour evolution. Their value was heuristically set to $\alpha=1.0$, $\beta=0.25$, $\gamma=2.0$. The outputs were refined level-set contours for both the lesions and the fuzzy boundaries. Once the contours were determined, lesions were discriminated from artifacts based on the prior morphological information: contours with a sphericity over $0.85$ were selected as lesions and included within the segmented lung (see \figurename~\ref{fig:chapt2/Figure_4} (d)).
\nomenclature[g-psi]{$\Psi$}{contour level set}
\nomenclature[x-partial]{$\frac{\partial f}{\partial t}$}{Partial derivative of $f$ with respect to time/step}
\nomenclature[x-abs]{$\lvert \cdot \rvert$}{absolute value}
\nomenclature[a-A]{$A$}{advection function}
\nomenclature[a-P]{$P$}{propagation function}
\nomenclature[a-Z]{$Z$}{spatial curvature modification function}
\nomenclature[g-k]{$\kappa$}{mean contour curvature}
\nomenclature[g-g]{$\nabla$}{gradient}

\subsection{Lung Segmentation Evaluation}
\label{ch2:ssec:GrounTruSeg}

The quality of the automatic segmentation for medical imaging applications is commonly estimated with respect to a manually or semi-automatically generated ground truth. The most commonly used evaluation measures are computed as an average of the intersected volumes between both segmentations (i.e., Dice Similarity Coefficient (DSC))~\cite{Mansoor2014ASegmentation,Noor2015PerformanceSegmentation}. For our application, suitable values of the measures could be misleading~\cite{Reinke2021CommonStory} if relatively small volumes at the fuzzy boundaries (i.e., lesions, respiratory motion artefacts) are incorrectly segmented. In those cases, the perceived decrease in quality given by the measure will be minor, but these errors in lung segmentation would generate considerable bias in the subsequent quantification of disease burden.\\

To mitigate this issue in evaluating the goodness of the proposed lung segmentation method, we use the procedure described below to select the slices that most probably have fuzzy boundaries. Rough segmentations of the lungs were semi-automatically computed in $63$ subjects using an in-house platform~\cite{Pascau2006MultimodalityAnalysys} created explicitly for the interactive segmentation of TB-infected lungs. To segment the lungs using the platform, the user specifies at least $1$ seed in the centre of the left lung and right lung. The segmentation then propagates employing a region-growing algorithm. The user can manually specify frontier surfaces to prevent the segmentation from reaching adjacent air-filled regions. The platform has added functionalities to enable manual correction of the results. Once the lungs are interactively segmented, the Hausdorff distances between the automatic lung segmentation obtained before and after the refinement step with respect to the semi-automatic segmentations are computed. The differences in the Hausdorff distances are due to the corrections performed by the refinement routine. The differences point out to those slices in which the segmentation is more uncertain due to the variability introduced by each subject and the disease course. We then choose the $156$ slices with the most considerable differences in the Hausdorff distance to build a surrogate ground truth, as described in detail in Appendix \ref{ap:sec:select}.\\

Three experts interactively segmented the selected slices, paying particular attention to the boundary delimitation. The very accurate segmentations obtained were then combined by consensus to provide a surrogate ground truth~\cite{Warfield2004SimultaneousSegmentation}. Characterization of the agreement, computing the intra-class correlation coefficient (ICC), between the lung segmentation performed by the experts showed excellent consistency (details can be found in Appendix \ref{app:sec:inter-exper-var}). 

The individual expert segmentations and the surrogate ground truth are compared with the proposed method (refined -Ref-) and two other approaches intended for healthy or slightly damaged lung segmentation. Namely, the aforementioned manual segmentation (referred to as semi-auto -Semi-) and the traditional fuzzy connectedness–based lung segmentation (referred to as FC), which has a publicly available open-source software lung segmentation tool (\url{http://www.nitrc.org/projects/nihlungseg/})~\cite{Mansoor2014CIDI-lung-seg:Scans}. For the latter, we used the best performing manual seeding mode, as recommended by the authors, for refining segmented region maps, namely, filling holes with a $0.44$ mm-diameter binary filter and checking fuzzy connectedness. \\

The similarity is measured as both volume overlap and distance between surfaces with the following metrics: Dice similarity coefficient ($DSC$)\nomenclature[z-DSC]{DSC}{Dice Similarity Coefficient}, Hausdorff distance ($HD$)\nomenclature[z-HD]{HD}{Hausdorff Distance}, Hausdorff distance averaged ($HDA$)\nomenclature[z-HDA]{HDA}{Hausdorff Distance Averaged}, false-positive error ($FPE$)\nomenclature[z-FPE]{FPE}{False Positive Error}, false-negative error ($FNE$)\nomenclature[z-FNE]{FNE}{False Negative Error} and volume dissimilarity ($VD$)\nomenclature[z-VD]{VD}{Volume Dissimilarity}. The $HD$ and $HDA$ measures are indicators of a given method’s ability to delineate the tissue boundaries. The $FPE$, $FNE$ and $VD$ indexes provide additional information for the volume overlap measured by the $DSC$. In particular, $FPE$ is related to over-segmentation, $FNE$ to under-segmentation and $VD$, evidently, to volume differences.

To better understand the measures dispersion, box plot charts for each similarity index are also obtained. The dispersion characterization of the similarity indexes is particularly interesting in our case, owing to the complexity of the dataset used. We refer to each comparison between a method and the surrogate ground truth for a given similarity index specifying the method as sub-index (e.g., $DSC_{Ref}$. refers to the median DSC of the comparison between the refined segmentation and the surrogate ground truth).

Finally, we studied the statistical significance of our results to assure the objectivity of our conclusions. For each evaluation metric and each reference segmentation, the outputs of the three segmentation methods were compared using a paired t-test.  A $p$ value below $0.05$ was considered statistically significant.  

\subsection{Results}
\label{ch2:sec:results}

\subsubsection{Qualitative Results}

\begin{figure*}[h!]
\centering
\captionsetup{justification=justified}
\includegraphics[width=\textwidth,height = 0.4\textheight]{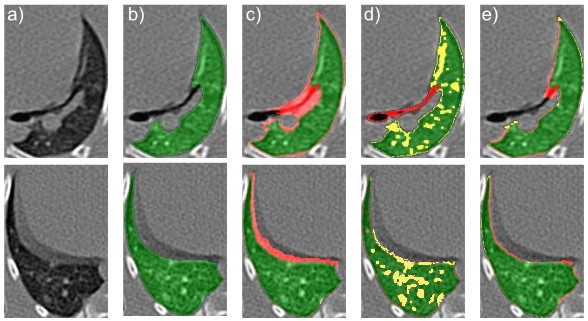}
\caption[Segmentation Examples]{Sample lung segmentations on a representative slice (a) corresponding with the surrogate ground truth (b), the semi-automatic segmentation (c), the fuzzy connectedness segmentation (d), and our proposed method (e). The regions in which there is overlap with the surrogate ground truth are colored in green, the false-positive errors in red and the false-negative errors in yellow.}
\label{fig:chapt2/Figure_7}
\end{figure*}

\figurename~\ref{fig:chapt2/Figure_7} illustrates the computed lung segmentation on a representative slice from those retained (i.e., those in which the segmentation is most uncertain).  The segmentations corresponding to the semi-automatic approach (panel c) are subject to over-segmentation: the delimitation of the lungs goes beyond the lung parenchyma, including respiratory movement artefacts. As per the FC approach (panel d), we observed that several lesions, independently of their localization, were not included in the segmentation due to the method's lack of sensitivity to those areas. The amount of over- and under-segmentation (highlighted in red and yellow, respectively) caused by the proposed method was reduced with respect to the other two approaches.

\subsubsection{Quantitative Results}

\begin{figure}[ht!]
\centering
\includegraphics[width=\textwidth,height=0.6\textheight,trim={0.5cm 0 0.5cm 0},clip]{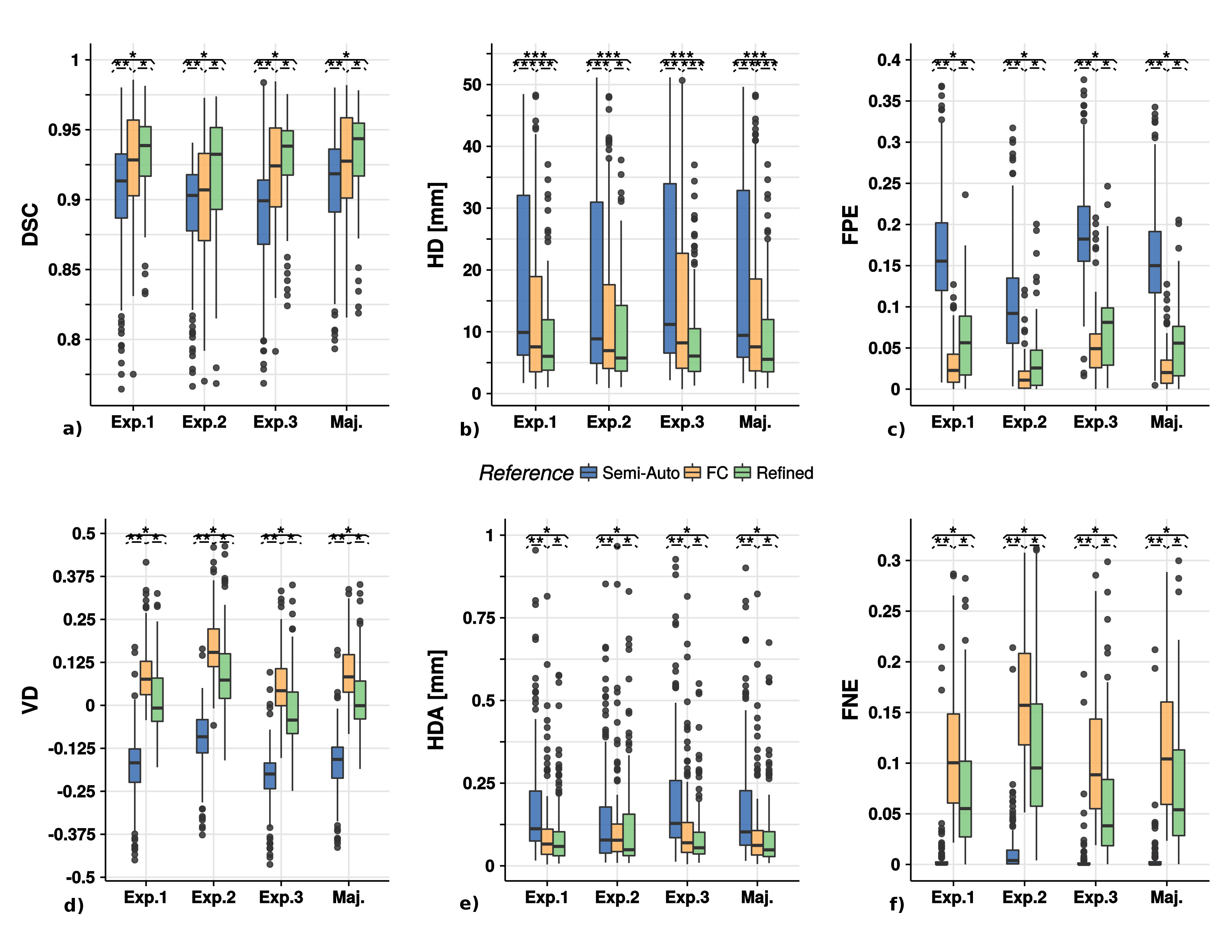}
\captionsetup{justification=justified}
\caption[Boxplot charts for the similarity indexes]{Boxplot charts for the similarity indexes: (a) Dice Similarity Coefficient ($DSC$); (b) Hausdorff Distance ($HD$); (c) False Positive Error ($FPE$); (d) Volume Dissimilarity (VD); (e) Hausdorff Distance Averaged ($HDA$);  (f) False Negative Error ($FNE$). The lung segmentation obtained with the proposed method (refined) is compared with the semi-automatic (semi-auto) and the fuzzy connectedness approaches in the individual expert annotations (Exp. 1, Exp. 2 and Exp. 3) and the surrogate ground truth obtained by the expert consensus as explained in the \ref{ap:sec:select} (\nameref{ap:sec:select}). The asterisks over each group of boxes indicate statistically significant differences between the lung segmentation methods compared: $p < 0.05 \equiv *$, $p < 0.01 \equiv **$ and $p < 0.001 \equiv ***$.} 
\label{fig:chapt2/Figure_5}
\end{figure}

\figurename~\ref{fig:chapt2/Figure_5} shows the box plot charts for each similarity index of the refined (Ref), the semi-automatic (Semi) and the fuzzy connectedness lung segmentation (FC) against the manual annotations performed by each expert (Exp. $\#$) and the consensus surrogate ground truth (Maj.). The numerical results are provided in \tablename~ \ref{ch2:tbl:means_res}. The refined segmentation provides the most similar results with respect to the experts' delimitation and, thus, with respect to the surrogate ground truth. In this sense, the proposed method achieves the largest volume overlap, as reflected by the $DSC$ (mean  $DSC_{Ref}=0.933$; median $DSC_{Ref}=0.943$). The second best-performing method, the FC, which was intended for the segmentation of slightly infected lungs, presents a close mean $DSC$ (mean $DSC_{FC}=0.926$) but more distant median $DSC$ (median $DSC_{FC}=0.922$). Our method achieves much lower distances ($HD$ and $HDA$) with respect to the surfaces of the surrogate ground truth than the others (between $1.2$ and $5.1$ mm with respect to the median ($HD_{Ref} = 5.537$ mm) and between $2.8$ and $11$ mm with respect to the average value ($HD_{Ref} = 8.642$ mm)). The method presents similar rates of under- and over-segmentation, around $6\%$. In contrast, the semi-auto approach achieve a larger over-segmentation rate (median $FPE_{Semi} = 15\%$, mean $FPE_{Semi}$ =16\%) but a much smaller under-segmentation rate (median $FNE_{Semi} = 0.2\%$, mean = 0.6\%) while the FC method provide the opposite results (mean $FPE_{FC} = 2.4\%$, median $FPE_{FC} = 2.2\%$, mean $FNE_{FC} = 11\%$ and median $FPE_{FC} = 10.4\%$). These imbalances make the differences between the volumes obtained by the experts (consensus) and those obtained with the semi-automatic and the fuzzy connectedness methods much higher than those measured for our approach. The volume dissimilarity index for the latter is close to zero in all cases (mean $VD_{Ref} = 0.026$, median $VD_{Ref} = -0.0009$). All the differences as illustrated in \figurename~\ref{fig:chapt2/Figure_5}, are statistically significant except for the $HDA$ index on the Refined and FC segmentations when Expert 2 is used as reference.

\begin{table}
\centering
\resizebox{\textwidth}{!}{%
\begin{tabular}{cccccccc}
\rowcolor[HTML]{9B9B9B} 
\multicolumn{1}{c|}{\cellcolor[HTML]{9B9B9B}Expert} & \multicolumn{1}{c|}{\cellcolor[HTML]{9B9B9B}Comparison} & $\overline{DSC} \pm \sigma_{DSC}$          & $\overline{HD} \pm \sigma_{HD}$              & $\overline{HDA} \pm \sigma_{HDA}$          & $\overline{FPE} \pm \sigma_{FPE}$          & $\overline{FNE} \pm \sigma_{FNE}$          & $\overline{VD} \pm \sigma_{Vol. Dis}$      \\ \hline
& Semi-Auto                                                & 0.904 $\pm$ 0.04 & 18.741 $\pm$  14.78 & 0.206 $\pm$ 0.25 & 0.167 $\pm$ 0.07 & 0.007 $\pm$ 0.03 &  -0.179 $\pm$ 0.10  \\
& FC                                                       & 0.926 $\pm$ 0.04 & 12.768 $\pm$  12.79 & 0.105 $\pm$ 0.14 & 0.028 $\pm$ 0.03 & 0.112 $\pm$ 0.07 & 0.092 $\pm$ 0.08  \\                                              
\multirow{-3}{*}{Exp. 1}                            
& Refined                                                  & 0.931 $\pm$ 0.03 & 8.801 $\pm$  7.37   & 0.093 $\pm$ 0.11  & 0.059 $\pm$ 0.04 & 0.074 $\pm$ 0.06 & 0.017 $\pm$ 0.10   \\ \hline
\rowcolor[HTML]{EFEFEF} 
\cellcolor[HTML]{EFEFEF}                            
& Semi-Auto                                                & 0.891 $\pm$ 0.04 & 17.448 $\pm$  14.66 & 0.159 $\pm$ 0.23 & 0.106 $\pm$ 0.07 & 0.013 $\pm$ 0.03  & -0.101 $\pm$ 0.09 \\
\rowcolor[HTML]{EFEFEF} 
\cellcolor[HTML]{EFEFEF}                            & FC                                                       & 0.901 $\pm$ 0.04 & 12.346 $\pm$  12.24 & 0.111 $\pm$ 0.13  & 0.015 $\pm$ 0.02 & 0.167 $\pm$ 0.07 & 0.170 $\pm$ 0.09   \\
\rowcolor[HTML]{EFEFEF} 
\multirow{-3}{*}{\cellcolor[HTML]{EFEFEF}Exp. 2}    
& Refined                                                  & 0.920 $\pm$ 0.04  & 9.576 $\pm$  7.95   & 0.117 $\pm$ 0.15 & 0.033 $\pm$ 0.04 & 0.118 $\pm$ 0.08  & 0.095 $\pm$ 0.11  \\ \hline                                             
& Semi-Auto                                                & 0.889 $\pm$ 0.04 & 19.835 $\pm$  15.19 & 0.235 $\pm$ 0.27 & 0.193 $\pm$ 0.07 & 0.005 $\pm$ 0.02  & -0.212 $\pm$ 0.09 \\
& FC                                                       & 0.919 $\pm$ 0.04 & 13.993 $\pm$  13.44 & 0.116 $\pm$ 0.15 & 0.052 $\pm$ 0.04 & 0.105 $\pm$ 0.06  & 0.059 $\pm$ 0.09  \\
\multirow{-3}{*}{Exp. 3}                            
& Refined                                                  & 0.931 $\pm$ 0.03 & 8.825 $\pm$  7.43   & 0.089 $\pm$ 0.09 & 0.059 $\pm$ 0.06 & 0.075 $\pm$ 0.05 & -0.016 $\pm$ 0.10  \\ \hline
\rowcolor[HTML]{EFEFEF} 
\cellcolor[HTML]{EFEFEF}                            
& Semi-Auto                                                & 0.909 $\pm$ 0.04 & 18.674 $\pm$  14.81 & 0.199 $\pm$ 0.25 & 0.16 $\pm$ 0.07 & \textbf{0.006 $\pm$ 0.02}   & -0.171 $\pm$ 0.09 \\
\rowcolor[HTML]{EFEFEF} 
\cellcolor[HTML]{EFEFEF}                            
& FC                                                       & 0.926 $\pm$ 0.04 & 12.786 $\pm$  12.85 & 0.103 $\pm$ 0.14 & \textbf{0.024 $\pm$ 0.02} & 0.116 $\pm$ 0.06  & 0.101 $\pm$ 0.08  \\
\rowcolor[HTML]{EFEFEF} 
\multirow{-3}{*}{\cellcolor[HTML]{EFEFEF}Maj.}      
& Refined                                                  & \textbf{0.933 $\pm$ 0.03} & \textbf{8.642 $\pm$  7.36}   & \textbf{0.091 $\pm$ 0.11} & 0.054 $\pm$ 0.04  & 0.077 $\pm$ 0.06 & \textbf{0.026 $\pm$ 0.09}  \\ \hline
\end{tabular}
}

\caption[Overall performance for the proposed model and comparisons]{Overall performance of the refined, the semi-automatic and the fuzzy connectedness (FC) lung segmentation against the manual annotations made by each expert (Exp. 1, Exp. 2 and Exp. 3) and the consensus surrogate ground truth (Maj.). Mean, and standard deviation are provided for each index. For the surrogate ground truth, the best performing method is highlighted in bold for each index. Note: Dice similarity coefficient (DSC), Hausdorff distance (HD), Hausdorff distance averaged (HDA), false-positive error (FPE), false-negative error (FNE) and volume dissimilarity (VD).}
\label{ch2:tbl:means_res}
\end{table}

\begin{figure}[h!]
\centering
\captionsetup{justification=justified}
\includegraphics[width=\textwidth]{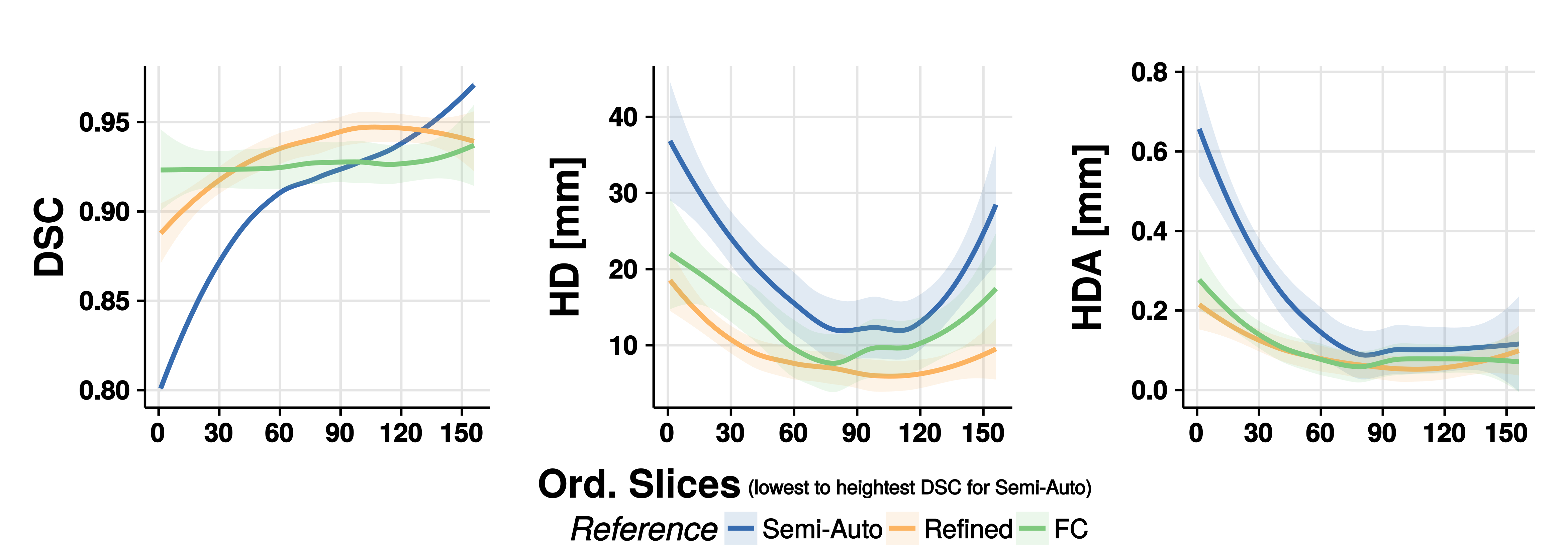}
\caption[Change of segmentation metrics per slice ]{Dice similarity coefficient (DSC), Hausdorff distance (HD), and Hausdorff distance averaged (HDA) plots along the slices sorted in ascending order based on the DSC of the semi-automatic segmentation with respect to the surrogate ground truth.  Data have been filtered with the locally weighted scatterplot smoothing (LOESS) model. The $95\%$ confidence interval is drawn as a shadow of the same colour as the corresponding line.}
\label{fig:chapt2/Figure_6}
\end{figure}

\figurename~\ref{fig:chapt2/Figure_6} displays $DSC$, $HD$ and $HDA$ plots over the slices arranged in ascending order as given by the $DSC$ of the semi-automatic segmentation with respect to the surrogate ground truth. The data have been filtered following the locally weighted scatterplot smoothing (LOESS)~\cite{Cleveland1988} model in order to achieve a better appreciation of the patterns and the differences between the approaches. The $DSC$ plot shows that the gap between the proposed and the semi-automatic method (about $10\%$ for the first slice) decreases as we move towards higher $DSC$ slice values, while the difference with the $FC$ method remains relatively stable. The $HD$ index corresponding to the proposed method is smaller than the other methods for all the slices. The improvement is $5-10$ mm with respect to the Semi-Automatic approach and $0.5-7.5$ mm with respect to the FC approach. Finally, the $HDA$ index exhibits an exponential decay for all the methods.

\subsection{Discussion}
\label{sec:discussion}
The experiments performed reveal substantial improvements when an input volume is processed through our pipeline. As expected from a method focused on improving boundary detection, the Hausdorff distance is significantly smaller than other methods while presenting reasonably good results for the volume overlap measures. This behaviour is explained by the ability to reject fuzzy boundary artefacts while retaining most of the damaged tissue (especially the lesions attached to the pleura). Since the Hausdorff distance computes the maximum among the minimal distances for all points in the two surfaces compared, small changes when delimiting a complex shape (such as those generated by the diseased lung) result in large Hausdorff distance values. Fortunately, the boundaries created by our method are consistent and stable, and inaccuracies in the boundary delimitation are less frequent. Moreover, improved delimitation enables the target volume to be filled more accurately, as reflected in the DSC values. 

For the macaque model context, where the lung segmentation is a preparatory step for quantifying the TB lesions burden during the disease, these small differences are vital. Especially, for relapse models in which small lesions are regularly the most often ones. High-quality segmentation is critical in the early stages. The sensitivity given by the radiological images is crucial when assessing latent tuberculosis due to the small parenchymal damage associated with this stage of the disease. Therefore, the fact that the advanced method achieves the lowest Hausdorff distance measured by far in almost all the slices (\figurename~\ref{fig:chapt2/Figure_6}) is a major step towards the proper quantification of disease burden, even with the current dispersion of the measure. This dispersion is mostly due to the intrinsic noise inherent in the delineation of complex slices. Thus, it is likely to appear in any segmentation method, including manual delineations~\cite{Udupa2006AAlgorithms}. In the Section \ref{app:sec:inter-exper-var}, the inter-agreement differences between the experts' delimitation are presented. They show a good intra-class correlation coefficient (ICC) \nomenclature[z-ICC]{ICC}{Intra-class Correlation Coefficient} for the overall surface delimitation ($HD = 0.88$,  $HDA = 0.85$) and lower values for the volume indicators of performance ($DSC = 0.74$, $FPE = 0.71$ and $FNE = 0.6$). The fact that small variations in delineation produce large dissimilarity values is even more obvious for the Hausdorff distance averaged. Although, as observed in \figurename~\ref{fig:chapt2/Figure_5}, the values of this measure are much smaller than the Hausdorff distance. Many outliers are present due to the relatively large distance between the surfaces corresponding to pairs of compared segmentations at several 
slices within the data set. 

The more conservative segmentations are those provided by the fuzzy connectedness–based method and our proposal. They perform better in terms of $HD$ and $HDA$ (\figurename~\ref{fig:chapt2/Figure_7}). Hence, the segmentations they provided are more suitable for subsequent quantification of the TB lesion burden.  

It is important to emphasize that our method achieves a good balance between false positive and false negative errors, in contrast to the semi-auto segmentation results, which show, on average, $15\%$ over-segmentation. The lung segmentation includes fuzzy regions, which will contaminate the subsequent analysis. In contrast, the FC segmentation is excessively conservative. It presents a tiny percentage of over-segmentation and $15\%$ false-negative errors on average for the most uncertain slices in the dataset. Thus, it potentially generates a misleading evaluation of TB infection. Although the advanced method balances out possible errors, it still exhibits $5\%$ false negative errors on average, which could still influence the quantification of disease burden, although less severely than the FC method. 

The information from the error types makes it possible to explain the volume dissimilarities shown in \figurename~\ref{fig:chapt2/Figure_5}. The semi-auto method presents the previously mentioned problems of over-segmentation, which account for the almost parabolic shape of the HD when the DSC increases in \figurename~\ref{fig:chapt2/Figure_6}. The method presents a limit (at the parabola vertex), from where the segmentation is unable to fill the region of interest without growing beyond. Thus, the method presents a few slices with better overlap (DSC) than the proposed approach at the expense of losing sensitivity at the boundaries. Consequently, the HD remains flat, between the 90th and the 120th slice, only to increase dramatically afterwards, while a suitable segmentation should decrease or, at least, keep a constant low distance. The HD plot for the FC method in \figurename~\ref{fig:chapt2/Figure_6} presents similar behaviour to the semi-auto method, albeit for different reasons. As illustrated with the examples in \figurename~\ref{fig:chapt2/Figure_7}, the FC method misses an important part of the volume-of-interest, resulting in considerable volume dissimilarity (see \figurename~\ref{fig:chapt2/Figure_5}). Although the DSC trend in \figurename~\ref{fig:chapt2/Figure_6} is flatter than the one corresponding to the semi-auto method, it also presents a parabola vertex, which indicates an inability to capture the intricate shape of the selected surrogate ground truth. In contrast, the refined method shows a negligible value of volume dissimilarity (see \figurename~\ref{fig:chapt2/Figure_5}) and a much less marked parabola shape (see \figurename~\ref{fig:chapt2/Figure_6}). To further improve the accuracy of the lung segmentation, it could be much more appropriate to use novel indicators of segmentation performance more closely associated with the ulterior quantification than the overlap and surface indicators. They are clearly of limited validity owing to the variability of human criteria during the segmentation process~\cite{Reinke2021CommonStory}. To this aim, we have introduced a quantification method, presented in the subsequent section \nameref{ch2:sec:quantication}, which makes use of the proposed pipeline for lung segmentation and that presents satisfactory results~\cite{Gordaliza2019ComputedTuberculosis}\\

The framework allows re-parametrization to other models (e.g., mice, humans) by fine-tuning of the parameters as shown in \tablename~\ref{tbl:chapt2/parameters}. As was mentioned at the beginning of the chapter, this hyper-tuning process would not probably achieve the best possible results given the SOTA DL methods available (see \nameref{ch2:sec:OutOfComf}). Nevertheless, the proper understanding of each parameter function provides instrumental knowledge for prospective segmentation models. Thus, to improve the results and extend the framework to the segmentation of extremely damaged lungs, within this thesis (see chapters \nameref{ch4:Intro} and \nameref{ch5:Intro}), AI/DL techniques are examined and developed. These implementations shown promising results for segmentation, besides further developments are introduced to cope with common limitations for DL models like the loss of resolution~\cite{Harrison2017ProgressiveImagesb}. These limitations impede the proper identification of boundaries, biases (Section \ref{ch1:shifts}), and results in the need for a large refined ground truth~\cite{Roy2018QuickNAT:Seconds}, which results quite challenging to obtain. To get the best labels (segmentations) to train the DL models introduced in the thesis, the unsupervised segmentation volumes obtained with the tool presented in this chapter are reviewed and corrected when necessary by experts. Sustantially minimizing in this way the time invested by them to create a good Ground Truth, especially in comparison with the use of the tools mentioned above (see Section \ref{ch2:ssec:GrounTruSeg}). 


\begin{table}
\caption[PLS Rule-based method parameters]{Pipeline process (first column), algorithm (second column), parameters (third column) and their values (fourth column).}
\label{tbl:chapt2/parameters}
\centering
\resizebox{\textwidth}{!}{
\arrayrulecolor[rgb]{1,0.361,0.361}
\begin{tabular}{cccc}
\rowcolor[rgb]{0.2,0.2,0.2} \multicolumn{2}{c}{\textbf{\textcolor{white}{Section}}}                                                                                                                                                                                                                                                      & \textbf{\textcolor{white}{Parameter}}                & \textbf{\textcolor{white}{Value}}                \\
{\cellcolor[rgb]{0,0,0.502}}                                                                                                                                                                                                                      & \textcolor[rgb]{0.2,0.2,0.6}{Adaptive Thresholding }                                 & \textcolor[rgb]{0.2,0.2,0.6}{Otsu Threshold }        & \textcolor[rgb]{0.2,0.2,0.6}{Auto. }             \\ 
\hhline{>{\arrayrulecolor[rgb]{0,0,0.502}}->{\arrayrulecolor[rgb]{0.2,0.2,0.6}}---}
{\cellcolor[rgb]{0,0,0.502}}                                                                                                                                                                                                                      & \textcolor[rgb]{0.2,0.2,0.6}{Rib Cage Extraction }                                   & \textcolor[rgb]{0.2,0.2,0.6}{Seeds }                 & \textcolor[rgb]{0.2,0.2,0.6}{$>900$ HU }         \\ 
\hhline{>{\arrayrulecolor[rgb]{0,0,0.502}}->{\arrayrulecolor[rgb]{0.2,0.2,0.6}}---}
\multirow{-3}{*}{{\cellcolor[rgb]{0,0,0.502}}\begin{tabular}[c]{@{}>{\cellcolor[rgb]{0,0,0.502}}c@{}}\textcolor{white}{Preliminary }\\\textcolor{white}{Lung }\\\textcolor{white}{Segmentation}\end{tabular}}                                     & \textcolor[rgb]{0.2,0.2,0.6}{Connectivity and Topological Analysis }                 & \textcolor[rgb]{0.2,0.2,0.6}{Min. Object size }      & \textcolor[rgb]{0.2,0.2,0.6}{$10\ mm^3$}         \\ 
\hhline{>{\arrayrulecolor[rgb]{1,0.361,0.361}}->{\arrayrulecolor[rgb]{0.2,0.2,0.6}}---}
{\cellcolor[rgb]{1,0.361,0.361}}                                                                                                                                                                                                                  & \multirow{2}{*}{\textcolor[rgb]{1,0.361,0.361}{Trachea detection }}                  & \textcolor[rgb]{1,0.361,0.361}{Expected Perimeter }  & \textcolor[rgb]{1,0.361,0.361}{$5.5 - 8.5\ mm$}  \\
{\cellcolor[rgb]{1,0.361,0.361}}                                                                                                                                                                                                                  &                                                                                      & \textcolor[rgb]{1,0.361,0.361}{Roundness }           & \textcolor[rgb]{1,0.361,0.361}{$>0.9$}           \\ 
\hhline{>{\arrayrulecolor[rgb]{1,0.361,0.361}}----}
{\cellcolor[rgb]{1,0.361,0.361}}                                                                                                                                                                                                                  & \multirow{4}{*}{\textcolor[rgb]{1,0.361,0.361}{Wavefront Propagation }}              & \textcolor[rgb]{1,0.361,0.361}{$Time\ Step$}         & \textcolor[rgb]{1,0.361,0.361}{$0.8$}            \\
{\cellcolor[rgb]{1,0.361,0.361}}                                                                                                                                                                                                                  &                                                                                      & \textcolor[rgb]{1,0.361,0.361}{$T_i$}                & \textcolor[rgb]{1,0.361,0.361}{$-625$ HU }       \\
{\cellcolor[rgb]{1,0.361,0.361}}                                                                                                                                                                                                                  &                                                                                      & \textcolor[rgb]{1,0.361,0.361}{$T_s$}                & \textcolor[rgb]{1,0.361,0.361}{$2.5$}            \\
{\cellcolor[rgb]{1,0.361,0.361}}                                                                                                                                                                                                                  &                                                                                      & \textcolor[rgb]{1,0.361,0.361}{$\alpha$}             & \textcolor[rgb]{1,0.361,0.361}{$1.4$}            \\ 
\hhline{----}
{\cellcolor[rgb]{1,0.361,0.361}}                                                                                                                                                                                                                  & \textcolor[rgb]{1,0.361,0.361}{Bifurcation Detection }                               & \textcolor[rgb]{1,0.361,0.361}{$\beta$}              & \textcolor[rgb]{1,0.361,0.361}{$2$}              \\ 
\hhline{----}
{\cellcolor[rgb]{1,0.361,0.361}}                                                                                                                                                                                                                  & \multirow{3}{*}{\textcolor[rgb]{1,0.361,0.361}{Leakage Detection }}                  & \textcolor[rgb]{1,0.361,0.361}{$T_{GR}$}             & \textcolor[rgb]{1,0.361,0.361}{$1$}              \\
{\cellcolor[rgb]{1,0.361,0.361}}                                                                                                                                                                                                                  &                                                                                      & \textcolor[rgb]{1,0.361,0.361}{$T_C$}                & \textcolor[rgb]{1,0.361,0.361}{$0.72$}           \\
\multirow{-10}{*}{{\cellcolor[rgb]{1,0.361,0.361}}\begin{tabular}[c]{@{}>{\cellcolor[rgb]{1,0.361,0.361}}c@{}}\textcolor{white}{Airway }\\\textcolor{white}{Tree }\\\textcolor{white}{Segmentation}\end{tabular}}                                 &                                                                                      & \textcolor[rgb]{1,0.361,0.361}{$T_W$}                & \textcolor[rgb]{1,0.361,0.361}{$10\%$}           \\ 
\hhline{>{\arrayrulecolor[rgb]{0.694,0.392,0.694}}->{\arrayrulecolor[rgb]{1,0.361,0.361}}---}
{\cellcolor[rgb]{0.694,0.392,0.694}}                                                                                                                                                                                                              & \textcolor[rgb]{0.694,0.392,0.694}{Morphological 3D Hole Filling }                   & \textcolor[rgb]{0.694,0.392,0.694}{$Kernel\ Radius$} & \textcolor[rgb]{0.694,0.392,0.694}{$1\ mm$}      \\ 
\hhline{>{\arrayrulecolor[rgb]{0.694,0.392,0.694}}----}
{\cellcolor[rgb]{0.694,0.392,0.694}}                                                                                                                                                                                                              & \multirow{4}{*}{\textcolor[rgb]{0.694,0.392,0.694}{Fuzzy Lung Border Segmentation }} & \textcolor[rgb]{0.694,0.392,0.694}{$\alpha$}         & \textcolor[rgb]{0.694,0.392,0.694}{$1.0$}        \\
{\cellcolor[rgb]{0.694,0.392,0.694}}                                                                                                                                                                                                              &                                                                                      & \textcolor[rgb]{0.694,0.392,0.694}{$\beta$}          & \textcolor[rgb]{0.694,0.392,0.694}{$0.25$}       \\
{\cellcolor[rgb]{0.694,0.392,0.694}}                                                                                                                                                                                                              &                                                                                      & \textcolor[rgb]{0.694,0.392,0.694}{$\gamma$}         & \textcolor[rgb]{0.694,0.392,0.694}{$2$}          \\
\multirow{-5}{*}{{\cellcolor[rgb]{0.694,0.392,0.694}}\begin{tabular}[c]{@{}>{\cellcolor[rgb]{0.694,0.392,0.694}}c@{}}\textcolor{white}{Closing }\\\textcolor{white}{and}\\\textcolor{white}{~Fuzzy}\\\textcolor{white}{~Boundaries}\end{tabular}} &                                                                                      & \textcolor[rgb]{0.694,0.392,0.694}{sphericity }      & \textcolor[rgb]{0.694,0.392,0.694}{$>0.85$}      \\
\hhline{----} 
\end{tabular}
\arrayrulecolor{black} }
\end{table}

\section{Quantifying trough correlation in a closed environment}\label{ch2:sec:quantication}

As mentioned several times before, automatic segmentation is a big step~\cite{Galban2012ComputedProgression}. However, the final goal is the implementation of techniques for the characterization of TB as a continuous spectrum employing radiological images given its sensitivity for findings TB manifestations~\cite{Nachiappan2017PulmonaryManagement,Pai2016Tuberculosis} (see \figurename~\ref{ch1:fig:Radiology_test}). Therefore, identifying biomarkers to ease the radiologists' evaluation of TB in extensive studies needs to be automatized.
However, the few methods dealing with TB damaged lungs do not contemplate quantification or are not automatic~\cite{Chen2014PET/CTTuberculosis,Via2012InfectionTomography,Via2013DifferentialJacchus, Wallis2013TuberculosisChallenges}.

To ease this task, within this section, our first approximation of a complete methodology to automatically extract biomarkers from CT images is presented (see Section \ref{ch1:ssec:LungSegSOTA}). Although with limitations, it is able to estimate the evolution of TB burden and could be used to assess the response to treatment of infected subjects when there is a causal relationship between the damaged lung tissue and TB burden~\cite{Chen2014PET/CTTuberculosis}. This scenario could be customary in several animal models, as was already pointed out in the our published work \textit{Computed Tomography-Based Biomarker for Longitudinal Assessment of Disease Burden in Pulmonary Tuberculosis} \cite{Gordaliza2019ComputedTuberculosis}. Most of the following sections were extracted from that manuscript and which \texttt{R} and \texttt{Python} based code can be found in \href{https://github.com/BIIG-UC3M/TLS-Piped}{https://github.com/BIIG-UC3M/TLS-Piped}.   

\subsection{Materials}
\label{sec:MandMs}

\subsubsection{Computer Tomography Images}
\label{ssec:materials}
The CT scans are already described in \nameref{ch2:ssec:CTScans}. It is important to remember that the macaques were treated with a different antibiotic cocktail of Isoniazid (H), Rifampicin (R), and Pyrazinamide (Z)~\cite{Sharpe2016UltraMacaques} in four phases, as is shown in the \tablename~\ref{ch2:tbl:treatments}.\\
All animal procedures and study designs were approved by the Public Health England Animal Welfare and Ethical Review Body, Porton Down, UK, and authorized under an appropriate UK Home Office project license.

\begin{table}
\centering
\arrayrulecolor[rgb]{0.843,0.843,0.843}
\begin{tabular}{c||c||c||c||c||c||c||c||c||c||c||c||c}
\rowcolor[rgb]{0.6,0.6,0.6} {\cellcolor[rgb]{0.6,0.6,0.6}}                           & \multicolumn{4}{c||}{Run-in ~}                                                                                                                                                                          & \multicolumn{2}{c||}{Phase 1}                                                                            & \multicolumn{2}{c||}{Phase 2}                                                     & \multicolumn{2}{c||}{Phase 3} & \multicolumn{2}{c}{Phase 4}                                                                                                                                          \\ 
\hhline{>{\arrayrulecolor[rgb]{0.6,0.6,0.6}\doublerulesepcolor[rgb]{0.6,0.6,0.6}}=>{\arrayrulecolor[rgb]{0.843,0.843,0.843}\doublerulesepcolor{white}}|:=:t:=:t:=:t:=::=:t:=::=:t:=::=:t:=::=:t:=}
\rowcolor[rgb]{0.804,0.804,0.804} \multirow{-2}{*}{{\cellcolor[rgb]{0.6,0.6,0.6}}ID} & 0$^{CT}$ & 3$^{CT}$ & ... & 12$^{CT}$                                                                                                                                                                                        & 14 & 16$^{CT}$                                                                                                  & 18 & 20$^{CT}$                                                                           & 22 & 24$^{CT}$                       & 26 & 28$^{CT}$                                                                                                                                                              \\ 
\hhline{=::=:b:=:b:=:b:=::=:b:=::=:b:=::=:b:=:b:=:b:=}
\rowcolor[rgb]{0.322,0.431,1} {\cellcolor[rgb]{0.804,0.804,0.804}}1                  & \multicolumn{4}{c||}{{\cellcolor[rgb]{0.322,0.431,1}}}                                                                                                                                                  & \multicolumn{2}{c||}{{\cellcolor[rgb]{0.361,0.678,0.361}}\textcolor{white}{HRZ}}                        & \multicolumn{2}{c||}{\textcolor{white}{No Treatment}}                            & \multicolumn{4}{c}{{\cellcolor[rgb]{0.89,0.051,0.051}}}\\ 
\hhline{=:|>{\arrayrulecolor[rgb]{0.322,0.431,1}\doublerulesepcolor[rgb]{0.322,0.431,1}}====>{\arrayrulecolor[rgb]{0.843,0.843,0.843}\doublerulesepcolor{white}}|:==::==:|>{\arrayrulecolor[rgb]{0.89,0.051,0.051}\doublerulesepcolor[rgb]{0.89,0.051,0.051}}====}\doublerulesepcolor{white}
\rowcolor[rgb]{0.322,0.431,1} {\cellcolor[rgb]{0.804,0.804,0.804}}2                  & \multicolumn{4}{c||}{{\cellcolor[rgb]{0.322,0.431,1}}}                                                                                                                                                  & \multicolumn{2}{c||}{{\cellcolor[rgb]{0.322,0.431,1}}}                                                   & \multicolumn{2}{c||}{{\cellcolor[rgb]{0.902,0.584,0}}\textcolor{white}{HR }}      & \multicolumn{4}{c}{\multirow{-2}{*}{{\cellcolor[rgb]{0.89,0.051,0.051}}\textcolor{white}{HZ }}}                                                                                                      \\ 
\hhline{>{\arrayrulecolor[rgb]{0.843,0.843,0.843}}=:|>{\arrayrulecolor[rgb]{0.322,0.431,1}\doublerulesepcolor[rgb]{0.322,0.431,1}}====>{\arrayrulecolor[rgb]{0.843,0.843,0.843}\doublerulesepcolor{white}}||>{\arrayrulecolor[rgb]{0.322,0.431,1}\doublerulesepcolor[rgb]{0.322,0.431,1}}==>{\arrayrulecolor[rgb]{0.843,0.843,0.843}\doublerulesepcolor{white}}|:==::====}
\rowcolor[rgb]{0.322,0.431,1} {\cellcolor[rgb]{0.804,0.804,0.804}}3                  & \multicolumn{4}{c||}{{\cellcolor[rgb]{0.322,0.431,1}}}                                                                                                                                                  & \multicolumn{2}{c||}{\multirow{-2}{*}{{\cellcolor[rgb]{0.322,0.431,1}}\textcolor{white}{No~Treatment }}} & \multicolumn{2}{c||}{{\cellcolor[rgb]{0.361,0.678,0.361}}\textcolor{white}{HRZ }} & \multicolumn{4}{c}{{\cellcolor[rgb]{0.902,0.584,0}}}                                                                                                                                                 \\ 
\hhline{=:|>{\arrayrulecolor[rgb]{0.322,0.431,1}\doublerulesepcolor[rgb]{0.322,0.431,1}}====>{\arrayrulecolor[rgb]{0.843,0.843,0.843}\doublerulesepcolor{white}}|:==::==:|>{\arrayrulecolor[rgb]{0.902,0.584,0}\doublerulesepcolor[rgb]{0.902,0.584,0}}====}\doublerulesepcolor{white}
\rowcolor[rgb]{0.322,0.431,1} {\cellcolor[rgb]{0.804,0.804,0.804}}4                  & \multicolumn{4}{c||}{{\cellcolor[rgb]{0.322,0.431,1}}}                                                                                                                                                  & \multicolumn{2}{c||}{{\cellcolor[rgb]{0.89,0.051,0.051}}\textcolor{white}{HZ }}                          & \multicolumn{2}{c||}{\textcolor{white}{No Treatment }}                            & \multicolumn{4}{c}{\multirow{-2}{*}{{\cellcolor[rgb]{0.902,0.584,0}}\textcolor{white}{HR }}}                                                                                                         \\ 
\hhline{>{\arrayrulecolor[rgb]{0.843,0.843,0.843}}=:|>{\arrayrulecolor[rgb]{0.322,0.431,1}\doublerulesepcolor[rgb]{0.322,0.431,1}}====>{\arrayrulecolor[rgb]{0.843,0.843,0.843}\doublerulesepcolor{white}}|:==::==::====}
\rowcolor[rgb]{0.322,0.431,1} {\cellcolor[rgb]{0.804,0.804,0.804}}5                  & \multicolumn{4}{c||}{{\cellcolor[rgb]{0.322,0.431,1}}}                                                                                                                                                  & \multicolumn{2}{c||}{\textcolor{white}{No~Treatment }}                                                   & \multicolumn{2}{c||}{{\cellcolor[rgb]{0.89,0.051,0.051}}\textcolor{white}{HZ }}   & \multicolumn{4}{c}{{\cellcolor[rgb]{0.361,0.678,0.361}}}                                                                                                                                             \\ 
\hhline{=:|>{\arrayrulecolor[rgb]{0.322,0.431,1}\doublerulesepcolor[rgb]{0.322,0.431,1}}====>{\arrayrulecolor[rgb]{0.843,0.843,0.843}\doublerulesepcolor{white}}|:==::==:|>{\arrayrulecolor[rgb]{0.361,0.678,0.361}\doublerulesepcolor[rgb]{0.361,0.678,0.361}}====}\doublerulesepcolor{white}
\rowcolor[rgb]{0.322,0.431,1} {\cellcolor[rgb]{0.804,0.804,0.804}}6                  & \multicolumn{4}{c||}{{\cellcolor[rgb]{0.322,0.431,1}}}                                                                                                                                                  & \multicolumn{2}{c||}{{\cellcolor[rgb]{0.902,0.584,0}}\textcolor{white}{HR }}                             & \multicolumn{2}{c||}{\textcolor{white}{No Treatment }}                            & \multicolumn{4}{c}{\multirow{-2}{*}{{\cellcolor[rgb]{0.361,0.678,0.361}}\textcolor{white}{HRZ }}}                                                                                                    \\ 
\hhline{>{\arrayrulecolor[rgb]{0.843,0.843,0.843}}=:|>{\arrayrulecolor[rgb]{0.322,0.431,1}\doublerulesepcolor[rgb]{0.322,0.431,1}}====>{\arrayrulecolor[rgb]{0.843,0.843,0.843}\doublerulesepcolor{white}}|:==::==::====}
\rowcolor[rgb]{0.322,0.431,1} {\cellcolor[rgb]{0.804,0.804,0.804}}7                  & \multicolumn{4}{c||}{{\cellcolor[rgb]{0.322,0.431,1}}}                                                                                                                                                  & \multicolumn{2}{c||}{{\cellcolor[rgb]{0.89,0.051,0.051}}\textcolor{white}{HZ }}                          & \multicolumn{2}{c||}{{\cellcolor[rgb]{0.902,0.584,0}}\textcolor{white}{HR }}      & \multicolumn{4}{c}{{\cellcolor[rgb]{0.322,0.431,1}}}                                                                                                                                                 \\ 
\hhline{=:|>{\arrayrulecolor[rgb]{0.322,0.431,1}\doublerulesepcolor[rgb]{0.322,0.431,1}}====>{\arrayrulecolor[rgb]{0.843,0.843,0.843}\doublerulesepcolor{white}}|:==::==:|>{\arrayrulecolor[rgb]{0.322,0.431,1}\doublerulesepcolor[rgb]{0.322,0.431,1}}====}\doublerulesepcolor{white}
\rowcolor[rgb]{0.322,0.431,1} {\cellcolor[rgb]{0.804,0.804,0.804}}8                  & \multicolumn{4}{c||}{{\cellcolor[rgb]{0.322,0.431,1}}}                                                                                                                                                  & \multicolumn{2}{c||}{{\cellcolor[rgb]{0.361,0.678,0.361}}\textcolor{white}{HRZ }}                        & \multicolumn{2}{c||}{{\cellcolor[rgb]{0.89,0.051,0.051}}\textcolor{white}{HZ }}   & \multicolumn{4}{c}{{\cellcolor[rgb]{0.322,0.431,1}}}                                                                                                                                                 \\ 
\hhline{>{\arrayrulecolor[rgb]{0.843,0.843,0.843}}=:|>{\arrayrulecolor[rgb]{0.322,0.431,1}\doublerulesepcolor[rgb]{0.322,0.431,1}}====>{\arrayrulecolor[rgb]{0.843,0.843,0.843}\doublerulesepcolor{white}}|:==::==:|>{\arrayrulecolor[rgb]{0.322,0.431,1}\doublerulesepcolor[rgb]{0.322,0.431,1}}====}\doublerulesepcolor{white}
\rowcolor[rgb]{0.322,0.431,1} {\cellcolor[rgb]{0.804,0.804,0.804}}9                  & \multicolumn{4}{c||}{\multirow{-9}{*}{{\cellcolor[rgb]{0.322,0.431,1}}\begin{tabular}[c]{@{}>{\cellcolor[rgb]{0.322,0.431,1}}c@{}}\textcolor{white}{ No~}\\\textcolor{white}{Treatment }\end{tabular}}} & \multicolumn{2}{c||}{{\cellcolor[rgb]{0.902,0.584,0}}\textcolor{white}{HR}}                             & \multicolumn{2}{c||}{{\cellcolor[rgb]{0.361,0.678,0.361}}\textcolor{white}{HRZ}} & \multicolumn{4}{c}{\multirow{-3}{*}{{\cellcolor[rgb]{0.322,0.431,1}}\begin{tabular}[c]{@{}>{\cellcolor[rgb]{0.322,0.431,1}}c@{}}\textcolor{white}{No}\\\textcolor{white}{Treatment}\end{tabular}}} 
\end{tabular}
\arrayrulecolor{black}
\caption[Antibiotic cocktail per week and subject]{Antibiotic cocktail per week and subject. Each color represents the treatment  (\textcolor{NavyBlue}{No Treatment}, \textcolor{BurntOrange}{HR ($Isoniazid + Rifampicin$)}, \textcolor{YellowGreen}{HRZ ($Isoniazid + Rifampicin + Pyrazinamide$)}, \textcolor{red}{ HZ ($Isoniazid + Pyrazinamide$)}) taken by a subject during each treatment phase. Weeks with the CT superindex indicate the acquisition of a computed tomography volume at that week.}
\label{ch2:tbl:treatments}
\end{table}

\subsection{Lungs Segmentation}
\label{sssec:lung_segment}
\begin{figure*}[t!]
\includegraphics[width=\textwidth]{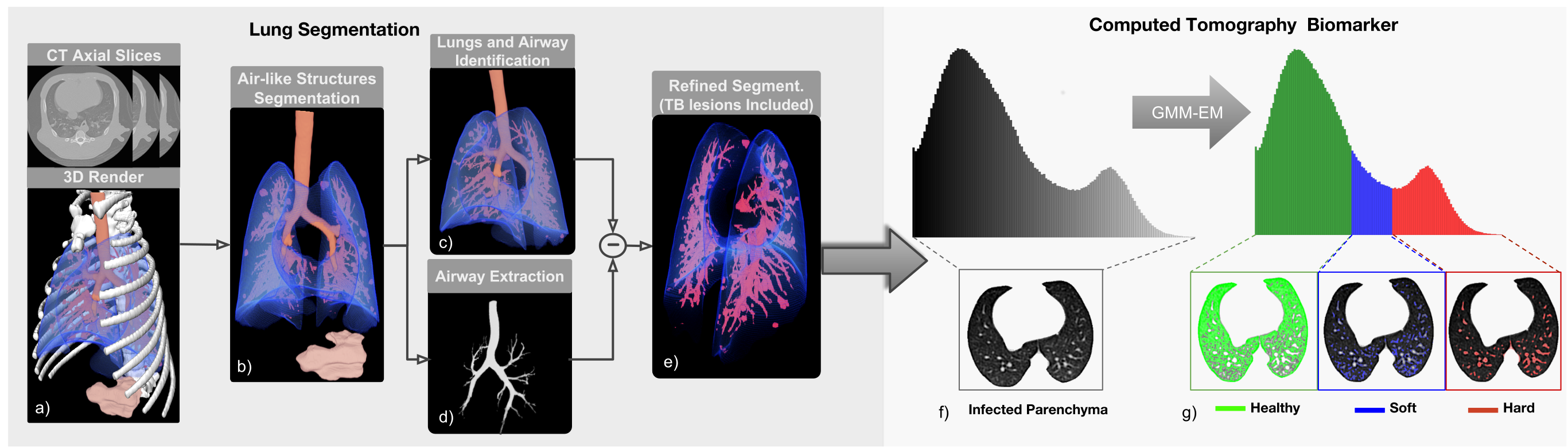}
\caption[Lung damage quantification]{\textit{Lung Segmentation:} From a chest CT scan (a) all aerated regions (b) are identified using an specific adapted algorithm~\cite{Hu2001AutomaticImages}. Successively, lungs are isolated, including the airways, by exploiting the topological information (c). The intricate structure formed by the airways (d) is extracted and eliminated from the lungs. Finally, to include the non-segmented damaged tissue, the lung segmentation is refined employing a hole-filling method based on active contours ~\cite{Caselles1997GeodesicContours}(e). \textit{CT Biomarkers:} The segmented lungs are separated into three TB-associated volumes as proposed by \textit{Chen et al.}~\cite{Chen2014PET/CTTuberculosis}. Each region comprises a grey-level range in the segmented lungs histogram (f) representing: Healthy Tissue: Parenchyma free of infection, Soft Tissue: Forming or healing lesions, Hard Tissue: Abnormal lung parenchyma (g).}
\label{fig:methods}
\end{figure*}

The entire procedure is detailed in Section \ref{ch2:sec:Methods}, therefore, if the reader is familiar with it can skip to Section \ref{ch2:ssec:biomExtrac} at this point. For those readers primarily interested in quantification, the previous segmentation step is illustrated in the left part of \figurename~\ref{fig:methods} and summarized in the following paragraph.\\

Initially, air-like organs (e.g., healthy lungs, airways tree, stomach) presented in the chest CT scans (\figurename~\ref{fig:methods}.a) are identified employing an adaptive thresholding method (\figurename~\ref{fig:methods}.b) to subsequently isolate the object formed by the lungs and airways studying the topology and connectivity of the organs (\figurename~\ref{fig:methods}.c). Next, the intricate airways tree structure is computed employing a region growing algorithm which propagates simulating a spherical wavefront ruled by active contours~\cite{Ceresa2010AutomaticImages} (\figurename~\ref{fig:methods}.d) and is removed from the segmented lungs. Finally, unsegmented pulmonary regions, corresponding to damaged parenchyma and TB lesion are included by a morphological hole filling process~\cite{Janaszewski2010HoleObjects} which is refined using \textit{Geodesic Active Contours}~\cite{Caselles1997GeodesicContours} to segment the most uncertain regions in the lungs boundary to include discarded lesions and expel previously included artefacts (\figurename~\ref{fig:methods}.e). 

\subsection{Computer Tomography Biomarker Extraction}\label{ch2:ssec:biomExtrac}
In order to automatically retrieve quantifiable information as a CT biomarker, the proposed method is inspired in ~\citet{Chen2014PET/CTTuberculosis} work. Within \textit{Chen's} work, the tissue belonging to the lungs is divided into three disease-associated volumes manually. This division depends on the grey level intensity of the voxels, measured employing Hounsfield Units (HU), and two thresholds selected by experts, which establish three regions in the lungs histogram corresponding with the following kind of tissues.
\begin{itemize}
    \item \textit{Healthy Tissue}, which corresponds to the voxels with the lower intensities in the lungs and free of TB
    \item \textit{Soft Tissue}, which match with voxels found in lower density of abnormal tissue, corresponding with forming or healing lesions.
    \item  \textit{Hard Tissue}, corresponding to intensity values in high density abnormal tissue.
\end{itemize}
In other words, \textit{Chen's} approach assigns a discrete class (\textit{healthy}, \textit{soft} or \textit{hard}) to a range of values distributed around an expected intensity given a variability that captures the subtle differences in the composition of each kind of tissue. The expected intensity value and variability of each class are intrinsically determined via the selection of thresholds by the experts. Fortunately, for the problem domain, this empirical approach can be computationally modelled employing the well known \textit{Gaussian Mixture Model} (GMM) and the more likelihood volumes separation obtained through the Expectation-Maximization (EM) algorithm~\cite{Bishop2006PatternLearning}. The GMM model allows us to represent a known histogram, like the one belonging to segmented lungs (see \figurename~\ref{fig:methods} right part), as a probability distribution composed of several overlapped Gaussian variables, in our case, the distribution of each kind of tissue.\\
Generally speaking is formulated as:
\begin{equation}\label{eq:gmm}
p(\textbf{x}) = \sum_{i=1}^K \pi_k \mathcal{N}(\boldsymbol{\mu_{k}},\boldsymbol\Sigma_k),
\end{equation}
\nomenclature[g-sum]{$\sum$}{summatory}
\nomenclature[g-boldpi]{$\boldsymbol{\Pi}$}{prior probability}
where $\textbf{x}$ is a vector of observed features (the intensity values of each voxel represented in the histogram), $K$ is the number of expected Gaussians ($K = 3$ corresponding to \textit{healthy}, \textit{soft} and \textit{hard} tissues), and $\mathcal{N}(\cdot)$ represent each one of the overlapped normal distributions ($k$) of the voxels grey level, being: $\pi_{k}$, the \textit{a priori} probability; $\boldsymbol{\mu_{k}}$, the mean; and $\boldsymbol\Sigma_k$ the covariance, respectively, for each distribution. These parameters are computed employing the EM algorithm, selecting those that set the Gaussians which overlapping is most similar to the known histogram (Eq. \ref{eq:gmm}). This way, each voxel is assigned to a lung tissue depending on which of the fitted Gaussians provides the biggest probability for a given voxel intensity.
\nomenclature[x-Normal]{$\mathcal{N}$}{Normal distribution}

\subsubsection{Gold Standard Computer Tomography Biomarker}
To measure the performance of the proposed automatic biomarker extraction method, the \textit{soft} and \textit{hard} volumes aforementioned were manually extracted by an expert from the original 63 CT scans comprised within the dataset.

\subsection{Evaluation Methods}
\label{ssec:Eval_Methods}
In order to provide a more general biomarker of the \textit{\textit{Mtb}} burden, besides the  volumes defined in the previous section, we include the total volume of diseased tissue for the comparison between the proposed method and the gold standard, the volume is defined as:
\begin{equation}
Diseased \ \ Vol. = Soft \ \  Vol. + Hard  \ \  Vol.
\end{equation}
Additionally, to avoid the effects of the changes in the whole lung volume due to the subjects' growth during the 28 weeks, the diseased volume is normalized to study the longitudinal disease behaviour. This volume is easily defined as follows:
\begin{equation}
Relative\ Diseased \ \ Vol. = \frac{Diseased \ \ Vol.}{Healthy \ \ Vol.}
\end{equation}
To evaluate the longitudinal change, we employ a multirow bar plot, known as waterfall (i.e., \figurename~ \ref{fig:waterfall}), in which the first row shows a relative volume at each subject in baseline time point and the rest of them the change in the volume at a concrete time point in a $log_2$ scale, therefore, the subject change is computed as:
\begin{equation}
change \ in \ vol. = \log_2\Bigg(\frac{Vol. \ at \ week \ of \ change}{Vol. \ at \ baseline}\Bigg)
\end{equation}

\subsection{Results}
\label{sec:results}

\begin{figure}[t!]
\includegraphics[width=\textwidth]{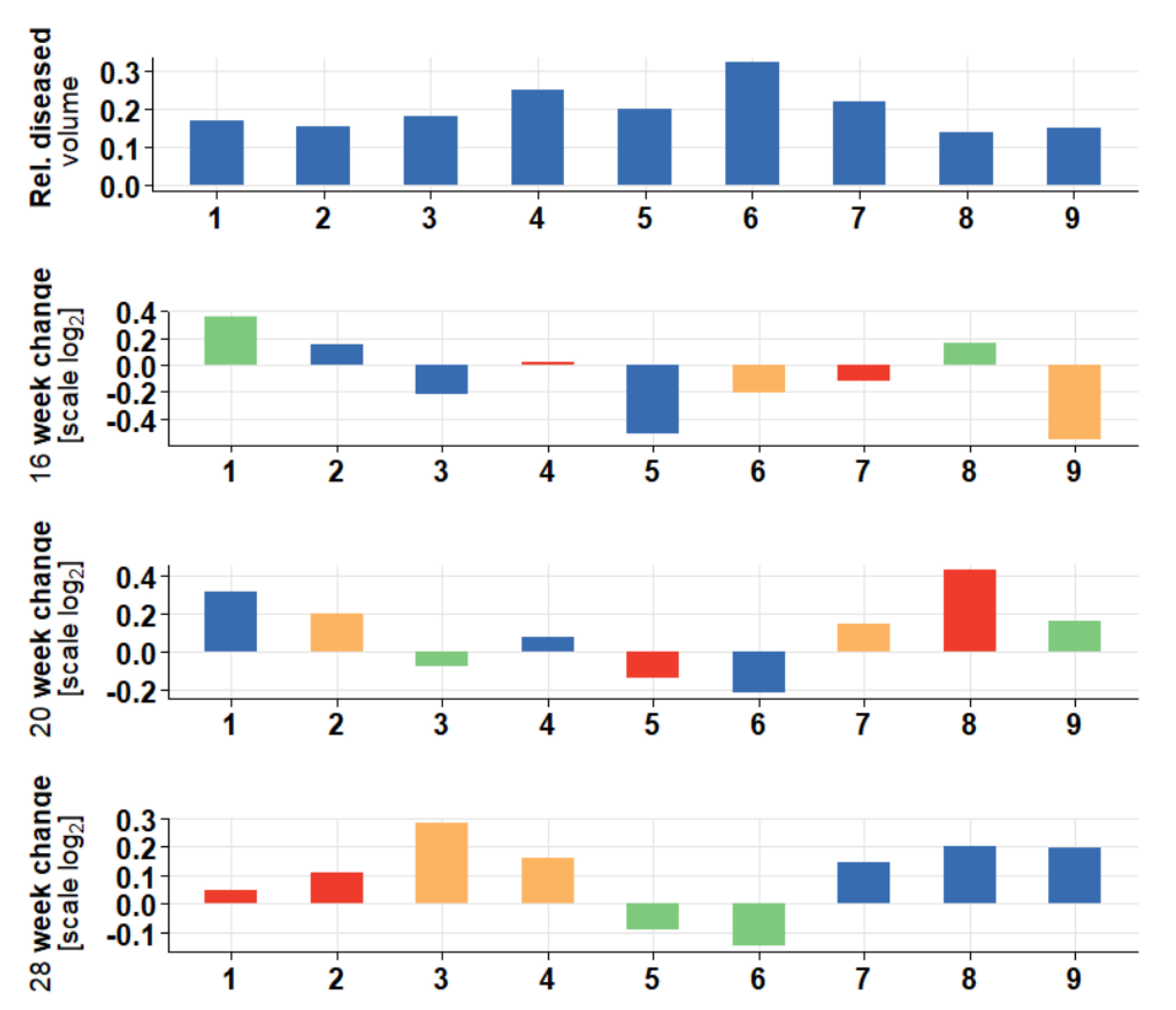}
\caption[Longitudinal evolution of TB infection]{Longitudinal evolution of TB infection: The waterfall plot, depicts the longitudinal change for diseased volume ($soft + hard vols.$) between the baseline week and the representative weeks as the $log_2$ fold change ($log_2(week \ change/baseline)$) and the correlation with treatments outcomes: \crule[NavyBlue]{.2cm}{.2cm} No-D (No Drugs), \crule[BurntOrange]{.2cm}{.2cm} HR ($Isoniazid + Rifampicin$), \crule[YellowGreen]{.2cm}{.2cm} HRZ ($Isoniazid + Rifampicin + Pyrazinamide$), \crule[red]{.2cm}{.2cm} HZ ($Isoniazid + Pyrazinamide$). Subjects with the same treatment in the final phase (week 28) present similar response ($diseased = hard + soft$) with respect to the baseline}
\label{fig:waterfall}
\end{figure}

\begin{figure}[t!]
\includegraphics[width=\textwidth]{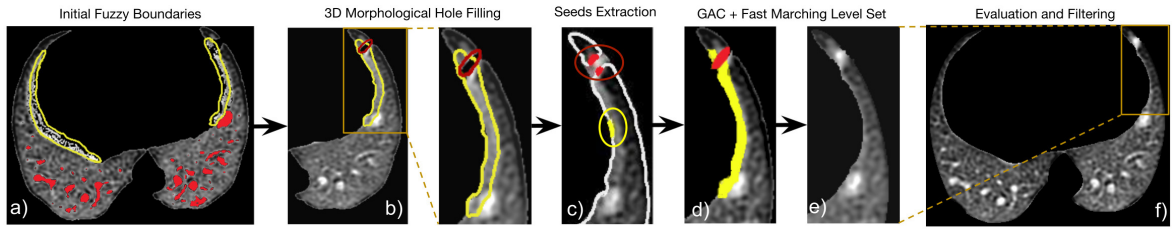}
\caption[Biomarker evaluation and Bland-Altman plot]{(Left) Biomarker evaluation: Correlation between the manual biomarker(relative diseased volume) and the proposed method ($R^2 \approx 0.8$, $p < 10^{-4}$). The $95\%$ confidence interval is drawn as the shadow of the regression line. (Right) The Bland- Altman plot presents a good agreement between measures with the regression bias of 0.47}
\label{fig:correlation}
\end{figure}

\figurename~\ref{fig:waterfall} depicts the longitudinal change of the \textit{Relative diseased volume}  through a waterfall plot for the nine subjects (horizontal axis) in four of the seven-time points for the shake of clarity. Concretely, the first row contains the relative volume of diseased tissue at week three after infection, while the rest represents the $log_2$ change (see Section \ref{ssec:Eval_Methods}) at weeks $16,20$ and $28$ with respect to the first row, the baseline. Beyond meaningful quantitative differences between equal treatments, it can be observed how subjects under the same drug cocktail (each treatment is shown with a different colour) at the end of the study (week 28) present a similar response to treatment to the baseline.\\
\figurename~\ref{fig:correlation} shows the diseased volume obtained employing the manual delimitation of regions against the volume obtained by the proposed automatic extraction method at each one of the 63 segmented lungs in the dataset (see Section \ref{ssec:materials}) together with the corresponding Bland-Altman plot in order to show the agreement between methods. The similarity between measures results is primarily independent of the subject, treatment and study time point (none of these factors shows a remarkable bias). The correlation coefficient was R $\approx$ 0.8 ($p < 1 \times 10^{-4}$), with a tendency to obtain higher values for the volumes obtained automatically. The Bland-Altman plot depicts all the values within the 95 \% limits of agreement. 

\subsection{Discussion}
\label{sec:discusion}
The results exhibit how the automatic biomarker extraction can provide good results by statistical modelling of the decision-making process carried out by an expert. Concretely, as an analogy between the experts' work and the proposed method, we can assert that our approach automatically assigns the thresholds established deterministically by the specialists. The method can fairly assess the longitudinal evolution of TB by showing significant similarities in the treatment response, as shown in \figurename~\ref{fig:waterfall}. As per the experimental design with a combination of antibiotics and as expected, differences are noticeable at the end-point (week 28). Besides, there is a correspondence between the results obtained automatically and the manual ones, as depicted by the $R^2 = 0.8$. This relation is biased by a factor of $0.47$ favouring the volumes obtained with the proposed method. Two causes can mainly explain these differences: a) The difficulty presented in the manual delimitation of complex three-dimensional structures, many times intricate in the healthy tissue (see \figurename~\ref{fig:chap2/Figure_1}), prevents from segmenting the whole region of interest which results in smaller volumes; b) The automatic extraction of the biomarker tend to include the unsegmented (in the lungs segmentation step) small vessels as diseased tissue, and therefore increase the obtained volumes. The inclusion of vessels as damaged lung tissue is undesirable. However, this side-effect is reduced by employing the normalized relative volume and assuming that the extra volume produced by the inclusion of vessels remains constant over time. Thus, the effect is not meaningful evaluating changes, as the trends presented in \figurename~\ref{fig:waterfall} seem to indicate. It is also essential to note that the proposed method is mainly intended to capture differences in the infection burden for animal models in which the subjects are not at the final stages of the disease. Namely, to establish a continuous spectrum between latent and active disease (\figurename~\ref{ch1:fig:Radiology_test}). Because of this, extra-large cavities (the manifestations proper of tissue destruction where the \textit{\textit{Mtb}} is not present anymore and the drugs cannot be effective) (see \figurename~\ref{ch1:fig:TB_lesions_cavidad}) are not included as damaged tissue, which can provoke small drifts in the data correlation corresponding with very particular high infected subjects. Some of the commented limitations are addressed in the next chapter, \nameref{ch3:Radiomics}, taking advantage of the Radiomics techniques~\cite{Lambin2017Radiomics:Medicine}, or in subsequent chapters by injecting part of qualitative information usually employed by the experts, and showed in this first approach, in DL models.          

%
%
%

\subsection{Conclusion}
\label{ch2:sec:Conclusion}
In this section, we introduce a complete methodology for the extraction of a biomarker to characterize the gradual change of \Mtb~infection. The proposed technique yields similar results to the ones obtained manually by a trained specialist. These facts highlight the capability of the method as a quantification tool in clinical assays devoted to design effective drugs against TB. Limitations of the framework, mainly caused by the lack of generalization, are shown in the following Section \ref{ch2:sec:OutOfComf}.

\section{Rule-Based Methods Under Unseen Domains \& Lack of Generalization}\label{ch2:sec:OutOfComf}

\begin{figure}[h!]
\centering
\captionsetup{justification=justified}
\includegraphics[width=\textwidth]{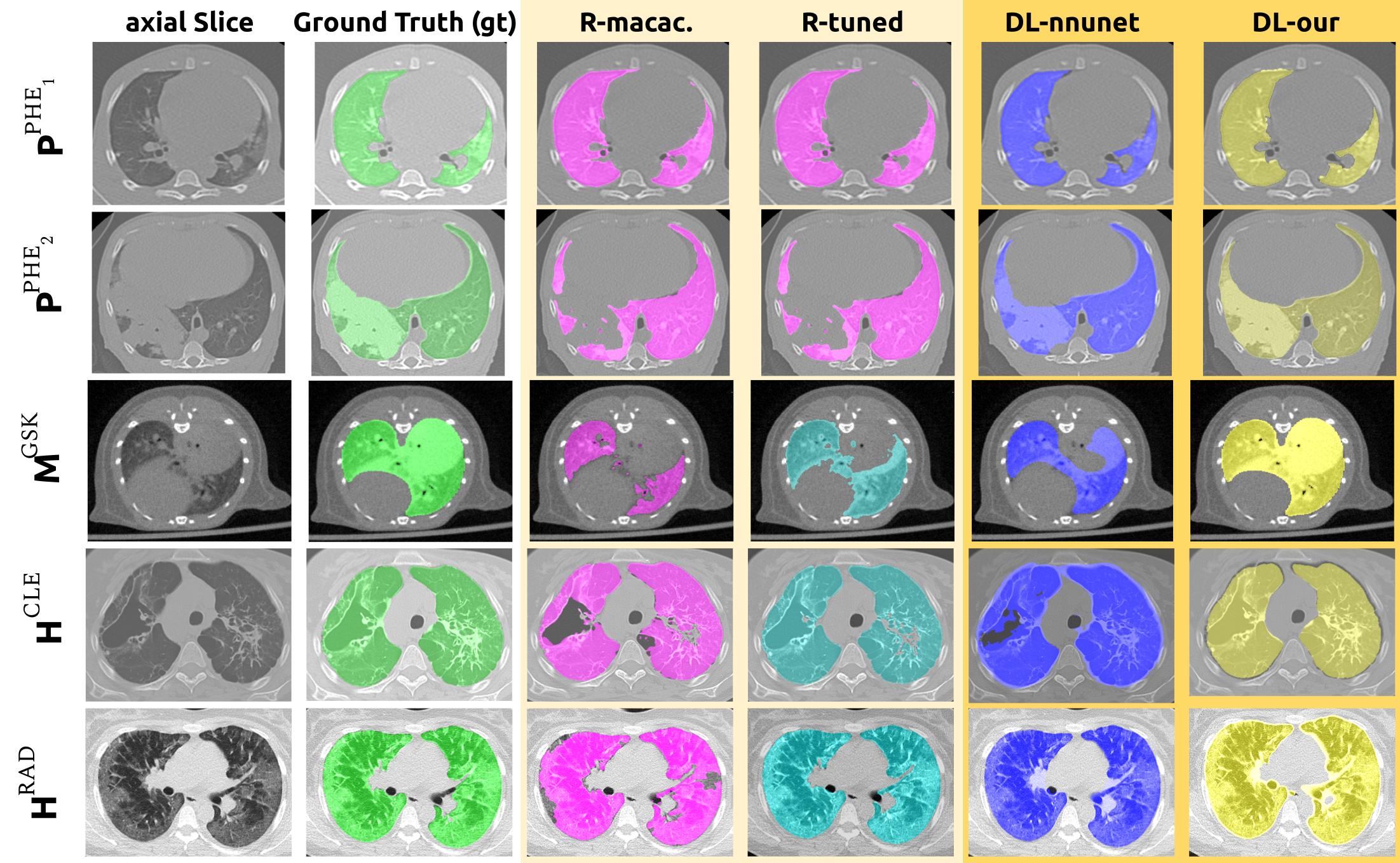}
\caption[Rules based segmentation limitations]{Pathological Lung Segmentation (PLS) masks examples obtained applying the method presented in this chapter and DL-based methods that will be in Chapter \ref{ch5:Intro}, employing different animal and disease models to those considered during the initial design. Each row shows an axial slice as an example of, respectively: 1) Initial dataset for which the presented model was designed, namely, mild TB macaque model ($P^{PHE_1}$, see Section \ref{ch2:sec:material}), 2) severe TB macaque model ($P^{PHE_2}$, see Section \ref{ch3:ssec:Matertials}), 3) severe TB mouse model ($M^{GSK}$) \cite{ERA4TBconsotium2021ERA4TB}, 4) human TB ($H^{CLE}$) \cite{DicenteCid2019OverviewAssessment} and 5) human COVID ($H^{RAD}$) \cite{Cohen2020COVID-19Collection} (description in \tablename~\ref{ch5:tbl:datasets}).\\ The columns correspond to a) the original chest CT axial slice, b) the \textcolor{green}{ground truth mask} delimited by experts (see details in Sections \ref{ch2:sec:material} and \ref{ch5:tbl:datasets}), c) the \textcolor{magenta}{mask obtained with the parameters given at the \tablename~\ref{tbl:chapt2/parameters} (\textit{R-macac.})}, d) the \textcolor{cyan}{mask obtained after tuning the parameters for each specific model (\textit{R-tuned.})}, e) \textcolor{blue}{mask obtained employing the SOTA DL-based method, nnU-Net \cite{Isensee2021NnU-Net:Segmentation} (\textit{DL-nnunet})} and f) the hybrid (discriminative + generative) \textcolor{darkyellow}{DL-based method proposed at Chapter \ref{ch5:Intro} of this work \cite{Gordaliza2021TranslationalRepresentations} (\textit{DL-our})}. }
\label{ch2:fig:segmetations_bad}
\end{figure}

We have referred before to the performance decay of traditional automation methods, especially when these receive inputs out-of-domain to those used during their design but whose similarity, however, allows experts to analyze them similarly. This section gives examples of the limited capability of rule-based algorithms to transfer knowledge to new domains and obtain acceptable lung mask delimitations. 
Thus, the first column of the \figurename~\ref{ch2:fig:segmetations_bad} shows chest CT axial slices corresponding to different mammals and diseases. Namely, the slice in the first row corresponds to the macaque model infected with mild TB as described in Section \ref{ch2:sec:material} (same domain, $P^{PHE_1}$). The slice at the second row also corresponds to the same macaque model but a different cohort ($P^{PHE_2}$). Such cohort models a much active TB infection (see Section \ref{ch3:ssec:Matertials}) \cite{Sharpe2016UltraMacaques}. As can be seen, the lesions, in this case, differ (lung vanishes), which supposes a domain shift.
The third image belongs to a dataset of a mouse infected by TB, $M^{GSK}$, (courtesy of \textit{GlaxoSmithKline} (GSK) a ERA4TB partner \cite{ERA4TBconsotium2021ERA4TB}, Section \nameref{ch1:sec:Erad:ERA4TB}). This example shows a domain shift due to the use of a different animal (i.e., different TB manifestations, change of CT scanner). The last two slices belong to images of human lungs extracted from publicly available clinical datasets \cite{Cohen2020COVID-19Collection,DicenteCid2019OverviewAssessment}. TB is the pathogen for $H^{CLE}$, while $H^{RAD}$ belongs to COVID-infected lungs, thus illustrating the limitations of classical methods in clinical practice \cite{Jacobs2019GooglesValidation, Topol2019High-performanceIntelligence}.

Ideally, automation mechanisms should perform segmentation on domain-shifted lung images exploiting information extracted from the dataset employed during design/learning. This fact holds while such information is similar for all datasets, in the same way, as experts can segment images of different animal models and diseases learning from particular datasets (Ground truth, GT). 

However, as illustrated in the third and fourth columns of \figurename~\ref{ch2:fig:segmetations_bad}, this only occurs when the input data have the same distribution as the training data.
Specifically, the \textit{R-macaq.} column shows the segmentation yield by the algorithm presented in this chapter with the parameters set for the macaque model dataset. The \textit{R-tuned} corresponds to the segmentation obtained when the parameters (see \tablename~\ref{tbl:chapt2/parameters}) are tuned to best fit other datasets. As shown, the algorithm performs excellently with data similar to those considered during the design. This approach is of enormous help in automating the analysis of hundreds or thousands of images. However, it is insufficient with severe infection models such as those in the second, third and fourth row of the figure corresponding to other animal and disease models.
This fact motivates the creation and implementation of methods, such as those studied in subsequent chapters, with the ability to perform the transfer of meaningful information between datasets. As a teaser, the fifth and sixth columns show the segmentation results obtained with two DL-based approaches; nnU-net \cite{Isensee2021NnU-Net:Segmentation} (\textit{DL-nnunet}), SOTA method and, an own approximation (\textit{DL-our}), that will be presented in Chapter \nameref{ch5:Intro}. 

\chapter{Radiomics for TB Manifestations Classification}
\label{ch3:Radiomics}

\ifpdf
    \graphicspath{{Chapter3/Figs/Raster/}{Chapter3/Figs/PDF/}{Chapter3/Figs/}}
\else
    \graphicspath{{Chapter3/Figs/Vector/}{Chapter3/Figs/}}
\fi

\section{Introduction}

The previous chapter, \nameref{ch2:chapter2}, illustrates how the more traditional methods for automating Pathological Lung Segmentation (PLS) \cite{Gordaliza2018} combined with a classic estimation algorithm such as Expectation-Maximization (EM) \cite{Gordaliza2019ComputedTuberculosis, Neal1998AVariants} allows quantifying infected lungs. As was commented, such an approach works as long as the input images belong to a specific domain, namely, a model of mild TB in macaques (see Sections \ref{ch1:sec:Erad:ERA4TB}). Even when this approach works as ideally expected, it may be insufficient to our goal of achieving a characterization of the continuous spectrum of the disease (see Section \ref{ch1:ssec:TBspectrum}) that requires the discrimination between the different types of lesions. 

Expert radiologists have claimed that TB lesions appear in high-resolution CT images at all disease stages, which radiological manifestations could be used as imaging biomarkers to provide information about the biological course of the disease. 
Thus, this chapter, presents a complete pipeline to detect TB lesions on thorax CT scans and extract informative features from them. In particular, the method infers the TB lesions after feeding with the texture features a \textit{Random Forest} classifier (see Section \nameref{ch1:ssec:list:rulesvsML}). The model can provide an adequate classification for a complex multi-label problem, distinguishing between five different TB lesions types: granulomas, conglomerations, trees in bud, consolidations and ground-glass opacities (see Section \ref{ch1:ssec:manifestList}). The work as previously published and presented orally in the \textit{International Symposium on Biomedical imaging} (ISBI) as \textit{Towards an informational model for tuberculosis lesion discrimination on X-ray CT images} \cite{Gordaliza2018TowardsImages} and in the \textit{European Molecular Imaging Meeting} (EMIC) as \textit{Radiomics for the Discrimination of Tuberculosis Lesions} \cite{Gordaliza2018RadiomicsLesionsb}.

\section{Materials and Methods}
\label{ch3:sec:MandMs}

The proposed methodology is summarized in the workflow shown in Figure \ref{ch3:fig:methods} and it is described in detail below:

\begin{figure}[htpb]
\includegraphics[width=\textwidth]{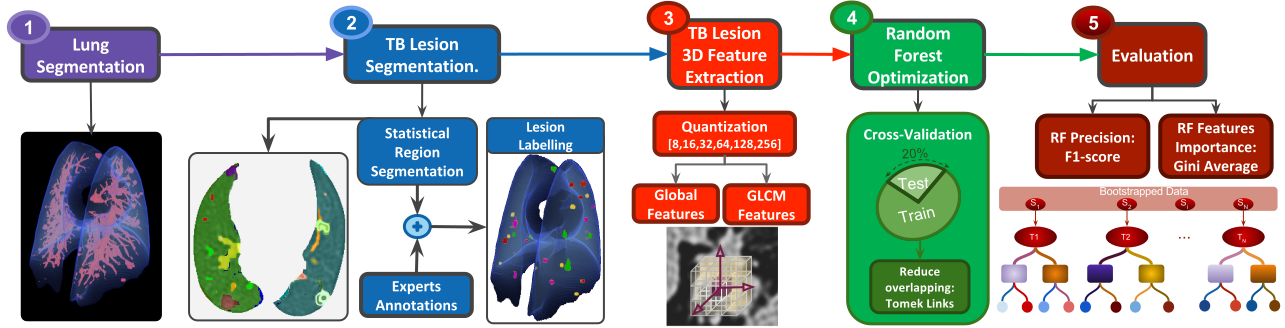}
\caption[Workflow for preliminary TB lesions discrimination]{Fully-automatic radiomics workflow for the extraction of informative features on the lung parenchyma: 1) Lung segmentation and airway  tree extraction; 2) Selection of relevant volumes employing the Statistical Region merging method \cite{Nock2004StatisticalMerging} matched with the expert annotations of lesions; 3) Extraction of texture features from each volume at $8$, $16$, $32$, $64$, $128$ and $256$ levels of quantization; $22$ features are extracted from the Grey Level Co-Occurrence Matrix (GLCM) and $4$ are global descriptor of the volume (Mean, Median, Maximum and Minimum); 4) Optimization of the \textit{Random Forest} (RF) hyperparameters (number of trees, minimum number of samples for split and the maximum number of features to evaluate per node);  The optimal RF classifier is computed per quantization level and number of features employed. The optimization employ a grid search process with $100$-fold cross validation where the training data ($80\%$ of the total) in each fold is filtered employing Tomek Links \cite{He2009LearningData} to handle class imbalance; 5) Two-fold evaluation: a) The weighted $F_1-score$ is employed as a measure of the classification quality of the most frequent TB lesion types; b) The importance of each feature is evaluated using as merit figure the Gini importance.
}
\label{ch3:fig:methods}
\end{figure}

\subsection{Materials: Computer Tomography Images}\label{ch3:ssec:Matertials}
In this chapter, forty-two thorax CT scans acquired on a medium size animal model of Tuberculosis were employed. Their voxel size is $0.26 \:\mbox{mm}\times 0.26\: \mbox{mm}\times 0.63\: \mbox{mm}$. In order to build a predictor, the identified lesions (Regions-of-Interest, ROIs) \nomenclature[z-ROI]{ROI}{Region Of Interest} were labeled by an expert distinguishing five types of lesions: $2140$ granulomas, $350$ conglomerations, $82$ trees in bud, $80$ consolidations and $53$ Ground Glass Opacities (see \nameref{ch1:ssec:manifestList}).\\
All animal procedures and study designs were approved by the Public Health England Animal Welfare and Ethical Review Body, Porton Down, UK, and authorized under an appropriate UK Home Office project license.\\

The reader should note that this dataset differs from the one presented in Section \ref{ch2:sec:material}. The former does not have the annotated lesions necessary for this chapter. However, the axial slice in the second row of \figurename~\ref{ch2:fig:segmetations_bad}, employed to illustrate a domain shift due to the present manifestations, belongs to the employed dataset.

\subsection{Lungs Segmentation}\label{ch3:ssec:lung_segment}
For completeness a small description of the process is included in this section, for further details see Section \ref{ch2:sec:Methods}.\\

\begin{figure}[htpb]
\centering
\includegraphics[width=0.85\textwidth]{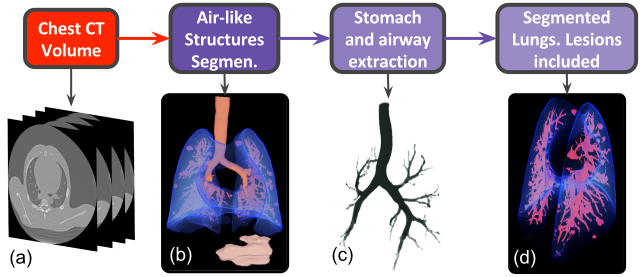}
\caption[Summary of lungs segmentation]{Summary of lungs segmentation methodology: From a CT scan a) aerated regions are identified via automatic thresholding. b) Then, lungs and airways together are isolated, based on topological priors. Next, the intricate airways are extracted (c). Finally, non-segmented damaged tissue is included by a hole filling method based on active contours (d).\\ For further details check \nameref{ch2:ssec:lungsegmentationGeneralDescr}.}
\label{ch3:fig:methods_lungseg}
\end{figure}

Air-like organs (e.g., healthy lungs, airways tree, stomach) are identified on a thorax CT scan. Next, the intricate airways tree structure is computed and removed from the segmented lungs. Finally, unsegmented pulmonary regions corresponding to damaged parenchyma and TB lesion are included by morphological hole filling. The whole procedure is summarised in \figurename~\ref{ch3:fig:methods_lungseg}.

\subsection{Lesions Segmentation}\label{ch3:ssec:lesion_segm}

The lung volume is divided into regions of similar appearance (i.e., lesions, vessels, healthy parenchyma) applying the statistical region merging (SRM) \nomenclature[z-SRM]{SRM}{Statistical Region Merging} approach ~\cite{Nock2004StatisticalMerging}.\\
The SRM works in two steps: sort and merge. Firstly, the voxels in each 26-connected neighbourhood are sorted based on their similitude. Subsequently, voxels are compared in the earlier order and merged into regions when a given condition is accomplished. The merging predicative is based on intensity similarity and region size (further details can be found in ~\cite{Nock2004StatisticalMerging}). The process achieves an automatic segmentation of the lesions on the extracted CT lung volumes as illustrated in \figurename~\ref{ch3:fig:methods}.2.

\subsection{Lesion Characterization \myfnt{As in the previous chapter the \texttt{C++/ITK} \cite{Johnson2015TheGuide} code implementation for the Tuberculosis Lung Segmentation (TLS) \href{https://github.com/BIIG-UC3M/TLS-Piped}{https://github.com/BIIG-UC3M/TLS-Piped} while the \texttt{Python} code for texture features extraction and analysis is in the following \href{https://github.com/BIIG-UC3M/TLS_Texture/tree/master/PET_Tumor_Analysis-master}{GitHub Repository}. This code was developed from scratch to enable feature extraction from three-dimensional neighbourhoods, reducing the time complexity of similar approaches. Indeed, this software is the choice for the radiomics application works to the clinical practice: \textit{Performance of ultra-high-frequency ultrasound in the evaluation of skin involvement in systemic sclerosis: a preliminary report} \cite{Naredo2019PerformanceReport} and \textit{Association of visual and quantitative heterogeneity of 18F-FDG PET images with treatment response in locally advanced rectal cancer: A feasibility study} \cite{Martin-Gonzalez2020AssociationStudy} thanks to its versatility and performance.}}\label{ch3:ssec:lesion_segm}

Texture features intend to formally define the spatial distribution of the pixel intensities perceived by experts during manual image evaluation. Numerous types of texture features (e.g., based on \textit{Grey-Level Run Length Matrix} (GLRLM) \nomenclature[z-GLRLM]{GLRLM}{Grey-Level Run Length Matrix} \cite{Galloway1975TextureLengths, Way2006Computer-aidedContours}, based on \textit{Local Binary Patterns} (LBP) \nomenclature[z-LBP]{LBP}{Local Binary Patterns} \cite{Ojala1994PerformanceDistributions, Song2013Feature-basedClassification}) can be found in a vast related literature \cite{Depeursinge2014Three-dimensionalOpportunities}. The nature of the problem to be solved designates the most appropriate set of characteristics to use. Namely, it depends on the imaging technique used, the region to be analyzed, and the specific manifestations of the disease to characterize.\\ 
Thus, among the variety of texture features, the ones based on the grey level co-occurrence matrix (GLCM) ~\cite{Haralick1979StatisticalTexture, Haralick1973TexturalClassification} are proved to be specially useful in our context ~\cite{Depeursinge2014Three-dimensionalOpportunities,Mansoor2014ASegmentation}. GLCM \nomenclature[z-GLCM]{GLCM}{Grey Level Co-occurrence Matrix} represents the joint frequency over all possible grey levels combinations in every pair of voxels separated by a predefined \textit{offset}, and it is computed as:\\

\vspace{-0.75cm}
\begin{equation}\label{eq:coOcurrenciaMatr}
\resizebox{\textwidth}{!}{${\displaystyle C_{\Delta x,\Delta y,\Delta z}(l_i,l_j)=\sum _{x=0}^{N_x-1}\sum _{y=0}^{N_y-1}\sum _{z=0}^{N_z-1}{\begin{cases}1,&{\text{if }}I(x,y,z)=l_i\text{ and }
I(x+\Delta x,y+\Delta y,z+\Delta z)=l_j\\0,&{\text{otherwise}}\end{cases}}}$}
\nonumber
\end{equation}
\nomenclature[g-Delta]{$\Delta$}{Change / Offset}
where $I(x,y,z)$ is the intensity at $(x,y,z)$, $(\Delta x, \Delta y, \Delta z)$ is the \textit{offset}, $N_x\times N_y\times N_z$ is the image size in voxels and $l_{[i,j]} \in L$ the grey level at the pair of voxels $i$ and $j$.

In this work, we restrict to unitary \textit{offsets} and compute GLCM matrices in the $13$ possible directions in $\mathbb{Z}^3$.  We further average them obtaining a unique GLCM matrix per ROI from which $26$ descriptors are extracted\footnote{For the full description of the 26 features see the Appendix \ref{ap:texture}.} as described in ~\cite{1011383,Depeursinge2014Three-dimensionalOpportunities}. These are added to the global ROI histogram descriptors: \textit{mean}, \textit{standard deviation}, \textit{minimum} and \textit{maximum} grey value.\\

Commonly, the GLCM is not computed on images with $2^{16}-1$ grey levels ($L$) \footnote{Note that, typically, not all the procured levels are used when using 16 bits, since the HU are usually set between -1024 and 3024} that can be found in a CT image, but a quantization is performed, that is, a reduction of the number of grey levels, using rounding and truncation techniques. In this way, it is possible to reduce the computational cost of the GLCM calculation while minimizing the noise ~\cite{Soh1999TextureMatrices}. However, it could mean a loss of detail if it is influenced by the artefacts present in the image (for example, the movement of the lungs). Therefore, it could not faithfully represent the tissue that \textit{a priori} characterizes.
For this reason, this work studies the effect of quantization to determine whether it is preferable to work with more smoothed images due to the noise or whether it is better to use all the available information.
With this objective, the classification study is performed using $8,16,32,64,128$ and $256$ grey levels independently.

\subsection{Random Forest Optimization and Evaluation}\label{ch3:ssec:RF}
It is well-known from the literature \cite{Breiman2001, Menze2009AData,Saeys2008} that the \nomenclature[z-RF]{RF}{Random Forest} (RF) classifiers provide a high precision due to their ability to discard irrelevant features. More formally, modelling with RF allows us to obtain a complex model estimator able to define a flexible manifold in high dimensional input spaces ~\cite{Breiman2001,Hastie2009ThePrediction}  without completely losing the interpretability of more classic statistical models (see \nameref{ch1:ssec:list:rulesvsML}).\\
Specifically, to estimate the importance of each feature, $\phi$, (the 26 GLCM descriptors in our case), we use the \textit{Gini importance} \cite{Ma2017CURE-SMOTEForests, Menze2009AData, Murphy2022ProbabilisticIntroduction}, $I_G$, which is computed as the averaged over all the trees $T$ and all the nodes $N$ of the \textit{Gini impurity}, $G(n) =\sum\limits_{\substack{k \in C }} p_k(1-p_k)$, change:

\begin{equation}
I_G(\phi) = \sum_{t \in T} \sum_{n \in \mathrm{N}}\Delta G(n),
\end{equation}

where $\Delta G(N) = G(N) - \sum_{s=0}^S G(S)$ ($s \in S$ set of nodes splited from $N$), $p_k$ is the probability of having an instance labeled with class $k$ in the set $C$. The Gini importance of each feature varies between $0$ and $1$ and the sum over the whole feature set adds to $1$.\\   

The RF hyper-parameters are the number of trees $T$ in each RF, the minimum number of samples for split and the maximum number of features to evaluate per node. The optimal RF hyper-parameters per quantization level are found by grid search employing a $100$-fold cross-validation (CV) per RF candidate.\\

The weighted $F_1$-score was used as the quality measure, which is defined as follows:

\begin{equation}
\frac{1}{\sum_{k \in C} \left|\hat{y}_k\right|} \sum_{k \in C} \left|\hat{y}_k\right| F_1(y_k, \hat{y}_k),
\end{equation}

where $y_k$ and $\hat{y}_k$ represent the predicted and the true labels. Namely, it is the weighted average of each class's $F_1$-score (harmonic mean of the precision and the recall).\\

Besides, due to the high imbalance of the annotated dataset, the \textit{Tomek Links} technique ~\cite{He2009LearningData} is applied to the training data to reduce the overlap in the feature space. 

\section{Results}\label{sec:results}

\begin{figure}[htpb]
\includegraphics[width=\textwidth]{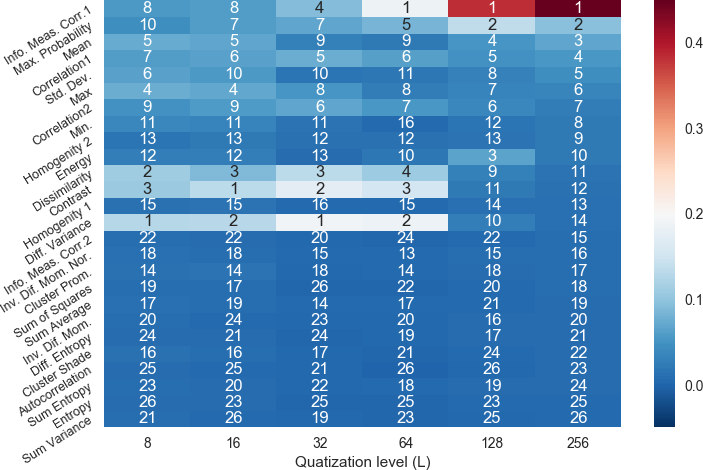}
\caption[Matrix of the Gini importances]{Matrix of the Gini importances estimating the importance of each feature for the optimal RF classifier. The number on each row corresponds to the feature ranking position and the colour, the averaged Gini index. Each column gives the results for a given quantization level.}
\label{ch3:fig:feats_ranking}
\end{figure}

Figure \ref{ch3:fig:feats_ranking} presents the Gini importance $I_G$ of each feature for a given quantization level $L$.\\
It can be observed that the \textit{difference variance, contrast} and \textit{information measure of correlation 1} rank first for at least one quantization level. The Gini importance of the most informative feature increases with $L$ (\textit{difference variance}, $L = 8$, $I_G=0.14$; \textit{information measure of correlation 1}, $L=256$, $I_G=0.45$). 

\begin{figure}[htpb]
\includegraphics[width=\textwidth]{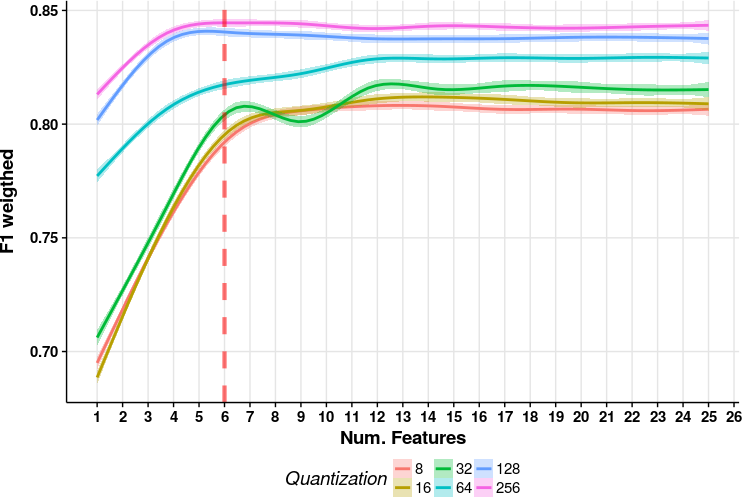}
\caption[Mean weighted $F_1$ score]{Per quantization level, mean weighted $F_1$ score obtained for groups  of sorted features (size $1$ to $26$) over the $100$ cross-validation folds of the best estimator. The features are sorted as per the ranking in Figure \ref{ch3:fig:feats_ranking}. The $95\%$ confidence interval is drawn as the line shadow.}\label{ch3:fig:feats_precision}
\end{figure}

Figure \ref{ch3:fig:feats_precision} shows the weighted $F_1$ score obtained for the optimal estimator at each $L$ in function of the number of features. The observed results confirm the ones presented above. At large quantization levels ($L=128, 256$), the precision reaches good values using just two or three features, while at small $L$, the increment in the weighted $F_1$-score with the number of features grows more slowly. However, all the estimators show symptoms of convergence when using $6$ features (red line in Figure \ref{ch3:fig:feats_precision}).  The weighted $F_1$-score at convergence (i.e., using $26$ features) increases as the quantization level does ($0.844\pm 0.01$, $L=256$) as shown in \tablename~\ref{ch3:tbl:results}.

\begin{table}
\centering

\resizebox{\textwidth}{!}{
\arrayrulecolor{white}
\begin{tabular}{|c|c|c|c|c|c|c|} 
\hhline{~------|}
\multicolumn{1}{c|}{} & {\cellcolor[rgb]{0.639,0,0}}\textbf{\textcolor{white}{8}} & {\cellcolor[rgb]{0.678,0.678,0.361}}\textbf{\textcolor{white}{16}} & {\cellcolor[rgb]{0.051,0.525,0.051}}\textbf{\textcolor{white}{32}} & {\cellcolor[rgb]{0.361,0.769,0.769}}\textbf{\textbf{\textcolor{white}{64}}} & {\cellcolor[rgb]{0.361,0.361,1}}\textcolor{white}{\textbf{128}} & {\cellcolor[rgb]{0.8,0.161,0.8}}\textcolor{white}{\textbf{256}}  \\ 
\hline
\textbf{\textbf{6}}   & \textcolor[rgb]{0.639,0,0}{0.788~$\pm$0.01 }              & \textcolor[rgb]{0.678,0.678,0.361}{0.792~$\pm$0.01 }               & \textcolor[rgb]{0.051,0.525,0.051}{0.803~$\pm$ 0.01 }              & \textcolor[rgb]{0.361,0.769,0.769}{0.818~$\pm$ 0.01}                        & \textcolor[rgb]{0.361,0.361,1}{0.841~$\pm$ 0.01}                & \textcolor[rgb]{0.8,0.161,0.8}{0.844~$\pm$ 0.01}                 \\ 
\arrayrulecolor[rgb]{0.753,0.753,0.753}\hline
\textbf{26}           & \textcolor[rgb]{0.639,0,0}{0.806~$\pm$0.01 }              & \textcolor[rgb]{0.678,0.678,0.361}{0.808~$\pm$ 0.01 }              & \textcolor[rgb]{0.051,0.525,0.051}{0.815~$\pm$ 0.01 }              & \textcolor[rgb]{0.361,0.769,0.769}{0.829~$\pm$ 0.01}                        & \textcolor[rgb]{0.361,0.361,1}{0.838~$\pm$ 0.01}                & \textcolor[rgb]{0.8,0.161,0.8}{0.844~$\pm$~0.01}                 \\
\arrayrulecolor{white}\hline
\end{tabular}
\arrayrulecolor{black}
}
\caption{Optimization results employing $6$ and $26$ most significant features}
\label{ch3:tbl:results}
\end{table}

\begin{figure}[htpb]
\centering
\includegraphics[width=0.7\textwidth]{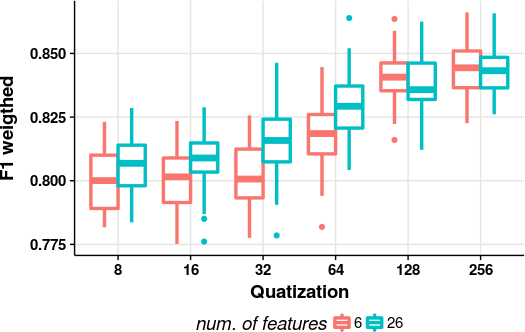}
\caption[Box plot for the weighted $F_1$]{Box plot representation of the weighted $F_1$ scores obtained on the $100$ folds of the cross validation process for the best RF estimator. The results are shown for each quantization level when employing the $6$ most relevant and all the $26$ features.}
\label{ch3:fig:optimization_boxes}
\end{figure}

In Figure \ref{ch3:fig:optimization_boxes}, we present the box plots of the weighted $F_1$-scores obtained for the best RF estimator at each quantization level for two interesting cases ($6$ and $26$ features). The differences between factors were assessed with the one-way analysis of variance test (ANOVA). The differences between quantization levels are statistically significant ($p < 2\times 10^{-7}$) while those between the number of features are not ($p>0.05$).

\section{Discussion}\label{ch3:ssec:discusion}

The results presented in this manuscript show that it is possible to obtain a good classification performance for a complex multi-label classification problem by training simple models that are still informative of the disease development.\\ 

It is crucial to notice that by using a unique \textit{offset}, averaging the GLCM directional descriptors and performing an independent treatment for each quantization level, we can keep the set of features small ($n = 26$). This is critical to facilitate the model comprehension and departs from the tendency to use models focused on achieving maximal precision at the cost of employing thousands of features. \\ 

As previously commented, identifying the lesions improves with the quantization level in parallel to the growth of the Gini importance of the most informative features. In particular, when employing $128$ and $256$ grey levels, the Gini importance of the first ranked feature,  \textit{Information Measure Correlation 1} or $r_0$, is particularly high, $0.38$ and $0.45$, respectively. This descriptor gives a general measure of the correlation between the intensity of adjacent voxels and can be interpreted as an information gain. Our results point out that when employing the proper level of detail, this descriptor can capture complex relations between the intensities of neighbour voxels characteristic of a particular type of lesion. \\

In general, we can confirm that our model can quantify changes in the lung parenchyma that are specific to a given type of lesion (e.g., density, heterogeneity). This affirmation gets strongly supported when analyzing the importance of the features at lower quantization levels. Namely, the most important features (\textit{contrast, difference of variance}) are strongly related to \textit{Information Measure Correlation 1}. This fact was expected as the RF classifiers discard features contributing with redundant information. For example, the descriptor, \textit{Information Measure Correlation 2}, defined as $r_1 = (1 - e^{2r_0})^{1/2}$, it is closely related to $r_0$ and consequently, provides similar information. So, it has a low Gini importance in our model for all the quantization levels appearing at the bottom of the ranking shown in Figure~\ref{ch3:fig:feats_ranking}. \\
 
\section{Conclusion}
The proposed framework gives promising results in its ability to extract informative biomarkers of tuberculosis development. Namely, we demonstrate that we can achieve a reasonable good classification of the most frequent TB lesion types.

This work will be the basis for the studies presented in the next chapters on the characterization of the biological changes induced by TB infection. We have selected the state-of-the-art (SOTA) methods that will hopefully lead to an improved understanding of the course of the disease. 
This work proves that machine learning (ML) can characterize segmented lesions, even employing a relatively medium/low capacity model. Thus, the next chapter, \nameref{ch4:Intro} \cite{Gordaliza2019AManifestations}, proposes a DL-based method capable of identifying lesions from a whole volume without relying on previously segmented lesions. Meanwhile, in Chapter \ref{ch5:Intro}, \nameref{ch5:Intro} \cite{Gordaliza2021TranslationalRepresentations}, we investigate the ability of a DL model to not only identify pathological lungs for different species and disease models (increase generalization capacities between domains). But as well, to synthesize realistic images of lungs damaged by Tuberculosis by introducing tools developed under the framework of graphical causality.
Such innovation initiated through the preliminary work presented in this chapter escalates the models' skill for characterization and quantification of TB to a new paradigm of infinite possibilities within clinical practice.  


\chapter{Deep Learning for TB Manifestation Classification}
\label{ch4:Intro}
The results obtained in the previous chapter show the possibility of extracting quantitative information from the Computed Tomography (CT) images through statistical descriptors (texture features) that allow us to formalize radiological descriptions of the manifestations of Tuberculosis (TB).\\ 
These features enable automatic manifestations identification using machine learning (ML) to characterize complex relationships among them. Such an approach requires previous manifestation delimitation. The difficulty of the delimitation process depends on the animal and disease model due to their biological variability. 
Indeed, experts face complications when performing this process manually in clinical practice or generating the ground truths needed to train supervised ML algorithms. Uncertainty in delimitation directly translates to the automatization performance.\\  

To tackle such issues, this chapter presents our previously published work as the \textit{A Multi-Task Self-Normalizing 3D-CNN to Infer Tuberculosis Radiological Manifestations} \cite{Gordaliza2019AManifestations}, in which we propose to identify the presence of lesions without prior segmentation.\\
We hypothesize that since a few handcrafted texture features can capture meaningful information to classify delimited lesions through a \textit{Random Forest}, much more powerful Deep Learning (DL) models should perform a suitable classification without the need to delimit lesions and extract handcrafted features.\\ 
As pointed out in the \nameref{ch1:ssec:list:rulesvsML}, DL models are end-to-end. Therefore, the features do not have to be defined. Oppositely these are learnt, and their effectiveness is well demonstrated in object localization tasks both in computer vision \cite{Dosovitskiy2020AnScale, Krizhevsky2012ImageNetNetworks} and medical imaging fields \cite{Esteva2017Dermatologist-levelNetworks, Litjens2017AAnalysis, Zhou2021APromises}.\\  
Specifically, we investigate this hypothesis by building a model to mimic the radiologist generated reports by inferring the presence of TB manifestations on thoracic Computer Tomography scans.\\
Our model exploits the well-known advantages of three-dimensional Convolutional Neural Networks (3D-CNNs). In particular, we adapt the \textit{V-Net} encoder to distinguish among five different radiological manifestations of TB at each lung lobe.\\
Specifically, since usually TB manifestations do not appear in the infected lungs isolated (i.e., nodules, conglomeration or cavities appear together in the lung parenchyma, see sections \ref{ch1:ssec:TBspectrum} and \ref{ch1:sec:MITB}), we propose a multi-task model (Section \ref{ch1:ssec:LungSegSOTA}) designed to identify single instances. A joint force strategy is established to overtake the issues (e.g., exploiding/vanishing gradients, lack of sensibility) that generally appear when training complicated deep 3D models with limited size datasets and large medical imaging volumes.\\

Our proposal employs: 1) At the network architecture level, the \textit{scaled exponential linear unit} (SELU) activation, which allows the self-normalization of the network, and 2) at the learning phase, multi-task learning with a loss function weighted by the task \textit{homoscedastic} uncertainty. This is performed independently of the binary or regressive nature of the task. 



\section{Introduction}

The automation of the radiologists' reports has been pursued during the last three decades. Most of the classical works in the literature approach the challenge as an image segmentation problem. Hand-crafted features are extracted to segment the lung parenchyma ~\cite{Mansoor2014ASegmentation} and to identify TB lesions ~\cite{Gordaliza2018TowardsImages, Xu2015Computer-aidedModels}. However, these techniques are usually limited to a particular application and are quite sensitive to the high implicit variance of medical images. Fortunately, in recent years, the use of DL techniques has drastically improved the automation of the  radiologist reports generation, reaching performances close to the  human error ~\cite{Esteva2017Dermatologist-levelNetworks,  Litjens2017AAnalysis, Shin2016DeepLearning, Wang2017ChestX-ray8:Diseases}. DL has opened a new paradigm in the medical imaging field \cite{Hinton2018DeepCare} with remarkable results when expert knowledge is properly incorporated into the models ~\cite{Bishop2013Model-basedLearning, Ghahramani2015ProbabilisticIntelligence}.\\
For the application at hand, knowledge is usually injected into the models in form of manually segmented masks of the lung-damaged Volumes-of-Interest (VOI). \nomenclature[z-VOI]{VOI}{Volume of Interest}
This technique yields to good results when 2D Convolutional Neural Networks are employed ~\cite{Harrison2017ProgressiveImagesb, Isensee2021NnU-Net:Segmentation, LaLonde2018CapsulesSegmentation, Ronneberger2015U-net:Segmentation} and promising ones ~\cite{Negahdar2018AutomatedNetwork}, when using complex 3D CNN models~\cite{Cicek20163DAnnotation, Hatamizadeh2021UNETR:Segmentation, Milletari2016V-Net:Segmentation}. However, to the best of our knowledge, there is a lack of studies directly employing the expertise acquired by radiologists through years of clinical practice as synthesized in tabular reports. For this reason, in this chapter, we aim to employ the information stored in the radiologists' reports.\\

With this aim, our methodology integrates the following techniques:

\begin{enumerate}
    \item A 3D-CNN based on \textit{V-Net} architecture \cite{Milletari2016V-Net:Segmentation} is employed to extract distinctive features from whole CT volumes. The extracted features are used to detect the presence or the quantity of specific TB manifestations employing as ground truth the radiologist reports. This approach makes use of network models and learning principles that are common in the literature
    \item The 3D-CNN architecture is modified to leverage a better regularization and act as \textit{Self-Normalizing Neural Networks} (SNNs)~\cite{Klambauer2017Self-NormalizingNetworks}. \nomenclature[z-SNN]{SNN}{Self-Normalizing Neural Network}
    \item Our deep network is configured as a multi-task model to perform multi-label classification, acknowledging that TB manifestations do not appear isolated. Moreover, uncertainty is used to weigh the influence of each task loss ~\cite{Kendall2018Multi-TaskSemantics} (see uncertainty definition at Section \ref{ch1:desc:uncert}). 
\end{enumerate}

\section{Material and Methods\myfnt{The \texttt{Python/Tensorflow} implemtation of the methods can be found under the following \href{https://github.com/PeterMcGor/Multitask_TB}{GitHub Repository}} }
\label{ch4:sec:M&Ms}

\subsection{Material}\label{ch4:sec:Materials}
The experiments (see \ref{ch4:sec:expAndResults}) were accomplished on a dataset constituted by $56$ chest CT-scans acquired from $14$ male Cynomolgus macaques at $3$, $7$, $11$ and $16$ weeks after TB aerosol exposure.  The voxel is $0.26\:\mbox{mm}~\times 0.26\:\mbox{mm}~\times 0.63\:\mbox{mm}$ with an in-plane resolution of $512~\mbox{pixels} \times 512~\mbox{pixels}$  and   201  to  270 slices,  which  are  preprocessed  to  feed  the  model (see  \ref{ch4:ssec:preproc}). We employ the quantitative report elaborated by a  radiologist with  $20$ years of experience as labelled data. These reports are tabular and contain the number of detected nodules (see \figurename~\ref{ch1:fig:TB_lesions_image}), which fluctuate between 0 and 15 depending on disease stage, and boolean annotations about the presence or absence of the most common TB manifestations ~\cite{Nachiappan2017PulmonaryManagement} -namely, cavitations (\figurename~\ref{ch1:fig:TB_lesions_cavidad}), conglomerations (\figurename~\ref{ch1:fig:TB_lesions_image}), consolidations (\figurename~\ref{ch1:fig:consolidacion}) and trees in bud (\figurename~\ref{ch1:fig:TB_lesions_TreeB}). The reports contain the disaggregated information per lung lobe (i.e., right superior, right middle, right inferior, left superior and left inferior) given a total of 25 manifestations to predict per subject.

\subsection{Model Architecture}\label{ch4:ssec:ModelArq}

Our model implementation aims to exploit the ability of state-of-the-art architectures to extract fine-grained features from chest CT-Images. We plan to use the extracted features as the input to Fully Connected Layers (FCLs)\nomenclature[z-FCL]{FCL}{Fully Connected Layer} for multi-label classification among the TB manifestations aforementioned. In this context, the \textit{V-Net}~\cite{Milletari2016V-Net:Segmentation} has been proven quite successful at segmentation tasks \cite{Litjens2017AAnalysis,Negahdar2018AutomatedNetwork} and even more importantly for our purpose, at the generation of synthetic lung nodules from these features \cite{Jin2018CT-RealisticSegmentation}.\\

\begin{figure}
  \includegraphics[width=\linewidth]{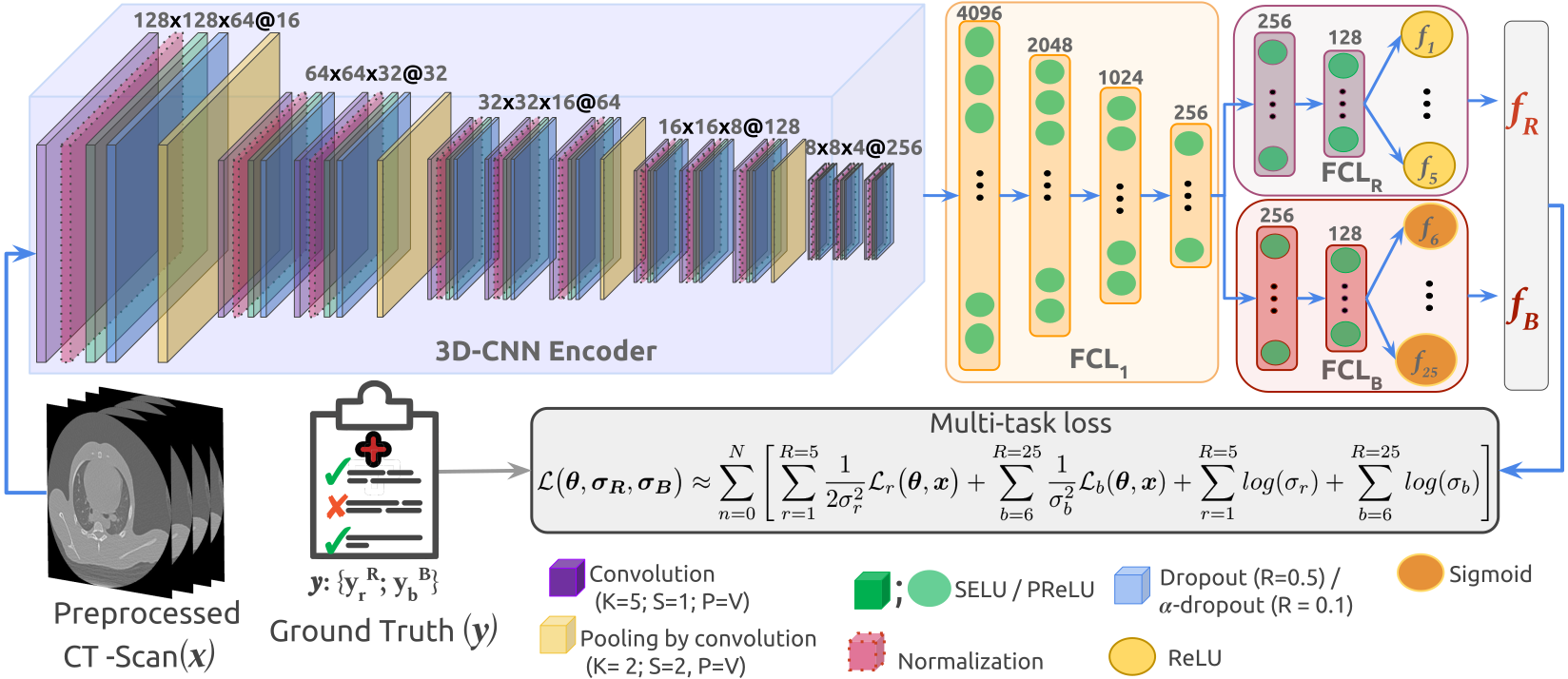}
  \caption[Multi-task Model Architecture]{The model is composed by a three-dimensional Covolutional Neural Network (3D-CNN) based on the V-Net ~\cite{Milletari2016V-Net:Segmentation} and three Fully Connected Layers (FCLs): $FCL_1$ with tasks-shared parameters and, $FCL_R$ and $FCL_B$ for prediction of regression and binary tasks, respectively. The encoder is built upon the following  layers: 1) Convolution characterized by their kernel size (K), stride (S) and the Valid (V) padding (P), in purple or in yellow for pooling; 2) Dropout (alpha-dropout for our proposed model) with a rate (R), in blue; 3) Activations: \textit{Scaled Exponential Linear Unit} (SELU) \cite{Klambauer2017Self-NormalizingNetworks} for our model, Parametric Rectified Linear Unit (PReLU) \cite{He2015DelvingClassification} for comparison experiments), in green; 4) Normalization (not employed in our model but for comparison), in pink. The final outputs units take \textit{Rectified Linear Unit}(ReLU) and \textit{sigmoid} activation to predict the labels used in the loss function~\cite{Kendall2018Multi-TaskSemantics}.}
  \label{ch4:fig:architecture}
\end{figure}

We adapt the \textit{V-Net} encoder to implement our Self-Normalizing Neural Network (SNN, explained in detail in the next section) or to include Batch Normalization layers~\cite{Ioffe2015BatchShift} for the comparison experiments.\\
The \textit{V-Net} encoder iteratively convolves inputs of preprocessed lung CT-Volumes (see \ref{ch4:ssec:preproc}) with a size of $128\times 128\times 64$ voxels with four codification stages, each of them halves the resolution and adds channels up to $256$, conducting to the generation of $1376256$ ($8\times 8\times 4\times 256$) features, which are flattened to feed our first FCL, referred as FCL$_{1}$.\\
This first FCL is task-shared. Therefore, it is in charge of characterizing the complex relationship among the low-level features obtained with the \textit{V-Net} in order to produce more abstract features common to the twenty-five tasks of our problem. FCL$_1$ is built up with four layers of $4096$, $2048$, $1024$ and $256$ units, respectively. All the parameters corresponding to the modified \textit{V-Net} encoder and FCL$_{1}$ are common to the prediction of each manifestation; this fact enables a more efficient training due to the possibility of modelling conditional relationships among the manifestations related.\\
However, to achieve a particular prediction for each task, we complete the architecture by employing the last $256$ features of FCL$_{1}$ as input of two independent FCLs, FCL$_R$ and FCL$_B$.  FCL$_R$  predicts the regression tasks (nodules  counting), while FCL$_B$ predicts the binary tasks. Each specific FCL is formed by two layers of $256$ and $128$ units and the final output units composed by $5$ \textit{Rectified Linear Units} (ReLU) for FCL$_R$ and $20$ \textit{sigmoid} activated units for FCL$_B$.\\
It is essential to mention that dropout or batch normalization layers are included where is needed, as shown in \figurename~\ref{ch4:fig:architecture}. When employing SNN, batch normalization is not needed (see next section for details).

\subsection{Self-Normalizing Neural Networks}\label{ch4:ssec:snn}

As explained above, we only implement half of the \textit{V-Net} architecture, the encoder, which results appropriate for our application but introduce a new issue. As \citet{Milletari2016V-Net:Segmentation} stated, \textit{V-Net} forwards the features extracted from the first levels of the encoder to feed the same levels of the decoder in order to improve the convergence of the model.\\
The vanishing/exploiding gradients effect is remarkably avoided trough the regularization effect produced by the partial deep Feed-forward Neural Networks (FNNs)\nomenclature[z-FNN]{FNN}{Feed-forward Neural Network} conforming such connections usually referred as \textit{skip connections}. Our model lacks these connections, making necessary the use of a different effective regularization technique.\\

In recent years, \textit{Batch Normalization} (BN) \cite{Ioffe2015BatchShift}\nomenclature[z-BN]{BN}{Batch Normalization} has been the most widely used regularization technique \cite{ Litjens2017AAnalysis}. The successful BN application depends on employing relatively large batch sizes during training since each BN layer needs to extract (ideally) unbiased statistics (mean and variance) from each input batch.\\
However, the usually small amount of available data and the large size of medical volumes force optimization of the network parameters on mini-batches. The use of mini-batches causes perturbations and high variance in the training error particularly, when the residual connections are not present in the network \cite{Klambauer2017Self-NormalizingNetworks}.\\

To alleviate this problem, we propose to include the \textit{Self-Normalizing Neural Networks} (SNNs) strategy~\cite{Klambauer2017Self-NormalizingNetworks} to our 3D-CNN + FCLs model.\\
SNNs leads to automatically normalized networks by preserving the activation close to zero mean and unit variance. Besides, \citet{Klambauer2017Self-NormalizingNetworks} proved that even in the worst-case scenarios, gradients could neither vanish nor explode.\\

Therefore, we implement our model to adjust it to the design conditions defined in \cite{Klambauer2017Self-NormalizingNetworks}. Namely, employing the following specific \textit{activation function}, \textit{weights initialization} and \textit{dropout}:
\begin{description}
\item \textit{Activation Function}: \textit{Scaled Exponential Linear Units} (SELUs)\nomenclature[z-SELU]{SELU}{Scaled Exponential Linear Units} given by

\begin{equation}\label{ch4:eq:selu}
\centering
    \mbox{SELU}(x)= \lambda
    \begin{cases}
      x &  if \hspace{10pt} x>0
      \\
      
      \alpha e^x-\alpha &  if \hspace{10pt} x \leqslant 1 
    \end{cases}
\end{equation}

being $\lambda$ = 1.0507 and $\alpha$ = 1.6733, which are fixed to assure and activation of unit variance and zero mean. A different configuration implies different normalization parameters.
\item \textit{Weights Initialization}:  It is needed to assure that at initialization, the first and second moments of the weights are zero and one, respectively. Because of this, the weights are drawn from the normal distribution,
\begin{equation}
    \mathcal{N}\bigg(0,\frac{1}{n}\bigg),
\end{equation}
being $n$ the number of units at each layer.
\item \textit{Alpha Dropout}: Standard dropout does not work properly with SELU activation (it is not possible to keep the mean and variance at the original values). The proposed alpha dropout technique changes the activation, $x$, as
  \begin{align}
  x &= a \big(\alpha\cdot d + \alpha' (1-d)\big) + b\\
  d &\sim\mathcal{B}(1,q), \qquad q:=\text{dropout rate}  \\
  a &= \big(q + \alpha'^2 q (1-q)  \big)^{-\frac{1}{2}} \\
  b &= - \big(q + \alpha'^2 q(1-q)^{-\frac{1}{2}}\big) \big( (1-q)\alpha' \big),
\end{align}
being $\alpha' = -\lambda \cdot \alpha = -1.7581$ and $\mathcal{B}$ a binomial distribution.  \nomenclature[x-Binomial]{$\mathcal{B}$}{binomial distribution}

\end{description}

\subsection{Learning Principle: Uncertainty Weighted Multi-task Loss }\label{ch4:ssec:Multi-loss}

The use of multi-task networks provides several significant advantages towards a more capable training since similar tasks modelled together leverage model convergence in contrast with single-task
models \cite{Ruder2016AnAlgorithms, Ruder2017AnNetworks}. However, the most employed learning strategies do not consider these facts explicitly promoting the appearance of the new methodology to exploit such property and tackle multi-task issues. \cite{ Bragman2018UncertaintyPlanning, Dorent2021LearningDatasets, Kendall2017WhatVision}.\\

In this work, we propose the inference principle of uncertainty weighting to cope with non-trivial aspect: the choice of the influence of each task in the final loss.\\
Traditionally, each task is associated with a weight selected either manually or after an exhaustive grid search. Then, the final loss is computed by the linear combination of each specific task loss and its weight, namely
\begin{equation}
    \mathbf{\mathcal{L}} = \sum_i w_i \mathcal{L}_i,
\end{equation}
being $w$ and $\mathcal{L}$ the weight and loss of each particular task, $i$.\\
This approach is highly influenced by the units and the scale of each task and is extremely computationally intensive and time-consuming.\\

Recently, this problem has been addressed by \citet{Kendall2018Multi-TaskSemantics} proposing the guidelines to compute the weight guiding each specific task loss by the \textit{task-dependent} or \textit{homoscedastic} uncertainty of the predictions (see further details at the description \ref{ch1:desc:uncert}). Besides, this work leverages the quantification of such uncertainty employing DL estimators, which is a hot topic in the literature due to the intractable nature of such models which do not respond to closed solutions. \cite{Mohamed2018PlantingLearning}.\\

Thus, following the success of other medical imaging applications of approaches embracing uncertainty \cite{Abdar2021AChallenges} (e.g., radiotherapy planning \cite{Bragman2018UncertaintyPlanning}, brain imaging \cite{Dorent2021LearningDatasets, Nair2020ExploringSegmentation}). In this work, we adapt to our multi-label classification problem.\\
For this, we apply the principles of estimation theory (see \equationautorefname~ \ref{ch1:eq:theoryRisk})  to  derive  the following  loss function for our model \cite{Kendall2018Multi-TaskSemantics}: 

\begin{equation}\label{ch4:eq:prob_model}
p\big(\mathbf{y} \big| \mathbf{x} \big) = p \big(\mathbf{y}\big|\mathbf{f(x)}; \mathbf{\theta}\big) = \prod_{t = 1}^{T = 25} p \big(y_t \big|f_t(\mathbf{x}); \mathbf{\theta}\big)  
\end{equation}
\nomenclature[g-prod]{$\prod$}{productory}
\nomenclature[a-Sigmoid]{$S$}{Sigmoid function}
\nomenclature[z-CE]{CE}{Cross Entropy}
\nomenclature[z-PReLU]{PReLU}{Parametric Rectified Linear Unit}
\nomenclature[z-CV]{CV}{Cross Validation}
\nomenclature[z-RMSE]{RMSE}{Root Mean Square Error}
where $\mathbf{y}$ are the $25$ labels assigned to each particular example (see \ref{ch4:sec:expAndResults}  below), $\mathbf{f(x)}$ are the outputs of our network when a chest CT image, $\mathbf{x}$, is employed as input. The parameters (to be learned) $\mathbf{\theta}$ are used as network weights. Therefore, the uncertainty in our predictions, $f_t$, for each task $t$ is measured by a probability distribution, Namely:\\
In the case of the nodule counting tasks, the probability takes the form: 

\begin{equation}\label{eq:normal}
    p\big(y \big |\mathbf{x};\mathbf{\theta} \big) = \mathcal{N}\big(f_t(\mathbf{x})\big|\mathbf{\theta},\sigma_t^2 \big) = \frac{1}{2\sigma_t^2} \exp\bigg(-\frac{\big(y - f_t(\mathbf{x},\mathbf{\theta})\big)^2}{\sigma_t}\bigg)
\end{equation}

While for the binary tasks, the squashed version of a \textit{sigmoid}, $S$, is employed 
\begin{equation}\label{eq:scale_sigmoid}
     p\big(y \big |\mathbf{x};\mathbf{\theta} \big) = S\bigg(\frac{f_t(\mathbf{x})\big | \mathbf{\theta}}{\sigma_t^2}   \bigg) = S\bigg( \frac{f_t(x,\mathbf{\theta})}{\sigma_t^2} \bigg)^y   \bigg(1 - S \bigg( \frac{f_t(x,\mathbf{\theta})}{\sigma_t^2} \bigg) \bigg)^{1 - y}
\end{equation}

The task-specific noise parameters, $\sigma_t$, are inferred to model the amount of noise in the ouputs. Taking this into consideration, Eq.~\ref{ch4:eq:prob_model} can be expressed as:

\begin{equation}\label{eq:prodEq}
     p \big(\mathbf{y} \big| \mathbf{f(x)}; \mathbf{\theta}\big) = \prod_{r=1}^{R=5} \mathcal{N}\big(f_r(\mathbf{x})\big|\mathbf{\theta},\sigma_r^2 \big) \prod_{b=6}^{B=25} S\bigg(\frac{f_b(\mathbf{x})\big | \mathbf{\theta}}{\sigma_b^2}   \bigg),
\end{equation}
where $r$ and $b$ correspond with the regression and binary tasks, respectively.

Given the model of Eq.~\ref{eq:prodEq}, we establish as loss to optimize, the following log-likelihood function of the parameters (see \cite{Kendall2018Multi-TaskSemantics} for details), $\mathcal{L}(\mathbf{\theta}, \mathbf{\sigma_R}, \mathbf{\sigma_B})$: 
\begin{equation} \label{ch4:eq:finall_loss}
\begin{split}
    \mathcal{\mathcal{L} \big(\mathbf{\theta}, \mathbf{\sigma_R}, \mathbf{\sigma_B}}\big) \approx \sum_{n=0}^N \Bigg[ \sum_{r=1}^{R=5}\frac{1}{2\sigma_r^{2}}\mathcal{L}_r \big(\mathbf{\theta},\mathbf{x} \big) + \sum_{b=6}^{R=25}\frac{1}{\sigma_b^{2}}\mathcal{L}_b(\mathbf{\theta},\mathbf{x})  + \sum_{r=1}^{R=5} log(\sigma_r) + \sum_{b=6}^{R=25} log(\sigma_b) \Bigg],
\end{split}
\end{equation}

being $N$ the number of examples and $\mathcal{L}_r(\mathbf{\theta},\mathbf{x}) = \big|\big| y_r - f_r(x, \mathbf{\theta}) \big|\big|^2$, the loss associated with the regression tasks. While the binary loss is defined as the \textit{Cross-Entropy} (CE), $\mathcal{L}_r(\mathbf{\theta},\mathbf{x}) = CE\big(y = [0,1], f(\mathbf{x},\mathbf{\theta}) \big)$. It is important to remark that the defined loss function (Eq.~\ref{ch4:eq:finall_loss}) provides an additional regularization term by including the logarithms of the noise factors for all the tasks, $\sigma$'s, which is added to the regularization implicit to the network (see section above). This way the weights of the network, $\mathbf{\theta}$, are preserved to zero mean and unit variance, which can be added to the model as $p(\mathbf{\theta}) = \mathcal{N}(\mathbf{0,1})$.

\subsection{Pre-procesing}\label{ch4:ssec:preproc}
The proposed architecture takes as inputs CT scans of $128\times 128\times 64$ voxels (see \ref{ch4:ssec:ModelArq}). Because of this, the original CT volumes are cropped. In this process, we extract the lungs by preserving the rib cage VOI as described in Section \ref{ch2:sec:Methods} \cite{Gordaliza2018}. Then, the resulting volume is sampled to the required size.\\

Besides, we augment the original data to amend the need of thousands of CT-scans to learn the model parameters in an environment with a minimal dataset. This process is performed online. During training, each input volume is augmented applying three transformations available in the \textit{DLTK} framework~\cite{Pawlowski2017DLTK:Images}: elastic transformation, the addition of Gaussian noise and flipping in the three spatial directions.

\section{Experiments and Results}\label{ch4:sec:expAndResults}

In order to measure the performance of the proposed model, a $5$-fold Cross-Validation (CV) is employed. Each fold is composed of $4$ CT volumes of each single subject, which lead to training sets of $13$ subjects and $52$ CT scans.\\
Besides, to compare the model behaviour against the state-of-the-art approaches, we perform the CV over the proposed model (\textit{SELU}) and a modified version which employs Parametric Rectified Linear Unit (PReLU) \cite{He2015DelvingClassification}, BN \cite{Ioffe2015BatchShift} and standard dropout \cite{Srivastava2014Dropout:Overfitting}, referred as \textit{BN+PReLU}.\\
In both cases, models are trained through $10.000$ iterations via ADAM optimizer, a learning rate of $10^{-5}$ and a mini-batch size of $n=1$. In the case of \textit{BN+PReLU}, we employed the standard parameters for ADAM~\cite{Kingma2014Adam:Optimization} and a dropout rate of $0.5$, while for the proposed model $\beta_2$ and $\epsilon$ were modified to $0.9$ and $0.01$ and alpha-dropout rate was settled to $0.1$ ~\cite{Klambauer2017Self-NormalizingNetworks}. \\

\begin{figure}
 \centering
  \includegraphics[width=0.8\linewidth,keepaspectratio]{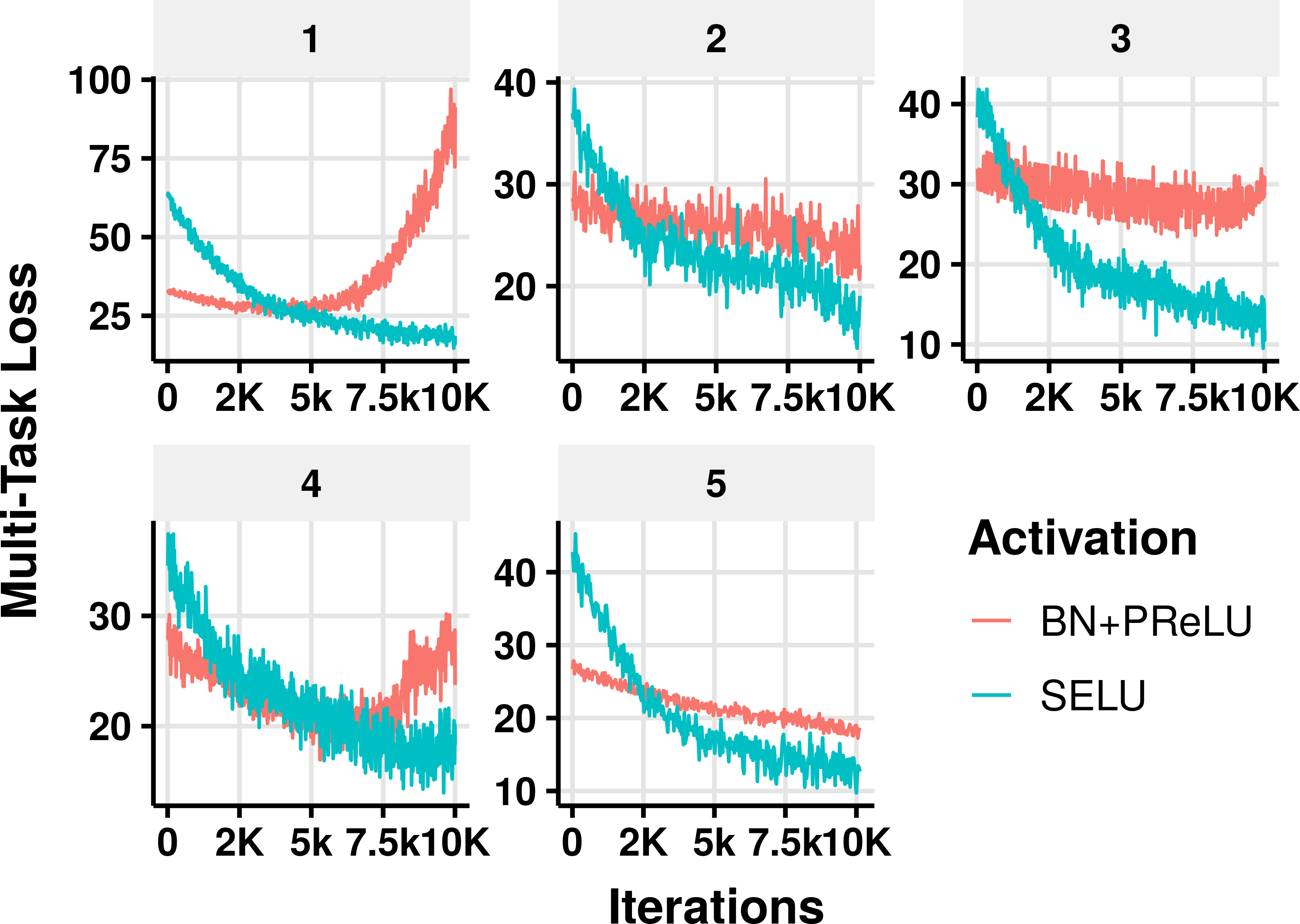}
  \caption[Multi-Task at validation]{Validation loss for $5$-fold Cross-Validation. In red, the standard model which employs \textit{Batch Normalization} \cite{Ioffe2015BatchShift} and  \textit{BN+PReLU}. In blue, our proposed model. \textit{SELU}.}
  \label{ch4:fig:loss_chart}
\end{figure}

In \figurename~\ref{ch4:fig:loss_chart}, we show the loss of the validation data per fold when employing the BN+PReLU (in red) and the proposed model, referred to as SELU (in blue). The SELU loss is always more extensive at the first iterations, although at the final iterations, it is always smaller than PReLU.\\

The inference error at convergence is estimated by the Root Mean Square Error (RMSE) for the five nodules count tasks and the $F_{1}$-score for the twenty binary tasks. Table \ref{ch4:tab:Lobe_results} presents the results per lung lobe and \tablename~\ref{ch4:tab:Fold_results}, by fold. 
 The five top rows represent lobes or folds; the last one, the average per column, shows the results for each manifestation when employing the \textit{BN+PReLU} or the proposed model.\\
 Results are very similar for both models: no significant statistical differences were found, $p \nleq 0.05$, for paired t-test of each kind of manifestation. Nevertheless, the average results are better for the proposed model, presenting a higher variance in most cases.\\  
The tables present a \textit{RMSE} around the unit for nodule counting and a good balance between the precision and the recall for the binary tasks. This way, the nodule counting task reaches \textit{RMSE}'s between $1.81-0.5$ ($2.21-0.92$) at lobe level and $1.09-0.45$ ($1.22-0.41$) at fold level for the proposed model (\textit{BN+PReLU}). The binary tasks reaches $F_1$-scores within $0.98-0.85$ ($0.98-0.74$) at lobe level and $0.98-0.79$ ($0.97-0.83$) for \textit{SELU} (\textit{BN+PReLU}). 

\begin{table}[h]

\centering\rowcolors{2}{gray!6}{white}

{

\resizebox{\linewidth}{!}{
\begin{tabular}{c|cc|cc|cc|cc|cc}
\rowcolor[HTML]{9B9B9B} 
\cellcolor[HTML]{343434}{\color[HTML]{FFFFFF} \textbf{Manifestation/} } & \multicolumn{2}{c|}{\cellcolor[HTML]{9B9B9B}{\color[HTML]{343434} \textbf{Nodules} [RMSE] }} & \multicolumn{2}{c|}{\cellcolor[HTML]{9B9B9B}{\color[HTML]{343434} \textbf{Cavitations} [$F_1$]}} & \multicolumn{2}{c|}{\cellcolor[HTML]{9B9B9B}{\color[HTML]{343434} \textbf{Conglomeration} [$F_1$]}} & \multicolumn{2}{c|}{\cellcolor[HTML]{9B9B9B}{\color[HTML]{343434} \textbf{Consolidations} [$F_1$]}} & \multicolumn{2}{c|}{\cellcolor[HTML]{9B9B9B}{\color[HTML]{343434} \textbf{Tree in bud} [$F_1$]}}  \\ 
\multirow{1}{*}{\cellcolor[HTML]{343434}{\color[HTML]{FFFFFF} \textbf{Lobe} }} & \cellcolor[HTML]{C0C0C0}{\color[HTML]{343434} BN+PReLU} & \cellcolor[HTML]{EFEFEF}{\color[HTML]{343434} SELU} & \cellcolor[HTML]{C0C0C0}{\color[HTML]{343434} BN+PReLU} & \cellcolor[HTML]{EFEFEF}{\color[HTML]{343434} SELU} & \cellcolor[HTML]{C0C0C0}{\color[HTML]{343434} BN+PReLU} & \cellcolor[HTML]{EFEFEF}{\color[HTML]{343434} SELU} & \cellcolor[HTML]{C0C0C0}{\color[HTML]{343434} BN+PReLU} & \cellcolor[HTML]{EFEFEF}{\color[HTML]{343434} SELU} & \cellcolor[HTML]{C0C0C0}{\color[HTML]{343434} BN+PReLU} & \cellcolor[HTML]{EFEFEF}{\color[HTML]{343434} SELU} \\ \hline
\cellcolor[HTML]{9B9B9B}{\color[HTML]{343434} \textbf{Left Inf.}} & $1.08_{0.66}$ & $0.94_{0.56}$ & $0.97_{0.13}$ & $0.95_{0.18}$ & $0.95_{0.15}$ & $0.95_{0.16}$ & $0.89_{0.18}$ & $0.88_{0.19}$ & $0.94_{0.14}$ & $0.93_{0.15}$ \\
\rowcolor[HTML]{EFEFEF} 
\cellcolor[HTML]{9B9B9B}{\color[HTML]{343434} \textbf{Left Sup.}} & $0.92_{0.64}$ & $1.25_{0.84}$ & $0.95_{0.15}$ & $0.95_{0.16}$ & $0.86_{0.22}$ & $0.91_{0.17}$ & $0.90_{0.19}$ & $0.89_{0.23}$ & $0.94_{0.16}$ & $0.99_{0.07}$ \\
\cellcolor[HTML]{9B9B9B}{\color[HTML]{343434} \textbf{Right Inferior}} & $2.21_{0.38}$ & $1.28_{0.89}$ & $0.98_{0.08}$ & $0.95_{0.13}$ & $0.94_{0.12}$ & $0.96_{0.13}$ & $0.91_{0.14}$ & $0.96_{0.1}$ & $0.93_{0.15}$ & $0.89_{0.2}$ \\
\rowcolor[HTML]{EFEFEF} 
\cellcolor[HTML]{9B9B9B}{\color[HTML]{343434} \textbf{Right Middle}} & $1.02_{0.77}$ & $0.5_{0.58}$ & $0.96_{0.11}$ & $0.98_{0.13}$ & $0.90_{0.12}$ & $0.91_{0.11}$ & $0.96_{0.13}$ & $0.93_{0.16}$ & $0.94_{0.13}$ & $0.85_{0.21}$ \\
\cellcolor[HTML]{9B9B9B}{\color[HTML]{343434} \textbf{Right Superior}} & $2.20_{1.01}$ & $1.81_{0.88}$ & $0.88_{0.20}$ & $0.94_{0.18}$ & $0.87_{0.19}$ & $0.93_{0.14}$ & $0.89_{0.16}$ & $0.93_{0.15}$ & $0.74_{0.25}$ & $0.89_{0.18}$ \\ \hline
\rowcolor[HTML]{EFEFEF} 
\cellcolor[HTML]{9B9B9B}{\color[HTML]{343434} \textbf{Total}} & {\color[HTML]{343434} $1.34_{0.69}$} & {\color[HTML]{343434} $\mathbf{1.16_{0.75}}$} & {\color[HTML]{343434} $0.95_{0.13}$} & {\color[HTML]{343434} $0.95_{0.16}$} & {\color[HTML]{343434} $0.91_{0.16}$} & {\color[HTML]{343434} $\mathbf{0.93_{0.14}}$} & {\color[HTML]{343434} $0.91_{0.16}$} & {\color[HTML]{343434} $\mathbf{0.92_{0.17}}$} & {\color[HTML]{343434} $0.90_{0.17}$} & {\color[HTML]{343434} $\mathbf{0.91_{0.16}}$}
\end{tabular}
}
}
\rowcolors{2}{white}{white}

  \caption[Manifestations predictions per lobe]{Predictions of the reported Tuberculosis manifestations per lung lobe (rows), and compared models, \textit{Batch Normalization and PReLU}(BN+PReLU) and our model, referred as SELU (columns). }%
\label{ch4:tab:Lobe_results}
\end{table}

\begin{table}[h]
 
\centering\rowcolors{2}{gray!6}{white}
{

\resizebox{\linewidth}{!}{
\begin{tabular}{ccccccccccc}
\rowcolor[HTML]{9B9B9B} 
\cellcolor[HTML]{343434}{\color[HTML]{FFFFFF} \textbf{Manifestation/}} & \multicolumn{2}{c}{\cellcolor[HTML]{9B9B9B}\textbf{Nodules} [RMSE]} & \multicolumn{2}{c}{\cellcolor[HTML]{9B9B9B}{\color[HTML]{343434} \textbf{Cavitations [$F_1$]}}} & \multicolumn{2}{c}{\cellcolor[HTML]{9B9B9B}{\color[HTML]{343434} \textbf{Conglomeration} [$F_1$]}} & \multicolumn{2}{c}{\cellcolor[HTML]{9B9B9B}{\color[HTML]{343434} \textbf{Consolidation} [$F_1$]}} & \multicolumn{2}{c}{\cellcolor[HTML]{9B9B9B}\textbf{Tree in bud} [$F_1$]} \\
\multirow{1}{*}{\cellcolor[HTML]{343434}{\color[HTML]{FFFFFF} \textbf{Fold}}} & \cellcolor[HTML]{C0C0C0}BN+PReLU & \cellcolor[HTML]{EFEFEF}SELU & \cellcolor[HTML]{C0C0C0}BN+PReLU & \cellcolor[HTML]{EFEFEF}SELU & \cellcolor[HTML]{C0C0C0}BN+PReLU & \cellcolor[HTML]{EFEFEF}SELU & \cellcolor[HTML]{C0C0C0}BN+PReLU & \cellcolor[HTML]{EFEFEF}SELU & \cellcolor[HTML]{C0C0C0}BN+PReLU & \cellcolor[HTML]{EFEFEF}SELU \\
\multicolumn{1}{c|}{\cellcolor[HTML]{9B9B9B}{\color[HTML]{343434} \textbf{1}}} & $0.73_{0.84}$ & \multicolumn{1}{c|}{$0.85_{0.35}$} & $0.88_{0.11}$ & \multicolumn{1}{c|}{$0.88_{0.12}$} & $0.90_{0.13}$ & \multicolumn{1}{c|}{$0.92_{0.12}$} & $0.88_{0.18}$ & \multicolumn{1}{c|}{$0.83_{0.22}$} & $0.83_{0.22}$ & $0.79_{0.23}$ \\
\rowcolor[HTML]{EFEFEF} 
\multicolumn{1}{c|}{\cellcolor[HTML]{9B9B9B}{\color[HTML]{343434} \textbf{2}}} & $1.15_{0.89}$ & \multicolumn{1}{c|}{\cellcolor[HTML]{EFEFEF}$1.09_{0.83}$} & $0.86_{0.23}$ & \multicolumn{1}{c|}{\cellcolor[HTML]{EFEFEF}$0.88_{0.22}$} & $0.94_{0.17}$ & \multicolumn{1}{c|}{\cellcolor[HTML]{EFEFEF}$0.93_{0.18}$} & $0.93_{0.15}$ & \multicolumn{1}{c|}{\cellcolor[HTML]{EFEFEF}$0.93_{0.18}$} & $0.97_{0.08}$ & $0.97_{0.08}$ \\
\multicolumn{1}{c|}{\cellcolor[HTML]{9B9B9B}{\color[HTML]{343434} \textbf{3}}} & $0.41_{0.34}$ & \multicolumn{1}{c|}{$0.23_{0.39}$} & $0.85_{0.12}$ & \multicolumn{1}{c|}{$0.87_{0.11}$} & $0.89_{0.19}$ & \multicolumn{1}{c|}{$0.97_{0.11}$} & $0.96_{0.11}$ & \multicolumn{1}{c|}{$0.98_{0.03}$} & $0.95_{0.14}$ & $0.96_{0.12}$ \\
\rowcolor[HTML]{EFEFEF} 
\multicolumn{1}{c|}{\cellcolor[HTML]{9B9B9B}{\color[HTML]{343434} \textbf{4}}} & $1.22_{0.6}$ & \multicolumn{1}{c|}{\cellcolor[HTML]{EFEFEF}$0.78_{0.74}$} & $0.94_{0.15}$ & \multicolumn{1}{c|}{\cellcolor[HTML]{EFEFEF}$0.9_{0.19}$} & $0.9_{0.19}$ & \multicolumn{1}{c|}{\cellcolor[HTML]{EFEFEF}$0.92_{0.15}$} & $0.88_{0.18}$ & \multicolumn{1}{c|}{\cellcolor[HTML]{EFEFEF}$0.87_{0.18}$} & $0.87_{0.21}$ & $0.94_{0.15}$ \\
\multicolumn{1}{c|}{\cellcolor[HTML]{9B9B9B}{\color[HTML]{343434} \textbf{5}}} & $0.41_{0.8}$ & \multicolumn{1}{c|}{$0.45_{0.8}$} & $0.93_{0.17}$ & \multicolumn{1}{c|}{$0.94_{0.17}$} & $0.94_{0.14}$ & \multicolumn{1}{c|}{$0.96_{0.12}$} & $0.90_{0.18}$ & \multicolumn{1}{c|}{$0.94_{0.14}$} & $0.91_{0.18}$ & $0.92_{0.17}$ \\ \hline
\rowcolor[HTML]{EFEFEF} 
\multicolumn{1}{c|}{\cellcolor[HTML]{9B9B9B}{\color[HTML]{343434} \textbf{Total}}} & $0.78_{0.69}$ & \multicolumn{1}{c|}{\cellcolor[HTML]{EFEFEF}$\mathbf{0.68_{0.62}}$} & $0.89_{0.16}$ & \multicolumn{1}{c|}{\cellcolor[HTML]{EFEFEF}$\mathbf{0.90_{0.17}}$} & $0.91_{0.16}$ & \multicolumn{1}{c|}{\cellcolor[HTML]{EFEFEF}$\mathbf{0.94_{0.16}}$} & $0.91_{0.16}$ & \multicolumn{1}{c|}{\cellcolor[HTML]{EFEFEF}$0.91_{0.15}$} & $0.91_{0.17}$ & $\mathbf{0.92_{0.15}}$
\end{tabular}}}
\rowcolors{2}{white}{white}
 \caption[Manifestations predictions per fold]{Predictions of the reported Tuberculosis manifestations per fold lobe (rows), and compared models, \textit{Batch Normalization and PReLU}(BN+PReLU) and our model, referred as SELU (columns).}%
\label{ch4:tab:Fold_results}%
\end{table}

\section{Discussion}

The models training in this work presents a promising inference of the radiologist reports. Although there are no significant statistical differences between the compared models for our reduced dataset experiment, our model presents some advantages.\\

By avoiding normalization layers, the number of parameters of the model is reduced, so it is the computational complexity. The novel loss function presents a better model convergence that assures a more robust training and avoids possible overfitting problems common to the state-of-the-art approaches.\\
Besides, the results are consistent with known facts about the disease. Specifically, TB spreads inside the lungs starting at the right superior lobe, usually more affected by diverse lesions and many nodules, creating difficulties in the report generation, which is reflected as poorer predictions at the region. Similarly, the inference of the severely diseased subjects' reports is poorer when compared to those moderately affected by the disease, as can be observed for subjects \#$2$ and  \#$3$ (\figurename~\ref{ch4:fig:loss_chart}).\\

We acknowledge that the number of subjects used to validate the study is limited, and further validation is needed.\\
Namely, prospective studies are a must for new applications to ease the black box effect produced by the DL models (learnt from limited samples) that limits their explainability and generalization (see \nameref{ch1:ssec:list:rulesvsML}).\\

The lack of generalization is not only due to the scarcity of observations but also to their simplified characterization as $p(y|x)$ instead of $p(x,y)$ since, as explained (see \ref{ch1:ssec:list:rulesvsML}), the latter is hardly tractable using the statistical learning framework.\\ 
Although it is not ideal due to the data scarcity, circumvention of the generalization problem is possible by implementing a model for each new domain after applying \textit{transfer learning} \cite{Raghu2019Transfusion:Imaging} or from scratch  (e.g., datasets belonging to other animal models in the context of ERA4TB or application to other lung diseases). Remarking such an approach is valid when the model just intends to achieve the highest possible predictive power on data belonging to the same distribution \cite{Breiman2001StatisticalCultures, Scholkopf2021TowardLearning}.\\
On the contrary, establishing informative (explainable) models can be vital for the correct understanding of the disease. However, the statistical learning framework is insufficient to establish causal relationships between the correlations found by the mere statistical characterization $p(y|x)$ or $p(x,y)$.\\

Not in vain, the following chapter presents an approach based on graphical causality embedding statistical learning.\\
The meeting of both modelling frameworks establishes a causal model governing $p(x,y)$, as shown by the results of the method in terms of generalization and ability to generate realistic images for the specific case of TB-infected lung imaging in different animal models.

\section{Conclusion}
The chapter introduces a novel methodology for multi-label classification, which enables the inclusion of \textit{Self-Normalizing Networks} within 3D CNNs. This approach allows an improved extraction of relevant features on large medical volumes through multi-task learning guided by the uncertainty in the model predictions. Therefore, demonstrating that DL models with designs adapted to the context of this thesis allow the extraction of essential information for the characterization of TB. This fact represents a significant advance in the field, even bearing in mind the limitations mentioned for \textit{explicability} and \textit{generalization} terms addressed in the next chapter.

\chapter{Translational Lung Imaging Analysis Through Disentangled Representations}
\label{ch5:Intro}
The development of new treatments often requires experiments with translational animal models using (pre)-clinical imaging to characterize inter-species pathological processes (see \nameref{ch1:sec:Erad:ERA4TB}).\\ 
To accelerate the development, this thesis presents a collection of novel methods to automatize related tasks and promote explainable Tuberculosis (TB) models for imaging biomarkers extraction.\\ 
Thus, the previous chapter (\nameref{ch4:Intro}) shows how Deep Learning (DL) models are commonly used to automate relevant information extraction from the images. However, as usual for automation tasks, the proposed DL model is specific for a domain (animal model) (see Section \ref{ch1:shifts}). This is due to low generalizability and explainability, a product of their entangled design (see \nameref{ch1:ssec:list:rulesvsML}).\\ 
Consequently, it is quite complicated to take advantage of the proven high capacity of DL to discover statistical relationships \cite{Arjovsky2020InvariantMinimization, Lopez-Paz2017DiscoveringImages} from inter-species images. Note that discovering such relationships would be essential to establish a shared disease marker beyond the particular manifestations of TB for each animal model.\\

In this chapter, we present a model to leverage such capacities. Concretely, we extract it from our publicly available work: \textit{Translational Lung Imaging Analysis Through Disentangled Representations} \cite{Gordaliza2021TranslationalRepresentations}.\\

This model extracts disentangled information from images of different animal models. Such an approach allows characterizing common mechanisms of the data generation, as is proven synthesizing realistic chest Computed Tomography (CT) volumes.\\

Our method stands at the intersection between deep generative models, disentanglement and causal representation learning. In this thesis context, it is optimized from images of pathological lung infected by Tuberculosis and is able: a) from an input slice, infer its position in a volume, the animal model to which it belongs, the damage present and even more, generate its lung delimitation mask (similar overlap measures to the \textit{nnU-Net} \cite{Isensee2021NnU-Net:Segmentation}), b) generate realistic lung images by setting the above variables and c) generate counterfactual images, as healthy versions of a damaged input slice.     
\section{Introduction}
Understanding disease progression is essential to develop new treatments ( \nameref{ch1:ssec:TBspectrum}). The longitudinal characterization of animal models in (pre-)clinical experiments is crucial \cite{Yang2021OneChemotherapeutics}. For this, we need to extract comparable biomarkers in similar phase of the pathology (\figurename~\ref{ch1:fig:Radiology_test}). We also need to prove the existence of similar pathophysiological mechanisms modulating common causal factors, that give rise to the variability of trial outcomes (see \nameref{ch1:sec:errad}). \\

In this context, medical imaging techniques enable the extraction of indicators (imaging biomarkers) from \textit{in vivo} studies \cite{Willmann2008MolecularDevelopment}. For example, the number of \textit{Mycobacterium tuberculosis} (\Mtb) colonies present in a subject can be approximated\footnote{The number of colonies correlates with some radiological manifestations of TB} from the damaged lung volume in an image of a human, primate, or mouse \cite{Yang2021OneChemotherapeutics} (see \ref{ch1:sec:MITB}).\\
The images contain meaningful information to interpret the mentioned physiological process. However, their analysis is tedious and automation is advantageous to process the vast amount of data produced during the trials. Thus, developing Artificial Intelligence (AI) systems that can not only automate the extraction of particular markers for each animal model (e.g., the damaged lung volume) but are also capable of inferring the common agents of such particular indicators (e.g., bacterial burden) is essential (see \nameref{ch1:sec:CAD}).\\
Although AI, has eased the process \cite{Hinton2018DeepCare, Zhou2021APromises}, some design premises has lessened its inference capabilities. In particular, DL models excel at extracting the statistical dependence between input-output pairs, i.e.,$(x_i, y_i) \in \mathcal{X,Y}$, from assumed \textit{independent and identically distributed (i.i.d.)} observational data \cite{Peters2017ElementsAlgorithmsb} (see \ref{ch1:shifts}). 
Such success has leaned the model designs towards an insufficient representation learning strategy \cite{Bengio2013RepresentationPerspectives}. Namely, the discovery of statistical dependence between specific data pair samples is priorized rather than the understanding of the physical model generating the whole data population (e.g., physiological mechanisms).\\

Since the i.i.d. assumption is fragile, data scarcity (especially for labelled data) and data mismatch are characteristic of the medical imaging field, well-known distribution shifts \cite{Castro2020CausalityImaging} between data employed at training, validation/test phases and ``real world" data are usual.\\
Under this scenario, the models tend to learn correlated representations that only hold for specific environments or domains, namely \textit{spurious correlations} \cite{Arjovsky2020InvariantMinimization}. 
Since (as a mantra)\textit{``correlation does not imply causation"}, such flaws cause ruinous effects \cite{DeGrave2021AISignal, Roberts2021CommonScans} for generalisation, transferability and explainability purposes \cite{Scholkopf2021TowardLearning}.\\

More formally, naive DL models maximize a joint distribution, $p(X, Y)$ or $p(X)$ (self-supervision), characterized by an entangled representation of the input. Namely, if $X$ and $Y$ correlate during training without necessarily derive from a causal representation ($X \rightarrow Y$), $p(X, Y)$ can adopt numerous (specific domain dependant) factorization forms  \cite{Goyal2021InductiveCognition}. Thus, there is a need to implement independent models to process the information in related domains (particularly, lung CT images of TB animal models). Such models are put in common through \textit{posthoc} analysis, losing possible data synergies.\\
In general, state-of-the-art learning strategies, mitigate this issue by shrinking the $p(X, Y)$ solutions space. To this aim, models are enriched injecting inductive biases (e.g., CNNs assume spatial correlation \cite{Dumoulin2018ALearning}, equivariant transformations \cite{Bronstein2021GeometricGauges}),
to facilitate the discovery of more meaningful and disentangled representations \cite{Liu2021ADomain}.\\
The above mentioned techniques simulate human cognition. Under the realms of causality \cite{Pearl2011Causality:Edition}, human cognition arranges the proper biases to extract a limited number of relevant factors related to a task holding among different environments.\\

Designing AI systems can follow a similar causal perspective. We can introduce specific biases to shrink the solution space. Thus, in this chapter, we consider: a) the strongly hierarchical nature of the human visual system and b) the data generation process. Such an approach intends to mimic the radiologists' tasks, who take into account specific patient factors (i.e., clinical history, sex, age) beyond the image \textit{per se}.\\
This approach yields more effective disentangled representations of the input \cite{Scholkopf2021TowardLearning}.\\

In particular, we intend to identify the unique mechanisms in the generation of translational imaging of lung Computed Tomography (CT) images and their corresponding segmentation masks (\figurename~\ref{ch5:fig:DAG}). We employ three different animal models (mouse, primate and human) infected by  \Mtb \cite{Pai2016Tuberculosis}.\\
From a simplified radiological point of view,  mammals' lungs share texture and shape features. We model these shared characteristics as an effect of the same causative factors, for example, the bacterial load (see \ref{ch1:fig:TB_tests}).\\
To prove the benefits of our strategy, we show how after optimizing the model employing a small limited number of volumes, our design can: 
\begin{itemize}
    \item Produce a very accurate reconstruction of the input images and generate  suitable segmentation masks (\figurename \ref{ch5:fig:lungs_seg},  \tablename~\ref{ch5:tbl:optimiza_results_DSC} and \tablename~\ref{ch5:tbl:optimiza_results_HD}). 
 
    \item Generate new realistic images of the three models controlling the lung damage on each, which implies the proper characterization of  the disentangled variables (\figurename~\ref{ch5:fig:generation}).

    \item Generate counterfactual images \cite{Cohen2021GifsplanationX-rays, Schutte2021UsingImages} of damaged lungs. Namely, the model is able to capture the meaningful representations of an input image to convert it into a healthy version by intervening on the damage variable value.  
\end{itemize}

\section{Methods}
\begin{figure}
 \centering
  \includegraphics[width=0.8\linewidth,keepaspectratio]{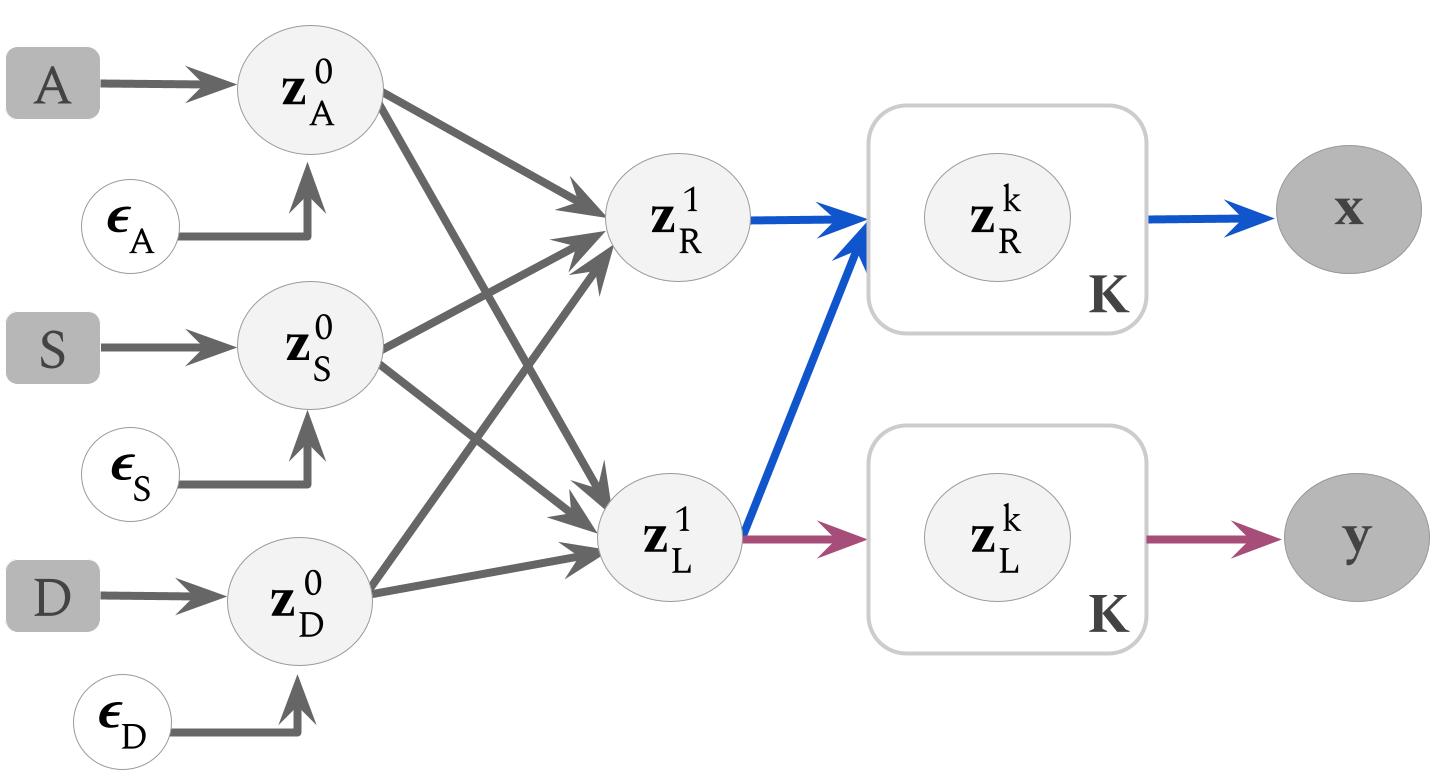}
  \caption[Generative DAG]{Direct Acyclic Graph (DAG) representing the generation of pathological lung CT images $\mbfx$, and their segmentation masks $\mbfy$. Both generated from a latent variables hierarchy at different resolutions scales, $K$, governed by three factors, i.e., animal model, $A$, the relative position of the axial slice, $S$, and the estimated lung damage caused by \Mtb, $D$.}
  \label{ch5:fig:DAG}
\end{figure}

We define a generative model in which the high dimensional texture and shape features that can be extracted from lung CT images and their corresponding segmentation masks are a result of the causal Direct Acyclic Graph (DAG) presented in \figurename~\ref{ch5:fig:DAG}.
\nomenclature[z-DAG]{DAG}{Direct Acyclic Graph}
\nomenclature[g-noise]{$\epsilon$}{noise}


\begin{figure}
 \centering
  \includegraphics[width=0.8\linewidth,keepaspectratio]{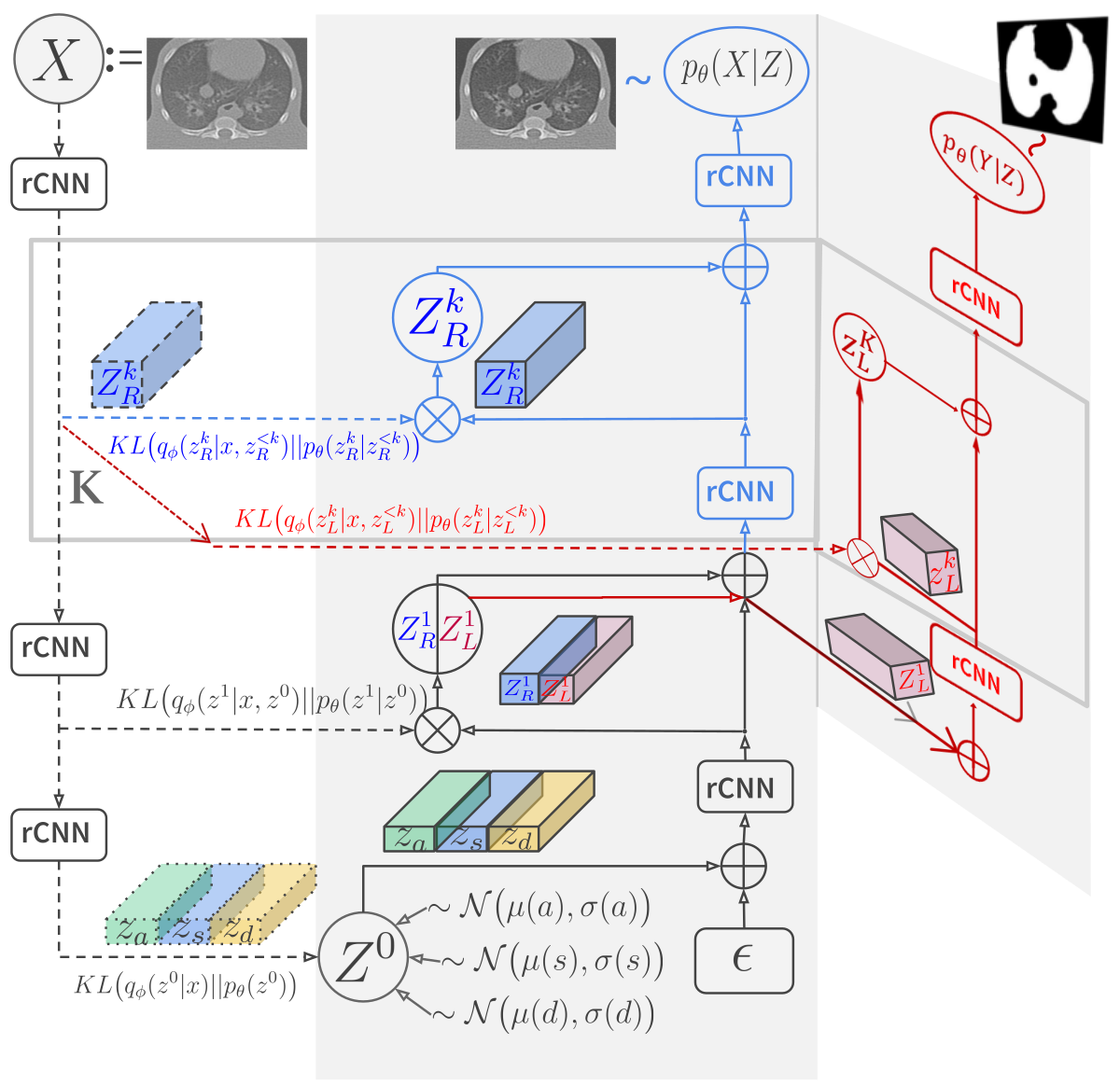}
  \caption[Multi-Task at validation]{ Summarized architecture: Blue and pink represent the image and mask generation branches, $\bigoplus$ features concatenation and $\bigotimes$ $p_\theta$, $q_\phi$ parameters combination during training and \texttt{rCNN} residual blocks.\\
  The encoder is not present for image generation (Section \ref{ssec:genration}). Counterfactual images arise inferring and setting some values at the deeper representation level $\mathbf{z}^0$ (Section \ref{ssec:counter})}
  \label{ch5:fig:arqu_sum}
\end{figure}

The proposed DAG simplify the physical image generation for obvious reasons. All the possible elementary causative factors (specific scanner, comorbidities, subject age, sex, etc.) are reduced to three: the animal model, $A$,  the observed lung axial slice, $S$, and the lung damage, $D$. The causative factors are modelled as three groups of independent variables, $\mathbf{z}^0$, under the noise term, $\epsilon_{\{A,S,D\}}$, which comprises noise and unconsidered variables. The primary variables govern the generative process which follows a part-whole hierarchy \cite{Hinton2021HowNetwork} from low-level representations of the texture and shape features, $\mathbf{z}^1$, to high dimensional ones, $\mathbf{z}^k$, the observed image, $\mbfx$ and the segmentation mask, $\mbfy$. This part-whole hierarchy resembles brain columns functioning \cite{Devlin2018BERT:Understanding, Locatello2020Object-CentricAttention}. Variable superscripts, $\mathbf{z}^k$, symbolize hierarchy levels at the DAG.\\
The plate notation at the DAG represents such upsampling generation. The DAG implements two paths diverging at the first hierarchy level (shared representation path), $\mathbf{z}^1$. The division forces, during optimization, to generate a disentangled representation of shape, $\mathbf{z}_L$ and texture, $\mathbf{z}_R$.  CT images depend on both shape and texture variables (blue path), while the segmentation masks only depend on shape variables (pink path). Then, assuming the independence of the noise terms,  the (\textit{independent causal mechanism} (ICM) principle is fulfilled \cite{Scholkopf2021TowardLearning}) and the following disentangled factorization arise: 
\nomenclature[z-ICM]{ICM}{Independent Causal Mechanism}
\nomenclature[z-KL]{KL}{ Kullback–Leibler divergence}
\begin{equation}\label{eq:factorization}
    p(\mbfx,\mbfy, \mbfz) = p(\mbfx|\mathbf{z}^K_R)p(\mbfy|\mathbf{z}^K_L)p(\mathbf{z}_R^k, )p(\mathbf{z}_L^k)p(\mathbf{z}_R^2|\mathbf{z}_R^1,\mathbf{z}_L^1)p(\mathbf{z}_{R}^1, \mathbf{z}_{L}^1|\mathbf{z}^0)p(\mathbf{z}^0)
\end{equation}
being
\begin{align}\label{eq:latent_scales}
p(\mathbf{z}_R^k)=\prod_{k=3}^{K}p(\mathbf{z}^k_R|\mathbf{z}^{k-1}_R); \quad  \qquad        &  p(\mathbf{z}_L^k)=\prod_{k=2}^{K}p(\mathbf{z}^k_L|\mathbf{z}^{k-1}_L); \quad   p(\mathbf{z^0}) = p(\mathbf{z}_A^0)p(\mathbf{z}_S^0)p(\mathbf{z}_D^0);          
\end{align}

\subsection{Model optimization}

For the above equations, each conditional distribution is parametrized by depthwise convolutional decoders with parameters $\theta$, leveraging a high capacity model (\figurename~\ref{ch5:fig:arqu_sum}) allowing to characterize the unobservable causes of variation ($\mathbf{\epsilon}$) consistent with the available data (lung CT images) \cite{Pawlowski2020DeepInference, Peters2017ElementsAlgorithmsb}. Once the model is optimized, it is possible to modify the disentangled variables to obtain new generated images (\ref{ssec:genration}) and counterfactual images \cite{Cohen2021GifsplanationX-rays, Schutte2021UsingImages} (see \ref{ssec:counter}).\\ 
The computation of the parameters requires optimization through training of the posterior probability, $p_\theta(\mbfz|\mbfx,\mbfy)$, which is intractable. To tackle this issue, we adapt the particular factorization in ~\ref{eq:factorization} to the methodology developed for deep Variational Autoencoders (deep VAEs) \cite{Child2020VeryImages, Kingma2016ImprovingFlow}. In this way, we obtain the best approximate amortized posterior distribution, $q_\phi(z|x)$, being $\phi$ the parameters of the encoder. Notice that the distribution is amortized just from $\mbfx$ (not from $\mbfy$), so we force the model to extract the meaningful mechanism to generate the segmentation masks just from the self-supervisory signal of the image \cite{LeCun2021Self-supervisedIntelligence}. Indeed, we add a segmentation branch in the architecture (\figurename~\ref{ch5:fig:arqu_sum}), dependent on the main branch.\\
Namely, we adopt the Noveau VAE (NVAE) \cite{Vahdat2020NVAE:Autoencoder}. This architecture is carefully designed for hierarchical models. Moreover, it has proven efficacy in approximating posteriors by introducing as inductive bias in the image generating process a deeply hierarchical architecture.  
To this aim, the set of $\mbfz$ variables at each representation level $k$, is divided in smaller sets, $m_{k}$, to get a total of $M$ groups of latent variables. They establish a hierarchical structure within each resolution too to help narrowing the solutions space, being $\mbfz$ the set:
\begin{equation}
\small
\mathbf{z} = \big\{ \{(\mathbf{z}_A, \mathbf{z}_S, \mathbf{z}_D)_0,\mathbf{z}_1,\mathbf{z}_2\,...,\mathbf{z}_{m_{k=0}}\}^0, \{(\mathbf{z}_L, \mathbf{z}_R)_{m+1},...,\mathbf{z}_{m_{k=1}}\big\}^1,..., \{ \mathbf{z}_{m+1}, ...,\mathbf{z}_{m_{k}}\}^k,\{\mathbf{z}_{m+1},...,\mathbf{z}_{M}\}^K \big\}
\end{equation}
Its prior and approximate posterior probability are given by:
\begin{equation}
    p_\theta(\mbfz) = \prod_m p_\theta(\mathbf{z}_m|\mathbf{z}_{m-1}) \qquad q_\phi(\mbfz|\mbfx) = \prod_m q_\phi(\mathbf{z}_m|\mathbf{z}_{m-1},\mbfx).
\end{equation}
Following this formulation, from marginalization of the $\log$ of \ref{eq:factorization} and rearranging terms, we obtain the variational lower bound to optimize (subscripts colors denote each optimization branch):
\begin{equation} \label{eq:loss}
\small
    \mathcal{L(\mbfx,\mbfy)} = \mathbb{E}_{q_\phi(\textcolor{blue}{\mbfz}|\mbfx)}\big[ log p_\theta(\mbfx|\textcolor{blue}{\mbfz}) \big] -KL(q_\phi(\mathbf{z}_0 | x)||p_\theta(\mathbf{z}_0)) + \mathbb{E}_{q_\phi(\textcolor{magenta}{\mbfz}|\mbfx)}\big[ log p_\theta(\mbfy|\textcolor{magenta}{\mbfz}) \big] - \mathbb{E}_{\textcolor{blue}{\mbfz}} \big[KL_{\textcolor{blue}{\mbfz}} \big] - \mathbb{E}_{\textcolor{magenta}{\mbfz}} \big[KL_{\textcolor{magenta}{\mbfz}} \big] 
\end{equation}
\normalsize
being $KL$ the Kullback–Leibler divergence and

\begin{equation}
    \mathbb{E}_{\mbfz} \big[KL_{\mbfz} \big] = \sum_{m}^{M} \mathbb{E}_{q_\phi(\mathbf{z_{m-1}}|\mbfx)} \big[ KL(q_\phi(\mathbf{z}_{m}| \mathbf{z}_{m-1}, \mbfx) || p_\theta(\mathbf{z}_m|\mathbf{z}_{m-1}) ) \big]
\end{equation}
Being $q_\phi(\mathbf{z}_{m-1}|\mbfx)$ the approximate posterior through the hierarchy of $m_{k-1}$ group.\\
Since, NVAE convergence depends on the proper approximation of KL terms (see \cite{Vahdat2020NVAE:Autoencoder}). To this aim, all priors and posterior probabilities are approximated as Normal distributions. Thus, we can write:
\begin{equation}\label{eq:prior_approx}
    p(\mathbf{z}_A^0) \sim \mathcal{N}(\mu(a), \sigma(a)); \qquad p(\mathbf{z}_S^0) \sim \mathcal{N}(\mu(s), \sigma(s));  \qquad   p(\mathbf{z}_D^0) \sim \mathcal{N}(\mu(d), \sigma(d));
\end{equation}






\section{Experiments and Results}

\subsection{Datasets description}\label{ssec:datasets}

The model is optimized employing small datasets: ten lung CT volumes per animal model ($\sim2000$ slices). 
The data used for the optimization phase (training) are axial slices from three models of pathological lungs infected by \Mtb~ The dataset names identify: the animal model, $A$, the data source and the phase as follows $A_{phase}^{source}$).  Namely, the human volumes, $H_{tr}^{CLE}$ , corresponds to the validation data of the 2019 ImageClefMed TB task \cite{DicenteCid2019OverviewAssessment}. The mice images, $M_{tr}^{GSK}$, are provided from \textit{GlaxoSmithKline plc.} (GSK) within the context of the ERA4TB project \cite{ERA4TBconsotium2021ERA4TB}, similarly to the primate ones, $P_{tr}^{PHE}$, from the \textit{Public Health of England} (PHE) \cite{Gordaliza2018, Gordaliza2019AManifestations}.
For testing (twenty volumes per model), $P_{ts}^{PHE}$ and $P_{ts}^{GSK}$, are selected from different cohorts of $P_{tr}^{PHE}$ and $P_{tr}^{GSK}$, while the human dataset, $H_{ts}^{CLE}$ is a partition of the mentioned data. The remaining sets are included to evaluate the model generalisation and transferability capabilities. $M_{ts}^{EXM}$ belongs to a publicly available dataset from the Institute for Experimental Molecular Imaging (ExMI) \cite{Rosenhain2018ASegmentations} which contains healthy subjects at low resolution. Finally, the human dataset, $H_{ts}^{RAD}$, presents subjects with lung damage caused by COVID-19 \cite{Cohen2020COVID-19Collection}.\\ 
Note that all datasets include segmentation masks produced/corrected by trained experts.\\
A detailed description of the different datasets is presented in Table \ref{ch5:tbl:datasets}.

\begin{table}[htpb]
\centering

\arrayrulecolor{white}
\scalebox{0.85}{
\begin{tabular}{c||c||c||c||c||c}
\rowcolor[rgb]{0.2,0.2,0.6} \textcolor{white}{\textbf{Dataset ID}} & \textcolor{white}{\textbf{Phase}}                                                               & \textcolor{white}{\textbf{Animal Model}}                                                           & \textcolor{white}{\textbf{Source}}                                                                   & \textcolor{white}{\textbf{Voxel Spacing [mm]}}                                                 & \textcolor{white}{\textbf{Resolution~}}                                 \\ 
\hhline{=::=::=::=::=::=}
\rowcolor[rgb]{0.945,0.945,0.945} $M^{GSK}_{tr}$                   & \textcolor[rgb]{0.259,0.259,0.259}{Training}                                                    & {\cellcolor[rgb]{0.945,0.945,0.945}}                                                               & {\cellcolor[rgb]{0.945,0.945,0.945}}                                                                 & {\cellcolor[rgb]{0.945,0.945,0.945}}                                                           & {\cellcolor[rgb]{0.945,0.945,0.945}}                                    \\ 
\hhline{-||-||>{\arrayrulecolor[rgb]{0.945,0.945,0.945}}->{\arrayrulecolor{white}}||>{\arrayrulecolor[rgb]{0.945,0.945,0.945}}->{\arrayrulecolor{white}}||>{\arrayrulecolor[rgb]{0.945,0.945,0.945}}->{\arrayrulecolor{white}}||>{\arrayrulecolor[rgb]{0.945,0.945,0.945}}-}
\rowcolor[rgb]{0.945,0.945,0.945} $M^{GSK}_{ts}$                   & {\cellcolor[rgb]{0.945,0.945,0.945}}                                                            & {\cellcolor[rgb]{0.945,0.945,0.945}}                                                               & \multirow{-2}{*}{{\cellcolor[rgb]{0.945,0.945,0.945}}\textcolor[rgb]{0.259,0.259,0.259}{GSK}}       & \multirow{-2}{*}{{\cellcolor[rgb]{0.945,0.945,0.945}}$0.087 \times 0.087$}                     & \multirow{-2}{*}{{\cellcolor[rgb]{0.945,0.945,0.945}}$500 \times 500$}  \\ 
\hhline{>{\arrayrulecolor{white}}-||>{\arrayrulecolor[rgb]{0.945,0.945,0.945}}->{\arrayrulecolor{white}}||>{\arrayrulecolor[rgb]{0.945,0.945,0.945}}->{\arrayrulecolor{white}}||-||-||-}
\rowcolor[rgb]{0.945,0.945,0.945} $M^{EXM}_{ts}$                   & \multirow{-2}{*}{{\cellcolor[rgb]{0.945,0.945,0.945}}\textcolor[rgb]{0.259,0.259,0.259}{Test }} & \multirow{-3}{*}{{\cellcolor[rgb]{0.945,0.945,0.945}}\textcolor[rgb]{0.259,0.259,0.259}{Mouse }}   & \textcolor[rgb]{0.259,0.259,0.259}{ExMI}                                                             & $0.282 \times 0.282$                                                                           & $144 \times 100$                                                        \\ 
\hhline{=::=::=::=::=::=}
\rowcolor[rgb]{0.867,0.867,0.867} $P^{PHE}_{tr}$                   & \textcolor[rgb]{0.259,0.259,0.259}{Training}                                                    & {\cellcolor[rgb]{0.867,0.867,0.867}}                                                               & {\cellcolor[rgb]{0.867,0.867,0.867}}                                                                 & {\cellcolor[rgb]{0.867,0.867,0.867}}                                                           & {\cellcolor[rgb]{0.867,0.867,0.867}}                                    \\ 
\hhline{-||-||>{\arrayrulecolor[rgb]{0.867,0.867,0.867}}->{\arrayrulecolor{white}}||>{\arrayrulecolor[rgb]{0.867,0.867,0.867}}->{\arrayrulecolor{white}}||>{\arrayrulecolor[rgb]{0.867,0.867,0.867}}->{\arrayrulecolor{white}}||>{\arrayrulecolor[rgb]{0.867,0.867,0.867}}-}
\rowcolor[rgb]{0.867,0.867,0.867} $P^{PHE}_{ts}$                   & \textcolor[rgb]{0.259,0.259,0.259}{Test}                                                        & \multirow{-2}{*}{{\cellcolor[rgb]{0.867,0.867,0.867}}\textcolor[rgb]{0.259,0.259,0.259}{Primate }} & \multirow{-2}{*}{{\cellcolor[rgb]{0.867,0.867,0.867}}\textcolor[rgb]{0.259,0.259,0.259}{PHE }}       & \multirow{-2}{*}{{\cellcolor[rgb]{0.867,0.867,0.867}}$0.235 \times 0.235$}                     & \multirow{-2}{*}{{\cellcolor[rgb]{0.867,0.867,0.867}}$512 \times 512$}  \\ 
\hhline{>{\arrayrulecolor{white}}=::=::=::=::=::=}
\rowcolor[rgb]{0.945,0.945,0.945} \textbf{$H^{CLE}_{tr}$~}         & \multicolumn{1}{l||}{\textcolor[rgb]{0.259,0.259,0.259}{Training}}                              & {\cellcolor[rgb]{0.945,0.945,0.945}}                                                               & {\cellcolor[rgb]{0.945,0.945,0.945}}                                                                 & {\cellcolor[rgb]{0.945,0.945,0.945}}                                                           & {\cellcolor[rgb]{0.945,0.945,0.945}}                                    \\ 
\hhline{-||-||>{\arrayrulecolor[rgb]{0.945,0.945,0.945}}->{\arrayrulecolor{white}}||>{\arrayrulecolor[rgb]{0.945,0.945,0.945}}->{\arrayrulecolor{white}}||>{\arrayrulecolor[rgb]{0.945,0.945,0.945}}->{\arrayrulecolor{white}}||>{\arrayrulecolor[rgb]{0.945,0.945,0.945}}-}
\rowcolor[rgb]{0.945,0.945,0.945} $H^{CLE}_{ts}$                   & {\cellcolor[rgb]{0.945,0.945,0.945}}                                                            & {\cellcolor[rgb]{0.945,0.945,0.945}}                                                               & \multirow{-2}{*}{{\cellcolor[rgb]{0.945,0.945,0.945}}\textcolor[rgb]{0.259,0.259,0.259}{ImageClef }} & \multirow{-2}{*}{{\cellcolor[rgb]{0.945,0.945,0.945}}$0.60\text{-}0.75\times0.60\text{-}0.75$} & \multirow{-2}{*}{{\cellcolor[rgb]{0.945,0.945,0.945}}$512 \times 512$}  \\ 
\hhline{>{\arrayrulecolor{white}}-||>{\arrayrulecolor[rgb]{0.945,0.945,0.945}}->{\arrayrulecolor{white}}||>{\arrayrulecolor[rgb]{0.945,0.945,0.945}}->{\arrayrulecolor{white}}||-||>{\arrayrulecolor[rgb]{0.945,0.945,0.945}}->{\arrayrulecolor{white}}||>{\arrayrulecolor[rgb]{0.945,0.945,0.945}}-}
\rowcolor[rgb]{0.945,0.945,0.945} $H^{RAD}_{ts}$                   & \multirow{-2}{*}{{\cellcolor[rgb]{0.945,0.945,0.945}}\textcolor[rgb]{0.259,0.259,0.259}{Test }} & \multirow{-3}{*}{{\cellcolor[rgb]{0.945,0.945,0.945}}\textcolor[rgb]{0.259,0.259,0.259}{Human }}   & \textcolor[rgb]{0.259,0.259,0.259}{Radiopedia}                                                       & $0.68\text{-}0.75\times0.68\text{-}0.75$                                                       & $512\text{-}630\times430\text{-}630$                                   
\end{tabular}
}
\arrayrulecolor{black}
\caption{Datasets description}
\label{ch5:tbl:datasets}
\end{table} 

\subsection{Implementation details}
The model is optimized employing six scales, $K=6$, with $18$ latent variables per scale, partitioned in $m_k$ groups per scale as follows, $m_k = [2,2,2,3,6,9]$ 
The three $\mu_A$, $\mu_S$ and $\mu_D$ per prior are known during training ($\mu_A=[-1,0,1]$, $\mu_D=(0,1)$, $\mu_S=(0,1)$), fix at image generation and inferred for image reconstruction and segmentation mask generation employing $KL\big(q_\phi(z^0) || \mathcal{N}(0,1)\big)$. Note that $\mu_D$ during optimization is given by the relative volume of the healthy lung (extracted by simple thresholding) with respect to the whole ground truth mask volume.

\subsection{Pathological Lungs Generation}
\label{ssec:genration}

\begin{figure}[htpb]
 \centering
  \includegraphics[width=\linewidth,keepaspectratio]{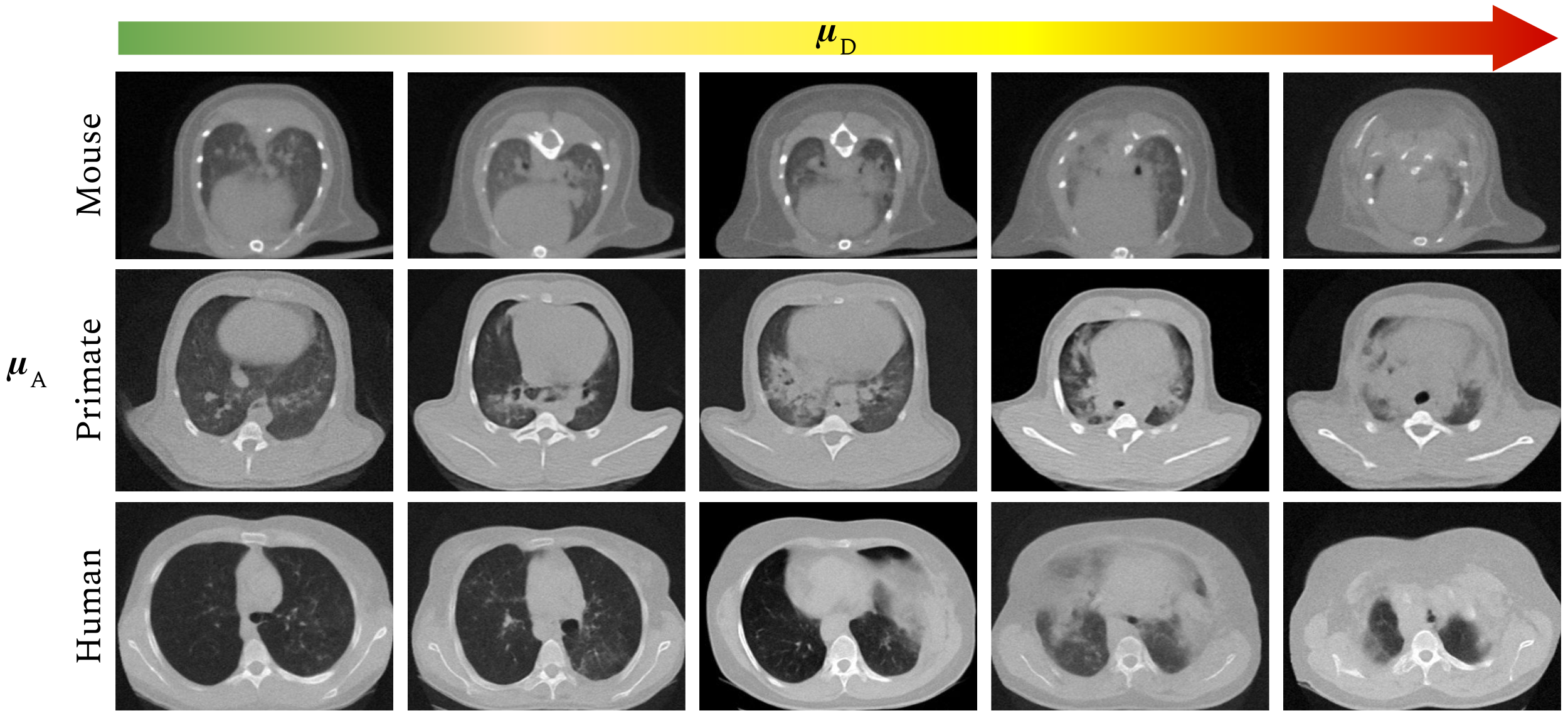}
  \caption[Synthetic lung CT images]{Synthetic lung CT images generated by our model. Images are generated with a fix slice relative position, ($\mu_S$). For each row, the animal model $\mu_A$ is fixed to $-1,0,1$, respectively, while for each column, the damage $\mu_D$ is increased [$0$-$1$].}
  \label{ch5:fig:generation}
\end{figure}

After optimization, 
the model is able to generate realistic images, such as those shown in ~\ref{ch5:fig:generation}, by choosing the mean values of $\mathbf{z}_A^0$, $\mathbf{z}_S^0$, $\mathbf{z}_D^0$ factors. To illustrate this capacity in \figurename~\ref{ch5:fig:generation}, we set a relative slice position of $0.5$, the animal model is fixed for each row and, the effect of the lung damage variable is modulated from lower to higher in each column. \\

\subsubsection{Pathological Lungs Generation: Varying the slice position}
This appendix shows generated slices instances fixing the damage and varying the relative slice position. This experiment extends the previous section, in which axial slices belong to a fixed relative slice position.

\begin{figure}[htpb]

 \centering
  \includegraphics[width=0.9\linewidth, height=0.3\textheight]{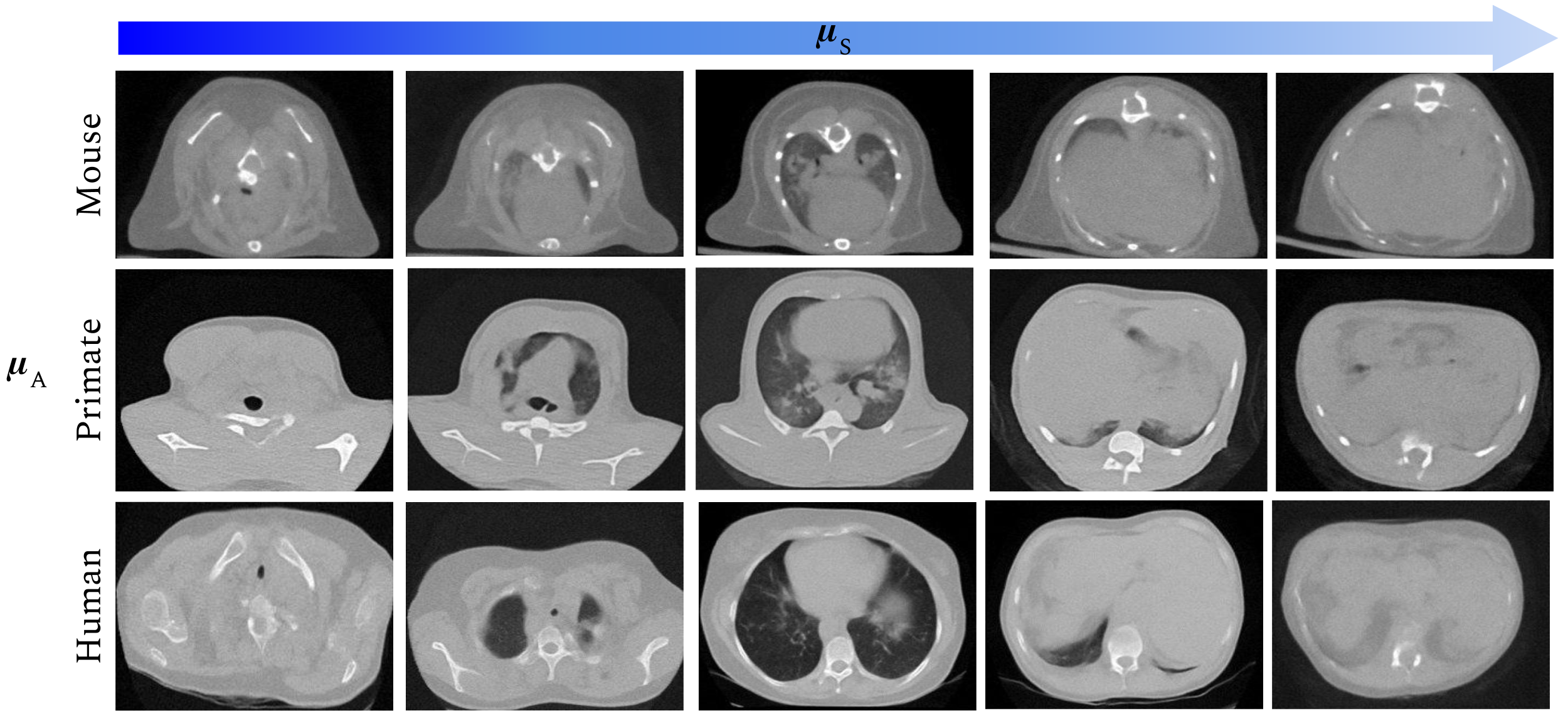}
  \caption{Synthetic lung CT images generated by our model. Images are generated with a fix relative damage, $\mu_D=0.5$. For each row, the animal model $\mu_A$ is fixed to $-1,0,1$, respectively, while for each column, the relative slice position $\mu_S$ is increased between $0$ and $1$.}
  \label{ch5:fig:var_slic}
\end{figure}

Since our chest CT volumes orientation is cephalic to caudal, the model generates axial images of the upper airways (trachea) and the corresponding per animal model surrounding tissues at the lowest slice position, as shown in the first column of the \figurename~\ref{ch5:fig:var_slic}. This way, the second column shows the corresponding generated anatomy for the superior lungs, while the third and fourth columns accordingly show the middle and inferior regions. Finally, the fifth column depicts the generated version at the beginning of the abdominal anatomy. 

\subsection{Counterfactual Images}\label{ssec:counter}
\begin{figure}[htpb]
 \centering
  \includegraphics[width=\linewidth,keepaspectratio]{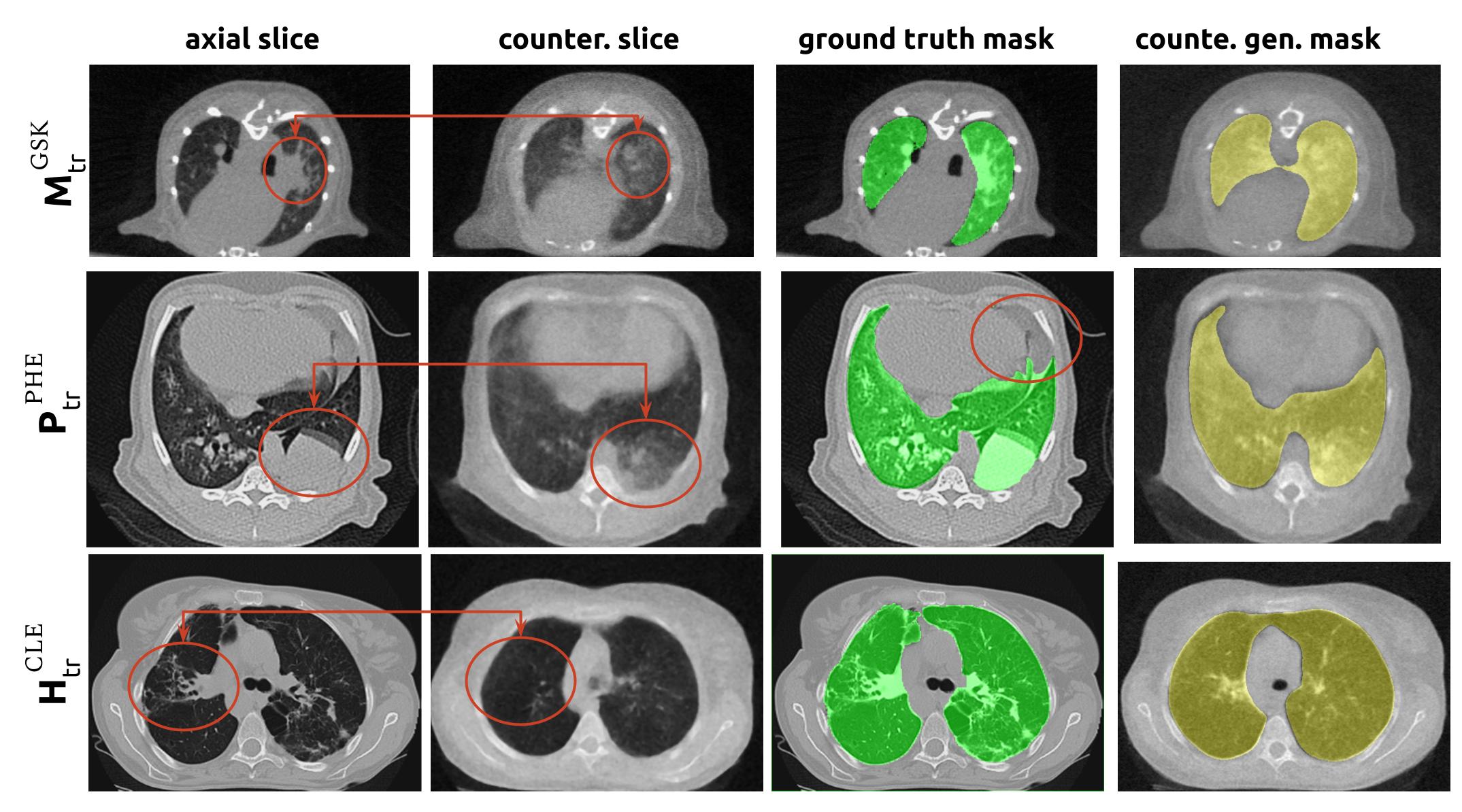}
  \caption{The encoder infers the real image (axial slice) disentangled representation, $\mathbf{z}_A^0$, $\mathbf{z}_S^0$, $\mathbf{z}_D^0$. By setting the damage variable $\mathbf{z}_D^0$ to $0$ the decoder generates the healthy counterfactual (counterfactual slice) and its respective mask (counterfactual mask).}
  \label{ch5:fig:conter}
\end{figure}

The first column of each row in \figurename~\ref{ch5:fig:conter} shows a real image of a damaged lung corresponding to a given animal model. When no actions are performed, the model infers the disentangled image representation of the causative variables ($\mathbf{z}_A^0$, $\mathbf{z}_S^0$, $\mathbf{z}_D^0$) through the encoder. Subsequently, the image is reconstructed and a segmentation mask is generated employing the optimized decoder (\figurename~\ref{ch5:fig:arqu_sum}).
The second column shows a healthy counterfactual of the input images. Each counterfactual image is generated after intervening on the inferred damage variable, $\mathbf{z_D^0}$. To this aim, its mean value is set to 0. The decoder is fed with the zero-mean counterfactual image and the rest (unaltered) inferred causal variables.

\subsubsection{Counterfactual Images: Extended Assessment}
This appendix extends the qualitative results presented in Section \ref{ssec:counter}. The former section shows the model capacity generating counterfactual images and their respective segmentation masks.

Here, we evaluate how realistic are the generated images. For that, we compare the Hounsfield Units (HU) of real CT slices with two cases: a) the reconstructed slice from the variable inferred by the encoder without modification of any of these values, and b) the counterfactual image, namely, after intervening on the inferred damage value. 
We compute the voxel-wise Root Mean Square Error (RMSE) for the reconstructed images per test dataset. \tablename~\ref{ch5:tbl:RMSE} shows these results with an average $RMSE=18.73\pm2.16$. 

Voxel-wise evaluation is not suitable for counterfactual images. Previous manual delimitation of comparable regions is needed, which is a priority for our future work.

\begin{figure}[htpb]
 \centering
  \includegraphics[width=\linewidth, height=0.55\textheight]{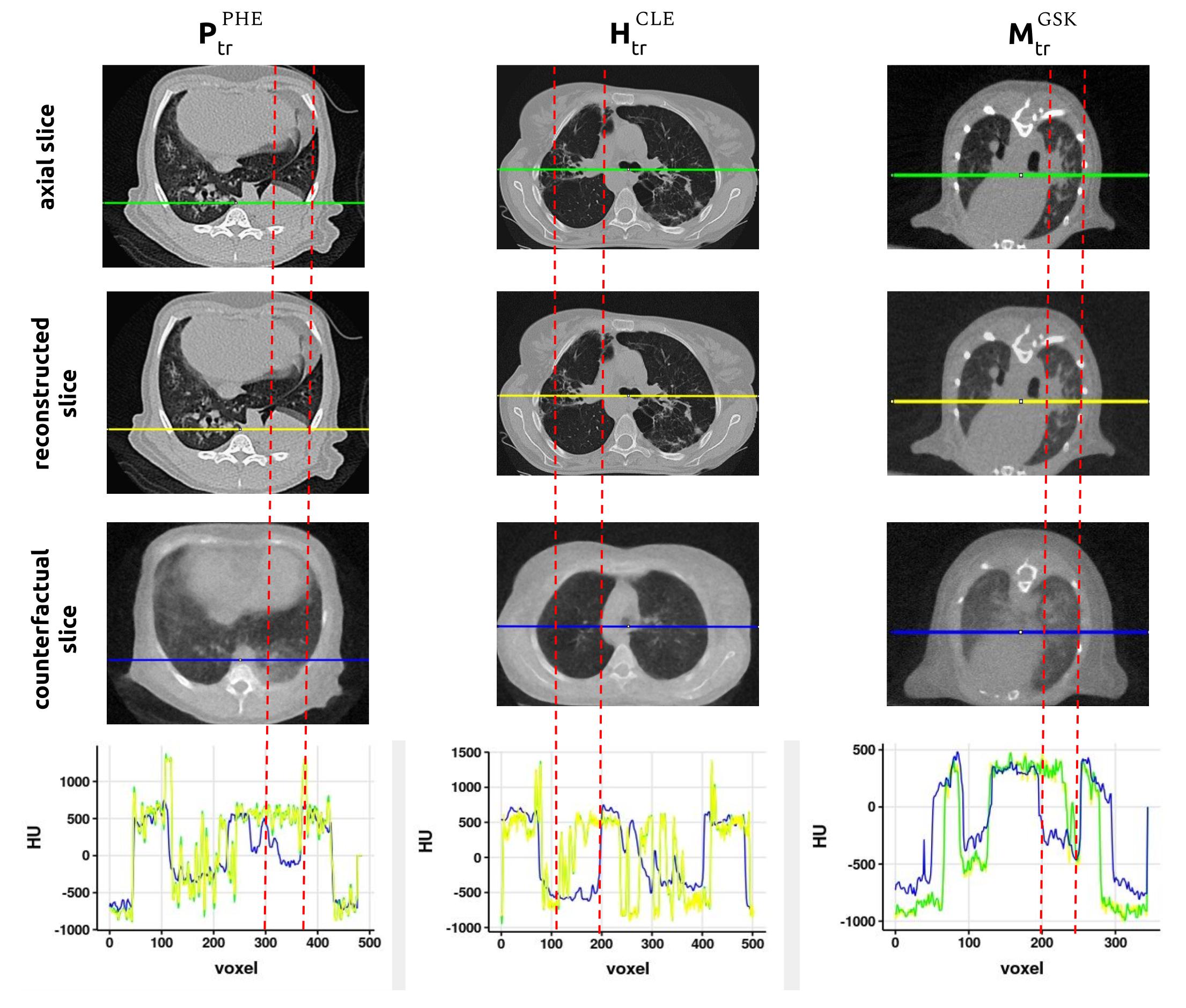}
  \caption{Hounsfield Units (HU) plots for profiles at regions damaged in original test axial slices. Each column contains instances of each dataset, previously employed in Section \ref{ssec:counter}. The first rows depict the original, reconstructed and counterfactual slices with the profile line green, yellow and blue, respectively. The last row draws the HU profiles per voxel. Vertical dashed lines highlight big differences between real/reconstructed and counterfactual slices.}
  \label{ch5:fig:profiles}
\end{figure}

To illustrate similarities and differences in the HU scale, in \figurename~\ref{ch5:fig:profiles}, we plot the HU profile belonging to the damaged regions shown in \figurename~\ref{ch5:fig:conter}. Respectively, the first three rows contain 1) the original axial slice from the different test datasets (the image is generated from the $\mu_a$, $\mu_s$ and $\mu_d$ inferred by our model), with the profile horizontal line in green, 2) the reconstructed slice (the image is generated maintaining $\mu_a$, $\mu_s$ inferred by our model and correcting $\mu_d$), with profile line in yellow and 3) the counterfactual after modifying the inferred expected damage, with the profile line in blue. 

The last row shows the HU plot for each profile-specific colour.   
HU values are similar for the three slices except for those regions where the slice counterfactual version replaces the damage with healthy tissue-like. We highlight such changes framing them in vertical dashed red lines.

Besides, it is important to note that the original and reconstructed images present more noisy patterns than the counterfactual version, as was expected from its blurrier appearance and the thickening of the soft tissue for the mice dataset. 

\begin{table}[htpb]
\centering
\caption{Root Mean Square Error (RMSE) between the real images and the image reconstructed from the $\mu_a$, $\mu_s$ and $\mu_d$ inferred by our model for the test datasets}
\label{ch5:tbl:RMSE}
\arrayrulecolor{white}
\begin{tabular}{c|c||c||c|c|}
\multicolumn{5}{c}{{\cellcolor[rgb]{0.322,0.322,0.322}}\begin{tabular}[c]{@{}>{\cellcolor[rgb]{0.322,0.322,0.322}}c@{}}\textcolor{white}{RMSE}\textcolor{white}{ [HU]}\\\end{tabular}} \\ 
\hhline{==:t:=:t:==|}
\rowcolor[rgb]{0.6,0.6,0.6} $M^{GSK}_{ts}$ & $M^{EXT}_{ts}$ & $P^{PHE}_{ts}$ & $H^{CLE}_{ts}$ & $H^{COV}_{ts}$ \\ 
\hline
$21.26$ & {\cellcolor[rgb]{0.973,0.973,0.973}}\textcolor[rgb]{0.2,0.2,0.2}{$18.75$} & {\cellcolor[rgb]{0.973,0.973,0.973}}\textcolor[rgb]{0.2,0.2,0.2}{$20.12$} & {\cellcolor[rgb]{0.973,0.973,0.973}}\textcolor[rgb]{0.2,0.2,0.2}{$17.89$} & {\cellcolor[rgb]{0.973,0.973,0.973}}\textcolor[rgb]{0.2,0.2,0.2}{$15.63$} \\
\hhline{~-|b|-|b|--|}
\end{tabular}
\arrayrulecolor{black}
\end{table}

\subsection{Segmentation employing counterfactual images}
Pathological lung segmentation is an important task for drug development studies. Unfortunately, it a complex task due to the difficulty of discrimination between lesions and other neighborhood tissues with the added difficulty of the diversity of the biological data  \cite{Hofmanninger2020AutomaticProblemb}. In this experiment, we retrain the optimized model with counterfactual images, such as the obtained in the previous experiment (see \ref{ssec:counter}), to generate the segmentation masks from the test datasets \ref{ssec:datasets}. 
To measure the strengths and weaknesses of this generative approach, we compare the obtained results, $our_{c}$, with the segmentation masks calculated by our method optimized before employing counterfactual images, $our_{nc}$, and the state of the art full supervised method, \textit{nnU-Nnet} \cite{Isensee2021NnU-Net:Segmentation}.\\

\begin{table}[htpb]
\centering

\arrayrulecolor{white}
\begin{tabular}{c||c|c||c||c|c|}
\multicolumn{1}{c|}{}                                                      & \multicolumn{5}{c|}{{\cellcolor[rgb]{0.322,0.322,0.322}}\textcolor{white}{DSC$_{\pm SD}$}}                                                                                                                                                                                          \\ 
\hhline{~==:t:=:t:==|}
\multicolumn{1}{c}{}                                                       & {\cellcolor[rgb]{0.6,0.6,0.6}}$M^{GSK}_{ts}$                                     & {\cellcolor[rgb]{0.6,0.6,0.6}}$M^{EXT}_{ts}$       & {\cellcolor[rgb]{0.6,0.6,0.6}}$P^{PHE}_{ts}$       & {\cellcolor[rgb]{0.6,0.6,0.6}}$H^{CLE}_{ts}$       & {\cellcolor[rgb]{0.6,0.6,0.6}}$H^{COV}_{ts}$        \\ 
\hline
\rowcolor[rgb]{0.973,0.973,0.973} \textcolor[rgb]{0.2,0.2,0.2}{nnU-Net }   & \textcolor[rgb]{0.2,0.2,0.2}{}$0.845_{ \pm 0.10}$\textcolor[rgb]{0.2,0.2,0.2}{} & \textcolor[rgb]{0.2,0.2,0.2}{$0.851_{\pm 0.11}$} & \textcolor[rgb]{0.2,0.2,0.2}{$0.957_{\pm 0.06}$} & \textcolor[rgb]{0.2,0.2,0.2}{$0.978_{\pm 0.04}$} & \textcolor[rgb]{0.2,0.2,0.2}{$0.973_{\pm 0.03}$}  \\ 
\hline
\rowcolor[rgb]{0.867,0.867,0.867} \textcolor[rgb]{0.2,0.2,0.2}{$our_{nc}$} & $0.849_{\pm 0.10}$                                                             & \textcolor[rgb]{0.2,0.2,0.2}{$0.843_{\pm 0.12}$} & \textcolor[rgb]{0.2,0.2,0.2}{$0.949_{\pm 0.06}$} & \textcolor[rgb]{0.2,0.2,0.2}{$0.963_{\pm 0.06}$} & \textcolor[rgb]{0.2,0.2,0.2}{$0.963_{\pm 0.06}$}  \\ 
\hline
\rowcolor[rgb]{0.973,0.973,0.973} \textcolor[rgb]{0.2,0.2,0.2}{$our_{c}$ } & \textcolor[rgb]{0.2,0.2,0.2}{$0.877_{\pm 0.08}$}                               & \textcolor[rgb]{0.2,0.2,0.2}{$0.859_{\pm 0.11}$} & \textcolor[rgb]{0.2,0.2,0.2}{$0.955_{\pm 0.06}$} & \textcolor[rgb]{0.2,0.2,0.2}{$0.977_{\pm 0.06}$} & \textcolor[rgb]{0.2,0.2,0.2}{$0.968_{\pm 0.04}$} 
\end{tabular}
\arrayrulecolor{black}
\caption[Mean and standard deviation of the DSC for segmentation comparison]{Mean and standard deviation (SD) of the Dice Similarity Coefficient (DSC) between the ground truth masks and mask obtained from the methods indicated at rows (\textit{nnU-Nnet}, proposed method before employing counterfactual images ($our_{nc}$), and after ($our_{c}$)) for each test dataset (columns).}
\label{ch5:tbl:optimiza_results_DSC}
\end{table}

\begin{table}
\centering

\arrayrulecolor{white}
\begin{tabular}{c||c|c||c||c|c}
\multicolumn{1}{c}{}                                                       & \multicolumn{5}{c}{{\cellcolor[rgb]{0.322,0.322,0.322}}\textcolor{white}{HD$_{\pm SD}$ [mm]}}                                                                                                                                                         \\ 
\hhline{~--|t|-|t|--}
\multicolumn{1}{c}{}                                                       & {\cellcolor[rgb]{0.6,0.6,0.6}}$M^{GSK}_{ts}$       & {\cellcolor[rgb]{0.6,0.6,0.6}}$M^{EXM}_{ts}$      & {\cellcolor[rgb]{0.6,0.6,0.6}}$P^{PHE}_{ts}$      & {\cellcolor[rgb]{0.6,0.6,0.6}}$H^{CLE}_{ts}$        & {\cellcolor[rgb]{0.6,0.6,0.6}}$H^{COV}_{ts}$         \\ 
\hhline{-|t:==::=::==}
\rowcolor[rgb]{0.973,0.973,0.973} \textcolor[rgb]{0.2,0.2,0.2}{nnU-Net }   & \textcolor[rgb]{0.2,0.2,0.2}{$1.737_{\pm 1.01}$} & \textcolor[rgb]{0.2,0.2,0.2}{$1.90_{\pm 1.52}$} & \textcolor[rgb]{0.2,0.2,0.2}{$3.30_{\pm 3.96}$} & $9.37_{\pm 15.14}$                                & \textcolor[rgb]{0.2,0.2,0.2}{$8.31_{\pm 10.71}$}   \\ 
\hline
\rowcolor[rgb]{0.867,0.867,0.867} \textcolor[rgb]{0.2,0.2,0.2}{$our_{nc}$} & \textcolor[rgb]{0.2,0.2,0.2}{$1.948_{\pm 1.11}$} & \textcolor[rgb]{0.2,0.2,0.2}{$2.06_{\pm 1.82}$} & \textcolor[rgb]{0.2,0.2,0.2}{$3.81_{\pm 4.10}$} & \textcolor[rgb]{0.2,0.2,0.2}{$10.12_{\pm 18.32}$} & \textcolor[rgb]{0.2,0.2,0.2}{$10.56_{\pm 10.77}$}  \\ 
\hline
\rowcolor[rgb]{0.973,0.973,0.973} \textcolor[rgb]{0.2,0.2,0.2}{$our_{c}$ } & \textcolor[rgb]{0.2,0.2,0.2}{$1.519_{\pm 0.89}$} & \textcolor[rgb]{0.2,0.2,0.2}{$1.88_{\pm 1.53}$} & \textcolor[rgb]{0.2,0.2,0.2}{$2.95_{\pm 3.54}$} & \textcolor[rgb]{0.2,0.2,0.2}{$8.78_{\pm 16.11}$}  & \textcolor[rgb]{0.2,0.2,0.2}{$9.48_{\pm 9.89}$}   
\end{tabular}
\arrayrulecolor{black}
\caption[Mean and standard deviation of the  HD for segmentation comparison]{Mean and standard deviation (SD) of the Hausdorff Distance (HD) between the ground truth masks and mask obtained from the methods indicated at rows (\textit{nnU-Nnet}, proposed method before employing counterfactual images ($our_{nc}$), and after ($our_{c}$)) for each test dataset (columns).}
\label{ch5:tbl:optimiza_results_HD}
\end{table}

\begin{figure}[htpb]

 \centering
  \includegraphics[width=\linewidth,keepaspectratio]{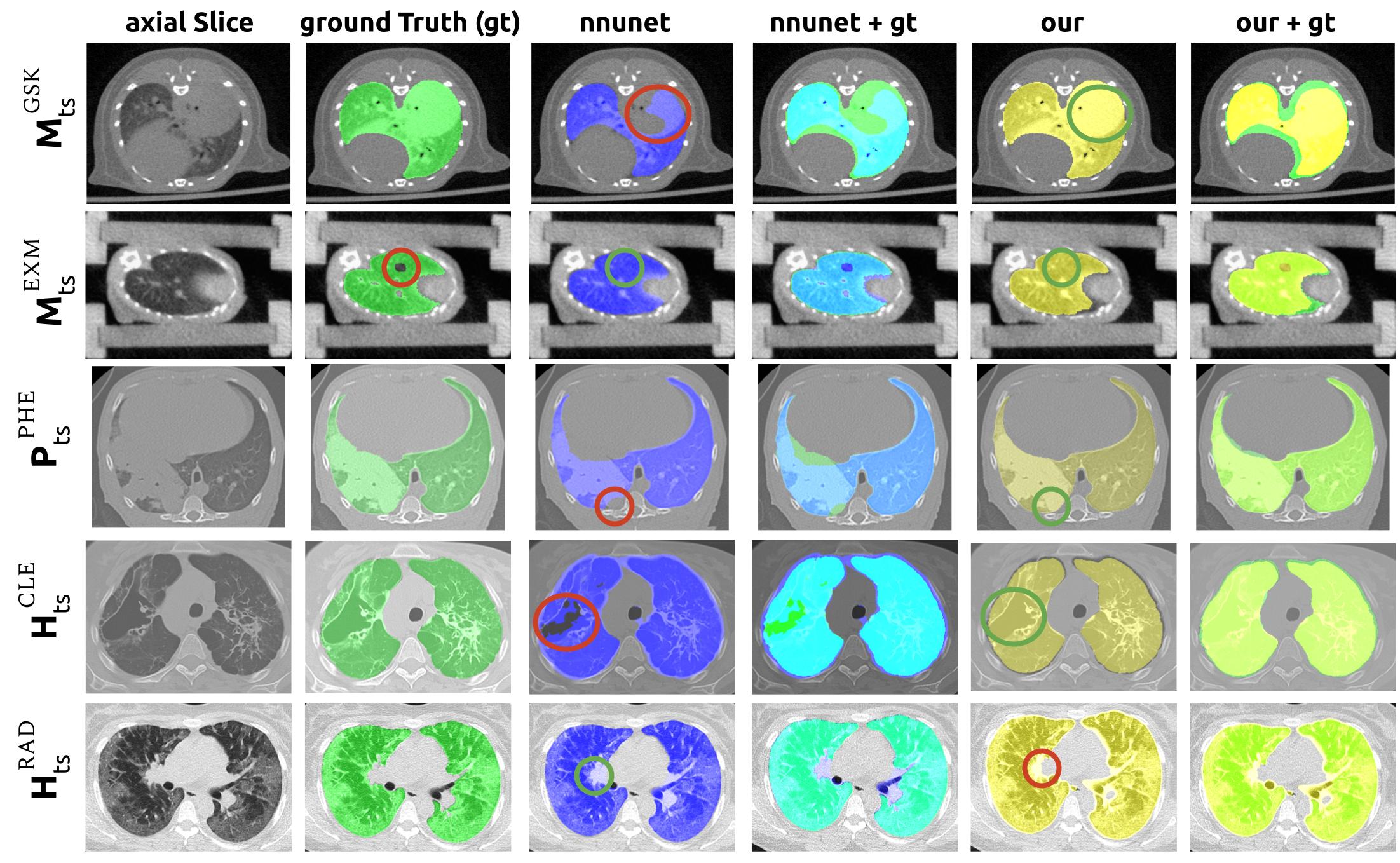}
  \caption[Comparison of Deep Learning methods (discriminative and hybrid) for PLS and different animal and disease models]{Comparison of methods. Each row contains axial slices and segmentation masks of each test dataset. Columns show the original CT image, \textcolor{green}{ground truth mask (gt)}, \textcolor{blue}{\textit{nnU-Net} mask}, \textcolor{cyan}{overlay of \textit{nnU-Net} and ground truth}, \textcolor{darkyellow}{the mask with our method employing counterfactual images during training} and \textcolor{lime}{the overlay with the ground truth}.}
  \label{ch5:fig:lungs_seg}
\end{figure}

The \tablename~\ref{ch5:tbl:optimiza_results_DSC} and \tablename~\ref{ch5:tbl:optimiza_results_HD} show the mean and standard deviation for Dice Similarity Coefficient (DSC) and Hausdorff Distance (HD) between each segmentation method and the ground truth masks for each test dataset. The results present an improvement for all measures and datasets when employing counterfactual images, yielding similar results to the \textit{nnU-Nnet}. The differences are due to subtle changes in most of the cases or even small imperfections in the ground truth masks as it is shown in \figurename~\ref{ch5:fig:lungs_seg}.

\subsubsection{Comparison with Chapter \ref{ch2:chapter2} Rule-Based Method}
For completeness, these paragraphs recover the results obtained with the traditional method of Chapter \ref{ch2:chapter2}. In that chapter, we show the qualitative performance of DL and the rules-based approaches for segmentation tasks to illustrate the issues of traditional methods under new environments.\\
Actually, the instance slices in the \figurename~\ref{ch2:fig:segmetations_bad} coincide with the ones in the \figurename~\ref{ch5:fig:lungs_seg} except for $M^{EXM}_{ts}$ (not suitable for the traditional approach) and the row for the mild TB primate model, $P^{PHE_1}$, in figure 2, already included in it.\\

Here, we complete the analysis for the $156$ slices specifically selected for Chapter \ref{ch2:chapter2} experiments. To this aim, the \tablename~\ref{ch5:tbl:adding_results} adds the DSC and HD to the previous chapter results (\tablename~\ref{ch2:tbl:means_res}) employing the DL methods, both the \textit{nnU-Net} and the two variants of our method.\\
As can be seen, the DL method overpasses the performance of traditional ones. 

\begin{table}[htpb]
\centering
\arrayrulecolor{white}
\begin{tabular}{c||c||c||}
\multicolumn{1}{c}{}                                                       & {\cellcolor[rgb]{0.322,0.322,0.322}}\textcolor{white}{DSC$_{\pm SD}$} & \multicolumn{1}{c}{{\cellcolor[rgb]{0.322,0.322,0.322}}\textcolor{white}{HD$_{\pm SD}$ [mm]}}  \\ 
\hhline{~=::=:t|}
\multicolumn{1}{c}{}                                                       & {\cellcolor[rgb]{0.6,0.6,0.6}}$P^{CH_2}$                              & {\cellcolor[rgb]{0.6,0.6,0.6}}$P^{CH_2}$                                                       \\ 
\hhline{=:t:=::=:|}
\rowcolor[rgb]{0.867,0.867,0.867} No DL (Section \ref{ch2:sec:Methods})                                    & \multicolumn{1}{l||}{$0.933_{\pm 0.03}$}                              & \multicolumn{1}{l||}{$8.642_{\pm 7.36}$}                                                       \\ 
\hline
\rowcolor[rgb]{0.973,0.973,0.973} \textcolor[rgb]{0.2,0.2,0.2}{nnU-Net }   & \textcolor[rgb]{0.2,0.2,0.2}{}$0.981_{\pm 0.05}$                      & \textcolor[rgb]{0.2,0.2,0.2}{$5.954_{\pm 5.11}$}                                               \\ 
\hline
\rowcolor[rgb]{0.867,0.867,0.867} \textcolor[rgb]{0.2,0.2,0.2}{$our_{nc}$} & $0.977_{\pm 0.06}$                                                    & \textcolor[rgb]{0.2,0.2,0.2}{$6.897_{\pm 5.87}$}                                               \\ 
\hline
\rowcolor[rgb]{0.973,0.973,0.973} \textcolor[rgb]{0.2,0.2,0.2}{$our_{c}$ } & \textcolor[rgb]{0.2,0.2,0.2}{$0.982_{\pm 0.06}$}                      & \textcolor[rgb]{0.2,0.2,0.2}{$6.142_{\pm 6.01}$}                                              
\end{tabular}
\arrayrulecolor{black}
\caption[DSC and HD for the $156$ slices of Chapter \ref{ch2:chapter2}]{DSC and HD for the $156$ slices defined at Appendix \ref{ap:slices} and employed at the experiments in Chapter \ref{ch2:chapter2}, $P^{CH_2}$ (see \ref{ch2:sec:Methods}) belonging to a mild TB primate model, which  results appears at \figurename~\ref{ch2:fig:segmetations_bad} and \tablename~\ref{ch2:tbl:means_res}.}\label{ch5:tbl:adding_results}
\end{table}



\section{Conclusions}
The methodology proposed in this work yields promising results for obtaining the shared factors between animal models that characterize the pathophysiological processes. Beyond the existing limitations, such as the use of isolated axial slices instead of more informative whole three dimensional images or the characterization of damage based simply on the damaged volume and not on the specific manifestations of the disease for each animal model, our model is capable of inferring meaningful disentangled representations.\\
The model can generate synthetic axial slices by setting the values of the modelled factors. Even more relevant, it produces counterfactual versions of existing axial slices by testing the effective disentanglement. This capability leverages potential benefits in the field of translational image analysis. Extending the diversity of existing data, essential for automatic segmentation, or providing the damage variable as a possible (to be validated) inter-species biomarker.

\chapter{Conclusions and Prospective work}\label{ch6:Intro}
\section{Conclusions}

This thesis develops different methods to automate fundamental CT image analysis tasks for TB infected lungs. We also facilitate the interpretation of CT images belonging to various animal models of TB infection, which is essential for the definition of helpful imaging biomarkers for anti-TB new drugs development. The works are presented from lower to a higher degree of complexity in response to the considered problems nature. The developed models embed in different Artificial Intelligence (AI) strategies and fulfil the thesis objectives specified in the introductory chapter. 

The method based on a traditional formulation presented in Chapter \ref{ch2:chapter2} allows segmenting damaged lungs automatically. This initial approach enables both subsequent quantification analyses and the extraction of fundamental knowledge in the implementation of higher capacity Deep Learning (DL) algorithms. The method presents limitations, such as the inability to provide lung segmentation for significant organ damage or animal models other than a macaque. However, it can assist experts during lung delimitation of other mammals by readjusting the adequate parameters (see Section \ref{tbl:chapt2/parameters}). The method validity in key experiments relies on the quantification results given in the second part of the same chapter. In Section \ref{ch2:sec:quantication}, we introduce an automatic quantification method of TB burden. The procedure, which may be considered simple in terms of specificity to TB manifestations, is vital in some translational analyses. Namely, when the animal models are not fully defined (i.e., there is no clear radiological taxonomy of the manifestations), the volume of the damaged lung can be used as a proxy. 

Even for models with sufficient image quality for radiological characterization of the lesions, their identification by experts is tedious, time-consuming and prone to errors. This fact motivates the development of the methods presented in Chapters \ref{ch3:Radiomics} and \ref{ch4:Intro} to help the task automatization. Thus, Chapter \ref{ch3:Radiomics} uses statistical descriptors that feed a machine learning (ML) model. These statistical descriptors result in lower interpretability than the previous approach but greater predictive power. The ML algorithms allow discerning between specific TB manifestations on the previously delimited whole lung of the macaque model. To avoid possible errors during delimitation, the model proposed in Chapter \ref{ch4:Intro} extends the ability to detect lesions. To this aim, we train a high-capacity Deep Learning model capable of extracting descriptors automatically from complete image volumes to infer the presence or the number of manifestation types. This way eliminates the need to delineate the lungs in a previous step. 

The methods in Chapters \ref{ch2:chapter2}, \ref{ch3:Radiomics} and \ref{ch4:Intro} allow classifying tissues. They have the ability to capture relevant information from images beyond human capabilities, as proved for the mild-TB macaque model. Such approaches are easily extendible to different datasets/animal models, as shown in the recent literature, allowing the experts to obtain specific dataset outcomes. Subsequently, multidisciplinary experts can combine individual outputs in the drug design context to provide model translational explanations. This idea reflects the traditional workflow for disease understanding and drug design relying on computer-assisted image analysis. It also reveals the lack of automation tools implemented to operate with multidomain data, mimicking human intelligence.

In this context, given the ability of current computational models to find relevant relationships or subtle information, introducing artificial intelligence systems that can exploit inter-species data is fundamental to overcoming the limitations mentioned above. Therefore, in the previous chapter, a method that assumes a causal structure in data generation from different animal models is employed to promote the formulated translational analyses.
The proposed DL model leverages shared features between images of distinct animal models. Apart from providing proper segmentation masks, the model infers similar levels of lung damage for comparable slices of the different animal models and alternatively generate such images when intervening in the model to fix a given level of damage.



\section{Prospective work and Perspectives}
Throughout this work, we have presented different methodologies based on Artificial Intelligence that enrich the extraction of relevant information from pathological images. Introducing the proposed methodologies in the drug assays pipelines is essential to facilitate diagnosis, disease longitudinal monitoring, and understanding disease etiology. From a mere technical point of view, what is missing in the current work for this inclusion is an extended validation. 

Fortunately, the context of ERA4TB (see \ref{ch1:sec:Erad:ERA4TB}) enables different ways to accomplish it. Prospective trials based on a higher subject number will allow us to extract further conclusions about the methods generalization capabilities. More importantly, such studies will enable to use of the benefits of computer assistance for understanding Tuberculosis pathophysiology.

In addition, the project context allows validation through triangulation, which is fundamental for qualitative research \cite{Carter2014TheResearch}. In ERA4TB, not only CT images are used for longitudinal analysis of infected animals, but also vital data produced from other imaging modalities (see \ref{ch1:sec:MITB}): Positron Emission Tomography (PET), pathological microscopy, Matrix-Assisted Laser Desorption Ionization (MALDI), etc., together with molecular and DNA analyses that greatly extend the description at the microscopic level of the bacteria interactions with the new drugs.

It is important to note that the method developed in Chapter \ref{ch5:Intro} is not only extensible to new animal models (e.g., rat, rabbit, pig) but also to all the information sources mentioned above. Therefore, extending the current state of the art (SOTA) computational methods towards mimicking much more human intelligence. Note that SOTA DL models usually rely on domain-specific assumptions. For instance, inference depends only on CT images. However, when a human expert makes decisions employing information extracted from images is acting in an \textit{imagined space}\footnote{\textit{Imagined space} concept was coined by \citet{Lorenz1973BehindKnowledge.}, and it is employed by \citet{Scholkopf2021TowardLearning} for AI}. Namely, the expert consciously or implicitly considers all previous knowledge, meaning, disease model, subject demographics, comorbidity, treatment, etc.
Therefore, our computational models need to resemble such mechanisms to assist during reasoning. The most meaningful works in recent literature already point in such a direction that will be a fundamental point in our future job \cite{Jaegle2021Perceiver:Attention, Jumper2021HighlyAlphaFold}. This prospective framework will enable the simulation of the longitudinal progress of the disease and determine the causal factors of treatments efficacy. Actually, for our future work, we have coined the term CaFE (Causal Factors of Efficacy) extended it to ICaFE (Imaging CaFE) when the factors are derived just from medical imaging data and TICaFE (Translational ICaFE) or TCaFE when the factors are translational. 

Undoubtedly the range of possible technical work that can arise in this context is extensive, especially if tools integration in the workflow is effective. However, this process goes beyond technicalities. Like any other new automation technology, there is an existing reluctance in many strata of society that strongly constrain its integration.

On the one hand, this rejection is due to the operation in itself of the new technologies \cite{Tsamados2021TheSolutions}. For example, it is unfortunately common to find cases in which Artificial Intelligence is employed without any control and let make biased decisions disfavoring certain social groups \cite{Bender2021OnFormat, Chinoy2019TheRecognition}. It is also of note the multiple instances in which the introduction of Artificial Intelligence can lead to the destruction of jobs without alternatives for workers \cite{2017TheWork}. Dealing with these problems necessarily involves making political decisions that contextualize the use of new tools within the normative ethical values that (\textit{a priori}) formalize societies \cite{Borenstein2020EmergingEducation}. Although legislation development usually lets behind, measures to introduce a control are becoming more common \cite{High-LevelExpertGrouponArtificialIntelligence2019EthicsAI}.

However, the rejection within the research context attains to different matters. 
The medical imaging field copes (for too long already), its exciting oversized version of a \textit{paradigm shift} \cite{Kuhn1970TheRevolutions}. It keeps an unnecessary struggle between the reluctance from the old "capos", the ones too embedded in the old fashioned clinical practices, and the necessity to boost the branch towards the Artificial Intelligence community direction \cite{Topol2019High-performanceIntelligence, Zhou2021APromises} supported by those
with \textit{"hands-on"} the actual predicament. It is horrific to witness how some self-interested negationists in the field take advantage of their political positions to badly exploit common resources that more than ever need to learn from the scientific community.
Therefore, the long term work must go beyond technical aspects and penetrate ethics and education. 




\begin{spacing}{0.9}


\bibliographystyle{plainnat} 
\cleardoublepage
\bibliography{References/references} 



\end{spacing}


\begin{appendices} 
\chapter{Surrogate Truth Extraction} 
\label{ap:slices}
\section{Selecting CT Slices With The More Uncertain Boundaries}\label{ap:sec:select}
Segmentation of lungs infected by Mycobacterium tuberculosis (Mtb) in chest computed tomography (CT) images is a complex task. Moreover, it is difficult to establish a suitable ground truth, as generation thereof is very time-consuming, subject to intra- and inter-expert variability, and prone to errors. As discussed, the commonly used measures of similarity do not represent well the unavoidable human variability inherent in a segmentation process. Therefore, in our workflow evaluation, we used a surrogate ground truth built as a consensus between three experts who performed detailed segmentations on 156 slices from our chest CT dataset. These slices were selected from the whole dataset using the procedure described in the next paragraph and were designed to ensure that the surrogate ground truth contains a representative sample of the most uncertain slices.

\begin{figure}
\centering
\includegraphics[width=0.7\textwidth,height = 0.35\textheight]{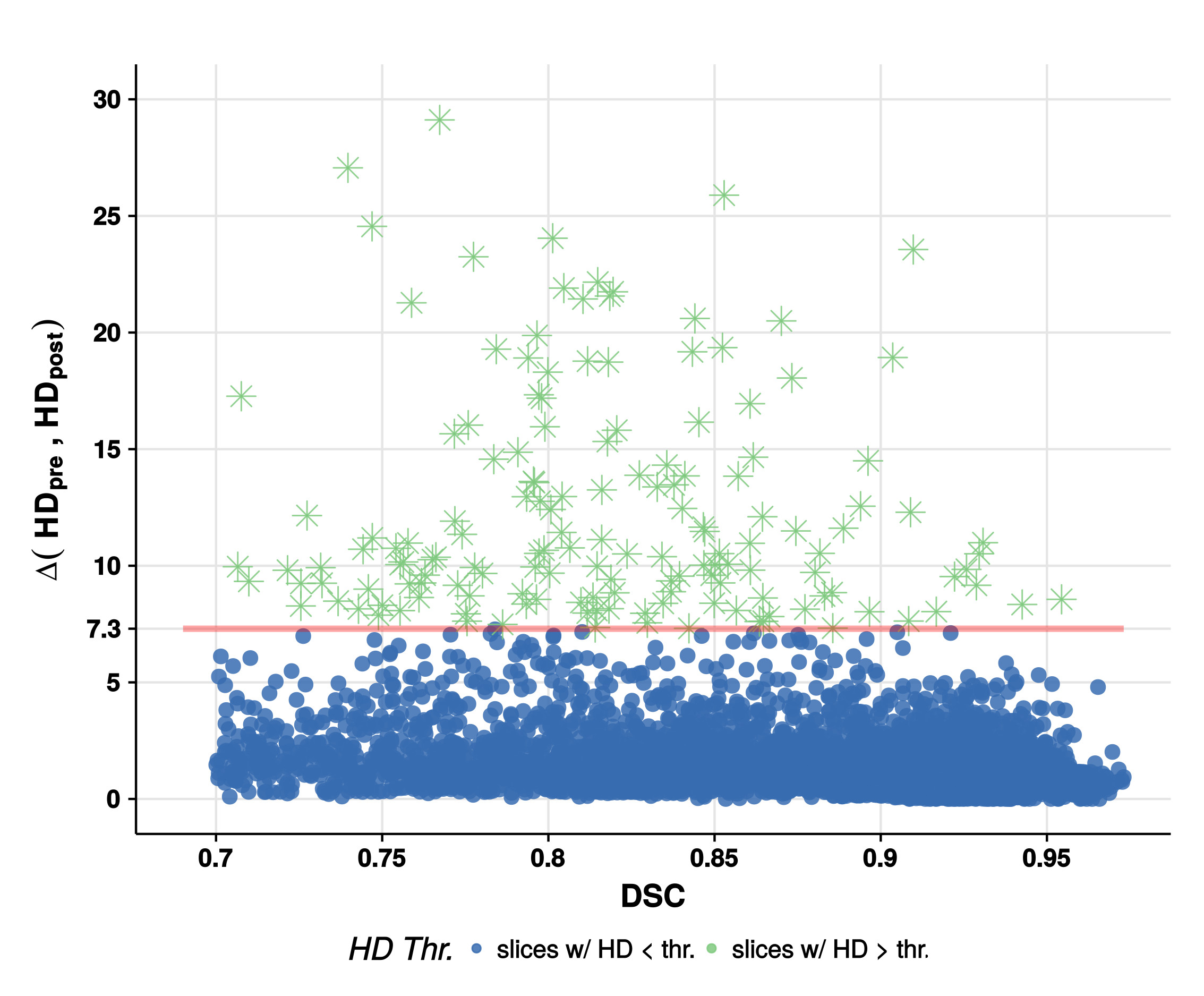}
\caption[HD differences for surrogate truth extraction]{Hausdorff  distance ($HD$) differences between those corresponding to the lung segmentation performed with our tool before and after the refinement process. $HD$ was measured using the semi-auto segmentations as a reference. Only those slices with a Dice similarity coefficient ($DSC$) over $0.7$ were included. $HD$ corresponding to the slices for which the $HD$ difference $(\Delta)$ is larger than  a given  threshold (thr.,  red line) are drawn in green, while points smaller than the threshold are shown in blue. In this case, the threshold is computed as $\mu_{\Delta (HD_{pre}, HD_{post})} + 3\sigma_{\Delta (HD_{pre}, HD_{post})}$ and the number of slices with an HD larger than the threshold is 156.}
\label{fig:SFigure_1}
\end{figure}

For the selection, we use the lung segmentation results obtained with the semi-automatic tool. This tool makes it possible to perform a simple interactive segmentation of each chest CT scan. Although the procedure is time-consuming and the results obtained are not ideal, they can be used as a reference to identify which of the lung segmentations computed with our tool have changed more with the refinement procedure. To work with reliable segmentations, we exclude slices for which the $DSC$ is below $0.7$. In the subset, we measure the Hausdorff distance (pre-refinement and post-refinement), using the semi-automatic lung segmentation as a reference. Finally, we select the slices for which the absolute differences between the Hausdorff distances are larger than  $\mu_{\Delta (HD_{pre}, HD_{post})} + 3\sigma_{\Delta (HD_{pre}, HD_{post})}$  (with $\mu$ and $\sigma$  being the mean and standard deviation of the HD differences, $\Delta$). In Figure \ref{fig:SFigure_1}, the $HD$ differences are plotted against the $DSC$ for all slices with a $DSC$ larger than $0.7$. The threshold is drawn in red, the slices with an $HD$ difference under the threshold are shown in blue and those above in green. As observed, the $DSCs$ of the latter are uniformly distributed among all the possible $DSC$ values, which is an indicator of disagreement at the surface delimitation and not at the complete filled volume. 

\section{Inter-Expert Variability}
\label{app:sec:inter-exper-var}

\begin{figure}
\centering
\includegraphics[width=\textwidth,height = 0.55\textheight,trim={0.35cm 0 0.5cm 0},clip]{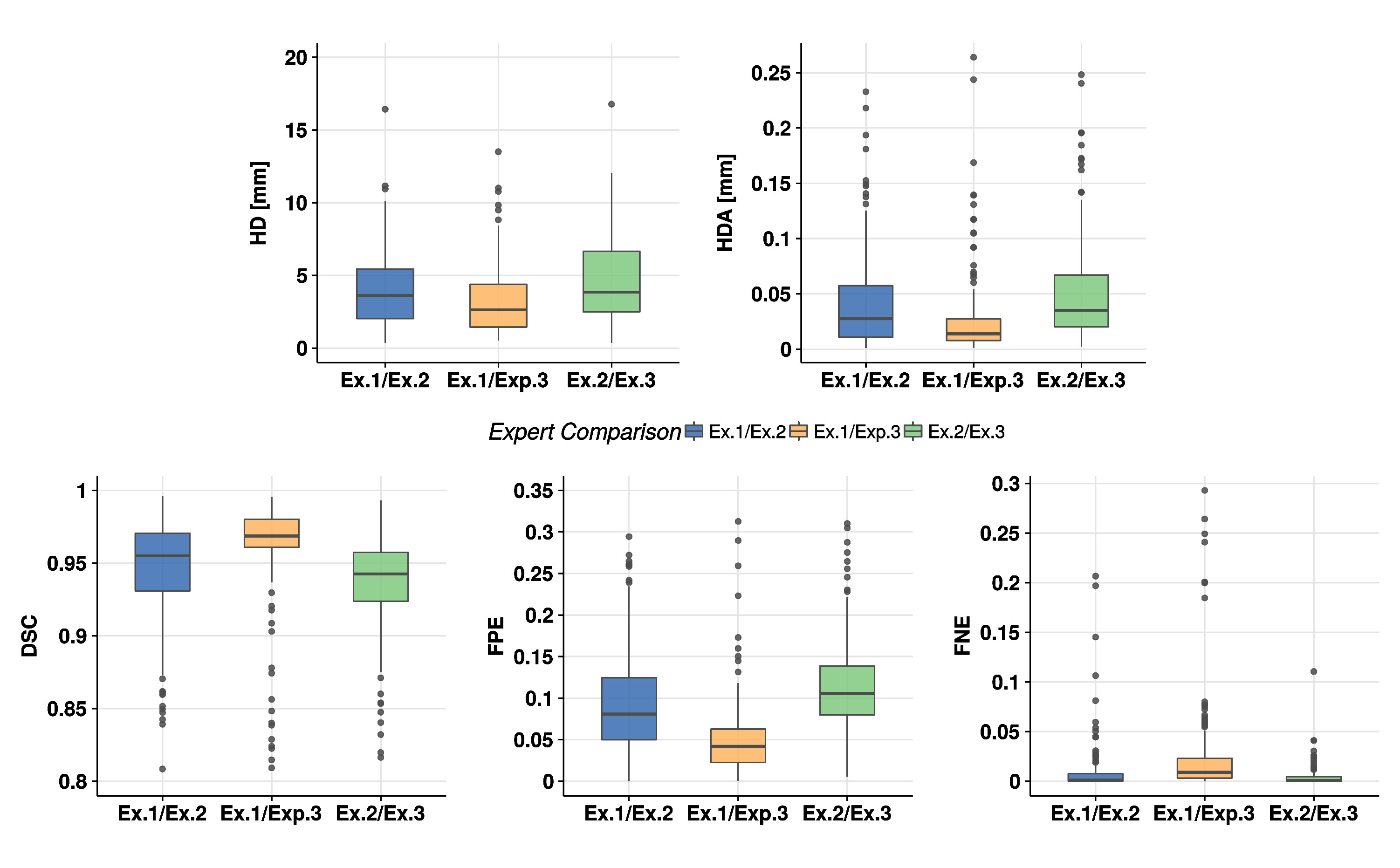}
\caption[Boxplot charts for the similarity coefficients]{Boxplot charts for the similarity coefficients obtained from the 156 most uncertain slices in the experts' delimitations.}
\label{apS:fig:SFigure_2}
\end{figure}

In order to characterize the agreement between the lung segmentations performed by the experts, the intraclass correlation coefficient ($ICC$) of each similarity measure was computed. The semi-automatic segmentation was used as reference. In Table \ref{tbl:Table_2}, the agreement coefficients are presented. We observed excellent consistency between the experts at the surface similarity measures (Hausdorff distance ($HD$), Hausdorff distance averaged ($HAD$), as was intended, and a good correlation for the volume overlap indicators (Dice similarity coefficient ($DSC$), false-positive error ($FPE$), false-negative error ($FNE$)). Figure \ref{apS:fig:SFigure_2} shows the boxplots corresponding to these results. 

\begin{table}
\centering
\begin{tabular}{lcll}
\hline
\rowcolor[HTML]{9B9B9B} 
\textbf{Coeff.} & \textbf{ICC} & \multicolumn{1}{c}{\cellcolor[HTML]{9B9B9B}\textbf{CI (95\%)}} & \multicolumn{1}{c}{\cellcolor[HTML]{9B9B9B}\textbf{p-val.}} \\ \hline
\textbf{HD} & 0.88 & 0.84 to 0.90 & \textless 0.001 \\
\textbf{HDA} & 0.85 & 0.79 to 0.88 & \textless 0.001 \\
\textbf{DSC} & 0.74 & 0.66 to 0.81 & \textless 0.001 \\
\textbf{FPE} & 0.71 & 0.27 to 0.86 & \textless 0.001 \\
\textbf{FNE} & 0.60 & 0.26 to 0.77 & \textless 0.001 \\ \hline
\end{tabular}
\caption[ICC and CI for the similarity coefficients between experts' delimitation]{Intra-class correlation coefficient (ICC) and $95\%$ confidence intervals (CI) for the similarity coefficients between the three experts' delimitation and the refined masks. Note: Haussdorff distance (HD), Haussdorff distance averaged (HDA), Dice similarity coefficient (DSC), false-positive error (FPE), false-negative error (FNE).}
\label{tbl:Table_2}
\end{table}

\chapter{Texture Features Definition}\label{ap:texture}

\begin{enumerate}
\item Maximum:\\
\begin{equation*}
f_1 = \max I(i,j)
\end{equation*}

\item Mean:\\
\begin{equation*}
f_2 =  \frac{1}{N+M}\sum_i^N\sum_j^M I(i,j) 
\end{equation*}

\item Minimum:\\
\begin{equation*}
f_3 = \min I(i,j)
\end{equation*}

\item Standard Deviation:\\
\begin{equation*}
f_4 = \frac{1}{N+M}(I(i,j) - f_2 )^{\frac{1}{2}}
\end{equation*}

\item Autocorrelation:\\
\begin{equation*}
f_5 = \sum_i\sum_j(ij)p(i,j)^2
\end{equation*}

\item \textit{Cluster Prominance}:\\
\begin{equation*}
f_6 = \sum_i\sum_j (i+j-\mu_x-\mu_y)^4p(i,j)
\end{equation*}

\item \textit{Cluster Shade}:\\
\begin{equation*}
f_7 = \sum_i\sum_j (i+j-\mu_x-\mu_y)^3p(i,j)
\end{equation*}

\item Contrast:\\
\begin{equation*}
f_8 =   \bigg |\sum_i\sum_j \bigg |^2p(i,j)
\end{equation*}

\item Correlation 1:\\
\begin{equation*}
f_9 = \sum_i\sum_j \frac{(i - \mu_x) (j - \mu_y)p(i,j)}{\sigma_x\sigma_y}
\end{equation*}

\item Correlation 2:\\
\begin{equation*}
f_{10} = \frac{\Large{\sum_{i=1}\sum_{j=1} } (ij)p(i,j) - \mu_x\mu_y}{\sigma_x\sigma_y}
\end{equation*}

\item Difference Entropy:\\
\begin{equation*}
f_{11} = -\sum_{i=0}^{L-1} p_{x-y}(i) \log \big (p_{x-y}(i) \big )
\end{equation*}

\item Difference Variance:\\
\begin{equation*}
f_{12} = \sum_{i=0}^{L-1} i^2 p_{x-y}(i)
\end{equation*}

\item Dissimilarity:\\
\begin{equation*}
f_{13} = \bigg |\sum_i\sum_j \bigg |p(i,j)
\end{equation*}

\item Energy:\\
\begin{equation*}
f_{14} =  \sum_i\sum_jp(i,j)^2
\end{equation*}

\item Entropy:\\
\begin{equation*}
f_{15} = -\sum_{i=1}^{L}\sum_{j=1}^{L} p(i,j) \log(p(i,j))
\end{equation*}

\item Homogenety 1:\\
\begin{equation*}
f_{16} = \sum_i\sum_j \frac{p(i,j)}{1+|i+j|}
\end{equation*}

\item Homogenety 2:\\
\begin{equation*}
f_{17} = \sum_i\sum_j \frac{p(i,j)}{1+|i+j|^2}
\end{equation*}

\item Information Measure Correlation 1:\\
\begin{equation*}
f_{18} = \frac{f_9 - HXY1}{\max(HX,HY)}
\end{equation*}

\item Information Measure Correlation 2:\\
\begin{equation*}
f_{19} = [1 - \exp(-2(HXY2 - f_9))]^{1/2}
\end{equation*}

\item Normalized Inverse Difference:\\
\begin{equation*}
f_{20} = \sum_{i=1}^{L}\sum_{j=1}^{L} \frac{1}{1 + |i - j|^2/L} p(i,j)
\end{equation*}

\item Normalized Moment Inverse Difference:\\
\begin{equation*}
f_{21} = \sum_{i=1}^{L}\sum_{j=1}^{L} \frac{1}{1 + (i - j)^2/L} p(i,j)
\end{equation*}

\item Maximum Probability:\\
\begin{equation*}
f_{22} = \max_{i,j}p(i,j)
\end{equation*}

\item Sum Average:\\
\begin{equation*}
f_{23} = \sum_{i=2}^{2L} i p_{x+y}(i)
\end{equation*}

\item Entropy Sum:\\
\begin{equation*}
f_{24} = -\sum_{i=2}^{2L} p_{x+y}(i) \log(p_{x+y}(i))
\end{equation*}

\item Sum Variance:\\
\begin{equation*}
f_{25} = \sum_{i=2}^{2L} (i - f_8)^2 p_{x+y}(i)
\end{equation*}

\item Sum of Squares:\\
\begin{equation*}
f_{26} = \sum_i\sum_j(i-\nu)^2p(i,j)
\end{equation*}

\end{enumerate}

\subsection*{Definitions}
\begin{itemize}
\item $L$: Quantization level
\item $p(i,j)$: Co-ocurrence matrix at position (i,j)
\item $\nu = \frac{1}{L} \sum_i^L\sum_j^L p(i,j)$
\item $p_x(i) = \sum_{j=1}^{L} p(i,j)$
\item $p_y(j) = \sum_{i=1}^{L} p(i,j)$
\item $p_{x+y}(k) = \sum_{i=1, i+j = k}^{L}\sum_{j=1}^{L}  p(i,j), \qquad k = 2,3,..., 2L$
\item $p_{x-y}(k) = \sum_{i=1, |i-j| = k}^{L}\sum_{j=1}^{L}  p(i,j), \qquad k = 0,1,..., L-1$
\item $HX = -\sum_i p_x(i) \log(p_x(i))$
\item $HY = -\sum_j p_y(j) \log(p_y(j))$
\item $HXY = -\sum_i\sum_j p(i,j) \log(p(i,j))$
\item $HXY1 = -\sum_i\sum_j p(i,j) \log(p_x(i)p_y(j))$
\item $HXY2 = -\sum_i\sum_j p_x(i)p_y(j) \log(p_x(i)p_y(j))$
\end{itemize}

\end{appendices}

\printthesisindex 

\end{document}